\documentclass[10pt]{article}
\usepackage{amsfonts}
\usepackage{amsmath}
\DeclareMathSymbol{\mlq}{\mathord}{operators}{``}
\DeclareMathSymbol{\mrq}{\mathord}{operators}{`'}
\usepackage{amssymb}
\usepackage[T1]{fontenc}
\usepackage{cancel}
\usepackage[all]{xy}
\usepackage{mathalfa}
\usepackage[mathscr]{euscript}
\usepackage{xcolor}
\usepackage{setspace}
\numberwithin{equation}{section}
\newtheorem{Thm}{Theorem}[section]
\newtheorem{Cor}{Corollary}[section]
\newtheorem{Def}{Definition}[section]
\newtheorem{Lemma}{Lemma}[section]
\newtheorem{Prop}{Proposition}[section]
\newtheorem{Remark}{Remark}[section]

\def\N{\mathbb{N}}
\def\Z{\mathbb{Z}}

\def\R{\mathbb{R}}
\def\C{\mathbb{C}}

\def\P{\mathbb{P}}
\def\fd{\hfill$\square$}
\newcommand{\prs}[2]{\mathop{\langle#1,#2\rangle}}
\date{}
\begin{document}
\title
{\emph{New  formula for the prepotentials associated with Hurwitz-Frobenius manifolds and generalized WDVV equations}}
\author
{{\small\bf Chaabane REJEB}\footnote{Universit\'{e} de Sherbrooke, CANADA.
Email: chaabane.rejeb@usherbrooke.ca, chaabane.rejeb@gmail.com}}
\maketitle

\begin{abstract}
We consider the Hurwitz spaces of ramified coverings of $\P^1$  with  prescribed ramification  profile over the point at infinity. By means of a particular symmetric bidifferential on a compact Riemann surface, we introduce  quasi-homogeneous differentials. By following Dubrovin, we  construct on  Hurwitz spaces a family of  Frobenius manifold structures associated with the quasi-homogeneous differentials.  We explicitly derive new generating formulas for the corresponding prepotentials. This produces quasi-homogeneous solutions to the following generalized WDVV associativity equations: $F_i\eta^{-1}F_j=F_j\eta^{-1}F_i$, where the invertible constant matrix $\eta$ is  a  linear combination of the matrices $F_j$. In particular, our approach provides another look at Dubrovin's construction of semi-simple Hurwitz-Frobenius manifolds and establishes an alternative practical  method to calculate their primary free energy functions. As applications, we use our formalism to obtain various explicit quasi-homogeneous solutions to the WDVV equations in genus zero and  one and give a new proof of Ramanujan's differential equations for  Eisenstein series.


\bigskip\noindent Key words: \scriptsize{WDVV equations, Frobenius manifold, Hurwitz space, Prepotential, Riemann surface, Ramified covering, Quasi-homogeneous solution.}
\end{abstract}
\tableofcontents

\section{Introduction}
The Witten-Dijkgraaf-Verlinde-Verlinde (WDVV) associativity  equations are the following  system of nonlinear partial differential equations for a function $F(t)=F(t^1,\dots,t^N)$:
\begin{equation}\label{WDVV-1}
F_{\alpha}F_1^{-1}F_{\beta}=F_{\beta}F_1^{-1}F_{\alpha}, \quad\quad \alpha,\beta=1,\dots,N,
\end{equation}
where for each $\alpha$, $F_{\alpha}$ denotes the matrix $\big(F_{\alpha}\big)_{\gamma\delta}:=\partial_{t^{\alpha}}\partial_{t^{\gamma}}\partial_{t^{\delta}}F$ such that the matrix  $F_1$ is constant and nondegenerate. The WDVV equations were discovered  in the context of 2D-topological field theory (2D-TFT)  at the end of 80's \cite{DVV, Witten}. \\
The concept  of a Frobenius manifold was first axiomatized and thoroughly studied by B. Dubrovin \cite{Dubrovin92, Dubrovin2D, Dubrovin93, Dubrovin99, Dubrovin2004}  as  a  geometric  reformulation  of the WDVV associativity  equations. Roughly speaking, a Frobenius manifold is a  manifold with a flat metric and a Frobenius algebra structure on the tangent bundle with two global vector fields, the unit and the Euler vector fields, satisfying some integrable conditions.
Each Frobenius manifold gives rise to a quasi-homogeneous solution to the WDVV equations (\ref{WDVV-1}), called the prepotential of the Frobenius manifold.
Besides 2D-TFT, Frobenius manifolds appear in a wide variety of mathematical physics contexts, including mirror symmetry \cite{Chen-KS, Sergyeyev, Pavlov}, theory of unfolding spaces of singularities \cite{Hertling, Strachan}, quantization theory \cite{Blackmore, Konopelchenko}, quantum cohomology \cite{Manin}, integrable systems \cite{Mokhov, M-Pavlov} and so on. \\
Various examples of Frobenius manifolds  have been constructed.  Some of them  are derived from   orbit spaces of certain groups, such as  finite irreducible Coxeter groups, affine Weyl groups and Jacobi groups \cite{Bertola PhD, Dubrovin2D, Dubrovin98}.\\
Hurwitz spaces provide  another  source of interesting examples of Frobenius manifold structures \cite{Dubrovin2D, Vasilisa, Vasilisa2, Romano}. A  ``simple'' Hurwitz space in genus $g$ is the space of ramified coverings $(C_g,\lambda)$ of the Riemann sphere $\P^1$, up to an equivalence relation (see Subsection 2.2), where $C_g$ is a surface of genus $g$ and $\lambda: C_g\longrightarrow\P^1$ is a meromorphic function of a fixed degree with $N$ simple branch points $\lambda_1,\dots,\lambda_N\in \C$  and a  prescribed ramification profile over the point at infinity. The $N$ branch points $\{\lambda_j\}$ serve as local coordinates on the ``simple'' Hurwitz space.\\
In \cite{Dubrovin2D} (Lecture 5), Dubrovin established that a family of $N$ specific  Darboux-Egoroff metrics  give rise to $N$ semi-simple Frobenius manifold structures on the same simple  Hurwitz space of dimension $N$. Such Darboux-Egoroff metrics were written  in terms of  the so-called primary differentials defined on the underlying Riemann surfaces. Moreover, in the system of local coordinates $\{\lambda_j\}_j$ consisting of simple branch points, the Frobenius algebra multiplication of the obtained family of  $N$ semi-simple Hurwitz-Frobenius manifolds is defined by  $\partial_{\lambda_i}\circ\partial_{\lambda_j}=\delta_{ij}\partial_{\lambda_i}$. \\
Let us emphasize that the crucial tool in Duborvin's construction was a bilinear  pairing defined on the space of differentials having some suitable properties.  Particularly, by means of such bilinear pairing, Dubrovin proposed a formula for the corresponding prepotentials, that is, quasi-homogeneous solutions to the WDVV equations (\ref{WDVV-1}). For more details, we refer to \cite{Dubrovin2D} (Lecture 5). In  Appendix 3 below, we recall the expression for the bilinear pairing and the formula for the prepotentials obtained in \cite{Dubrovin2D}.\\
By employing a generalized  bilinear pairing and viewing the above  described ``simple'' Hurwitz spaces  as real manifolds, Shramchenko  \cite{Vasilisa} gave another class of Darboux-Egoroff metrics  in terms of the Schiffer and Bergman kernels and obtained  $``\text{real double}"$ semi-simple Hurwitz-Frobenius manifolds.
Furthermore, by introducing some parameters, Shramchenko  extended in \cite{Vasilisa2} Dubrovin's construction and built deformed semi-simple Hurwitz-Frobenius manifold structures. The solutions to the WDVV equations that are associated with the ``real double'' and deformed Hurwitz-Frobenius manifold are also described by means of a Dubrovin type bilinear pairing.\\
Notwithstanding the fact that a general formula for the prepotential is known for all the already mentioned Hurwitz-Frobenius manifolds, their doubles and deformations, it seems still be tricky to apply Dubrovin's bilinear pairing method and to produce explicit solutions to the WDVV equations. Moreover, at least to our knowledge, the only known examples come from three and four  dimensional Hurwitz spaces in genus zero and one \cite{Miguel, Dubrovin2D, Vasilisa, Vasilisa2}.

\medskip

The main purpose of the current work is to develop a new formula for the prepotentials associated with (a generalized) Hurwitz-Frobenius manifolds.   The idea of our approach is the following. First, by following \cite{Dubrovin2D} (see also \cite{Vasilisa}), we construct new Frobenius manifold structures on simple Hurwitz spaces induced by quasi-homogeneous differentials defined on the underlying Riemann surfaces.  A differential $\omega_0$ on a Riemann surface $C_g$ is called quasi-homogeneous of degree $d_0\geq0$ if $\omega_0$ admits the following integral representation with respect to the canonical symmetric bidifferential $W(P,Q)$ on  $C_g$:
\begin{equation}\label{Intro-quasi}
\omega_0(P)=c_0\int_{\ell_0}\big(\lambda(Q)\big)^{d_0}W(P,Q),
\end{equation}
where $c_0$ is a nonzero constant and $\ell_0$ is a contour on the surface $C_g$  satisfying some conditions. For  properties of the symmetric bidifferential $W(P,Q)$, we refer to \cite{Fay, Fay92, Yamada} and Subsection 2.3 below.  Note that the language of the bidifferential $W(P,Q)$ in the context of Frobenius manifolds was first employed by Shramchenko in \cite{Vasilisa}. Interestingly, it was observed in \cite{Vasilisa} that Dubrovin's primary differentials  can be conveniently rewritten in terms  of the symmetric bidifferential $W(P,Q)$ as in (\ref{Intro-quasi}). This means that each primary differential $\phi_{t^A}$ of Dubrovin is quasi-homogeneous and can be characterized, through formula (\ref{Intro-quasi}),  by the triplet  $(d_0,c_0,\ell_0)=(d_A,c_A,\ell_A)$ where $d_A$ is the degree of $\phi_{t^A}$, $c_A$ is a constant and $\ell_A$ is a suitable contour on $C_g$.\\
Let us denote by $t^A(\omega_0)$ the function defined by
\begin{equation}\label{Intro-tA}
t^A(\omega_0):=c_A\int_{\ell_A}\big(\lambda(P)\big)^{d_A}\omega_0(P),
\end{equation}
with $(d_A,c_A,\ell_A)$ being the triplet determined by the primary differential $\phi_{t^A}$ (when $\ell_A$ is a curve joining two poles of $\lambda$,  the integral in the right hand side of (\ref{Intro-tA}) is defined by an appropriate principal value (see \cite{Dubrovin2D} and  Subsection 3.3 of this paper). Then the set consisting of the $N$ functions  $\big\{t^A(\omega_0)\big\}$ provides a system of flat coordinates of  the flat metric $\eta(\omega_0)$ of the Frobenius manifold associated with the differential  $\omega_0$.  Analogously to \cite{Dubrovin2D}, the metric $\eta(\omega_0)$ is defined by
\begin{equation}\label{Intro-eta}
\eta(\omega_0)=\sum_{j=1}^N\bigg(\underset{P_j}{{\rm res}}\frac{\omega_0(P)^2}{d\lambda(P)}\bigg)(d\lambda_j)^2,
\end{equation}
where $P_1,\dots,P_N$ are the simple ramification points of the covering $(C_g,\lambda)$. Let us denote by $\eta$ the Gram constant matrix of the flat metric $\eta(\omega_0)$ in  flat coordinates $\big\{t^A(\omega_0)\big\}$.  Then, by  using a convenient choice of the constants $c_A$ in the integral representation formula (\ref{Intro-quasi})  for the primary and quasi-homogeneous differentials $\phi_{t^A}$, we prove that the constant Gram matrix $\eta$ can be obtained in such a way  that all the rows  of the  constant matrix $\eta$ contain exactly one nonzero entry which is 1. This leads us to introduce a useful duality relations between flat coordinates $\big\{t^A(\omega_0)\big\}$ of the metric $\eta(\omega_0)$ (this duality was already implicitly present in \cite{Dubrovin2D}). More precisely, two flat coordinates $t^{A}(\omega_0)$ and $t^B(\omega_0)$ are called $\eta(\omega_0)$-dual to each other if $\eta^{AB}=1$ and in this case $t^B$ is denoted by $t^{A^{\prime}}$. Here $\eta^{AB}$ are the constant entries of the contravariant metric $\eta^*(\omega_0)$ induced by $\eta(\omega_0)$. Note that, with the help of the bidifferential $W(P,Q)$, the contravariant metric $\eta^*(\omega_0)$, as well as the introduced duality relations, we present new arguments to prove some results obtained by Dubrovin \cite{Dubrovin2D} and used as key  ingredients in the construction of Hurwitz-Frobenius manifolds. The details of  these results are discussed  in Sections 3.1, 3.2, 3.3 and 4.1.

\medskip

Next, we establish  the following new formula for the  prepotentials associated with the Hurwitz-Frobenius manifolds induced by the quasi-homogeneous differential $\omega_0$ (\ref{Intro-quasi}):
\begin{equation}\label{Intro-Prep}
\begin{split}
\mathcal{F}_{\omega_0}&=\frac{1}{2(1+d_0)}\sum_{A,B}\frac{\big((d_A+d_0)t^A(\omega_0)+r_{0,A}\big)\big((d_B+d_0)t^B(\omega_0)+r_{0,B}\big)}
{1+d_0+d_{A^{\prime}}}t^{A^{\prime}}\big(\phi_{t^{B^{\prime}}}\big),\\
\end{split}
\end{equation}
where $d_0, d_A, d_{A^{\prime}}$ are respectively the degrees of the quasi-homogeneous  differentials $\omega_0$, $\phi_{t^A}$, $\phi_{t^{A^{\prime}}}$ and $t^{A^{\prime}}$ is the dual coordinate of $t^A$, while $\{r_{0,A}\}_A$ are certain constants depending on the differentials $\omega_0$ and $\phi_{t^A}$ and $\big\{t^{A^{\prime}}(\phi_{t^{B^{\prime}}})\big\}$ denotes the set of flat coordinates of the flat metric $\eta\big(\phi_{t^{B^{\prime}}}\big)$ (\ref{Intro-eta}) induced by the primary differential $\phi_{t^{B^{\prime}}}$ (see Theorem \ref{Prep-eta-thm} and formulas (\ref{Prep-eta}), (\ref{Prep-eta2}) and (\ref{Prep-eta3}) below).\\
The prepotential $\mathcal{F}_{\omega_0}$ (\ref{Intro-Prep}) is a quasi-homogeneous  solution to the following generalized WDVV equations:
\begin{equation}\label{WDVV equation}
\left\{
\begin{array}{ll}
F_{\alpha}\eta^{-1}F_{\beta}=F_{\beta}\eta^{-1}F_{\alpha}, \quad \alpha,\beta=1,\dots,N,\\
\\
\eta=\sum_jf_{\alpha}(t)F_{\alpha},
\end{array}
\right.
\end{equation}
where $F_{\alpha}=\partial_{t^{\alpha}(\omega_0)}H$  and $H$ is the Hessian matrix of $\mathcal{F}_{\omega_0}$, $\eta$ is the constant $N\times N$-matrix of the metric $\eta(\omega_0)$ and the coefficients $f_{\alpha}$ are some functions of flat coordinates $\{t^{\alpha}(\omega_0)\}$.\\
Note that when $\omega_0$ is among the $N$ primary differentials of Dubrovin,  formula (\ref{Intro-Prep})  provides a novel alternative  method to calculate   prepotentials   associated with the aforementioned  semi-simple Hurwitz-Frobenius manifolds developed  by Dubrovin and the function  $\mathcal{F}_{\omega_0}$ becomes a solution to the WDVV equations (\ref{WDVV-1}).\\
As a first application,  we use our approach  to reconstruct  the Frobenius manifold structure  associated with the Chazy equation satisfied by the Eisenstein series $E_2$ and recalculate the corresponding prepotential $\mathcal{F}_{\phi_1}$. This Frobenius manifold was first obtained by Dubrovin \cite{Dubrovin2D}. It is  constructed on the three dimensional Hurwitz space of Weierstrass elliptic curves and induced by the normalized holomorphic differential $\phi_1$. Note that, in this case, the two remaining  primary differentials $\phi_2,\phi_3$ are such that $\phi_2$ is an Abelian differential of the second kind and $\phi_3$ is a multivalued differential.   Moreover, by showing that flat coordinates of  the flat metrics $\eta(\phi_2)$ and $\eta(\phi_3)$ can be  computed directly from the Hessian matrix  of the obtained prepotential $\mathcal{F}_{\phi_1}$,  we propose a new proof of the  Ramanujan differential equations for the Eisenstein series $E_2$, $E_4$ and $E_6$ \cite{Ramanujan1916}. \\
As a second application of formula (\ref{Intro-Prep}), we prove that for any positive integer $m$, the following function $\mathcal{F}_{\omega_0}=\mathcal{F}_{\omega_0}(t_0,t_1,\dots,t_{2m+1})$ involving the Jacobi odd $\theta_1$-function
\begin{equation}\label{Intro-ex}
\begin{split}
\mathcal{F}_{\omega_0}
&=\frac{(t_{2m+1})^2}{2}t_0+t_{2m+1}\sum_{k=1}^mt_kt_{2m+1-k}+\frac{1}{2}\sum_{k=1}^m\big(t_{2m+1-k}\big)^2\log(t_{2m+1-k})\\
&\quad +\frac{1}{2}\bigg(\sum_{k=1}^mt_{2m+1-k}\bigg)^2\log\bigg(\sum_{j=1}^mt_{2m+1-j}\bigg)
 -\sum_{k=1}^m\big(t_{2m+1-k}\big)^2\log\left(\frac{\theta_{1}\big(t_k\big|{2{\rm i}\pi}t_0\big)}{\theta_1'\big(0\big|2{\rm i}\pi t_0\big)}\right)\\
&\quad +\frac{1}{2}\sum_{j,k=1, j\neq k}^mt_{2m+1-j}t_{2m+1-k}
\log\bigg(\frac{\theta_1'\big(0\big|{2{\rm i}\pi}t_0\big)\theta_1\big(t_j-t_k\big|{2{\rm i}\pi}t_0\big)}{\theta_1\big(t_j\big|{2{\rm i}\pi}t_0\big)
\theta_1\big(t_k\big|{2{\rm i}\pi}t_0\big)}\bigg)
\end{split}
\end{equation}
is a genus one quasi-homogeneous solution to the WDVV equations (\ref{WDVV equation}), with $\omega_0$ being the normalized holomorphic differential. This explicit solution is directly  suggested by our formula (\ref{Intro-Prep})  and a  particular choice of the combinatorial parameters of the Hurwitz space (here $g=1$ and the $m+1$ poles of the ramified coverings are all simple).    We refer the reader to Section 4.1.5 for details. \\
Note that the case $m=1$ of the  prepotential (\ref{Intro-ex}) was obtained  in \cite{Dubrovin2009, Miguel}   through  a long  computation based on Dubrovin's bilinear pairing method.\\
In our case, most of the terms of expression (\ref{Intro-ex}) are obtained by simple specialization of the general formula (1.5) to the chosen
Hurwitz space. Moreover, it is the structure of the general formula (1.5) that suggests which Hurwitz space and primary differential
to look at in order to obtain an explicit solution while avoiding lengthy calculations. Note that all known explicit examples obtained through
Dubrovin’s bilinear pairing method require an explicit parametrization of the Hurwitz space. Such a parametrization is not necessary for
applying formula (1.5) which allows to construct prepotentials for a larger class of Hurwitz spaces, of which (\ref{Intro-ex}) is an example.
To our knowledge this is the first example of a prepotential of a Dubrovin-Hurwitz-Frobenius manifold depending on a large (greater than 4)
number of variables.\\
In addition,  the following function depending on  $2m$-variables:
$$
\mathcal{F}=\sum_{k=1}^mt_{2m+1-k}e^{t_k}+\frac{1}{2}\sum_{k=1}^m\big(t_{2m+1-k}\big)^2\Big(t_k+\log\big(t_{2m+1-k}\big)\Big)
+\frac{1}{2}\sum_{k,s=1, s\neq k}^mt_{2m+1-k}t_{2m+1-s}\log\big(e^{t_s}-e^{t_k}\big)
$$
provides an example of explicit solution to the generalized WDVV equations  (\ref{WDVV equation}), where  $\eta$ is the anti-diagonal matrix $\eta_{\alpha\beta}=\delta_{2m+1,\alpha+\beta}$ and the coefficients $f_{\alpha}$ are given by
$$
f_{\alpha}=\frac{e^{-t_{\alpha}}}{1-\sum_{r=1}^mt_{2m+1-r}e^{-t_r}},\quad\quad  f_{2m+1-\alpha}=-\frac{t_{2m+1-\alpha}e^{-t_{\alpha}}}{1-\sum_{r=1}^mt_{2m+1-r}e^{-t_r}},
\quad\quad\alpha=1,\dots,m.
$$
This example comes from the  genus zero Hurwitz-Frobenius manifold structure determined by a quasi-homogeneous differential not belonging to the family of Dubrovin primary differentials.\\
On the other hand, we extend formula (\ref{Intro-Prep}) to prepotentials associated with the already mentioned deformations of Hurwitz-Frobenius manifolds and obtained in  \cite{Vasilisa2}. This brings us to give new explicit quasi-homogeneous solutions to the WDVV equations by considering  the deformations of the obtained prepotentials.

\medskip

This paper is organized as follows. In Section 2, we recall the definitions of a Frobenius manifold structure,  simple and double Hurwitz spaces and  that of the symmetric bidifferential on a Riemann surface. In Section 3, we  introduce a family of flat metrics in terms of the symmetric bidifferential and construct their systems of flat  coordinates. The study here  includes the intersection forms regarded as flat metrics on double Hurwitz spaces.    The goal  of Section 4 is to describe  Frobenius manifold structures induced by quasi-homogeneous differentials and to give a formula the corresponding prepotentials as well as examples in genus one. Sections 5 is  devoted to extending our approach to the framework of deformed  Frobenius manifold structures.
In Section 6, we construct examples of Frobenius manifold structures on $2m$-dimensional Hurwitz space in genus zero and  compute the corresponding explicit prepotentials. Ramanujan's differential equations and other examples of genus one quasi-homogeneous solutions to the WDVV equations are discussed in Section 7. Finally, in the Appendix, we  give  details of some  results used in this work for the reader’s convenience.

\section{Preliminaries}
\subsection{Frobenius manifolds}
\begin{Def}\label{F algebra} A $\C$-algebra $(\mathcal{A},\circ,e)$ supplied with a nondegenerate and symmetric inner product  $\prs{\cdot}{\cdot}:\mathcal{A}\times \mathcal{A}\longrightarrow\C$ is a Frobenius algebra if
\begin{description}
\item[i)] the algebra $(\mathcal{A},\circ,e)$  is associative, commutative with unity $e$;
\item[ii)] the multiplication  $``\circ"$ is compatible  with  $\prs{\cdot}{\cdot}$, in the sense that
$$
\prs{x\circ y}{z}=\prs{x}{y\circ z},\quad \quad \forall\ x,y,z\in \mathcal{A}.
$$
\end{description}
A Frobenius algebra $(\mathcal{A},\circ,e, \prs{\cdot}{\cdot})$ is called semi-simple if it contains no nilpotent element.
\end{Def}

The next definition of a Frobenius manifold differs slightly from the original one  given by Dubrovin. More precisely, in order to be consistent  with the generalized WDVV equations (\ref{WDVV equation}), the unit vector field  is not required to be flat.
\begin{Def}\label{Frob Man def} Let $\mathcal{M}$ be a manifold of dimension N. A Frobenius manifold structure on $\mathcal{M}$ is a data of $\big(\mathcal{M},\circ,e,\prs{\cdot}{\cdot},E\big)$ such that each tangent space $T_t\mathcal{M}$ is a Frobenius algebra varying smoothly or analytically over $\mathcal{M}$ with the additional properties:
\begin{itemize}
\item[\textbf{\emph{1.}}] the metric $\prs{\cdot}{\cdot}_t$ on $\mathcal{M}$ is flat  (but not necessarily real and  positive);
\item[\textbf{\emph{2.}}] the tensor $\big(\nabla_{v}\textbf{c}\big)(x,y,z)$ is symmetric in four vector fields $v,x,y,z\in T_t\mathcal{M}$, where $\nabla$ is the Levi-Civita connection of the metric $\prs{\cdot}{\cdot}$ and $\textbf{c}$ is the symmetric 3-tensor
$$
\textbf{c}(x,y,z):=\prs{x\circ y}{z},\quad \forall\ x,y,z\in T_t\mathcal{M};
$$
\item[\textbf{\emph{3.}}] the vector field $E$ is covariantly linear, i.e. $\nabla\nabla E=0$, and
\begin{align}
\begin{split}\label{Euler vector def}
&[E,e]=-e,\\
&[E,x\circ y]-[E,x]\circ y-x\circ[E,y]=x\circ y,\\
&\big(Lie_E\prs{\cdot}{\cdot}\big)(x,y):=E\prs{x}{y}-\prs{[E,x]}{y}-\prs{x}{[E,y]}=(2-D)\prs{x}{y},
\end{split}
\end{align}
for some constant $D$, called the charge of the Frobenius manifold. The vector field $E$ is called Euler vector field.
\end{itemize}
\end{Def}
\begin{Def}\label{Flatness-e} Let $\big(\mathcal{M},\circ,e,\prs{\cdot}{\cdot},E\big)$ be a Frobenius manifold. \\
We shall call it  normalized if the unit vector field $e$ is covariantly constant with respect to $\nabla$, the Levi-Civita connection of $\prs{\cdot}{\cdot}$, namely $\nabla e=0$.\\
A Frobenius manifold is called semi-simple if the Frobenius algebra in the tangent space at any point of $\mathcal{M}$ is semi-simple.
\end{Def}

The flatness of the metric $\prs{\cdot}{\cdot}$ implies that locally there exist flat coordinates $\{t^{\alpha}: \ 1\leq\alpha\leq N\}$ so that the matrix  $\big(\prs{\partial_{t^{\alpha}}}{\partial_{t^{\beta}}}\big)_{\alpha\beta}$ is constant. On the other hand, from the symmetry of the tensor $\nabla\textbf{c}$, it follows that there is a function $F=F(t^1,\dots,t^{N})$ such that
\begin{equation}\label{Introd-prep def}
\partial_{t^{\alpha}}\partial_{t^{\beta}}\partial_{t^{\gamma}}F=\textbf{c}\big(\partial_{t^{\alpha}},\partial_{t^{\beta}},\partial_{t^{\gamma}}\big).
\end{equation}
The function $F$  is unique up to additional  quadratic and linear  polynomial functions of $t^1,\dots,t^N$ and it is known as the prepotential of the Frobenius manifold. Physicists call it the primary free energy function of the  2D-TFT \cite{DVV}.  \\
Note that (\ref{Euler vector def}) yields that the Lie derivative, along the Euler vector field $E$, of the symmetric 3-tensor $\textbf{c}$ is such that:
\begin{equation}\label{Intro-Lie deriv-c}
\begin{split}
\big(Lie_E.\textbf{c}\big)(x,y,z)&:=E.\textbf{c}(x,y,z)-\textbf{c}\big([E,x],y,z\big)-\textbf{c}\big(x,[E,y],z\big)-\textbf{c}\big(x,y,[E,z]\big)\\
&=(3-D)\textbf{c}(x,y,z).
\end{split}
\end{equation}
Let us assume that $E=\sum_{\epsilon}E^{\epsilon}(t)\partial_{t^{\epsilon}}$, with $t=(t^1,\dots,t^N)$. Then the requirement  $\nabla\nabla E=0$ implies that $\partial_{t^{\alpha}}\partial_{t^{\beta}}E^{\epsilon}=0$, for all $\alpha,\beta,\epsilon$. This fact, (\ref{Introd-prep def}) and (\ref{Intro-Lie deriv-c}) allow us to write
$$
\big(Lie_E.\textbf{c}\big)\big(\partial_{t^{\alpha}},\partial_{t^{\beta}},\partial_{t^{\gamma}}\big)=\partial_{t^{\alpha}}\partial_{t^{\beta}}\partial_{t^{\gamma}}E.F=
(3-D)\partial_{t^{\alpha}}\partial_{t^{\beta}}\partial_{t^{\gamma}}F.
$$
Thus the prepotential $F$ is a quasi-homogeneous function of degree $3-D$ (modulo quadratic and/or linear terms). Following Dubrovin \cite{Dubrovin2D} (Lecture 1),  we use a generalized quasi-homogeneity property is considered in our context. More precisely,  a function $f$ is called quasi-homogeneous  of degree $\nu_f$ with respect to the Euler vector field $E=\sum_{\alpha}\big(d_{\alpha}t^{\alpha}+r_{\alpha}\big)\partial_{t^{\alpha}}$ if
\begin{equation}\label{quasihomog-def}
\textstyle E.f=\nu_ff+\sum_{\alpha,\beta}A_{\alpha\beta}t^{\alpha}t^{\beta}+\sum_{\alpha}B_{\alpha}t^{\alpha}+C.
\end{equation}
On the other hand, we emphasize that the flatness of the unit vector field $e$ (i.e. $\nabla e=0$) implies that $e$ can be taken of the form $e=\partial_{t^1}$ (by making a linear change of flat coordinates). In this case, the metric $\prs{\cdot}{\cdot}$ satisfies
$$
\prs{\partial_{t^{\alpha}}}{\partial_{t^{\beta}}}=\partial_{t^1}\partial_{t^{\alpha}}\partial_{t^{\beta}}F.
$$
Otherwise, we have $e=\sum_{\epsilon}e^{\epsilon}(t)\partial_{t^{\epsilon}}$ and
$$
\prs{\partial_{t^{\alpha}}}{\partial_{t^{\beta}}}=\textstyle\sum_{\epsilon}e^{\epsilon}(t)\partial_{t^{\epsilon}}\partial_{t^{\alpha}}\partial_{t^{\beta}}F.
$$
Lastly,  we mention some results concerning a semi-simple Frobenius manifold $\big(\mathcal{M},\circ,e,\prs{\cdot}{\cdot},E\big)$ that were derived by Dubrovin \cite{Dubrovin2D} (Lecture 3):
In a neighborhood of each   point of $\mathcal{M}$, there is  a system of canonical coordinates $\lambda_1,\dots,\lambda_N$ such that the Frobenius algebra multiplication is given by $\partial_{\lambda_i}\circ\partial_{\lambda_j}=\delta_{ij}\partial_{\lambda_i}$, the unit and Euler vector fields are respectively represented by $e=\sum_j\partial_{\lambda_j}$
and $E=\sum_j\lambda_j\partial_{\lambda_j}$ and, with respect to the canonical coordinates, the flat metric $\prs{\cdot}{\cdot}$ becomes of Darboux-Egoroff type (diagonal with diagonal terms generated by a potential).

\subsection{Hurwitz spaces}
The Hurwitz space $\mathcal{H}_{g,L}$  is the moduli space  of equivalence classes of ramified (or branched) coverings $[(C_g,\lambda)]$, where $C_g$ is a compact Riemann surface of genus $g$ and $\lambda: C_g\longrightarrow\P^1$ is a meromorphic function on $C_g$ of degree $L\geq 2$. Two ramified coverings $(C_g,\lambda)$ and $(\widetilde{C}_g,\widetilde{\lambda})$ are equivalent if there is a biholomorphic map $h:C_g\longrightarrow \widetilde{C}_g$ such that the following diagram commutes:
$$
 \xymatrix{C_g \ar[rr]^h  \ar[rd]_{\lambda} && \widetilde{C}_g \ar[ld]^{\widetilde{\lambda}} \\& \P^1}
$$
The ramification  points of the covering $(C_g,\lambda)$ are the zeros  of the differential $d\lambda(P)$ (here $\lambda$ is regarded as a holomorphic map) and its  branch points are the images of the ramification points by the projection $\lambda$. \\
A branch point is called simple if the corresponding ramification index is  $r=1$. Here we mention that the ramification index at a point is a nonnegative integer $r$ such that the number of sheets of the covering which are glued at this point is $r+1$. In particular, the fiber of a simple branch point contains  exactly $L-1$ points, with $L$ being the degree of $\lambda$. \\
The (classical) Hurwitz space of simply branched coverings (i.e. all its  branching points are simple)  of the Riemann sphere of degree $L$ and genus $g$ appeared first in the works  of Clebsch \cite{Clebsch} and Hurwitz \cite{Hurwitz}, where they showed that such space is a connected complex manifold and the simple branch points can be taken  as local coordinates on it.

\medskip

Let $m,n$ be two positive integers and $\mathbf{m}=(m_0,\dots,m_n)$ and $\mathbf{n}=(n_0,\dots,n_m)$ be two fixed vectors of nonnegative integers such that
\begin{equation*}\label{Partition}
\textstyle \sum_{i=0}^n(m_i+1)=\sum_{j=0}^m(n_j+1)=L.
\end{equation*}
The simple Hurwitz space $\mathcal{H}_{g, L}(n_0,\dots,n_m)=\mathcal{H}_{g, L}(\mathbf{n})$ parameterizes  isomorphism classes of ramified coverings $[(C_g,\lambda)]\in \mathcal{H}_{g,L}$ where the branched covering $(C_g,\lambda)$ has $N$ simple distinct branch points $\lambda_1=\lambda(P_1),\dots,\lambda_N=\lambda(P_N)\in \C$ and   $m+1$ prescribed poles $\infty^0,\dots,\infty^m$ of order $n_0+1,\dots,n_m+1$, respectively. In particular, the divisor of the differential $d\lambda$ has  the following form:
$$
(d\lambda)=\textstyle \sum_{j=1}^NP_j-\sum_{i=0}^m(n_i+2)\infty^i.
$$
According to Fulton \cite{Fulton}, the simple Hurwitz space $\mathcal{H}_{g, L}(n_0,\dots,n_m)$ is still a connected complex manifold of dimension $N$.
Let us emphasize  that, due to the Riemann-Hurwitz formula, the number $N$ of  simple and finite  branch points  of any covering
$[(C_g,\lambda)]\in \mathcal{H}_{g, L}(n_0,\dots,n_m)$  can be expressed as follows:
$$
N= 2g+2L-2-\textstyle\sum_{i=0}^mn_i=2g+2m+\textstyle\sum_{i=0}^mn_i.
$$
As previously indicated, simple Hurwitz spaces were considered by Dubrovin  \cite{Dubrovin2D} and Shramchenko \cite{Vasilisa, Vasilisa2} in the framework of  Frobenius manifolds. Note that the Frobenius manifold structures were constructed on open subsets of a covering $\widehat{\mathcal{H}}_{g, L}(n_0,\dots,n_m)$ of the simple Hurwitz space $\mathcal{H}_{g, L}(n_0,\dots,n_m)$ whose points are pairs of $\big\{\big(C_g,\lambda\big),\{a_k,b_k\}\big\}$, where $[(C_g,\lambda)]$ is a point in
$\mathcal{H}_{g, L}(n_0,\dots,n_m)$ and $\{a_k,b_k\}$ is a canonical homology  basis  of cycles on $C_g$. Here, we will also  work on a fixed open set $\widehat{\mathcal{H}}_{g, L}(n_0,\dots,n_m)$ and we continue to call it simple Hurwitz space for brevity.

\medskip

As a generalization of  simple Hurwitz spaces, we can consider equivalence classes of ramified  coverings  in $\mathcal{H}_{g,L}$ with two special fibers over $0$ and the point at infinity. These are called double Hurwitz spaces.  More precisely, the double Hurwitz space $\mathcal{H}_{g,L}(\mathbf{n}; \mathbf{m})$ consists  of the classes $[(C_g,\lambda)]\in \mathcal{H}_{g,L}$  such that  the following  conditions are satisfied:
\begin{description}
\item[1.] The covering $\lambda$ has $N$ simple ramification  points $P_1,\dots,P_N$ with distinct images $\lambda_1,\dots,\lambda_N\in \P^1\setminus \{0,\infty\}$;
\item[2.] The  divisor of $\lambda$ is as follows:
$$
(\lambda)=\textstyle\sum_{i=0}^n(m_i+1)\mathbf{z}^i-\sum_{i=0}^m(n_i+1)\infty^i.
$$
\end{description}
In this case, the Riemann-Hurwitz formula tells us  that the number $N$ of simple branch points belonging to $\C\setminus \{0\}$ is given by
$$
N=2g+2L-2-\textstyle\sum_{i=0}^nm_i-\sum_{i=0}^mn_i=2g+m+n.
$$
Contrary to simple Hurwitz spaces, double Hurwitz spaces are not in general connected.
Note that when $\mathbf{m}=(0,\dots,0)$, the double Hurwitz space $\mathcal{H}_{g, L}(\mathbf{n};\mathbf{0})$ becomes an open subset of  the simple Hurwitz space  $\mathcal{H}_{g, L}(n_0,\dots,n_m)$ determined by the conditions $\lambda_1\neq0,\dots, \lambda_N\neq 0$.\\
As above, we shall work with a connected (component of a) covering  $\widehat{\mathcal{H}}_{g,L}(\mathbf{n}; \mathbf{m})$ of the considered double Hurwitz space, whose elements are pairs: a point of the space $\mathcal{H}_{g,L}(\mathbf{n}; \mathbf{m})$ and  a canonical homology  basis  of cycles on the underlying surface.

\medskip

Now, we shall introduce some notation that will be frequently used throughout this work.\\
\noindent\textbf{Standard local parameters near ramification points} Let $[(C_g,\lambda)]\in \mathcal{H}_{g, L}(n_0,\dots,n_m)\cup \mathcal{H}_{g,L}(\mathbf{n}; \mathbf{m})$. The  complex structure on the surface $C_g$ is defined by the covering as follows.\\
The distinguished  local parameter near a simple ramification point $P_j\in C_g$ is defined by  $x_j(P):=\sqrt{\lambda(P)-\lambda_j}$ provided that $P_j$
is  not a pole, nor a zero of the meromorphic function $\lambda$.  \\
Near a pole $\infty^i$ of $\lambda$ of order $n_i+1$, the standard local parameter $z_i(P)$ is defined by  $z_i(P):=\big(\lambda(P)\big)^{-\frac{1}{n_i+1}}$.
Lastly, in a neighborhood of a zero $\mathbf{z}^i$ of $\lambda$, the local parameter is denoted by $\mathfrak{z}_i(P)$ and  given by $\mathfrak{z}_i(P):=\big(\lambda(P)\big)^{\frac{1}{m_i+1}}$, with $m_i$ being the ramification index of $\lambda$ at the point $\mathbf{z}^i$.

\medskip

\noindent\textbf{Evaluation of a differential at a ramification point} By following Dubrovin \cite{Dubrovin2D}, we shall introduce an appropriate  evaluation of a differential at a ramification point. It is defined with respect  to the previously  mentioned standard local parameters. \\
Let $\omega$ be a meromorphic differential on the surface $C_g$. Assume that  $\omega$ is  holomorphic near a simple ramification point $P_j$, with
$$
\omega(P)\underset{P\sim P_j}=\textstyle\Big(\sum_{k=0}^{\infty}a_kx_j^k(P)\Big)dx_j(P),\quad \quad x_j(P)=\sqrt{\lambda(P)-\lambda_j}.
$$
Define the evaluation $\omega(P_j)$ of the differential $\omega(P)$ at $P=P_j$ with respect to the fixed  local parameter $x_j(P)$ by
\begin{equation}\label{notation1}
\omega(P_j)=a_0=\frac{\omega(P)}{dx_j(P)}\Big|_{P=P_j}.
\end{equation}
Since $d\lambda(P)=2x_j(P)dx_j(P)$ near the point $P_j$, it follows that the quantity $\omega(P_j)$ satisfies
\begin{equation}\label{notation1-1}
\frac{\big(\omega(P_j)\big)^2}{2}=\underset{P_j}{{\rm res}}\ \frac{\big(\omega(P)\big)^2}{d\lambda(P)}.
\end{equation}
We shall define the evaluation of $\omega$ at a pole $\infty^i$ or at a zero $\mathbf{z}^i$ of $\lambda$ as follows
\begin{equation}\label{notation3}
\begin{split}
&\omega^{(k)}(\infty^i)=k!\underset{\infty^i}{{\rm res}}\ \lambda(P)^{\frac{k+1}{n_i+1}}\omega(P), \quad \forall\ k\in \N;\\
&\omega^{(k)}(\mathbf{z}^i)=k!\underset{\mathbf{z}^i}{{\rm res}}\ \lambda(P)^{-\frac{k+1}{m_i+1}}\omega(P), \quad \forall\ k\in \N.
\end{split}
\end{equation}
Here, $\omega$ is not supposed to be holomorphic near $\infty^i$ and $\mathbf{z}^i$, but if it is, then the terms $\omega^{(k)}(\infty^i)/k!$ (resp. $\omega^{(k)}(\mathbf{z}^i)/k!$) are nothing but the coefficients of the Taylor expansion of $\omega$ near the pole $\infty^i$ (resp. the zero $\mathbf{z}^i$).
\subsection{Symmetric bidifferentials on  compact Riemann surfaces}
Let $C_g$ be a compact Riemann surfaces of genus $g\geq1$  equipped with a canonical homology  basis $\{a_k,b_k\}$. Corresponding to the fixed basis of cycles $\{a_k,b_k\}$, there is a uniquely determined basis of holomorphic differentials $\omega_1,\dots,\omega_g$, normalized by the conditions $\oint_{a_k}\omega_j=\delta_{jk}$. Denote by $\mathbb{B}$ the Riemann matrix of $b$-periods, i.e. $\mathbb{B}_{jk}=\oint_{b_k}\omega_j$.\\
Let $\alpha$ and $\beta$ be two arbitrary real $g$-dimensional vectors. The Riemann theta function with  characteristic $\Delta=[\alpha,\beta]$ and modulus  $\mathbb{B}$ is defined by
\begin{equation}\label{Theta}
\Theta_{\Delta}(z)=\Theta_{\Delta}(z|\mathbb{B}):=\sum_{n\in \Z^g}e^{{\rm{i}}\pi\prs{\mathbb{B}(n+\alpha)}{n+\alpha}+2{\rm{i}}\pi\prs{z+\beta}{n+\alpha}},\quad \quad z\in \C^g.
\end{equation}
For any integer vectors $p,q\in \Z^g$, the function $\Theta_{\Delta}$ enjoys the following quasi-periodicity properties:
\begin{equation}\label{theta-quasi}
\Theta_{\Delta}(z+p+\mathbb{B}q)=e^{-{\rm{i}}\pi\prs{\mathbb{B}q}{q}-2{\rm{i}}\pi\prs{z+\beta}{q}+2{\rm{i}}\pi\prs{p}{\alpha}}\Theta_{\Delta}(z).
\end{equation}

In what follows we will  work with an odd and non-singular half-integer characteristic  $\Delta$  where by this we mean that $\alpha,\beta\in \frac{1}{2}\Z^g$,  $4\prs{\alpha}{\beta}\in 1+2\Z$ and the  gradient of the function $\Theta_{\Delta}$ at the origin is nonzero:
\begin{equation}\label{Theta gradient}
\big(\partial_{z_1}\Theta_{\Delta}(0),\dots,\partial_{z_g}\Theta_{\Delta}(0)\big)\neq (0,\dots,0).
\end{equation}
Particularly, $\Theta_{\Delta}$ becomes an odd function and the holomorphic differential
\begin{equation*}\label{holomorphic diff}
\omega_{\Delta}(P):=\textstyle\sum_{k=1}^g\partial_{z_k}\Theta_{\Delta}(0)\omega_k(P)
\end{equation*}
does  not vanish identically. For more detailed treatment on the Riemann theta function, we refer to \cite{Farkas, Fay, Lewittes}.\\

The (canonical) symmetric bidifferential $W$ on $C_g$  is the meromorphic differential on $C_g\times C_g$ defined by means  of the $\Delta$-characteristic Riemann theta function $\Theta_{\Delta}$ by \cite{Fay, Fay92, Yamada}
\begin{equation}\label{W-def}
W(P,Q):=d_Pd_Q\log\Big(\Theta_{\Delta}\big(\mathcal{A}(P)-\mathcal{A}(Q)\big)\Big)=d_Pd_Q\log\big(E(P,Q)\big),
\end{equation}
where $\mathcal{A}(P):=\big(\int^P\omega_1,\dots,\int^P\omega_g\big)$ stands for the Abel map on $C_g$ and $E$ is the prime form.\\
By using the properties of Riemann theta function \cite{Farkas, Fay, Lewittes}, we can observe that the  bidifferential $W(P,Q)$ satisfies the following properties:
\begin{description}
\item[i)] it is a symmetric meromorphic differential on $C_g\times C_g$: $W(P,Q)=W(Q,P)$;
\item[ii)] it has a pole of second order on the diagonal with biresidue 1:
\begin{equation}\label{W-asymptotic}
W(P,Q)\underset{P\sim Q}=\left(\frac{1}{\big(\xi(P)-\xi(Q)\big)^2}+\text{reg}\right)d\xi(P)d\xi(Q),
\end{equation}
where $\xi$ is a local parameter near $P\sim Q$;
\item[iii)] its $a$-periods with respect to either of the arguments vanish:
\begin{equation}\label{W-periods}
\oint_{a_k}W(P,Q)=0, \quad\quad \forall\ k=1,\dots,g.
\end{equation}
\end{description}
We point out that  properties \textbf{i)}, \textbf{ii)} and \textbf{iii)} provide a characterization of the canonical  symmetric bidifferential $W(P,Q)$.
Indeed, the difference of two kernels that fulfill these three conditions  would have no pole and would have vanishing $a$-periods, and therefore it would vanish identically.\\
In genus zero, the symmetric  bidifferential on the Riemann sphere $\P^1$ is given by the following rational expression:
\begin{equation}\label{g=0-W}
W(P,Q)=\frac{dz_Pdz_Q}{(z_P-z_Q)^2}.
\end{equation}

\noindent\textbf{Rauch variational formulas}  Let $[(C_g,\lambda)]$ be a point in the already described (covering of the) simple or double Hurwitz space. Then,  for each fixed $P,Q\in C_g$, the quantity $W(P,Q)$ becomes a function of simple branch points $\lambda_1,\dots,\lambda_N$. The dependence of the bidifferential $W(P,Q)$ on $\{\lambda_j\}_j$ is specified  by Rauch's variational formulas \cite{Fay92, Rauch, Yamada}:
\begin{equation}\label{Rauch}
\partial_{\lambda_j}W(P,Q)=\frac{1}{2}W(P,P_j)W(P_j,Q)=\underset{P_j}{{\rm res}}\frac{W(P,S)W(S,Q)}{d\lambda(S)}.
\end{equation}
Here $W(P,P_j)$ denotes the evaluation of the bidifferential $W(P,Q)$ (the point $P$ is fixed) at $Q=P_j$ with respect to the standard local parameter $x_j(Q)=\sqrt{\lambda(Q)-\lambda_j}$ near the ramification point $P_j$:
$$
W(P,P_j):=\frac{W(P,Q)}{dx_j(Q)}\Big|_{Q=P_j}.
$$

\section{Flat metrics and flat coordinates in terms of the bidifferential on simple and double Hurwitz spaces}
In this section, we consider a family of  flat metrics  associated with  differential forms  $\omega$ with respect to the bidifferential $W$, i.e., let $\omega$ be the differential form corresponding to the covering $(C_g,\lambda)$ defined by:
\begin{equation}\label{Diff model}
\omega(P)=\int_{\gamma}h(\lambda(Q))W(P,Q).
\end{equation}
Here  $\gamma$  and $h$ are assumed to  satisfy the following conditions:
\begin{itemize}
\item[\texttt{C1)}] $\gamma$ is an arbitrary smooth contour on the compact Riemann surface $C_g$ not passing through any of the ramification points $P_j$ and such that  its projection $\lambda(\gamma)\subset\P^1$ is independent of the branch points $\{\lambda_j\}_j$, where by this we mean that the contour $\lambda(\gamma)$ does not change under small variations of $\lambda_1,\dots,\lambda_N$;
\item[\texttt{C2)}] $h$ is a  function defined in a neighborhood of $\gamma$ such that $h$ does not depend on  the branch points $\lambda_1,\dots,\lambda_N$.
\end{itemize}
Motivated by the work of Dubrovin \cite{Dubrovin2D} (formulas (5.35a)-(5.35d)) (see also \cite{Vasilisa, Romano}), the pair $(h,\gamma)$ will be chosen in such a manner that the differential $\omega$ on the surface $C_g$ is holomorphic outside the assigned poles $\infty^0,\dots,\infty^m$ and zeros $\mathbf{z}^0,\dots,\mathbf{z}^n$  of $\lambda$.\\
Due to the properties of the symmetric bidifferential $W(P,Q)$ and those of the contour $\gamma$, we observe that $\omega$ is holomorphic around each ramification point $P_j$.
Moreover, in view of Rauch's formulas (\ref{Rauch}), the partial derivatives of $\omega(P)$ with respect to the branch points $\{\lambda_j\}_j$ of the covering $(C_g,\lambda)$ are expressed by means of the evaluations $\big\{\omega(P_j)\big\}_j$ as follows:
\begin{equation}\label{Rauch-2}
\partial_{\lambda_j}\omega(P)=\frac{1}{2}W(P,P_j)\int_{\gamma}h(\lambda(Q))W(Q,P_j)=\frac{\omega(P_j)}{2} W(P,P_j).
\end{equation}
In particular, for each $j$ we observe that if $\omega(P_j)\neq 0$, then  $\partial_{\lambda_j}\omega(P)$ is an Abelian differential of the second kind with a unique double pole at the point $P_j$. Moreover, their $a$-periods are all zero.\\
For a fixed  differential $\omega$ as in (\ref{Diff model}), we will work in the open domain $\mathcal{M}_{\omega}$ in the simple Hurwitz space
$\widehat{\mathcal{H}}_{g, L}(n_0,\dots,n_m)$ or double Hurwitz space  $\widehat{\mathcal{H}}_{g,L}(\mathbf{n}; \mathbf{m})$  determined by the conditions
\begin{equation}\label{Assymptions}
 \omega(P_j)\neq0,\quad \forall\ j=1,\dots,N.
\end{equation}
For $\beta_1,\beta_2\in \C$, with $(\beta_1,\beta_2)\neq(0,0)$, consider the  diagonal metric $\mathbf{ds^2_{\beta_1,\beta_2}}(\omega)$ depending on the chosen  differential $\omega$ defined by:
\begin{equation}\label{ds}
\mathbf{ds^2_{\beta_1,\beta_2}}(\omega):=\sum_{j=1}^N\bigg(\frac{\omega(P_j)^2}{2\big(\beta_1\lambda_j+\beta_2\big)}\bigg)(d\lambda_j)^2=
\sum_{j=1}^N\bigg(\underset{P_j}{{\rm res}}\frac{\omega(P)^2}{\big(\beta_1\lambda(P)+\beta_2\big)d\lambda(P)}\bigg)(d\lambda_j)^2.
\end{equation}
Consider  the M\"{o}bius transformation:
$$
\textsf{M}.z:=\frac{\delta{z}-\beta_2}{\delta+\beta_1},\quad \quad \text{with}\quad \delta(\beta_1+\delta)=1.
$$
We can observe that when $\beta_1=0$, then $\textsf{M}.z=z\pm\beta_2$ and thus $\infty$ is the unique  fixed point of $\textsf{M}$. In this case, the metric $\mathbf{ds^2_{0,\beta_2}}(\omega)$ is well defined in $\mathcal{M}_{\omega}$ (\ref{Assymptions}) seen as an open subset of the simple Hurwitz space $\widehat{\mathcal{H}}_{g, L}(n_0,\dots,n_m)$.\\  When $\beta_1\neq0$, then $\textsf{M}$ has two fixed points which are $\infty$ and $\xi=-\beta_2/\beta_1$.
Therefore  we regard $\mathbf{ds^2_{\beta_1,\beta_2}}(\omega)$ as a metric defined in $\mathcal{M}_{\omega}$ viewed as an open subset of the double Hurwitz space $\widehat{\mathcal{H}}_{g,L}(\mathbf{n}; \mathbf{m}){(\xi)}$. Here the space $\widehat{\mathcal{H}}_{g,L}(\mathbf{n}; \mathbf{m}){(\xi)}$ is the image of the already described  double Hurwitz  space $\widehat{\mathcal{H}}_{g,L}(\mathbf{n}; \mathbf{m})$ by the biholomorphic mapping
$$
[(C_g,\lambda)]\longmapsto[(C_g,\lambda+\xi)],\quad \quad \xi=-\beta_2/\beta_1
$$
(that is the two special fibers become over the two fixed points $\infty$ and $\xi=-\beta_2/\beta_1$ of $\textsf{M}$).\\
It is   known that  the Christoffel symbols of a diagonal metric $\sum_ig_{ii}(d\lambda_i)^2$ satisfy
\begin{equation}\label{Christoffel symbols}
\Gamma_{ij}^k=0,\quad \Gamma_{ii}^k=-\frac{\partial_kg_{ii}}{2g_{kk}},\quad \Gamma_{ii}^i=
\frac{\partial_ig_{ii}}{2g_{ii}},\quad\Gamma_{ij}^i=\frac{\partial_jg_{ii}}{2g_{ii}},\quad \quad \text{$i,j,k$ are distinct}.
\end{equation}
Therefore, by (\ref{Rauch-2}), (\ref{Christoffel symbols}) and a straightforward computation,  we deduce that the  nonvanishing Christoffel symbols of the diagonal metric $\mathbf{ds^2_{\beta_1,\beta_2}}(\omega)$ are
\begin{equation}\label{CS}
\begin{split}
&\Gamma_{ij}^{i}=\frac{\partial_{\lambda_j}\omega(P_i)}{\omega(P_i)}=\frac{\omega(P_{j})}{2\omega(P_{i})}W(P_i,P_j),\quad\quad  i\neq j;\\
&\Gamma_{ii}^{j}=-\frac{\beta_1\lambda_j+\beta_2}{\beta_1\lambda_i+\beta_2}\Gamma_{ij}^{j},\quad\quad \quad  \quad i\neq j;\\
&\Gamma_{ii}^{i}=-\frac{\beta_1}{2\beta_1\lambda_i+2\beta_2}+\frac{\partial_{\lambda_{i}}\omega(P_i)}{\omega(P_i)}.
\end{split}
\end{equation}
In  Appendix 1, we mainly use (\ref{CS}) as well as  identity (\ref{Auxi1}) below  to prove that the Riemann curvature tensor of $\nabla$ (the Levi-Civita connection of $\mathbf{ds^2_{\beta_1,\beta_2}}(\omega)$) vanishes identically.  This establishes the flatness property of the metric $\mathbf{ds^2_{\beta_1,\beta_2}}(\omega)$. We refer to Theorem \ref{flatness-W} and its proof for more details.\\
In particular, the diagonal metrics $\mathbf{ds^2_{0,\beta_2}}(\omega)$ are all  Darboux-Egoroff metrics  with common rotation coefficients given in terms of the symmetric bidifferential $W(P,Q)$ by
$$
\kappa_{ij}:=\frac{W(P_i,P_j)}{2}=\frac{W(P,Q)}{2dx_i(P)dx_j(Q)}\Big|_{P=P_i,Q=P_j},\quad i\neq j.
$$
\subsection{An auxiliary result}
The following useful result describes the action of the family of the vector fields
$$
\textstyle \sum_{j=1}^N\big(\beta_1\lambda_j+\beta_2\big)\partial_{\lambda_j},\quad \quad (\beta_1,\beta_2)\neq(0,0),
$$
on the quantities $W(P_i,P_j)$ and $\omega(P_j)$ seen as functions of the branch points $\{\lambda_1,\dots,\lambda_N\}$ of a covering $(C_g,\lambda)$.
\begin{Thm}\label{E-M-action} Let  $(C_g,\lambda)$ be a branched covering of $\P^1$ with $N$ simple critical points $P_1,\dots,P_N$ and  $\omega$ be a differential on $C_g$
defined by (\ref{Diff model}). Then for all $i,j=1,\dots,N$, $i\neq j$,  we have
\begin{align}
&\label{Auxi1}\sum_{k=1}^N(\beta_1\lambda_k+\beta_2)\partial_{\lambda_k}W(P_i,P_j)=-\beta_1W(P_i,P_j);\\
&\label{Auxi2}\sum_{k=1}^N(\beta_1\lambda_k+\beta_2)\partial_{\lambda_k}\omega(P_j)=-\frac{\beta_1}{2}\omega(P_j)
+\int_{\gamma}\Big(\big(\beta_1\lambda(Q)+\beta_2\big)h'(\lambda(Q))W(Q,P_j)\Big).
\end{align}
\end{Thm}
\emph{Proof:} We distinguish two cases depending on wether $\beta_1$ is zero or not.\\
\emph{First case: $\beta_1=0$}.  Consider the family of branched coverings
$\big\{(C_g, \lambda_{\varepsilon})\big\}_{\varepsilon}$ with
\begin{equation}\label{Translation}
\lambda_{\varepsilon}(P):=\lambda(P)+\varepsilon.
\end{equation}
We observe that  all the branched coverings $(C_g,\lambda_{\varepsilon})$ share the set of ramification points and then, by (\ref{Translation}) we deduce that the standard local parameter $x_j^{\varepsilon}(P)$ near the point $P_j$, induced by the covering $\lambda_{\varepsilon}$,  is given by
$$
x_j^{\varepsilon}(P):=\sqrt{\lambda_{\varepsilon}(P)-\lambda_{\varepsilon}(P_j)}=x_j(P).
$$
Moreover,  we have a one-parameter group $(\chi_{\varepsilon})_{\varepsilon\in\C}$, a subgroup of the automorphism group of the simple Hurwitz space  $\mathcal{H}_{g, L}(n_0,\dots,n_m)$,  defined by
$$
\chi_{\varepsilon}\big([(C_g,\lambda)]\big)=[(C_g, \lambda_{\varepsilon})].
$$
Denote by $\big((\chi_{-\varepsilon})^*W\big)(P,Q)$ the pullback of the function $W(P,Q)$ by the map $\chi_{-\varepsilon}=\chi_{\varepsilon}^{-1}$
(when $P,Q$ are fixed, we view $W(P,Q)$ as a function on the Hurwitz space). Then  the quantity $\big((\chi_{-\varepsilon})^*W\big)(P,Q)$ defines a bidifferential on the same surface $C_g$ equipped with complex structure induced by the covering $(C_g, \lambda_{-\varepsilon})$.
Since for any $\varepsilon$, the complex structure on surface $C_g$ defined by the coverings $(C_g, \lambda_{\varepsilon})$ and $(C_g, \lambda)$ is the same, by uniqueness of the symmetric bidifferential it follows that $\big((\chi_{-\varepsilon})^*W\big)(P,Q)=W(P,Q)$, $\forall\ \varepsilon$.\\
On the other hand, given that the differential $\omega$ in (\ref{Diff model})  is defined by the pair $(h(\lambda),\gamma)$ and the contour $\gamma$ is  on the surface $C_g$ (hence $\gamma$ is independent of $\lambda$ and $\lambda_{\varepsilon}$), we deduce that  the $\chi_{-\varepsilon}$-pullback of $\omega$ is defined by $(h(\lambda_{-\varepsilon}),\gamma)$ and thus the differential $(\chi_{-\varepsilon})^*\omega$ is given by
$$
\big((\chi_{-\varepsilon})^*\omega\big)(P)=\int_{\gamma}h\big(\lambda_{-\varepsilon}(Q)\big)W(P,Q)=
\int_{\gamma}h\big(\lambda(Q)-\varepsilon\big)W(P,Q).
$$
This brings us to write:
\begin{align*}
&\frac{\big((\chi_{-\varepsilon})^*W\big)(P,Q)}{dx_i^{\varepsilon}(P)dx_j^{\varepsilon}(Q)}=\frac{W(P,Q)}{dx_i(P)dx_j(Q)};\\
&\frac{\big((\chi_{-\varepsilon})^*\omega\big)(P)}{dx_j^{\varepsilon}(P)}=\int_{\gamma}h\big(\lambda(Q)-\varepsilon\big)\frac{W(P,Q)}{dx_j(P)}.
\end{align*}
Recalling that the ramification points of $\lambda$ and $\lambda_{-\varepsilon}$ coincide, we perform the evaluation at $(P,Q)=(P_i,P_j)$ in the first equality and $P=P_j$ in the second one and recalling notation (\ref{notation1}) we get
\begin{align*}
&\big((\chi_{-\varepsilon})^*W\big)(P_i,P_j)=W(P_i,P_j),\quad i\neq j;\\
&\big((\chi_{-\varepsilon})^*\omega\big)(P_j)=\int_{\gamma}h\big(\lambda(Q)-\varepsilon\big)W(P_j,Q).
\end{align*}
Now differentiating these two relations with respect to $\varepsilon$ at $\varepsilon=0$ and using the fact that the flow associated with the one-parameter group $(\chi_{\varepsilon})_{\varepsilon}$ is generated by the vector field $\sum_{k=1}^N\partial_{\lambda_k}$, we obtain
\begin{align*}
&\textstyle\sum_{k=1}^N\partial_{\lambda_k}W(P_i,P_j)=-\frac{d}{d\varepsilon} \big((\chi_{-\varepsilon})^*W\big)(P_i,P_j)\Big|_{\varepsilon=0}=0;\\
&\textstyle\sum_{k=1}^N\partial_{\lambda_k}\omega(P_j)=-\frac{d}{d\varepsilon}\big((\chi_{-\varepsilon})^*\omega\big)(P_j)\Big|_{\varepsilon=0}
=\int_{\gamma}h'(\lambda(Q))W(P_j,Q).
\end{align*}
\noindent \emph{Second case: $\beta_1\neq0$.} Let us consider the family of branched coverings $(C_g, \lambda_{\varepsilon})$  of  $\P^1$ defined by
\begin{equation}\label{Dilation-translation}
\lambda_{\varepsilon}(P)=e^{\varepsilon\beta_1}\big(\lambda(P)-\xi\big)+\xi, \quad \quad \text{with}\quad \xi:=-\beta_2/\beta_1.
\end{equation}
As in the first case, we can see that the family of  transformations $\{\varphi_{\varepsilon}\}_{\varepsilon}$, with
$$
\varphi_{\varepsilon}\big([(C_g, \lambda)]\big):=[(C_g, \lambda_{\varepsilon})],
$$
gives rise to a local one-parameter subgroup of automorphisms of the Hurwitz space.\\
Similarly to the first case, we have $\big((\varphi_{-\varepsilon})^*W\big)(P,Q)=W(P,Q)$ and
$$
\big((\varphi_{-\varepsilon})^*\omega\big)(P)=\int_{\gamma}h\big(\lambda_{-\varepsilon}(Q)\big)W(P,Q)
=\int_{\gamma}h\big(e^{-\varepsilon\beta_1}\big(\lambda(Q)-\xi\big)+\xi\big)W(P,Q).
$$
In addition, (\ref{Dilation-translation}) implies that $d\lambda_{\varepsilon}(P)=0$ if and only if $d\lambda(P)=0$ and thus the ramification points of the coverings
$(C_g, \lambda_{\varepsilon})$ coincide with those of $(C_g, \lambda)$. Consequently, the standard local parameter $x_j^{\varepsilon}(P)$ near the critical point $P_j$ and induced by $\lambda_{-\varepsilon}$ becomes
$$
x_j^{\varepsilon}(P):=\sqrt{\lambda_{-\varepsilon}(P)-\lambda_{-\varepsilon}(P_j)}=e^{-\varepsilon\frac{\beta_1}{2}}x_j(P).
$$
Therefore $dx_j^{\varepsilon}(P)=e^{-\varepsilon\frac{\beta_1}{2}}dx_j(P)$ and
\begin{align*}
&\frac{\big((\varphi_{-\varepsilon})^*W\big)(P,Q)}{dx_i^{\varepsilon}(P)dx_j^{\varepsilon}(Q)}=e^{\varepsilon\beta_1}\frac{W(P,Q)}{dx_i(P)dx_j(Q)};\\
&\frac{\big((\varphi_{-\varepsilon})^*\omega\big)(P)}{dx_j^{\varepsilon}(P)}
=e^{\varepsilon\frac{\beta_1}{2}}\int_{\gamma}h\big(e^{-\varepsilon\beta_1}\big(\lambda(Q)-\xi\big)+\xi\big)\frac{W(P,Q)}{dx_j(P)}.
\end{align*}
Plugging $(P,Q)=(P_i,P_j)$ and $P=P_j$ in these equalities, respectively, and using introduced notation (\ref{notation1}), we deduce that they can be rewritten as follows
\begin{align*}
&\big((\varphi_{-\varepsilon})^*W\big)(P_i,P_j)=e^{\varepsilon\beta_1}W(P_i,P_j);\\
&\big((\varphi_{-\varepsilon})^*\omega\big)(P_j)=e^{\varepsilon\frac{\beta_1}{2}}\int_{\gamma}h\big(e^{-\varepsilon\beta_1}\big(\lambda(Q)-\xi\big)+\xi\big)W(P_j,Q).
\end{align*}
Let us now differentiate both sides with respect to $\varepsilon$ at $\varepsilon=0$. Since the vector field $\sum_j\big(\beta_1\lambda_j-\beta_1\xi\big)\partial_{\lambda_j}$ generates the local flow induced by the local one-parameter group $(\varphi_{\varepsilon})_{\varepsilon}$, it follows that
\begin{align*}
& \sum_k\big(\beta_1\lambda_k-\beta_1\xi\big)\partial_{\lambda_k}W(P_i,P_j)=-\frac{d}{d\varepsilon} \big((\varphi_{-\varepsilon})^*W\big)(P_i,P_j)\Big|_{\varepsilon=0}=
-\beta_1W(P_i,P_j);\\
&\sum_k\big(\beta_1\lambda_k-\beta_1\xi\big)\partial_{\lambda_k}\omega(P_j)
=-\frac{\beta_1}{2}\omega(P_j)+\int_{\gamma}\bigg(\big(\beta_1\lambda(Q)-\beta_1\xi)\big)h'(\lambda(Q))W(P_j,Q)\bigg).
\end{align*}
Finally replacing $\xi$ by $-\beta_2/\beta_1$, we conclude that formulas (\ref{Auxi1}) and (\ref{Auxi2}) hold true.

\fd

\subsection{Characterization of flat functions}
Flat coordinates of a flat metric $\Lambda$ are a set of coordinates $\{t^A\}$ in which coefficients of the metric are all constant. In these coordinates the Christoffel symbols vanish and the covariant derivative $\nabla_{\partial_{t^A}}$ is nothing but  the usual partial derivative $\partial_{t^A}$ in the direction of the flat coordinate $t^A$. We refer to the monograph \cite{Spivak} for details.\\
It is known that the flat coordinates are solutions to the partial differential equations
$$
\partial_{\lambda_i}\partial_{\lambda_j}t=\textstyle\sum_k\Gamma_{ij}^k\partial_{\lambda_k}t,
$$
where $\Gamma_{ij}^k$ are the Christoffel symbols of the flat metric $\Lambda$.\\
In the case where the considered flat metric $\Lambda$ is diagonal, by (\ref{Christoffel symbols})  we see that these equations can be rewritten in the form:
\begin{align}
&\label{system flat coordinates1}\partial_{\lambda_i}\partial_{\lambda_j}t=\Gamma_{ij}^i\partial_{\lambda_i}t+ \Gamma_{ij}^j\partial_{\lambda_j}t,\quad \quad i\neq j,\\
&\label{system flat coordinates2}\partial_{\lambda_j}^2t=\displaystyle\sum_s\Gamma_{jj}^s\partial_{\lambda_s}t.
\end{align}
We shall call a solution to equations (\ref{system flat coordinates1})-(\ref{system flat coordinates2})  a flat function with respect to  the metric $\Lambda$ or more shortly a $\Lambda$-flat function.\\

Consider  the diagonal flat metric $\mathbf{ds^2_{\beta_1,\beta_2}}(\omega)$ defined by (\ref{ds}), where $\omega$ is a fixed   differential with respect to the kernel $W$  given by   (\ref{Diff model}) and (\ref{Assymptions}).  We are going  to construct flat coordinates $t(\omega)$ for this metric of the following specific type:
\begin{equation}\label{FF-model}
t(\omega):=\int_{\ell}f(\lambda(P))\omega(P),
\end{equation}
where the contour $\ell$ and the function $f$ enjoy the conditions \texttt{C1)} and \texttt{C2)} in (\ref{Diff model}), i.e., $\lambda(\ell)$ and $f$ are independent of the branch points $\lambda_1,\dots,\lambda_N$.\\
Note the pair $(h,\gamma)$ defining the differential $\omega$  in (\ref{Diff model}) gives rise to a function of type (\ref{FF-model}). Conversely,
given that a function $t(\omega)$ of type (\ref{FF-model}), then the properties of $(f,\ell)$ allow us to attach it to the following  differential   $\phi_{t}(P)$ of the form(\ref{Diff model}):
\begin{equation}\label{phi-t}
\phi_t(P):=\int_{\ell}f(\lambda(Q))W(P,Q).
\end{equation}
Below a characterization of the $\mathbf{ds^2_{\beta_1,\beta_2}(\omega)}$-flatness of the function $t(\omega)$ will be given with the help  of its associated differential form $\phi_t$.\\

Our choice of  functions of (\ref{FF-model}) is firstly motivated by the following result.
\begin{Prop}
The function $t(\omega)$  in (\ref{FF-model}) is a solution to  equations (\ref{system flat coordinates1}) for the flat  metrics $\mathbf{ds^2_{\beta_1,\beta_2}}(\omega)$ for all $(\beta_1,\beta_2)\neq(0,0)$.
\end{Prop}
\emph{Proof:} Since by (\ref{CS}) the Christoffel  coefficients $\Gamma_{ij}^i$ are independent of $\beta_1,\beta_2$ when $i\neq j$,
 the system (\ref{system flat coordinates1}) is the same for all the considered  flat metrics $\mathbf{ds^2_{\beta_1,\beta_2}}(\omega)$. \\
Now, the properties of $\ell$ permit us to differentiate  under the integral sign and thanks to Rauch formulas (\ref{Rauch-2}) we have
\begin{equation}\label{partial-t-W}
\forall\ j,\quad \partial_{\lambda_j}t(\omega)=\frac{\omega(P_j)}{2}\phi_t(P_j)=\underset{P_j}{{\rm res}}\frac{\omega(P)\phi_t(P)}{d\lambda(P)}.
\end{equation}
Here, the second equality is an immediate consequence of the holomorphic  property of the differentials $\omega$ and $\phi_t$ near the point $P_j$ (a simple zero of $d\lambda(P)$).\\
Differentiating  in (\ref{partial-t-W}) once again with respect to  $\lambda_i$, $i\neq j$,  we obtain
\begin{align*}
\partial_{\lambda_i}\partial_{\lambda_j}t(\omega)&=\frac{1}{4}W(P_i,P_j)\phi_t(P_i)\omega(P_j)+\frac{1}{4}W(P_i,P_j)\phi_t(P_j)\omega(P_i).
\end{align*}
On the other hand, using (\ref{CS}) and (\ref{partial-t-W}), we deduce that
\begin{align*}
\Gamma_{ij}^i\partial_{\lambda_i}t(\omega)+ \Gamma_{ij}^j\partial_{\lambda_j}t(\omega)
=\frac{1}{4}\phi_t(P_i)W(P_i,P_j)\omega(P_j)+\frac{1}{4}\phi_t(P_j)W(P_i,P_j)\omega(P_i),\quad i\neq j.
\end{align*}
This shows that the function $t(\omega)$ obeys  equations (\ref{system flat coordinates1}) as desired.

\fd

\begin{Prop}\label{CNS-flat function} Let $(\beta_1,\beta_2)\neq(0,0)$ and $t(\omega)$ be the function defined by (\ref{FF-model}) and $\phi_t$ be the corresponding differential (\ref{phi-t}), where the pair $(f,\ell)$ enjoys the two conditions \texttt{C1)} and \texttt{C2)} as in (\ref{Diff model}).\\
 Then the  following are equivalent:
\begin{description}
\item[i)] The function $t(\omega)$ satisfies  equations (\ref{system flat coordinates2}) for the metric $\mathbf{ds^2_{\beta_1,\beta_2}}(\omega)$;
\item[ii)] for all $j=1,\dots, N$, we have
$$
\textstyle \sum_{k=1}^N\big(\beta_1\lambda_k+\beta_2\big)\partial_{\lambda_k}\phi_t(P_j)=-\frac{\beta_1}{2}\phi_t(P_j);
$$
\item[iii)] for all $j=1,\dots, N$, the following holds:
\begin{equation}\label{CNS-FF}
\int_{\ell}\big(\beta_1\lambda(P)+\beta_2\big)f'(\lambda(P))W(P,P_j)=0.
\end{equation}
\end{description}
\end{Prop}
\noindent\emph{Proof:} The equivalence between  items \textbf{ii)} and \textbf{iii)} follows immediately from (\ref{Auxi2}) applied to the differential form $\phi_t$ (\ref{phi-t}).\\
On the other hand,  we have
\begin{align*}
&\sum_s\Gamma_{jj}^s\partial_{\lambda_s}t(\omega)
=\sum_{s\neq j}\Gamma_{jj}^s\partial_{\lambda_s}t(\omega)+\Gamma_{jj}^j\partial_{\lambda_j}t(\omega)\\
&=-\frac{\omega(P_j)}{4\beta_1\lambda_j+4\beta_2}\sum_{s\neq j}\big(\beta_1\lambda_s+\beta_2)W(P_s,P_j)\phi_t(P_s)
+\left(-\frac{\beta_1\omega(P_j)}{4\beta_1\lambda_j+4\beta_2}+\frac{1}{2}\partial_{\lambda_j}\omega(P_j)\right)\phi_t(P_j)\\
&=-\frac{\omega(P_j)}{2\beta_1\lambda_j+2\beta_2}\sum_{s\neq j}\big(\beta_1\lambda_s+\beta_2)\partial_{\lambda_s}\phi_t(P_j)
+\left(-\frac{\beta_1\omega(P_j)}{4\beta_1\lambda_j+4\beta_2}+\frac{1}{2}\partial_{\lambda_j}\omega(P_j)\right)\phi_t(P_j)\\
&=-\frac{\omega(P_j)}{2\beta_1\lambda_j+2\beta_2}\sum_{s=1}^N\big(\beta_1\lambda_s+\beta_2)\partial_{\lambda_s}\phi_t(P_j)
+\frac{\omega(P_j)}{2}\partial_{\lambda_j}\phi_t(P_j)
+\left(-\frac{\beta_1\omega(P_j)}{4\beta_1\lambda_j+4\beta_2}+\frac{1}{2}\partial_{\lambda_j}\omega(P_j)\right)\phi_t(P_j)\\
&=\frac{\beta_1\omega(P_j)\phi_t(P_j)}{4\beta_1\lambda_j+4\beta_2}
-\frac{\omega(P_j)}{2\beta_1\lambda_j+2\beta_2}\int_{\ell}\bigg(\big(\beta_1\lambda(P)+\beta_2\big)f'(\lambda(P))W(P,P_j)\bigg)
+\frac{\omega(P_j)}{2}\partial_{\lambda_j}\phi_t(P_j)\\
&\quad +\left(-\frac{\beta_1\omega(P_j)}{4\beta_1\lambda_j+4\beta_2}+\frac{1}{2}\partial_{\lambda_j}\omega(P_j)\right)\phi_t(P_j)\\
&=\partial_{\lambda_j}^2t(\omega)-\frac{\omega(P_j)}{2\beta_1\lambda_j+2\beta_2}\int_{\ell}\bigg((\beta_1\lambda(P)+\beta_2\big)f'(\lambda(P))W(P,P_j)\bigg),
\end{align*}
where we have used relations (\ref{CS}) and (\ref{partial-t-W}) in the second equality,  Rauch's formulas (\ref{Rauch-2}) in the third equality and  relation (\ref{Auxi2}) in the fifth one.\\
From the last equality we conclude that $t(\omega)$ is a solution to  the system of  partial differential equations $\partial_{\lambda_j}^2t(\omega)=\sum_s\Gamma_{jj}^s\partial_{\lambda_s}t(\omega)$ if and only if   conditions (\ref{CNS-FF})  are satisfied.  This finishes  the proof of the proposition.

\fd

\medskip

Note that the $N$ conditions (\ref{CNS-FF}) do not depend  on  the choice of the  differential $\omega$ that we work with.
In addition, by making use of the properties of the symmetric bidifferential $W(P,Q)$ (\ref{W-def}), Proposition \ref{CNS-flat function} offers a method to construct flat coordinates of the flat metric $\mathbf{ds^2_{\beta_1,\beta_2}}(\omega)$. More precisely, this method consists in finding  functions of type (\ref{FF-model}) and  adequately choosing  the corresponding pairs $(f,\ell)$ in such a way that the $N$ conditions in (\ref{CNS-FF})  are fulfilled. This method will be applied in the next two subsections. \\

The next useful result gives  practical techniques  to calculate the entries of the constant Gram  matrix of the contravariant metric $\left(\mathbf{ds^2_{\beta_1,\beta_2}}(\omega)\right)^*$.\\
Here we recall that if $\Lambda=\sum_{i,j}\Lambda_{ij}dx^idx^j$ is a metric on a manifold $\mathcal{M}$, then $\Lambda$ induces a dual metric $\Lambda^*$  in the cotangent spaces. The components of $\Lambda^*$ are given  by
$$
\Lambda^*(dx^i,dx^j)=\Lambda^{ij},
$$
where $\Lambda^{ij}$ are the entries of the inverse matrix of $(\Lambda_{ij})$.
\begin{Thm}\label{Practical F}
Let $t^j(\omega)$ and $\phi_{t^j}(P)$ be respectively the functions and  the corresponding differentials defined  by
$$
t^j(\omega):=\int_{\ell_j}f_j(\lambda(P))\omega(P),\quad \quad  \phi_{t^j}(P):=\int_{\ell_j}f_j(\lambda(Q))W(P,Q),\quad j=1,2.
$$
Here, we assume that the pairs $(f_j,\ell_j)$  obey the same conditions  \texttt{C1)} and \texttt{C2)} satisfied by the pair $(h,\gamma)$ defining  the differential $\omega$  (\ref{Diff model}), but the functions $t^1(\omega)$ and $t^2(\omega)$ are not necessarily  supposed to be flat with respect to the flat metric $\mathbf{ds^2_{\beta_1,\beta_2}}(\omega)$.\\
Then we have the following results:
\begin{align}
\left(\mathbf{ds^2_{\beta_1,\beta_2}}(\omega)\right)^*\left(dt^1(\omega),dt^2(\omega)\right)
\label{dual metric-R1}&=\frac{1}{2}\sum_{j=1}^N\big(\beta_1\lambda_j+\beta_2\big)\phi_{t^1}(P_j)\phi_{t^2}(P_j)\\
&\label{dual metric-R2}=\sum_{j=1}^N\big(\beta_1\lambda_j+\beta_2\big)\partial_{\lambda_j}t^1\left(\phi_{t^2}\right)\\
&\label{dual metric-R4}=\sum_{j=1}^N\underset{P_j}{{\rm res}}\bigg(\big(\beta_1\lambda(P)+\beta_2\big)\frac{\phi_{t^1}(P)\phi_{t^2}(P)}{d\lambda(P)}\bigg).
\end{align}
where $t^1\left(\phi_{t^2}\right)$ is the function  defined as in (\ref{FF-model}).
\end{Thm}
\emph{Proof:}  We have
\begin{align*}
\left(\mathbf{ds^2_{\beta_1,\beta_2}}(\omega)\right)^*(dt^1(\omega),dt^2(\omega))
&=\sum_{i,j}\partial_{\lambda_i}t^1(\omega)\partial_{\lambda_j}t^2(\omega)\left(\mathbf{ds^2_{\beta_1,\beta_2}}(\omega)\right)^*(d\lambda_i,d\lambda_j)\\
&=\sum_{j}\frac{2\beta_1\lambda_j+2\beta_2}{\omega(P_j)^2}\partial_{\lambda_j}t^1(\omega)\partial_{\lambda_j}t^2(\omega)\\
&=\frac{1}{2}\sum_{j}\big(\beta_1\lambda_j+\beta_2\big)\phi_{t^1}(P_j)\phi_{t^2}(P_j),
\end{align*}
where we have used the chain rule formula in the first equality, (\ref{ds}) in the second one and (\ref{partial-t-W}) in the third equality.\\
Formula (\ref{dual metric-R2}) is direct consequence  of  (\ref{dual metric-R1}) combined with the  Rauch partial derivatives (\ref{partial-t-W}) for $\omega=\phi_{t_2}$:
$$
\partial_{\lambda_j}t^1\left(\phi_{t^2}\right)=\frac{1}{2}\phi_{t^1}(P_j)\phi_{t^2}(P_j).
$$
Moreover, by using (\ref{dual metric-R1}), we see that relation (\ref{dual metric-R4}) holds.

\fd

\begin{Remark}
Formula (\ref{dual metric-R1}) has been derived by Romano \cite{Romano} in the  particular case where $\beta_1=1$ and $\beta_2=0$.
\end{Remark}


\subsection{Flat coordinates of  Dubrovin's flat metrics and duality relations}
In this subsection, we assume that $\beta_1=0$ and $\beta_2=1$ and then  we shall work in open  subsets of the simple Hurwitz space
$\widehat{\mathcal{H}}_{g, L}(n_0,\dots,n_m)$. \\
Inspired by Dubrovin's work \cite{Dubrovin2D} (see also \cite{Vasilisa}), we shall give a system of flat coordinates of the Darboux-Egoroff  metric
\begin{equation}\label{eta-def}
\eta(\omega):=\mathbf{ds^2_{0,1}}(\omega)=\frac{1}{2}\sum_j\big(\omega(P_j)\big)^2(d\lambda_j)^2=
\sum_j\bigg(\underset{P_j}{{\rm res}}\frac{\omega(P)^2}{d\lambda(P)}\bigg)(d\lambda_j)^2,
\end{equation}
where $\omega$ is a differential satisfying (\ref{Diff model}) and (\ref{Assymptions}).
As already mentioned in the introduction, we allow $\omega$ to belong to a family of differentials  larger  than  the finite list of primary differentials given by
Dubrovin in \cite{Dubrovin2D} (Lecture 5).\\
As in the Dubrovin case, among the list of flat coordinates of the metric $\eta(\omega)$,  there is a particular subfamily of flat functions involving an appropriate  principal value. More precisely, the principal value in the function
$$
s^i(\omega):= p.v.\int_{\infty^0}^{\infty^i}\omega(P)
$$
is defined by omitting the divergent part of the integral as a function of the local parameters $z_0(P):=\lambda(P)^{-\frac{1}{n_0+1}}$, $z_i(P):=\lambda(P)^{-\frac{1}{n_i+1}}$ near the points $\infty^0$ and $\infty^i$, respectively.\\
In Theorem \ref{s-ij} below, we pay particular attention  to  the contribution of this specific principal value when $\omega$ is the Abelian differential of the third kind $\Omega_{\infty^0\infty^j}$.
\begin{Prop} Let $\omega$ be a differential satisfying  (\ref{Diff model}) and (\ref{Assymptions}) and $\eta(\omega)$ be the corresponding Dubrovin metric (\ref{eta-def}).\\
Then the following $N$ functions are all $\eta(\omega)$-flat :
\begin{equation}\label{FBP-W}
\begin{array}{lllll}
t^{i,\alpha}(\omega):=\displaystyle\frac{\sqrt{n_i+1}}{\alpha} \underset{\infty^i}{{\rm res}\ } \lambda(P)^{\frac{\alpha}{n_i+1}}\omega(P),&\quad i=0,\dots,m,\quad \alpha=1,\dots,n_i;\\
\\
v^i(\omega):=\displaystyle \underset{\infty^i}{{\rm res}\ }\lambda(P)\omega(P),&\quad i=1,\dots,m;\\
\\
s^i(\omega):=\displaystyle p.v.\int_{\infty^0}^{\infty^i}\omega(P),&\quad i=1,\dots,m;\\
\\
\rho^k(\omega):=\displaystyle \frac{1}{2{\rm{i}}\pi}\oint_{b_k}\omega(P),&\quad k=1,\dots,g;\\
\\
u^k(\omega):=\displaystyle \oint_{a_k}\lambda(P)\omega(P),&\quad k=1,\dots,g.
\end{array}
\end{equation}
\end{Prop}
\emph{Proof:} The functions $t^{i,\alpha}(\omega)$, $v^i(\omega)$, $\rho^k(\omega)$, $u^k(\omega)$ in (\ref{FBP-W}) are of the form (\ref{FF-model}). Furthermore, by employing the properties of the symmetric  bidifferential, we can directly  check  that  necessary and sufficient conditions (\ref{CNS-FF}) are  satisfied. This establishes their flatness property.\\
Now let us deal with the functions $s^i(\omega)$, $i=1,\dots,m$. If the differential $\omega$  has no poles (within a fundamental polygon of the surface $C_g$), then
we have $s^i(\omega)=\int_{\infty^0}^{\infty^i}\omega(P)$ (here the integral converges). Hence $s^i(\omega)$ is of the form (\ref{FF-model}) and obeys the conditions in (\ref{CNS-FF}).\\
Assume now that  the poles of $\omega$ are among the  assigned ramification points $\infty^0,\dots,\infty^m$. Note that (\ref{CNS-FF}) is valid for $\omega$ such that   Rauch's formula (\ref{partial-t-W}) holds.  In order to  ensure  the validity of  formula (\ref{partial-t-W}) for the functions $s^i(\omega)= p.v.\int_{\infty^0}^{\infty^i}\omega(P)$, we  always assume that the singular  parts of $\omega$ near its poles do not depend on the branch points $\lambda_1,\dots, \lambda_N$ (which serve as local coordinates in the considered simple Hurwitz space). Therefore the differential (\ref{phi-t}) corresponding to the function $s^i(\omega)$ is $\phi_{s^i}(P)=\int_{\infty^0}^{\infty^i}W(P,Q)$ and thus  (\ref{CNS-FF}) is valid for the function $s^i(\omega)$.

\fd

\medskip

We now move on to study  some characteristic properties of the differentials $\phi_{t^A}$ associated   with the flat functions  $t^A$ (\ref{FBP-W}) in the sense of (\ref{FF-model}) and (\ref{phi-t}).
These properties will be used as a crucial ingredient   to calculate  the entries of the Gram matrix of the contravariant  metric $\eta^*(\omega)$. \\
According to the correspondence  between (\ref{FF-model}) and (\ref{phi-t}), the family of the $\eta(\omega)$-flat functions
\begin{equation}\label{S-eta}
\mathcal{S}(\omega):=\Big\{t^{i,\alpha}(\omega),\ v^i(\omega),\ s^i(\omega),\ \rho^k(\omega),\  u^k(\omega)\Big\}
\end{equation}
defined by (\ref{FBP-W}) brings us to the following differentials:
\begin{equation}\label{Primary-e}
\begin{array}{lllll}
\textbf{\emph{1.}}\quad \phi_{t^{i,\alpha}}(P):=\displaystyle \frac{\sqrt{n_i+1}}{\alpha}\underset{\infty^i}{{\rm res}\ }\lambda(Q)^{\frac{\alpha}{n_i+1}}W(P,Q),&i=0,\dots,m; \quad \alpha=1,\dots,n_i;\\
\\
\textbf{\emph{2.}}\quad \phi_{v^i}(P):=\displaystyle \underset{\infty^i}{{\rm res}\ }\lambda(Q)W(P,Q), &i=1,\dots,m;\\
\\
\textbf{\emph{3.}}\quad \phi_{s^i}(P):=\displaystyle \int_{\infty^0}^{\infty^i}W(P,Q), &i=1,\dots,m;\\
\\
\textbf{\emph{4.}}\quad \phi_{\rho^k}(P):=\displaystyle\frac{1}{2{\rm{i}}\pi}\oint_{b_k}W(P,Q),&k=1,\dots,g;\\
\\
\textbf{\emph{5.}}\quad \phi_{u^k}(P):=\displaystyle\oint_{a_k}\lambda(Q)W(P,Q), &k=1,\dots,g.
\end{array}
\end{equation}
\begin{Remark}\label{Remark-primary}
\end{Remark}
\textbf{i)} For later use, we emphasize  that all the  differentials listed in (\ref{Primary-e}) share  the following particular integral representation with respect to the kernel $W(P,Q)$:
\begin{equation}\label{Int-rep-e}
\phi_{t^A}(P)=c_A\int_{\ell_A}\big(\lambda(Q)\big)^{d_A}W(P,Q),
\end{equation}
where $c_A$ is a nonzero  complex number and $d_A\geq0$. As we will see later, the nonnegative constants $d_A$ are of particular importance and play the role of the quasi-homogeneous degrees of  the differentials $\phi_{t^A}$ (see  Subsections 4.1.3 and 4.1.4). \\
As an example, if $t^A=t^{i,\alpha}$ we have $c_{i,\alpha}=\frac{\sqrt{n_i+1}}{\alpha}$, $d_{i,\alpha}=\frac{\alpha}{n_i+1}$ and $\ell_{i,\alpha}$ is a small contour
around the pole $\infty^i$.\\
\textbf{ii)} Consider a flat function  $t^A(\omega)\in \mathcal{S}(\omega)$ from the list (\ref{FBP-W})-(\ref{S-eta}).  If $t^A(\omega)\neq s^i(\omega)$, then  $t^A(\omega)$ also admits  the following integral representation:
$$
t^A(\omega)=c_A\int_{\ell_A}\big(\lambda(P)\big)^{d_A}\omega(P),
$$
where $c_A$ and $d_A$ are as above.\\
\textbf{iii)} The list (\ref{Primary-e}) coincides (up to multiplicative constants) with Dubrovin's primary differentials given by Shramchenko \cite{Vasilisa} (Theorem 2) in terms of the bidifferential $W(P,Q)$. We have chosen different multiplicative constants in order to introduce a convenient duality relation between flat coordinates (see Definition \ref{duality def} below).

\begin{Prop}\label{Primary-e-prop} The following statements hold:
\begin{itemize}
\item[\textbf{\emph{1)}}] For $i=0,\dots,m$ and $\alpha=1,\dots,n_i$, the differential $\phi_{t^{i,\alpha}}$ is an Abelian differential of the second kind.
It has the only pole at $\infty^i$ with the principal part:
\begin{equation}\label{type1}
\phi_{t^{i,\alpha}}(P)\underset{P\sim \infty^i}=\left(\frac{\sqrt{n_i+1}}{z_i(P)^{\alpha+1}}+O(1)\right)dz_i(P),
\quad \quad \lambda(P)=z_i(P)^{-n_i-1}.
\end{equation}
\item[\textbf{\emph{2)}}] For $i=1,\dots,m$, the differential $\phi_{v^i}$ is an Abelian differential of the second kind having
the only pole at $\infty^i$ with principal part:
\begin{equation}\label{type2}
\phi_{v^i}(P)\underset{P\sim \infty^i}=\left(\frac{n_i+1}{z_i(P)^{n_i+2}}\ +\ O(1)\right)dz_i(P)
=-d\lambda(P)+\ \text{holomorphic}
,\quad \quad \lambda(P)=z_i(P)^{-n_i-1}.
\end{equation}
\item[\textbf{\emph{3)}}] For $i=1,\dots,m$,  the differential $\phi_{s^i}$ is an Abelian differential of the third kind having simple poles at the points $\infty^i$
 and $\infty^0$ with residues $1$, $-1$ respectively. Furthermore, the following items hold true:
\begin{itemize}
\item[\textbf{i)}] In genus zero, we have
\begin{equation}\label{third kind-g0}
\phi_{s^i}(P)=\Omega_{\infty^0\infty^i}(P)=\left(\frac{1-\delta_{\infty^i,\infty}}{z_P-\infty^i}-\frac{1-\delta_{\infty^0,\infty}}{z_P-\infty^0}\right)dz_P,
\end{equation}
with $\infty\in \P^1$ being the point at infinity and $\delta_{\infty^i,\infty}$  the Kronecker symbol.
\item[\textbf{ii)}] If $g\geq 1$, then the differential $\phi_{s^i}$ is given by
\begin{equation}\label{third kind}
\begin{split}
\phi_{s^i}(P)&=\Omega_{\infty^0\infty^i}(P)=d_P\log\left(\frac{\Theta_{\Delta}\big(\mathcal{A}(P)-\mathcal{A}(\infty^i)\big)}
{\Theta_{\Delta}\big(\mathcal{A}(P)-\mathcal{A}(\infty^0)\big)}\right)\\
&=\sum_{k=1}^g\omega_k(P)\bigg(\frac{\partial_{z_k}\Theta_{\Delta}\big(\mathcal{A}(P)-\mathcal{A}(\infty^i)\big)}{\Theta_{\Delta}\big(\mathcal{A}(P)-\mathcal{A}(\infty^i)\big)}
-\frac{\partial_{z_k}\Theta_{\Delta}\big(\mathcal{A}(P)-\mathcal{A}(\infty^0)\big)}{\Theta_{\Delta}\big(\mathcal{A}(P)-\mathcal{A}(\infty^0)\big)}\bigg).
\end{split}
\end{equation}
Here  $\mathcal{A}$ denotes the Abel map and  $\Theta_{\Delta}$ is  the Riemann theta function (\ref{Theta}) corresponding to an odd and non-singular  half integer characteristic $\Delta$.
\end{itemize}
\item[\textbf{\emph{4)}}] For $k=1,\dots,g$, the differentials $\phi_{\rho^k}=\omega_k$ are holomorphic.
\item[\textbf{\emph{5)}}] For $k=1,\dots,g$,  $\phi_{u^k}(P)$ is a multivalued  differential with a jump along the cycle $b_k$ as follows:
\begin{equation}\label{type5}
\phi_{u^k}(P+b_k)-\phi_{u^k}(P)=-2{\rm{i}}\pi d\lambda(P).
\end{equation}
\end{itemize}
Furthermore, the  $a$-periods of $\phi_{\rho^k}$ are as follows $\displaystyle \oint_{a_l}\phi_{\rho^k}(P)=\delta_{kl}$, while the other differentials
 have zero $a$-periods.
\end{Prop}

\medskip

These analytic properties  of the differentials (\ref{Primary-e}) were essentially stated in \cite{Dubrovin2D} (Lecture 5) and \cite{Vasilisa} (Theorem 2).  Since they will be important in several subsequent results, we include detailed proofs for the reader’s convenience.

\medskip

\noindent \emph{Proof:}  \textbf{1)} and \textbf{2)} Let $P\in C_g\setminus\{\infty^i\}$ be fixed, where $\infty^i$ is among the assigned poles $\infty^0,\dots,\infty^m$ of the covering $(C_g,\lambda)$. Since the differential $W(P,\cdot)$ is holomorphic on $C_g\setminus \{P\}$,  by notation (\ref{notation3}), we can write
\begin{align*}
\frac{1}{\alpha}\underset{\infty^i}{{\rm res}}\lambda(Q)^{\frac{\alpha}{n_i+1}}W(P,Q)
&=\frac{1}{\alpha!}W^{(\alpha-1)}(P,\infty^i).
\end{align*}
In particular, $\phi_{t^{i,\alpha}}$  is a holomorphic differential on $C_g\setminus\{\infty^i\}$.\\
On the other hand, from (\ref{W-asymptotic}) we know that
$$
W(P,Q)\underset{Q\sim \infty^i}{\underset{P\sim \infty^i}=}\bigg(\frac{1}{(z_i(P)-z_i(Q))^2}+O(1)\bigg)dz_i(P)dz_i(Q).
$$
Differentiating $(\alpha-1)$-times this relation w.r.t. $z_i(Q)$ and evaluating it at $Q=\infty^i$ we obtain
$$
W^{(\alpha-1)}(P,\infty^i)=\Big(\frac{\alpha!}{z_i(P)^{\alpha+1}}+O(1)\Big)dz_i(P).
$$
This gives us the stated result of the first item. The second one can be proven in the same way  taking $n_i+1$ instead of $\alpha$.\\
\textbf{3)} When $g=0$, we obtain  formula (\ref{third kind-g0}) for the differential $\phi_{s^i}$ by integrating the genus zero bidifferential (\ref{g=0-W}) along a path from $\infty^0$ to $\infty^i$. If $g\geq 1$, then formula (\ref{W-def}) defining  the symmetric bidifferential $W(P,Q)$ in terms of the Abel map and Riemann theta function brings us to write:
\begin{align*}
\phi_{s^i}(P)&=d_P\log\left(\frac{\Theta_{\Delta}\big(\mathcal{A}(P)-\mathcal{A}(\infty^i)\big)}
{\Theta_{\Delta}\big(\mathcal{A}(P)-\mathcal{A}(\infty^0)\big)}\right)\\
&=\sum_{k=1}^g\omega_k(P)\bigg(\frac{\partial_{z_k}\Theta_{\Delta}\big(\mathcal{A}(P)-\mathcal{A}(\infty^i)\big)}{\Theta_{\Delta}\big(\mathcal{A}(P)-\mathcal{A}(\infty^i)\big)}
-\frac{\partial_{z_k}\Theta_{\Delta}\big(\mathcal{A}(P)-\mathcal{A}(\infty^0)\big)}{\Theta_{\Delta}\big(\mathcal{A}(P)-\mathcal{A}(\infty^0)\big)}\bigg).
\end{align*}
We thus arrive at  (\ref{third kind}).
Moreover the function $\Theta_{\Delta}\big(\mathcal{A}(P)-\mathcal{A}(\infty^j)\big)$ is holomorphic near the point $\infty^j$ and has the following behavior:
$$
\Theta_{\Delta}\big(\mathcal{A}(P)-\mathcal{A}(\infty^j)\big)\underset{P\sim\infty^j}{=}
\bigg(\sum_{l=1}^g\omega_l(\infty^j)\partial_{z_l}\Theta_{\Delta}(0)\bigg)z_j(P)+o\big(z_j(P)\big),\quad j=0,i.
$$
This shows the desired analytic properties of the differential $\phi_{s^i}$.\\
\textbf{4)} The fact that $\phi_{\rho^k}$ coincides with the normalized holomorphic differential $\omega_k$ follows directly from  the quasi-periodicity property (\ref{theta-quasi}) of the function $\Theta_{\Delta}$:
$$
\phi_{\rho^k}(P)=
\frac{1}{2i\pi}d_P\Big[\log\Big(\Theta_{\Delta}\big(\mathcal{A}(P)-\mathcal{A}(Q)-\mathbb{B}e_k\big)\Big)
-\log\Big(\Theta_{\Delta}\big(\mathcal{A}(P)-\mathcal{A}(Q)\big)\Big)\Big]
=\omega_k(P).
$$
\textbf{5)} Now we are going to prove the last assertion. Our arguments here are different from those used in \cite{Vasilisa}.  By (\ref{W-def}), we can write
\begin{align*}
\lambda(Q)W(P,Q)&=\lambda(Q)d_Pd_Q\log\Big(\Theta_{\Delta}\big(\mathcal{A}(P)-\mathcal{A}(Q)\big)\Big)\\
&=d_Q\Big[\lambda(Q)d_P\log\Big(\Theta_{\Delta}\big(\mathcal{A}(P)-\mathcal{A}(Q)\big)\Big)\Big]
-d\lambda(Q) d_P\log\Big(\Theta_{\Delta}\big(\mathcal{A}(P)-\mathcal{A}(Q)\big)\Big).
\end{align*}
Therefore
$$
\phi_{u^k}(P)=-\oint_{a_k}d_P\log\Big(\Theta_{\Delta}\big(\mathcal{A}(P)-\mathcal{A}(Q)\big)\Big)d\lambda(Q).
$$
Consider the function
\begin{equation*}\label{primitive-uk}
F_{u^k}(P):=\int^P\phi_{u^k}=-\oint_{a_k}\log\Big(\Theta_{\Delta}\big(\mathcal{A}(P)-\mathcal{A}(Q)\big)\Big)d\lambda(Q).
\end{equation*}
Using the quasi-periodicity properties (\ref{theta-quasi}) of the $\Theta_{\Delta}$-function, we conclude that  $F_{u^k}$ is a multivalued function with
\begin{align*}
F_{u^k}(P+a_j)-F_{u^k}(P)&=\oint_{a_j}\phi_{u^k}=\oint_{a_k}\lambda(Q)\left(\oint_{P\in a_j}W(P,Q)\right)=0;\\
F_{u^k}(P+b_j)-F_{u^k}(P)&=\oint_{b_j}\phi_{u^k}=-\oint_{a_k}\log\bigg(\frac{\Theta_{\Delta}\big(\mathcal{A}(P+b_j)-\mathcal{A}(Q)\big)}
{\Theta_{\Delta}\big(\mathcal{A}(P)-\mathcal{A}(Q)\big)}\bigg)d\lambda(Q)\\
&=-\oint_{a_k}\bigg(-{\rm{i}}\pi \prs{\mathbb{B}e_j}{e_j}-2{\rm{i}}\pi\prs{\beta}{e_j}-2{\rm{i}}\pi\int_Q^P\omega_j\bigg)d\lambda(Q)\\
&=2{\rm{i}}\pi \oint_{a_k}\Big(\int_Q^P\omega_j\Big)d\lambda(Q)\\
&=2{\rm{i}}\pi \oint_{a_k}\bigg(d_Q\Big(\lambda(Q)\int_Q^P\omega_j\Big)+\lambda(Q) \omega_j(Q)\bigg)\\
&=2{\rm{i}}\pi\bigg(\lambda(Q)\int_Q^P\omega_j\bigg)\bigg|_{Q=P_0}^{Q=P_0+a_k}+2{\rm{i}}\pi \oint_{a_k}\lambda(Q)\omega_j(Q),
\end{align*}
where $P_0$ is a starting point of the contour $a_k$ which we keep arbitrary for now.\\
Thus we arrive at
\begin{equation}\label{Fubini1}
F_{u^k}(P+b_j)-F_{u^k}(P)=\oint_{b_j}\phi_{u^k}=-2{\rm{i}}\pi\lambda(P_0)\delta_{jk}+2{\rm{i}}\pi \oint_{a_k}\lambda(Q)\omega_j(Q).
\end{equation}
On the other hand, note that when considering the difference $F_{u^k}(P+b_j)-F_{u^k}(P)$ we implicitly consider the point $P$ as a starting point of the cycle $b_j$.
In the case $j=k$, let us take the starting points of $a_k$ and $b_k$ to coincide with their intersection point. That is let us choose $P_0=P$. Now, given that $\phi_{u^k}(P)=dF_{u^k}(P)$, by (\ref{Fubini1}) we deduce that  $\phi_{u^k}$ has jumps only along the cycle $b_k$ and we have
$$
\phi_{u^k}(P+b_k)-\phi_{u^k}(P)=dF_{u^k}(P+b_k)-dF_{u^k}(P)=-2{\rm{i}}\pi{d\lambda(P)}.
$$

\fd

\medskip

The goal of the following theorem is to elucidate  the principal value of the integral $\int_{\infty^0}^{\infty^j}$ of the Abelian differential of the third kind
$\phi_{s^i}(P)=\Omega_{\infty^0\infty^i}(P)$ given by (\ref{third kind-g0})-(\ref{third kind}). In other words, according to the third type of  flat functions (\ref{FBP-W}), we are going to calculate the functions $s_{ij}:=s^j(\phi_{s^i})$.
\begin{Thm}\label{s-ij}
Let  $(C_g,\lambda)$ be a covering of $\P^1$ with prescribed poles $\infty^0,\dots,\infty^m$ of order $n_0+1,\dots,n_m+1$, respectively.
\medskip

\noindent \textbf{\emph{i)}} Let $g=0$ and choose the point $\infty^0$ in such a way that $\infty^0\in \C$ (without loss of generality, this may always be assumed).\\
Then
\begin{equation*}
\begin{array}{lllll}
s_{ij}=\displaystyle\log\left(\frac{\infty^i-\infty^j}{(\infty^j-\infty^0)(\infty^i-\infty^0)}\right)
+\frac{\log\big(\mu_0(\infty^0)\big)}{n_0+1}, & \hbox{if}\quad  i\neq j\ \text{and}\ \infty^i,\infty^j\in \C;\\
\\
s_{ij}=\displaystyle-\log(\infty^j-\infty^0)+\frac{\log\big(\mu_0(\infty^0)\big)}{n_0+1}, &\hbox{if}\quad  i\neq j\ \text{and}\ \infty^i=\infty;\\
\\
s_{ij}=\displaystyle-\log(\infty^0-\infty^i)+\frac{\log\big(\mu_0(\infty^0)\big)}{n_0+1}, &\hbox{if}\quad  i\neq j\ \text{and}\ \infty^j=\infty;\\
\\
s_{ij}=\displaystyle\frac{\log\big(\mu_i(\infty^i)\big)}{n_i+1}+\frac{\log\big(\mu_0(\infty^0)\big)}{n_0+1}
-2\log(\infty^i-\infty^0)-\log(-1),&\hbox{if}\quad  i=j\ \text{and}\ \infty^i\in \C;\\
\\
s_{ij}=\displaystyle\frac{\log\big(\mu_i(\infty^i)\big)}{n_i+1}+\frac{\log\big(\mu_0(\infty^0)\big)}{n_0+1},&\hbox{if}\quad  i=j\ \text{and}\ \infty^i=\infty,
\end{array}
\end{equation*}
where, for $i=0,\dots,m$, $\mu_i$ denotes  the meromorphic function on $\P^1$ defined by
\begin{equation}\label{mu-i}
\mu_i(P)=
\left\{
\begin{array}{ll}
(z_P-\infty^i)^{n_i+1}\lambda(P),& \quad \hbox{if}\quad \infty^i\in\C;\\
\\
z_P^{-n_i-1}\lambda(P),&\quad \hbox{if}\quad \infty^i=\infty.
\end{array}
\right.
\end{equation}
\textbf{\emph{ii)}} Assume that $g\geq1$. Then
\begin{equation}\label{s-ij Formula}
\begin{split}
&s_{ij}=\log\left(\frac{\Theta_{\Delta}\big(\mathcal{A}(\infty^j)-\mathcal{A}(\infty^i)\big)}
{\Theta_{\Delta}\big(\mathcal{A}(\infty^j)-\mathcal{A}(\infty^0)\big)\Theta_{\Delta}\big(\mathcal{A}(\infty^i)-\mathcal{A}(\infty^0)\big)}\right)
+\log\left(\sum_{k=1}^g\omega_k(\infty^0)\partial_{z_k}\Theta_{\Delta}\big(0\big)\right)-\log(-1),\quad i\neq j;\\
&s_{ii}=\log\left(\sum_{k=1}^g\omega_k(\infty^i)\partial_{z_k}\Theta_{\Delta}\big(0\big)\right)
+\log\left(\sum_{k=1}^g\omega_k(\infty^0)\partial_{z_k}\Theta_{\Delta}\big(0\big)\right)
-2\log\left(\Theta_{\Delta}\big(\mathcal{A}(\infty^i)-\mathcal{A}(\infty^0)\big)\right)-\log(-1).
\end{split}
\end{equation}
Here, as above, $\Theta_{\Delta}$ denotes   the Riemann theta function (\ref{Theta}) corresponding to an odd and non-singular  half integer characteristic $\Delta$.
\end{Thm}
\begin{Remark}
By definition, the function $\mu_i$ in (\ref{mu-i}) is holomorphic near $\infty^i$ and $\mu_i(\infty^i)\in \C\setminus\{0\}$.\\
When $g\geq1$, the k-th component of the  vector $\mathcal{A}(\infty^i)-\mathcal{A}(\infty^0)\in \C^g$ is $\int_{\infty^0}^{\infty^i}\omega_k=s^i(\omega_k)$.
\end{Remark}
\emph{Proof:}  \textbf{i)} Below, for any $i_0=0,\dots,m$, the point $\infty^{i_0,\varepsilon}$ is taken  near the pole $\infty^{i_0}$ of $\lambda$  such that
$$
\infty^{i_0,\varepsilon}=
\left\{
\begin{array}{ll}
\infty^{i_0}+\varepsilon,& \quad \hbox{if}\quad \infty^{i_0}\in\C;\\
\\
1/{\varepsilon},&\quad \hbox{if}\quad \infty^{i_0}=\infty,
\end{array}
\right.
$$
where $\varepsilon=\varepsilon(i_0)$ is a complex number such that $|\varepsilon|$ is small enough.\\
This yields  that the image of $\infty^{i_0,\varepsilon}$ by $\mu_{i_0}$ is given by
\begin{equation}\label{claim g=0}
\mu_{i_0}(\infty^{i_0,\varepsilon})=\varepsilon^{n_i+1}\lambda(\infty^{i_0,\varepsilon})=\Big(\varepsilon/{z_{i_0}(\infty^{i_0,\varepsilon})}\Big)^{n_i+1},
\end{equation}
with $z_{i_0}(P)$ being  the standard local parameter near the pole $\infty^{i_0}$.\\
To demonstrate our result, we  consider two cases, depending on wether $\infty^i$ is finite or infinite.\\
\textbf{First case: $\infty^i\neq\infty$}. This implies that $\Omega_{\infty^0\infty^i}(P)$  is as follows:
$$
\Omega_{\infty^0\infty^i}(P)=\frac{dz_P}{z_P-\infty^i}-\frac{dz_P}{z_P-\infty^0}.
$$
If $\infty^j\neq\infty$ and $i\neq j$, then by (\ref{claim g=0}) we have
\begin{align*}
\int_{\infty^{0,\varepsilon}}^{\infty^j}\Omega_{\infty^0\infty^i}(P)
&=\int_{\varepsilon+\infty^0}^{\infty^j}\frac{dz_P}{z_P-\infty^i}-\int_{\varepsilon+\infty^0}^{\infty^j}\frac{dz_P}{z_P-\infty^0}\\
&=\log(\infty^j-\infty^i)-\log(\infty^0+\varepsilon-\infty^i)+\log(\varepsilon)-\log(\infty^j-\infty^0)\\
&=\log\Big(\frac{\infty^j-\infty^i}{\infty^j-\infty^0}\Big)-\log(\infty^0+\varepsilon-\infty^i)
+\frac{\log\big(\mu_0(\infty^{0,\varepsilon})\big)}{n_0+1}+\log\big(z_0(\infty^{0,\varepsilon})\big).
\end{align*}
Omitting the term $\log\big(z_0(\infty^{0,\varepsilon})\big)$ which diverges and letting $\varepsilon\longrightarrow0$, we arrive at
$$
p.v.\int_{\infty^0}^{\infty^j}\Omega_{\infty^0\infty^i}(P)=\log\bigg(\frac{\infty^j-\infty^i}{(\infty^j-\infty^0)(\infty^0-\infty^i)}\bigg)
+\frac{\log\big(\mu_0(\infty^0)\big)}{n_0+1}.
$$
If $\infty^j=\infty$, then by applying the above case to a point $R$ belonging to a small neighborhood of $\infty$, we find
$$
p.v. \int_{\infty^0}^{R}\Omega_{\infty^0\infty^i}(P)
=\log\bigg(\frac{z_{R}-\infty^i}{(z_R-\infty^0)(\infty^0-\infty^i)}\bigg)+\frac{\log\big(\mu_0(\infty^0)\big)}{n_0+1}.
$$
Moreover
\begin{align*}
\int_{R}^{\infty^j}\Omega_{\infty^0\infty^i}(P)&=\int_0^{1/{z_{R}}}\bigg(\frac{1}{\tau(1-\tau\infty^i)}-\frac{1}{\tau(1-\tau\infty^0)}\bigg)d\tau
=\log\Big(\frac{z_{R}-\infty^0}{z_{R}-\infty^i}\Big).
\end{align*}
Therefore
$$
p.v. \int_{\infty^0}^{\infty^j}\Omega_{\infty^0\infty^i}(P)=-\log\big(\infty^0-\infty^i)+\frac{\log\big(\mu_0(\infty^0)\big)}{n_0+1}.
$$
We now treat the case $i=j$. Using twice (\ref{claim g=0}), we get
\begin{align*}
\int_{\infty^{0,\varepsilon}}^{\infty^{i,\epsilon}}\Omega_{\infty^0\infty^i}(P)
&=\int_{\varepsilon+\infty^0}^{\epsilon+\infty^i}\frac{dz_P}{z_P-\infty^i}-\int_{\varepsilon+\infty^0}^{\epsilon+\infty^i}\frac{dz_P}{z_P-\infty^0}\\
&=\log(\epsilon)-\log(\infty^0+\varepsilon-\infty^i)+\log(\varepsilon)-\log(\epsilon+\infty^i-\infty^0)\\
&=\frac{\log\big(\mu_i(\infty^{i,\epsilon})\big)}{n_i+1}+\log\big(z_i(\infty^{i,\varepsilon})\big)-\log(\infty^0+\varepsilon-\infty^i)\\
&\quad +\frac{\log\big(\mu_0(\infty^{0,\varepsilon})\big)}{n_0+1}+\log\big(z_0(\infty^{0,\varepsilon}))-\log(\epsilon+\infty^i-\infty^0).
\end{align*}
Accordingly, we arrive at
$$
p.v. \int_{\infty^0}^{\infty^i}\Omega_{\infty^0\infty^i}(P)=\frac{\log\big(\mu_i(\infty^i)\big)}{n_i+1}+\frac{\log\big(\mu_0(\infty^0)\big)}{n_0+1}
-2\log(\infty^i-\infty^0)-\log(-1).
$$
\textbf{Second case: $\infty^i=\infty$}.  In this case, we have $\displaystyle\Omega_{\infty^0\infty^i}(P)=-\frac{dz_P}{z_P-\infty^0}$.\\
If $\infty^j\in \C$, then
\begin{align*}
\int_{\infty^{0,\varepsilon}}^{\infty^j}\Omega_{\infty^0\infty^i}(P)
&=-\int_{\varepsilon+\infty^0}^{\infty^j}\frac{dz_P}{z_P-\infty^0}
=\log(\varepsilon)-\log(\infty^j-\infty^0)\\
&=\frac{\log\big(\mu_0(\infty^{0,\varepsilon})\big)}{n_0+1}+\log\big(z_0(\infty^{0,\varepsilon})\big)-\log(\infty^j-\infty^0)
\end{align*}
and thus
$$
p.v.\int_{\infty^0}^{\infty^j}\Omega_{\infty^0\infty^i}(P)=\frac{\log(\mu_0(\infty^0)\big)}{n_0+1}-\log(\infty^j-\infty^0).
$$
If $\infty^j=\infty=\infty^i$, then we have
\begin{align*}
\int_{\infty^{0,\varepsilon}}^{\infty^{i,\epsilon}}\Omega_{\infty^0\infty^i}(P)&=-\int_{\varepsilon+\infty^0}^{1/\epsilon}\frac{dz_P}{z_P-\infty^0}
=\log(\epsilon)-\log(1-\epsilon\infty^0)+\log(\varepsilon)\\
&=\frac{\log\big(\mu_i(\infty^{i,\epsilon})\big)}{n_i+1}+\log\big(z_i(\infty^{i,\varepsilon})\big)-\log(1-\epsilon\infty^0) +\frac{\log\big(\mu_0(\infty^{0,\varepsilon})\big)}{n_0+1}+\log\big(z_0(\infty^{0,\varepsilon})\big).
\end{align*}
Therefore
$$
p.v. \int_{\infty^0}^{\infty^i}\Omega_{\infty^0\infty^i}(P)=\frac{\log\big(\mu_i(\infty^i)\big)}{n_i+1}+\frac{\log\big(\mu_0(\infty^0)\big)}{n_0+1}.
$$
\textbf{ii)} Let $i=0,\dots,m$ and  $\infty^{i,\varepsilon}$ be a point belonging to a small neighborhood of $\infty^i$  such that $z_i(\infty^{i,\varepsilon}):=\lambda(\infty^{i,\varepsilon})^{-\frac{1}{n_i+1}}=\varepsilon$. Here, as usual, $z_i(P)$ denotes the standard local parameter near the point $\infty^i$. We first claim the following asymptotic behavior:
\begin{equation}\label{theta-asymp}
\begin{split}
\Theta_{\Delta}\big(\mathcal{A}(\infty^{i,\varepsilon})-\mathcal{A}(\infty^i)\big)
&=  \varepsilon\sum_{k=1}^g\omega_k(\infty^i)\partial_{z_k}\Theta_{\Delta}\big(0\big)+
\frac{\varepsilon^2}{2}\sum_{k=1}^g\omega_k'(\infty^i)\partial_{z_k}\Theta_{\Delta}\big(0\big)\\
&\quad +\frac{\varepsilon^2}{2}\sum_{k,l=1}^g\omega_k(\infty^i)\omega_l(\infty^i)\partial_{z_k}\partial_{z_l}\Theta_{\Delta}\big(0\big)+
o(\varepsilon^2),
\end{split}
\end{equation}
where $\omega_k(\infty^i)$ and $\omega_k'(\infty^i)$ are the evaluations defined by (\ref{notation3}).\\
In particular, the following holds:
\begin{equation}\label{claim}
\log\Big(\Theta_{\Delta}\big(\mathcal{A}(\infty^{i,\varepsilon})-\mathcal{A}(\infty^i)\big)\Big)=\log(\varepsilon)+
\log\left(\sum_{k=1}^g\omega_k(\infty^i)\partial_{z_k}\Theta_{\Delta}\big(0\big)+O(\varepsilon)\right).
\end{equation}
Indeed, for all $k=1,\dots,g$, we observe that
\begin{align*}
\chi_k(\varepsilon)&:=\int_{\infty^i}^{\infty^{i,\varepsilon}}\omega_k(P)=\int_{0}^{\varepsilon}\Big(\omega_k(\infty^i)+\omega_k'(\infty^i)z_i(P)+\dots\Big)dz_i(P)\\
&=\varepsilon\omega_k(\infty^i)+\frac{\varepsilon^2}{2}\omega_k'(\infty^i)+o(\varepsilon^2),
\end{align*}
and we can  write
\begin{align*}
f(\varepsilon)&:=\Theta_{\Delta}\big(\mathcal{A}(\infty^{i,\varepsilon})-\mathcal{A}(\infty^i)\big)
=\Theta_{\Delta}\Big(\chi_1(\varepsilon),\ \dots\ , \chi_g(\varepsilon)\Big).
\end{align*}
Differentiating $f$    with respect to $\varepsilon$ at $\varepsilon=0$, we obtain
\begin{align*}
&f'(0)=\sum_{k=1}^g\omega_k(\infty^i)\partial_{z_k}\Theta_{\Delta}\big(0\big);\\
&f''(0)=\sum_{k=1}^g\omega_k'(\infty^i)\partial_{z_k}\Theta_{\Delta}\big(0\big)
+\sum_{k,l=1}^g\omega_k(\infty^i)\omega_l(\infty^i)\partial_{z_k}\partial_{z_l}\Theta_{\Delta}\big(0\big).
\end{align*}
On the other hand, since $\Theta_{\Delta}$ is odd, it follows that $f(0)=\Theta_{\Delta}(0)=0$. We thus get (\ref{theta-asymp}) expanding $f$ in the Taylor series near $\varepsilon=0$.\\
We now move on to prove the stated formulas for the functions $s_{ij}$.\\
\textbf{First case: $j\neq i$.} Take a point $\infty^{0,\varepsilon}$ belonging to a small neighborhood of $\infty^0$
with  $z_0(\infty^{0,\varepsilon}):=\lambda(\infty^{0,\varepsilon})^{-\frac{1}{n_0+1}}=\varepsilon$. \\
In view of  (\ref{third kind}) and (\ref{claim}), we obtain
\begin{align*}
\int_{\infty^{0,\varepsilon}}^{\infty^j}\Omega_{\infty^0\infty^i}(P)
&=\log\left(\frac{\Theta_{\Delta}\big(\mathcal{A}(\infty^j)-\mathcal{A}(\infty^i)\big)}{\Theta_{\Delta}\big(\mathcal{A}(\infty^j)-\mathcal{A}(\infty^0)\big)}\right)
-\log\left(\frac{\Theta_{\Delta}\big(\mathcal{A}(\infty^{0,\varepsilon})-\mathcal{A}(\infty^i)\big)}
{\Theta_{\Delta}\big(\mathcal{A}(\infty^{0,\varepsilon})-\mathcal{A}(\infty^0)\big)}\right)\\
&=\log\left(\frac{\Theta_{\Delta}\big(\mathcal{A}(\infty^j)-\mathcal{A}(\infty^i)\big)}{\Theta_{\Delta}\big(\mathcal{A}(\infty^j)-\mathcal{A}(\infty^0)\big)}\right)
-\log\Big(\Theta_{\Delta}\big(\mathcal{A}(\infty^{0,\varepsilon})-\mathcal{A}(\infty^i)\big)\Big)\\
&\quad +\log(\varepsilon)+\log\bigg(\sum_{k=1}^g\omega_k(\infty^0)\partial_{z_k}\Theta_{\Delta}\big(0\big)+O(\varepsilon)\bigg).
\end{align*}
Therefore by subtracting  the divergent term $\log(\varepsilon)=\log\big(z_0(\infty^{0,\varepsilon})\big)$
and letting $\varepsilon\longrightarrow0$, we deduce that
\begin{align*}
p.v.\int_{\infty^0}^{\infty^j}\Omega_{\infty^0\infty^i}(P)
&=\log\bigg(\frac{\Theta_{\Delta}\big(\mathcal{A}(\infty^j)-\mathcal{A}(\infty^i)\big)}
{\Theta_{\Delta}\big(\mathcal{A}(\infty^j)-\mathcal{A}(\infty^0)\big)\Theta_{\Delta}\big(\mathcal{A}(\infty^i)-\mathcal{A}(\infty^0)\big)}\bigg)\\
&\quad +\log\bigg(\sum_{k=1}^g\omega_k(\infty^0)\partial_{z_k}\Theta_{\Delta}\big(0\big)\bigg)-\log(-1).
\end{align*}
\textbf{Second case: $j=i$.} Take $\infty^{0,\varepsilon}$ as in the first case and  choose $\infty^{i,\epsilon}$  in neighborhood of $\infty^i$ such that $z_i(\infty^{i,\epsilon}):=\lambda(\infty^{i,\epsilon})^{-\frac{1}{n_i+1}}=\epsilon$. \\
Again by virtue of (\ref{third kind}) and (\ref{claim}), we have
\begin{align*}
\int_{\infty^{0,\varepsilon}}^{\infty^{i,\epsilon}}\Omega_{\infty^0\infty^i}(P)
&=\log\Big(\Theta_{\Delta}\big(\mathcal{A}(\infty^{i,\epsilon})-\mathcal{A}(\infty^i)\big)\Big)
-\log\Big(\Theta_{\Delta}\big(\mathcal{A}(\infty^{0,\varepsilon})-\mathcal{A}(\infty^i)\big)\Big)\\
&\quad -\log\Big(\Theta_{\Delta}\big(\mathcal{A}(\infty^{i,\epsilon})-\mathcal{A}(\infty^0)\big)\Big)+
\log\Big(\Theta_{\Delta}\big(\mathcal{A}(\infty^{0,\varepsilon})-\mathcal{A}(\infty^0)\big)\Big)\\
&=\log(\epsilon)+ \log\bigg(\sum_{k=1}^g\omega_k(\infty^i)\partial_{z_k}\Theta_{\Delta}\big(0\big)+O(\epsilon)\bigg)
-\log\Big(\Theta_{\Delta}\big(\mathcal{A}(\infty^{0,\varepsilon})-\mathcal{A}(\infty^i)\big)\Big)\\
&\quad -\log\Big(\Theta_{\Delta}\big(\mathcal{A}(\infty^{i,\epsilon})-\mathcal{A}(\infty^0)\big)\Big)
 +\log(\varepsilon)+\log\bigg(\sum_{k=1}^g\omega_k(\infty^0)\partial_{z_k}\Theta_{\Delta}\big(0\big)+O(\varepsilon)\bigg).
\end{align*}
Thus, by arguing as above we can see that
\begin{align*}
s_{ii}:=p.v.\int_{\infty^0}^{\infty^i}\Omega_{\infty^0\infty^i}(P)
&=\log\bigg(\sum_{k=1}^g\omega_k(\infty^i)\partial_{z_k}\Theta_{\Delta}\big(0\big)\bigg)
+\log\bigg(\sum_{k=1}^g\omega_k(\infty^0)\partial_{z_k}\Theta_{\Delta}\big(0\big)\bigg)\\
&\quad-2\log\Big(\Theta_{\Delta}\big(\mathcal{A}(\infty^i)-\mathcal{A}(\infty^0)\big)\Big)-\log(-1).
\end{align*}
The proof of the theorem is now complete.

\fd

\begin{Remark} Assume that the surface $C_g$ is of genus $g\geq 1$ and consider the holomorphic differentials $\omega_k$ and the multivalued differentials $\phi_{u^k}$ (\ref{type5}). Similarly to (\ref{s-ij Formula}),  we can also express the flat functions $s^i(\omega_k)$ and $s^i\big(\phi_{u^k})$ in terms of the Riemann Theta function and the Abel map. More precisely, for any $k=1,\dots,g$, we have
\begin{align*}
&s^i(\omega_k)=\int_{\infty^0}^{\infty^i}\omega_k=\mathcal{A}_k(\infty^i)-\mathcal{A}_k(\infty^0);\\
&s^i(\phi_{u^k})=-\oint_{a_k}\log\bigg(\frac{\Theta_{\Delta}\big(\mathcal{A}(\infty^i)-\mathcal{A}(Q)\big)}
{\Theta_{\Delta}\big(\mathcal{A}(\infty^i)-\mathcal{A}(Q)\big)}\bigg)d\lambda(Q).
\end{align*}
In addition, if we make use of (\ref{third kind}) as well as the quasi-periodicity properties of the Riemann theta function, then we observe that:
$$
s^i(\omega_k)=s^i(\phi_{\rho^k})=\rho^k(\phi_{s^i})\quad\quad \text{and}\quad\quad  s^i(\phi_{u^k})=u^k(\phi_{s^i}), \quad \quad \forall\ i,k.
$$
\end{Remark}

\medskip

Let us define the  $N\times N$-matrix $H_{AB}:=t^A(\phi_{t^B})$ for the $N$ operations $\{t^A\}$ in (\ref{FBP-W}) and (\ref{S-eta}) and the corresponding differentials $\{\phi_{t^A}\}$ given by (\ref{Primary-e}). The next result studies the symmetry property of this matrix.
\begin{Prop} Let $P_0$ be a marked point on $C_g$ such that $\lambda(P_0)=0$. Assume that all the cycles $\{a_k,b_k\}$ start at the point $P_0$. Then the following symmetry property holds:
\begin{equation}\label{Fubini-flat}
t^A(\phi_{t^B})=t^B(\phi_{t^A})+\log(-1)\sum_{i,j=1}^m(1-\delta_{ij})\delta_{s^i,t^A}\delta_{s^j,t^B},
\end{equation}
where $t^A$ and $t^B$ are from the list of flat functions (\ref{FBP-W})-(\ref{S-eta}).
\end{Prop}
\emph{Proof:} According to Theorem \ref{s-ij},  we know that
$$
s^j(\phi_{s^i})=s_{ij}=s_{ji}+\log(-1)=s^i(\phi_{s^j})+\log(-1),\quad \forall\ i\neq j.
$$
When $t^A$ or $t^B$ is not of type $s^i$, then it suffices to justify (\ref{Fubini-flat}) when the contours defining the functions $t^A$ and $t^B$ have an intersection point. Thus we need to discuss the following three cases
$$
\rho^k\big(\phi_{u^k}\big)=u^k\big(\phi_{\rho^k}\big),\quad \quad v^i\big(\phi_{s^i}\big)=s^i\big(\phi_{v^i}\big)\quad \quad\text{and} \quad \quad t^{i,\alpha}\big(\phi_{s^i}\big)=s^i\big(\phi_{t^{i,\alpha}}\big).
$$
Equality (\ref{Fubini1}) tells us that
$$
\rho^j\big(\phi_{u^k}\big)=u^k\big(\phi_{\rho^j}\big)-\lambda(P_0)\delta_{jk},\quad j,k=1,\dots,g,
$$
with $P_0$ being the intersection point of the cycles $a_k$ and $b_k$.  Thus the first equality holds true under the added assumption on the cycles $a_k$ and $b_k$. \\
Let us prove  the second equality. Denote by $\gamma_i$ a curve on the surface $C_g$ from $\infty^0$ to $\infty^i$ not passing through any of the ramification points $P_j$ or any of the $\infty^j$ except of extremities. \\
Let\\
\textrm{i)} $U$ be a small connected open subset of $C_g$ containing  the point $\infty^i$ such that
$$
\forall\ P,Q\in U, \quad W(P,Q)=\Big(\frac{1}{(z_i(P)-z_i(Q))^2}+\sum_{j,l\geq0}c_{j,l}z_i^{j}(P)z_i^{l}(Q)\Big)dz_i(P)dz_i(Q),
$$
where $z_i(P)$ is the standard local parameter near $\infty^i$ defined by $\lambda(P)=z_i^{-n_i-1}(P)$; \\
\textrm{ii)}  $\mathfrak{c}_i$ be a cycle inside  $U$ such that
$$
\underset{Q=\infty^i}{{\rm res}}\lambda(Q)W(P,Q)=\frac{1}{2{\rm{i}}\pi}\oint_{Q\in \mathfrak{c}_i}\lambda(Q)W(P,Q);
$$
\textrm{iii)} $P_{0,i}$ be a point in $U\cap \gamma_i$ chosen in such a way that there is no intersection point between the cycle $\mathfrak{c}_i$ and  the part of the curve $\gamma_i$ from $\infty^0$ to $P_{0,i}$. \\
With these assumptions, it suffices  to prove that
$$
p.v. \int_{P_{0,i}}^{\infty^i}\bigg(\underset{\infty^i}{{\rm res}}\lambda(Q)W(P,Q)\bigg)=\underset{\infty^i}{{\rm res}}\bigg(\lambda(Q)\int_{P_{0,i}}^{\infty^i}W(P,Q)\bigg).
$$
We have, using the notation introduced in (\ref{notation3}),
\begin{align*}
\forall\ P\in U,\quad \underset{\infty^i}{{\rm res}}\lambda(Q)W(P,Q)&=\frac{1}{n_i!}W^{(n_i)}(P,\infty^i)\\
&=\Big(\frac{n_i+1}{z_i(P)^{n_i+2}}+\sum_{j\geq0}c_{j,n_i}z_i^{j}(P)\Big)dz_i(P)\\
&=-d\lambda(P)+\Big(\sum_{j\geq0}c_{j,n_i}z_i^{j}(P)\Big)dz_i(P)
\end{align*}
and then
\begin{align*}
\int_{P_{0,i}}^{\infty^{i,\varepsilon}}\underset{\infty^i}{{\rm res}}\lambda(Q)W(P,Q)
&=\lambda(P_{0,i})-\lambda(\infty^{i,\varepsilon})+\int_{z_i(P_{0,i})}^{\varepsilon}\Big(\sum_{j\geq0}c_{j,n_i}z_i^{j}(P)\Big)dz_i(P)\\
&=\lambda(P_{0,i})-\lambda(\infty^{i,\varepsilon})+\sum_{j\geq0}\frac{c_{j,n_i}}{j+1}\big(\varepsilon^{j+1}-z_i^{j+1}(P_{0,i})\big),
\end{align*}
where $\varepsilon=z_i(\infty^{i,\varepsilon})=\Big(\lambda(\infty^{i,\varepsilon})\Big)^{-\frac{1}{n_i+1}}$.\\
Letting $\varepsilon\to0$ and omitting the divergent term, we arrive at
$$
p.v. \int_{P_{0,i}}^{\infty^i}\bigg(\underset{\infty^i}{{\rm res}}\lambda(Q)W(P,Q)\bigg)
=\lambda(P_{0,i})-\sum_{j\geq0}\frac{c_{j,n_i}}{j+1}z_i^{j+1}(P_{0,i}).
$$
On the other hand
\begin{align*}
\int_{P_{0,i}}^{\infty^i}W(P,Q)&=\left(\int_{z_i(P_{0,i})}^{0}\bigg(\frac{1}{(z_i(P)-z_i(Q))^2}+\sum_{j,l\geq0}c_{j,l}z_i^{j}(P)z_i^{l}(Q)\bigg)dz_i(P)\right)dz_i(Q)\\
&=\bigg(\frac{1}{z_i(Q)}+\frac{1}{z_i(P_{0,i})-z_i(Q)}-\sum_{j,l\geq0}\frac{c_{j,l}}{j+1}z_{i}^{j+1}(P_{0,i})z_i^{l}(Q)\bigg)dz_i(Q).
\end{align*}
Therefore
\begin{align*}
&\underset{\infty^i}{{\rm res}}\bigg(\lambda(Q)\int_{P_{0,i}}^{\infty^i}W(P,Q)\bigg)\\
&=\underset{0}{\rm res}\left(z_i^{-n_i-1}(Q)\bigg(\frac{1}{z_i(Q)}+\frac{1}{z_i(P_{0,i})-z_i(Q)}-\sum_{j,l\geq0}\frac{c_{j,l}}{j+1}z_i^{j+1}(P_{0,i})z_i^{l}(Q)\bigg)dz_i(Q)\right)\\
&=\frac{1}{z_i^{n_i+1}(P_{0,i})}-\sum_{j\geq0}\frac{c_{j,n_i}}{j+1}z_i^{j+1}(P_{0,i})\\
&=\lambda(P_{0,i})-\sum_{j\geq0}\frac{c_{j,n_i}}{j+1}z_{i}^{j+1}(P_{0,i}).
\end{align*}
Thus we have established that   $v^i\big(\phi_{s^i}\big)=s^i\big(\phi_{v^i}\big)$. The symmetry  $t^{{i,\alpha}}\big(\phi_{s^i}\big)=s^i\big(\phi_{t^{i,\alpha}}\big)$ can be shown similarly.

\fd

\begin{Thm}\label{FC-thm-eta} Let $\big\{\big(C_g,\lambda\big),\{a_k,b_k\}\big\}$ be a point of (a fixed covering of) the simple Hurwitz space
$\widehat{\mathcal{H}}_{g, L}(n_0,\dots,n_m)$ and  $P_0$ be a marked point on $C_g$ such that $\lambda(P_0)=0$. Assume that all the cycles $\{a_k,b_k\}$ start at the point $P_0$.\\
Then the  $N$ flat functions of the set $\mathcal{S}(\omega)$ defined by (\ref{FBP-W})-(\ref{S-eta})  give  a system  of flat coordinates of the Dubrovin  metric $\eta(\omega)$.\\
Moreover, in these coordinates, the nonzero entries of the constant  matrix of the contravariant metric $\eta^*(\omega)$ are as follows:
\begin{equation}\label{Entries eta}
\begin{split}
&\eta^*(\omega)(dt^{i,\alpha}(\omega),dt^{j,\beta}(\omega))=\delta_{ij}\delta_{\alpha+\beta,n_j+1};\\
&\eta^*(\omega)(dv^{i}(\omega),ds^{j}(\omega))=\delta_{ij};\\
&\eta^*(\omega)(d\rho^{i}(\omega),du^{j}(\omega))=\delta_{ij}.
\end{split}
\end{equation}
\end{Thm}
\noindent\emph{Proof:} We already know that the functions of the set $\mathcal{S}(\omega)$ are all flat with respect to the metric $\eta(\omega)$.
Let us calculate the matrix of $\eta^*(\omega)$ in these coordinates. The proof  relies  on the following tools:\\
- the characteristic properties of the differentials (\ref{Primary-e}) that were described  in Proposition \ref{Primary-e-prop};\\
- the following particular versions of formulas (\ref{dual metric-R2})-(\ref{dual metric-R4}) for $\eta^*(\omega)=\left(\mathbf{ds^2_{0,1}}(\omega)\right)^*$:
\begin{equation}\label{dual metric-e}
\eta^*(\omega)\left(dt^A(\omega),dt^B(\omega)\right)=e.t^A(\phi_{t^B})=\sum_{k=1}^N\underset{P_k}{{\rm res}}\frac{\phi_{t^A}(P)\phi_{t^B}(P)}{d\lambda(P)},
\end{equation}
where $e$ is the vector field $e:=\sum_{j=1}^N\partial_{\lambda_j}$.\\
Note that  the symmetry property (\ref{Fubini-flat}) implies that $e.t^A(\phi_{t^B})=e.t^B(\phi_{t^A})$.\\
From their analytic  properties, the differentials $\phi_{t^{i,\alpha}}$, $\phi_{v^i}$, $\phi_{s^i}$ and  $\phi_{\rho^k}=\omega_k$ are all Abelian differentials on the surface $C_g$ and they are all holomorphic near any simple ramification point $P_k$. Therefore, by analysing the poles of the differential $\frac{\phi_{t^A}(P)\phi_{t^B}(P)}{d\lambda(P)}$ and applying the residue theorem, it follows that the right  hand side of (\ref{dual metric-e}) is straightforwardly  computed provided that $t^A$ and $t^B$ are among the four types of operations $\big\{t^{i,\alpha}, v^i, s^i, \rho^k\big\}$.\\
For example if we consider the case $t^A=t^{i,\alpha}$ and $t^B=t^{i,\beta}$, with $i=0,\dots, m$ and $\alpha,\beta=1,\dots,n_i$, then the residue theorem and the behavior
$$
\phi_{t^{i,\alpha}}(P)\underset{P\sim \infty^i}=\Big(\sqrt{n_i+1}z_i(P)^{-\alpha-1}+ \text{reg}\Big)dz_i(P),\quad \quad \lambda(P)=z_i(P)^{-n_i-1},
$$
imply that
\begin{align*}
&\eta^*(\omega)\left(dt^A(\omega),dt^B(\omega)\right)
=e.t^{i,\alpha}(\phi_{t^{i,\beta}})=-\underset{\infty^i}{{\rm res}}\frac{\phi_{t^{i,\alpha}}(P)\phi_{t^{i,\beta}}(P)}{d\lambda(P)}\\
&=\frac{1}{n_i+1}\underset{0}{{\rm res}}\bigg(z_i(P)^{n_i+2}\Big(\sqrt{n_i+1}z_i(P)^{-\alpha-1}+ \text{reg}\Big)\Big(\sqrt{n_i+1}z_i(P)^{-\beta-1}+ \text{reg}\Big)\bigg)dz_i(P)\\
&=\delta_{\alpha+\beta,n_i+1}.
\end{align*}
However, the use of (\ref{dual metric-e}) in the case where $\phi_{t^A}=\phi_{u^k}$ is the multivalued differential having jumps along the cycle $b_k$ needs some details.\\
For $k=1,\dots,g$, let us consider the differential
\begin{equation}\label{Phi-k}
\Phi_k(P):=\frac{\phi(P)}{d\lambda(P)}\phi_{u^k}(P),
\end{equation}
where $\phi$ is  one of the differentials listed in (\ref{Primary-e}). Then $\Phi_k$ is a multivalued differential and by (\ref{type5}), its jumps are as follows:
\begin{align*}
\begin{array}{llll}
\Phi_k(P+a_j)-\Phi_k(P)=0, \quad & \hbox{if}\quad j=1,\dots,g;\\
\\
\Phi_k(P+b_k)-\Phi_k(P)=-2{\rm{i}}\pi\phi(P), \quad & \hbox{if}\quad \phi\neq \phi_{u^k};\\
\\
\Phi_k(P+b_k)-\Phi_k(P)=-4{\rm{i}}\pi\phi_{u^k}(P)-4\pi^2d\lambda(P), \quad & \hbox{if}\quad \phi=\phi_{u^k};\\
\\
\Phi_k(P+b_j)-\Phi_k(P)=-2{\rm{i}}\pi\phi_{u^k}(P), \quad & \hbox{if}\quad \phi=\phi_{u^j}\quad \text{and}\quad  j\neq k;\\
\\
\Phi_k(P+b_j)-\Phi_k(P)=0, \quad & \hbox{if}\quad \text{$\phi$ is Abelian and $j\neq k$}.
\end{array}
\end{align*}
In addition, since the multivalued differentials $\phi_{u^1},\dots,\phi_{u^g}$ are all single valued and holomorphic inside the fundamental polygon $F_g$ associated with  a compact surface $C_g$, it follows that the differential $\Phi_k$ is  single valued and meromorphic inside $F_g$ and has simple poles at the ramifications points $P_1,\dots,P_N$.\\
Furthermore, by using the distinctive analytic properties of the differentials (\ref{Primary-e}) (see Proposition \ref{Primary-e-prop}) and keeping in mind the presence of the factor $d\lambda(P)$, we deduce that $\Phi_k$ has no other poles inside the fundamental polygon $F_g$.\\
Thus, by applying the residue theorem and recalling that $\partial F_g=\sum_{j=1}^g \big(a_j+b_j+a_j^{-1}+b_j^{-1}\big)$ and taking into account the mentioned jumps of $\Phi_k$, we obtain
\begin{align*}
e.u^k(\phi)&=\sum_{s=1}^N\underset{P_s}{{\rm res}}\Phi_k(P)=\frac{1}{2{\rm{i}}\pi}\oint_{\partial{F_g}}\Phi_k(P)\\
&=\frac{1}{2{\rm{i}}\pi}\sum_{j=1}^g\bigg(\oint_{a_j}\Phi_k(P)+\oint_{a_j^{-1}}\Phi_k(Q)\bigg)
+\frac{1}{2{\rm{i}}\pi}\sum_{j=1}^g\bigg(\oint_{b_j}\Phi_k(P)+\oint_{b_j^{-1}}\Phi_k(Q)\bigg)\\
&=\frac{1}{2{\rm{i}}\pi}\sum_{j=1}^g\bigg(\oint_{a_j}\Phi_k(P)-\oint_{a_j}\Phi_k(P+b_j)\bigg)
+\frac{1}{2{\rm{i}}\pi}\sum_{j=1}^g\bigg(\oint_{b_j}\Phi_k(P)-\oint_{b_j}\underbrace{\Phi_k(P+a_j^{-1})}_{=\Phi_k(P)}\bigg)\\
&=\bigg(\sum_{j=1,j\neq k}^g\delta_{\phi,\phi_{u^j}}\oint_{a_j}\phi_{u^k}\bigg) +\big(1-\delta_{\phi,\phi_{u^k}}\big)\oint_{a_k}\phi +\delta_{\phi,\phi_{u^k}}\oint_{a_k}\Big(2\phi_{u^k}(P)-2{\rm{i}}\pi d\lambda(P)\Big)\\
&= \big(1-\delta_{\phi,\phi_{u^k}}\big)\delta_{\phi,\phi_{\rho^k}}=\delta_{\phi,\phi_{\rho^k}},
\end{align*}
where $\delta_{\phi,\phi_{t^B}}$ denotes the Kronecker delta function.

\fd

\begin{Remark} Observe that if $g\geq 1$, then by using  explicit formulas (\ref{s-ij Formula}) for the functions $s_{ij}$, we see  that the action of the vector field $e:=\sum_{j=1}^N\partial_{\lambda_j}$ on $s_{ij}$ is completely determined by knowing  its action on the following functions:
$$
\mathbb{B}_{kl}=2{\rm{i}}\pi\rho^l(\omega_k),\quad \omega_k(\infty^i)=\frac{t^{i,1}(\omega_k)}{\sqrt{n_i+1}},\quad \mathcal{A}_k(\infty^i)-\mathcal{A}_k(\infty^0)=\int_{\infty^0}^{\infty^i}\omega_k=s^i(\omega_k).
$$
This provides an alternative way to show that  $e.s_{ij}=0$.
\end{Remark}

\medskip

Using (\ref{Entries eta}), we remark  that all the rows  of the  constant matrix of the Darboux-Egoroff metric $\eta^*(\omega)$ contain exactly one nonzero
entry which is 1.This permits us to introduce the following natural duality relation  between  flat coordinates of $\eta(\omega)$:
\begin{Def}\label{duality def} For each $t^A\in \mathcal{S}(\omega):=\Big\{t^{i,\alpha}(\omega), v^i(\omega), s^i(\omega), \rho^k(\omega), u^k(\omega)\Big\}$, let us  denote by $t^{A^{\prime}}$ the unique flat function belonging to the set $\mathcal{S}(\omega)$ such that
\begin{equation}\label{duality}
\eta^*(\omega)\big(dt^A(\omega), dt^{A^{\prime}}(\omega)\big)=1.
\end{equation}
The flat coordinates  $t^A$ and $t^{A^{\prime}}$ are called dual to each other with respect to the metric $\eta(\omega)$. \\
\end{Def}
Let us mention the following remarks:
\begin{itemize}
\item From (\ref{Entries eta}) and (\ref{dual metric-e}), it follows that the introduced $\eta(\omega)$-duality does not depend on the choice of the differential $\omega$ that we work with.
\item For any $A$, we have  $t^{(A^{\prime})^{\prime}}(\omega)=t^{A}(\omega)$.
\item Due to (\ref{Entries eta}), the five families of operations $t^{i,\alpha}, v^i,s^i,\rho^k$ and $u^k$ are connected by:
\begin{equation}\label{duality-picture}
\begin{split}
&t^{({i,\alpha})^{\prime}}(\omega)=t^{i,n_i+1-\alpha}(\omega),\quad \quad \quad \quad v^{{i}^{\prime}}(\omega)=s^i(\omega),\quad \quad \ \quad \rho^{k^{\prime}}(\omega)=u^k(\omega);\\
&\phi_{t^{({i,\alpha})^{\prime}}}(P)=\phi_{t^{i,n_i+1-\alpha}}(P),\quad \quad \quad \phi_{v^{i^{\prime}}}(P)=\phi_{s^i}(P),\quad \quad \phi_{\rho^{k^{\prime}}}(P)=\phi_{u^k}(P).
\end{split}
\end{equation}
\end{itemize}

\begin{Prop} The derivatives of the canonical coordinates $\{\lambda_j\}_{1\leq j\leq N}$ with respect to $\eta(\omega)$-flat coordinates $t^A(\omega)$ belonging to the system (\ref{FBP-W}) are as follows
\begin{align}\label{derivatives lambda-j}
\frac{\partial \lambda_j}{\partial t^A(\omega)}=\frac{\phi_{t^{A^{\prime}}}(P_j)}{\omega(P_j)},
\end{align}
where  $\phi_{t^{A^{\prime}}}$ is the  differential  (\ref{phi-t}) associated with the $\eta(\omega)$-flat function $t^{A^{\prime}}(\omega)$.
\end{Prop}
\emph{Proof:}  In \cite{Dubrovin2D}, Dubrovin  has proven (\ref{derivatives lambda-j}) by using a thermodynamical type identity. We suggest a different proof based  on Rauch's formulas, (\ref{Entries eta}) as well as some basic facts from linear algebra.\\
 Let us denote by $(\Lambda_{AB})$ the constant matrix of $\eta(\omega)$ in flat coordinates $\{t^A(\omega)\}$ and $(\eta_{jj})$ the diagonal matrix of $\eta(\omega)$ in the coordinates $\{\lambda_j\}$.\\
On  one hand, Rauch's formula (\ref{partial-t-W}) yields that
$$
dt^A(\omega)=\sum_j\partial_{\lambda_j}t^A(\omega)d\lambda_j=\frac{1}{2}\sum_j\omega(P_j)\phi_{t^A}(P_j)d\lambda_j.
$$
On the other hand, using twice the fact that $\eta(\omega)$ induces a linear  isomorphism from the tangent space onto the cotangent space, namely regarding $(\Lambda_{AB})$ and $(\eta_{jj})$ as the change of basis matrices,  we can write
\begin{align*}
dt^A(\omega)&=\sum_B\Lambda^{AB}\eta(\omega)\Big(\partial_{t^B(\omega)},\cdot\Big)\\
&=\sum_B\Lambda^{AB}\eta(\omega)\Big(\sum_j\big(\partial_{t^B(\omega)}\lambda_j\big)\partial_{\lambda_j},\cdot\Big)\\
&=\sum_j\sum_B\Lambda^{AB}\big(\partial_{t^B(\omega)}\lambda_j\big)\eta(\omega)(\partial_{\lambda_j},\cdot)\\
&=\sum_j\bigg(\eta_{jj}\sum_B\Lambda^{AB}\big(\partial_{t^B(\omega)}\lambda_j\big)\bigg)d\lambda_j.
\end{align*}
Thus we must have
$$
\sum_B\Lambda^{AB}\big(\partial_{t^B(\omega)}\lambda_j\big)=\frac{1}{2}\eta^{jj}\omega(P_j)\phi_{t^A}(P_j)=\frac{\phi_{t^A}(P_j)}{\omega(P_j)}.
$$
Since $\Lambda^{AB}=\eta^*(\omega)\big(dt^A(\omega),dt^B(\omega)\big)=\delta_{t^B,t^{A^{\prime}}}$, we deduce that
$$
\sum_B\Lambda^{AB}\big(\partial_{t^B(\omega)}\lambda_j\big)=\partial_{t^{A^{\prime}}(\omega)}\lambda_j.
$$
This establishes the desired formula (\ref{derivatives lambda-j}).

\fd

\begin{Cor} The following assertions hold true:
\begin{align}
\begin{split}\label{flat-canonical relation}
&\partial_{t^A(\omega)}=\sum_j\frac{\phi_{t^{A^{\prime}}}(P_j)}{\omega(P_j)}\partial_{\lambda_j};
\end{split}\\
\begin{split}\label{Entries-eta2}
&\eta(\omega)\left(\partial_{t^A(\omega)},\partial_{t^B(\omega)}\right)=\frac{1}{2}\sum_j\phi_{t^{A^{\prime}}}(P_j)\phi_{t^{B^{\prime}}}(P_j)
=\eta^*(\omega)\left(dt^{A^{\prime}}(\omega),t^{B^{\prime}}(\omega)\right).
\end{split}
\end{align}
In particular, the flat metric $\eta(\omega)$ can be written as
\begin{equation}\label{eta-tA}
\eta(\omega)=\textstyle \sum_Adt^A(\omega)\otimes dt^{A^{\prime}}(\omega).
\end{equation}

\end{Cor}
\emph{Proof:}  The relation (\ref{flat-canonical relation}) is a direct consequence of (\ref{derivatives lambda-j}) combined with the chain rule formula. The first equality in (\ref{Entries-eta2}) follows from (\ref{flat-canonical relation}) and (\ref{eta-def}) and the second one from  (\ref{dual metric-R1}).
Lastly, in view of (\ref{Entries-eta2}) and (\ref{Entries eta}), we obtain the stated expression for the metric $\eta(\omega)$.

\fd

\begin{Remark}\end{Remark}
In the case where $\omega=\phi$ is a  primary differential, Dubrovin's method to calculate the constant  matrix of the flat metric $\eta(\phi)$ in flat coordinates uses his bilinear pairing (see formula (5.60) in \cite{Dubrovin2D}).  In \cite{Vasilisa}, p. 562-564, Shramchenko employed  the same method and gave a detailed computation of constant matrices of the Daroubx-Egoroff metrics in the real double setting of Frobenius manifolds.

\medskip

We close this subsection  by mentioning  the following duality-exchange relation:
\begin{Prop} Let $\omega$ be a given differential satisfying (\ref{Diff model}) and (\ref{Assymptions}). Let $t^A(\omega), t^B(\omega)$ be two $\eta(\omega)$-flat coordinates from the set (\ref{FBP-W})-(\ref{S-eta}) and $\phi_{t^A}$, $\phi_{t^B}$ be the corresponding differentials belonging to the list (\ref{Primary-e}).\\
Then the following holds:
$$
\frac{\partial\phi_{t^B}(P)}{\partial{t^A(\omega)}}=\frac{\partial\phi_{t^{A^{\prime}}}(P)}{\partial{t^{B^{\prime}}(\omega)}}.
$$
\end{Prop}
\emph{Proof:} We have
\begin{align*}
\partial_{t^A(\omega)}\phi_{t^B}(P)
&=\sum_j\frac{\phi_{t^{A^{\prime}}}(P_j)}{\omega(P_j)}\partial_{\lambda_j}\phi_{t^B}(P)
=\frac{1}{2}\sum_j\frac{\phi_{t^{A^{\prime}}}(P_j)}{\omega(P_j)}\phi_{t^B}(P_j)W(P,P_j)\\
&=\sum_j\frac{\phi_{t^B}(P_j)}{\omega(P_j)}\partial_{\lambda_j}\phi_{t^{A^{\prime}}}(P)
=\partial_{t^{B^{\prime}}(\omega)}\phi_{t^{A^{\prime}}}(P),
\end{align*}
where the first and  fourth equalities follow from (\ref{flat-canonical relation}), while  the second and third ones are a direct consequence of  Rauch's formulas
(\ref{Rauch-2}).

\fd


\subsection{Flat coordinates of  intersection forms}
According to Dubrovin \cite{Dubrovin2D}, given a semi-simple Frobenius manifold $(\mathcal{M}, \prs{\cdot}{\cdot})$,  there is a second flat metric $\mathfrak{g}$, called the intersection form, defined on an open subset of $\mathcal{M}$. The metric $\mathfrak{g}$ and the initial metric $\prs{\cdot}{\cdot}$ give rise to the so-called flat pencil.\\
In the case of a semi-simple Hurwitz-Frobenius manifolds, the (covariant metric induced by the) intersection form is defined by:
\begin{equation}\label{Inter-form}
\mathbf{g}(\omega):=\mathbf{ds^2_{1,0}}(\omega)=\sum_{j=1}^N\frac{\omega(P_j)^2}{2\lambda_j}(d\lambda_j)^2,
\end{equation}
where, as above, $\omega$ denotes a  differential satisfying (\ref{Diff model}) and (\ref{Assymptions}).\\
In this subsection,  we focus on flat coordinates of the intersection form (\ref{Inter-form}) considered as a flat metric on a connected  open subset  $\widehat{\mathcal{H}}_{g,L}(\mathbf{n}; \mathbf{m})$ of the double Hurwitz space.   Recently, Romano  used  a Dubrovin type bilinear pairing  and gave a set of flat coordinates of the metric $\mathbf{g}(\omega)$ \cite{Romano}.  However, he did not  calculate the corresponding constant matrix.\\
 In our case, by employing the language of the symmetric bidifferential $W(P,Q)$, we show that the  approach used in Section  3.3  can be adapted to  study of the metric $\mathbf{g}(\omega)$.

\begin{Thm}\label{Flat coord-g}  Let $\big\{\big(C_g,\lambda\big),\{a_k,b_k\}\big\}$ be a point of the double Hurwitz space
$\widehat{\mathcal{H}}_{g, L}(\mathbf{n}; \mathbf{m})$ and  Let $P_0$ be a marked point on $C_g$ such that $\lambda(P_0)=1$. Assume that all the cycles $\{a_k,b_k\}$ start at the point $P_0$ and the differential $\omega$ has zero $a$-periods. \\
Then the following $N=2g+m+n$ functions provide a system  of flat coordinates of the metric $\mathbf{g}(\omega)$:
\begin{equation}\label{flat basis-g}
\begin{array}{llll}
s^i(\omega):=\displaystyle p.v.\int_{\infty^0}^{\infty^i}\omega(P),&\quad i=1,\dots,m;\\
\\
y^j(\omega):=\displaystyle p.v.\int_{\mathbf{z}^0}^{\mathbf{z}^j}\omega(P),&\quad j=1,\dots,n;\\
\\
\rho^k(\omega):=\displaystyle \frac{1}{2{\rm{i}}\pi}\oint_{b_k}\omega(P),&\quad k=1,\dots,g;\\
\\
q^k(\omega):=\displaystyle \oint_{a_k}\log\big(\lambda(P)\big)\omega(P),&\quad k=1,\dots,g.
\end{array}
\end{equation}
In addition, in these coordinates, the nonzero entries of the contravariant  metric $\mathbf{g}^*(\omega)$ are
\begin{equation}\label{Entrie-g}
\begin{split}
&\mathbf{g}^*(\omega)\big(ds^i(\omega),ds^j(\omega)\big)=\frac{1}{n_0+1}+\delta_{ij}\frac{1}{n_i+1};\\
&\mathbf{g}^*(\omega)\big(dy^i(\omega),dy^j(\omega)\big)=-\frac{1}{m_0+1}-\delta_{ij}\frac{1}{m_i+1};\\
&\mathbf{g}^*(\omega)\big(d\rho^i(\omega),dq^j(\omega)\big)=\delta_{ij}.
\end{split}
\end{equation}
Here the positive integers $n_i+1$ and $m_j+1$ denote respectively the multiplicities  of the prescribed  ramifications points $\infty^i$ and $\mathbf{z}^j$, with $i=0,\dots,m$ and $j=0,\dots,n$.
\end{Thm}

\bigskip

\noindent Before we start to prove this theorem, let us mention some remarks and preliminary results.\\

\noindent $\bullet$  Similarly to the function $s^i(\omega)$, the principal value in the formula of $y^j(\omega)$ is defined  by subtraction of the divergent part of the integral as a function of the local parameters $\mathfrak{z}_0(P):=\lambda(P)^{\frac{1}{m_0+1}}$ and  $\mathfrak{z}_i(P):=\lambda(P)^{\frac{1}{m_i+1}}$ near the points $\mathbf{z}^0$ and $\mathbf{z}^i$, respectively.

\medskip

\noindent $\bullet$ Let us denote by $\Omega_{\mathbf{z}^0\mathbf{z}^i}(P)$ the normalized  Abelian differential of the third kind having simple poles at  the points $\mathbf{z}^0$ and $\mathbf{z}^i$, i.e.
$$
\underset{\mathbf{z}^i}{{\rm res}}\ \Omega_{\mathbf{z}^0\mathbf{z}^i}=1,\quad \underset{\mathbf{z}^0}{{\rm res}}\ \Omega_{\mathbf{z}^0\mathbf{z}^i}=-1,\quad \oint_{a_k}\Omega_{\mathbf{z}^0\mathbf{z}^i}=0, \quad \forall\ k=1,\dots,g.
$$
We emphasize that formulas analogous to (\ref{third kind-g0}) and (\ref{third kind})  and a direct adaptation of Theorem \ref{s-ij} allow us to calculate the functions
$$
y_{ij}:= p.v.\int_{\mathbf{z}^0}^{\mathbf{z}^j}\Omega_{\mathbf{z}^0\mathbf{z}^i}(P),\quad \quad \forall\ i,j=1,\dots,n.
$$
More precisely, we obtain the following statements:
\begin{description}
\item[1)] Assume that $g=0$ and $\mathbf{z}^0\in \C$. For $i=0,\dots,n$, consider the meromorphic function $\nu_i$ defined by
\begin{equation}\label{nu-i}
\nu_i(P)=
\left\{
\begin{array}{ll}
(z_P-\mathbf{z}^i)^{-m_i-1}\lambda(P),& \quad \hbox{if}\quad \mathbf{z}^i\in\C,\\
\\
z_P^{m_i+1}\lambda(P),&\quad \hbox{if}\quad \mathbf{z}^i=\infty.
\end{array}
\right.
\end{equation}
\begin{itemize}
\item[\textbf{i)}] If $i\neq j$ and $\mathbf{z}^i,\mathbf{z}^j\in \C$, then
\begin{equation*}
y_{ij}=\log\left(\frac{\mathbf{z}^j-\mathbf{z}^i}{\big(\mathbf{z}^j-\mathbf{z}^0\big)\big(\mathbf{z}^0-\mathbf{z}^i\big)}\right)
-\frac{\log\big(\nu_0(\mathbf{z}^0)\big)}{m_0+1}.
\end{equation*}
When $\mathbf{z}^r=\infty$, with $r=i$ or $j$, we deduce the desired formula from the previous one by letting $\mathbf{z}^r\to\infty$.
\item[\textbf{ii)}] If $i=j$ we have
\begin{equation*}
\begin{array}{ll}
y_{ii}=\displaystyle-\frac{\log\big(\nu_i(\mathbf{z}^i)\big)}{m_i+1}-\frac{\log\big(\nu_0(\mathbf{z}^0)\big)}{m_0+1}
-2\log\big(\mathbf{z}^i-\mathbf{z}^0\big)-\log(-1),\quad &\hbox{if}\quad  \mathbf{z}^i\in \C;\\
\\
y_{ii}=\displaystyle-\frac{\log\big(\nu_i(\mathbf{z}^i)\big)}{m_i+1}-\frac{\log\big(\nu_0(\mathbf{z}^0)\big)}{m_0+1},\quad &\hbox{if}\quad \mathbf{z}^i=\infty.
\end{array}
\end{equation*}
\end{itemize}
\item[2)] Assume that $g\geq1$.
\begin{itemize}
\item[\textbf{i)}] When $i\neq j$, we have
\begin{equation*}
y_{ij}=\log\left(\frac{\Theta_{\Delta}\big(\mathcal{A}(\mathbf{z}^j)-\mathcal{A}(\mathbf{z}^i)\big)}
{\Theta_{\Delta}\big(\mathcal{A}(\mathbf{z}^j)-\mathcal{A}(\mathbf{z}^0)\big)\Theta_{\Delta}\big(\mathcal{A}(\mathbf{z}^i)-\mathcal{A}(\mathbf{z}^0)\big)}\right)
+\log\left(\sum_{k=1}^g\omega_k(\mathbf{z}^0)\partial_{z_k}\Theta_{\Delta}\big(0\big)\right)-\log(-1).
\end{equation*}
\item[\textbf{ii)}] If $i=j$ we have
\begin{equation*}
y_{ii}=\log\left(\sum_{k=1}^g\omega_k(\mathbf{z}^i)\partial_{z_k}\Theta_{\Delta}\big(0\big)\right)
+\log\left(\sum_{k=1}^g\omega_k(\mathbf{z}^0)\partial_{z_k}\Theta_{\Delta}\big(0\big)\right)
-2\log\left(\Theta_{\Delta}\big(\mathcal{A}(\mathbf{z}^i)-\mathcal{A}(\mathbf{z}^0)\big)\right)-\log(-1).
\end{equation*}
\end{itemize}
\end{description}
\medskip

\noindent $\bullet$ Consider the differentials connected with  the four families of flat functions $s^i,y^i,\rho^k,q^k$
(as above, the link here is in the sense of (\ref{FF-model}) -(\ref{phi-t})):
\begin{equation}\label{Primary-E}
\begin{split}
&\psi_{s^i}(P)=\int_{\infty^0}^{\infty^i}W(P,Q)=\Omega_{\infty^0\infty^i}(P);\\
&\psi_{y^i}(P)=\int_{\mathbf{z}^0}^{\mathbf{z}^i}W(P,Q)=\Omega_{\mathbf{z}^0\mathbf{z}^i}(P);\\
&\psi_{\rho^k}(P)=\frac{1}{2{\rm{i}}\pi}\oint_{b_k}W(P,Q)=\omega_k(P);\\
&\psi_{q^k}(P)=\oint_{a_k}\log\big(\lambda(Q)\big)W(P,Q).
\end{split}
\end{equation}
The analytic properties of the Abelian differentials $\psi_{s^i}, \psi_{y^i}, \psi_{\rho^k}$ have previously been specified.   Let us deal with the differential $\psi_{q^k}$. Since the differential $W(P,\cdot)$ has vanishing $a$-periods (\ref{W-periods}),  the definition of $\psi_{q^k}$ does not depend on the choice of the branch of the complex logarithm  and its $a$-periods are all zero.  Now, we claim that $\psi_{q^k}(P)$ is a multivalued  differential with jumps along the cycle $b_k$ and
\begin{equation}\label{jump-E}
\psi_{q^k}(P+b_k)-\psi_{q^k}(P)=-2{\rm{i}}\pi \frac{d\lambda(P)}{\lambda(P)}.
\end{equation}
Indeed, using (\ref{W-def}) and integrating by parts, we can write
$$
\psi_{q^k}(P)=-\oint_{a_k}d_P\log\Big(\Theta_{\Delta}\big(\mathcal{A}(P)-\mathcal{A}(Q)\big)\Big)\frac{d\lambda(Q)}{\lambda(Q)}=dF_{q^k}(P),
$$
where $F_{q^k}$ is   given by
$$
F_{q^k}(P):=\int^P\psi_{q^k}=-\oint_{a_k}\log\Big(\Theta_{\Delta}\big(\mathcal{A}(P)-\mathcal{A}(Q)\big)\Big)\frac{d\lambda(Q)}{\lambda(Q)}.
$$
In view of relation (\ref{theta-quasi}), we obtain
\begin{align*}
F_{q^k}(P+a_j)-F_{q^k}(P)&=\oint_{a_j}\psi_{q^k}=0;\\
F_{q^k}(P+b_j)-F_{q^k}(P)&=\oint_{b_j}\psi_{q^k}=-\oint_{a_k}\log\bigg(\frac{\Theta_{\Delta}\big(\mathcal{A}(P+b_j)-\mathcal{A}(Q)\big)}
{\Theta_{\Delta}\big(\mathcal{A}(P)-\mathcal{A}(Q)\big)}\bigg)\frac{d\lambda(Q)}{\lambda(Q)}\\
&=-\oint_{a_k}\bigg(-{\rm{i}}\pi \prs{\mathbb{B}e_j}{e_j}-2{\rm{i}}\pi\prs{\beta}{e_j}-2{\rm{i}}\pi\int_Q^P\omega_j\bigg)\frac{d\lambda(Q)}{\lambda(Q)}\\
&=2{\rm{i}}\pi \oint_{a_k}\Big(\int_Q^P\omega_j\Big)\frac{d\lambda(Q)}{\lambda(Q)}\\
&=2{\rm{i}}\pi \oint_{a_k}\bigg(d_Q\Big(\log\big(\lambda(Q)\big)\int_Q^P\omega_j\Big)+\log\big(\lambda(Q)\big)\omega_j(Q)\bigg)\\
&=2{\rm{i}}\pi\bigg(\log\big(\lambda(Q)\big)\int_Q^P\omega_j\bigg)\bigg|_{Q=Q_0}^{Q=Q_0+a_k}+2{\rm{i}}\pi \oint_{a_k}\log\big(\lambda(Q)\big)\omega_j(Q)\\
&=-2{\rm{i}}\pi\delta_{jk}\log\big(\lambda(Q_0)\big)+2{\rm{i}}\pi q^k(\omega_j),
\end{align*}
where $Q_0$ is a starting point of the cycle $a_k$. Here, we are going to study general properties of the differential $\psi_{q^k}$ and so (contrary to the assumption of Theorem \ref{Flat coord-g}) we don't suppose that the cycles $\{a_k,b_k\}$ start at a particular point $P_0$ such that $\lambda(P_0)=1$.\\
From the above calculation, we see that $\psi_{q^k}(P+b_j)-\psi_{q^k}(P)=0$ for all $j\neq k$. Moreover, regarding $P$ as a starting point of the cycle $b_k$ in the expression
$F_{q^k}(P+b_j)-F_{q^k}(P)$ and  choosing $Q_0=P$ to be the intersection points of $a_k$ and $b_k$, we arrive at (\ref{jump-E}).

\medskip

\noindent $\bullet$ Let $u^A(\omega),u^B(\omega)\in \Big\{s^i(\omega), y^j(\omega), \rho^k(\omega), q^k(\omega)\Big\}$. The above computation shows that
$$
q^k(\psi_{\rho^j})=\oint_{a_k}\log\big(\lambda(Q)\big)\omega_j(Q)=\frac{1}{2{\rm{i}}\pi}\oint_{b_j}\psi_{q^k}=\rho^j(\psi_{q^k}),\quad \quad  \forall\ j,k=1,\dots,g;
$$
provided that  the cycles $\{a_k,b_k\}$ start at a marked point $P_0$ so that  $\lambda(P_0)=1$.\\
Analogously  to the symmetry property (\ref{Fubini-flat}), using  the latter assumption as well as the explicit formulas for the functions $s_{ij}$ and $y_{ij}$, we see that the interchanging of the order of two operations $u^A,u^B$ among $\big\{s^i, y^j, \rho^k, q^k\big\}$ only results in the addition of a constant. More concretely, the following holds:
\begin{equation}\label{Fubini-E}
u^A(\psi_{u^B})=u^B(\psi_{u^A})+\log(-1)\sum_{i,j=1}^m(1-\delta_{ij})\delta_{s^i,u^A}\delta_{s^j,u^B}
+\log(-1)\sum_{i,j=1}^n(1-\delta_{ij})\delta_{y^i,u^A}\delta_{y^j,u^B}.
\end{equation}

\bigskip

Now we are in position to prove  Theorem \ref{Flat coord-g}.\\

\noindent\emph{Proof of Theorem \ref{Flat coord-g}:} By applying (\ref{CNS-FF}) with $\beta_1=1$, $\beta_2=0$ and taking into account  the properties of the symmetric bidifferential $W(P,Q)$, we deduce that the functions $s^i(\omega)$, $y^j(\omega)$, $\rho^k(\omega)$ and $q^k(\omega)$ in (\ref{flat basis-g}) are all flat with respect to the intersection form $\mathbf{g}(\omega)$.\\
Now, in order to calculate the entries of the constant matrix of the  metric $\mathbf{g}^*(\omega)$, we will adapt the scheme of the  proof of  Theorem \ref{FC-thm-eta}.
More precisely, the crucial ingredient is a formula analogous to (\ref{dual metric-e}):
\begin{equation}\label{dual metric-E}
\mathbf{g}^*(\omega)\left(du^A(\omega),du^B(\omega)\right)=E.u^A(\psi_{u^B})=\sum_{k=1}^N\underset{P_k}{{\rm res}}\frac{\lambda(P)\psi_{u^A}(P)\psi_{u^B}(P)}{d\lambda(P)},
\end{equation}
where $E$ is the vector field $E:=\sum_{j=1}^N\lambda_j\partial_{\lambda_j}$ and $\psi_{u^A}(P)$, $\psi_{u^B}(P)$ are among the differentials in (\ref{Primary-E}). Note that (\ref{dual metric-E}) is  an immediate consequence of (\ref{dual metric-R2}) and (\ref{dual metric-R4}) where we take $(\beta_1,\beta_2)=(1,0)$. \\
Due to the residue theorem and (\ref{dual metric-E}) we have
\begin{align*}
E.s^i(\psi_{s^j})&=\sum_{k=1}^N\underset{P_k}{{\rm res}}\frac{\lambda(P)\Omega_{\infty^0\infty^i}(P)\Omega_{\infty^0\infty^j}(P)}{d\lambda(P)}\\
&=-\underset{\infty^0}{{\rm res}}\frac{\lambda(P)\Omega_{\infty^0\infty^i}(P)\Omega_{\infty^0\infty^j}(P)}{d\lambda(P)}
-\delta_{ij}\underset{\infty^i}{{\rm res}}\frac{\lambda(P)\Omega_{\infty^0\infty^i}(P)\Omega_{\infty^0\infty^j}(P)}{d\lambda(P)}\\
&=\frac{1}{n_0+1}+\delta_{ij}\frac{1}{n_i+1}.
\end{align*}
Similarly we get
\begin{align*}
&E.s^i(\psi_{y^j})=E.s^i(\psi_{\rho^k})=E.y^i(\psi_{\rho^k})=E.\rho^k(\psi_{\rho^j})=0;\\
&E.y^i(\psi_{y^j})=-\frac{1}{m_0+1}+\delta_{ij}\frac{1}{m_i+1}.
\end{align*}
It remains to make precise  the action of vector field $E$ on the functions $q^k(\psi)$ with $\psi\in \big\{\psi_{s^i}, \psi_{y^j}, \psi_{\rho^k}, \psi_{q^k}\big\}$. To this end, let us consider the differential
$$
\Psi_k(P):=\frac{\lambda(P)\psi(P)}{d\lambda(P)}\psi_{q^k}(P),\quad k=1,\dots,g,
$$
where $\psi$ is  one of the differentials belonging to (\ref{Primary-E}).\\
Then $\Psi_k$ is a multivalued differential and its jumps are as follows:
\begin{align*}
\begin{array}{llll}
\Psi_k(P+a_j)-\Psi_k(P)=0, \quad & \hbox{if}\quad j=1,\dots,g;\\
\\
\Psi_k(P+b_k)-\Psi_k(P)=-2{\rm{i}}\pi\psi(P), \quad & \hbox{if}\quad \psi\neq \psi_{q^k};\\
\\
\Psi_k(P+b_k)-\Psi_k(P)=-4{\rm{i}}\pi\psi_{q^k}(P)-4\pi^2\frac{d\lambda(P)}{\lambda(P)}, \quad & \hbox{if}\quad \psi=\psi_{q^k};\\
\\
\Psi_k(P+b_j)-\Psi_k(P)=-2{\rm{i}}\pi\psi_{q^k}(P), \quad & \hbox{if}\quad \psi=\psi_{q^j}\quad \text{and}\quad  j\neq k;\\
\\
\Psi_k(P+b_j)-\Psi_k(P)=0, \quad & \hbox{if}\quad \text{$\psi$ is Abelian and $j\neq k$}.
\end{array}
\end{align*}
To summarize, the jumps along the cycles $b_j$ can be rewritten as
\begin{align*}
\Psi_k(P+b_j)-\Psi_k(P)&=\big(1-\delta_{\psi,\psi_{q^k}}\big)\big(-2{\rm{i}}\pi\psi(P)\big)+\delta_{\psi,\psi_{q^k}}
\Big(-4{\rm{i}}\pi\psi_{q^k}(P)-4\pi^2\frac{d\lambda(P)}{\lambda(P)}\Big)\\
&\quad+(1-\delta_{jk})\delta_{\psi,\psi_{q^j}}\big(-2{\rm{i}}\pi\psi_{q^k}(P)\big).
\end{align*}
On the other hand, observing that the multivalued differentials $\psi_{q^1},\dots,\psi_{q^g}$ are all single valued and holomorphic inside the fundamental polygon $F_g$ associated with $C_g$, we deduce that the differential $\Psi_k$ becomes  meromorphic inside $F_g$ and the ramification points $P_1,\dots,P_N$ are its unique simple poles belonging to $F_g$.  \\
Therefore, from (\ref{dual metric-E}) we have
\begin{align*}
E.q^k(\psi)&=\sum_{s=1}^N\underset{P_s}{{\rm res}}\Psi_k(P)=\frac{1}{2{\rm{i}}\pi}\oint_{\partial{F_g}}\Psi_k(P)\\
&=\frac{1}{2{\rm{i}}\pi}\sum_{j=1}^g\bigg(\oint_{a_j}\Psi_k(P)-\oint_{a_j}\Psi_k(P+b_j)\bigg)
+\frac{1}{2{\rm{i}}\pi}\sum_{j=1}^g\bigg(\oint_{b_j}\Psi_k(P)-\oint_{b_j}\Psi_k(P+a_j^{-1})\bigg)\\
&=\bigg(\sum_{j=1,j\neq k}^g\delta_{\psi,\psi_{q^j}}\oint_{a_j}\psi_{q^k}\bigg) +\big(1-\delta_{\psi,\psi_{q^k}}\big)\oint_{a_k}\psi +\delta_{\psi,\psi_{q^k}}\oint_{a_k}\bigg(2\psi_{q^k(P)}-2{\rm{i}}\pi \frac{d\lambda(P)}{\lambda(P)}\bigg)\\
&=\big(1-\delta_{\psi,\psi_{q^k}}\big)\delta_{\psi,\psi_{\rho^k}}=\delta_{\psi,\psi_{\rho^k}},
\end{align*}
with $\delta_{\psi,\psi_{\rho^k}}$ being the Kronecker delta. This finishes the proof of the theorem.

\fd

\begin{Cor} The nonzero entries of constant matrix of the intersection form $\mathbf{g}(\omega)$ are given by
\begin{equation}\label{g-entries}
\begin{split}
&\mathbf{g}(\omega)\big(\partial_{s^i(\omega)},\partial_{s^j(\omega)}\big)=(n_i+1)\delta_{ij}-\frac{(n_i+1)(n_j+1)}{L};\\
&\mathbf{g}(\omega)\big(\partial_{y^i(\omega)},\partial_{y^j(\omega)}\big)=-(m_i+1)\delta_{ij}+\frac{(m_i+1)(m_j+1)}{L};\\
&\mathbf{g}(\omega)\big(\partial_{\rho^i(\omega)},\partial_{q^j(\omega)}\big)=\delta_{ij}.
\end{split}
\end{equation}
Here $L$ denotes the common number of sheets of the coverings $[(C_g,\lambda)]\in \widehat{\mathcal{H}}_{g,L}(\mathbf{n}; \mathbf{m})$.
\end{Cor}
\emph{Proof:} Let $\mathbf{A}=(\mathbf{A}_{ij})$ be a $m\times m$ square matrix of the form
$$
\mathbf{A}_{ij}=\alpha_0+\delta_{ij}\alpha_i,
$$
where $\alpha_0,\alpha_1,\dots,\alpha_m$ are all nonzero complex numbers.  Then one can check that the matrix $\mathbf{A}$ is invertible and the entries of its inverse are:
$$
\mathbf{A}^{ij}=\frac{1}{\alpha_i}\delta_{ij}-\frac{1}{\alpha_i\alpha_j\sigma},\quad \quad \text{with}\quad \sigma:=\sum_{k=0}^m\frac{1}{\alpha_k}.
$$
When $\alpha_0,\dots,\alpha_m$ are the positive real numbers:
$$
\alpha_i=\frac{1}{n_i+1}, \quad \forall\ i=0,1,\dots,m,
$$
we have
$$
\sigma=\sum_{k=0}^m\frac{1}{\alpha_k}=L.
$$
This and (\ref{Entrie-g}) imply  (\ref{g-entries}).

\fd

\begin{Cor} Let $u^A(\omega)\in \Big\{s^i(\omega), y^j(\omega), \rho^k(\omega), q^k(\omega)\Big\}$. Then the partial derivative $\partial_{u^A(\omega)}$ has the following form
\begin{equation}\label{g-partial der-FC}
\partial_{u^{A}(\omega)}=\sum_{j}\bigg(\frac{\lambda_j}{\omega(P_j)}\sum_{\alpha}\textsf{G}_{A\alpha}\psi_{u^{\alpha}}(P_j)\bigg)\partial_{\lambda_j},
\end{equation}
where $\textsf{G}_{A\alpha}=\mathbf{g}(\omega)\big(\partial_{u^A(\omega)},\partial_{u^{\alpha}(\omega)}\big)$ are the entries (\ref{g-entries}) of the constant matrix of the metric $\mathbf{g}(\omega)$ and $\psi_{u^A}$ is among the differentials (\ref{Primary-E}) corresponding to the function $u^A$.
\end{Cor}
\emph{Proof:} By arguing as in the proof of (\ref{derivatives lambda-j}), we can check  that the partial derivatives of $\lambda_j$ with respect to flat coordinates $\{u^A(\omega)\}_A$ are given by
$$
\partial_{u^A(\omega)} \lambda_j=\frac{\lambda_j}{\omega(P_j)}\sum_{\alpha}\textsf{G}_{A\alpha}\psi_{u^{\alpha}}(P_j).
$$
This  shows that  (\ref{g-partial der-FC}) holds true.

\fd

\medskip

At the end of this subsection, we shall  describe the matrix of the intersection form  $\mathbf{g}^*(\omega)$ in flat coordinates (\ref{FBP-W})-(\ref{S-eta})   of the Darboux-Egoroff metric $\eta(\omega)$ defined by (\ref{eta-def}). Then we shall view $\mathbf{g}(\omega)$ as a metric on an open subset of the particular double Hurwitz space $\widehat{\mathcal{H}}_{g,L}(\mathbf{n}; \mathbf{0})$. The latter is  an open subset of the simple Hurwitz space $\widehat{\mathcal{H}}_{g, L}(n_0,\dots,n_m)$ determined by the conditions $\lambda_j\neq 0$, for all $j=1,\dots,N$.\\
From \cite{Dubrovin2D} (see also next section), it is known that the vector field $E=\sum_{j}\lambda_j\partial_{\lambda_j}$ plays the role of the Euler vector field of the semi-simple Hurwitz-Frobenius manifold structures induced by Dubrovin's primary differentials (\ref{Primary-e}). Thus the obtained matrix of $\mathbf{g}^*(\omega)$ describes, through formula (\ref{Entries-g-eta}) below,  the quasi-homogeneity properties  of  flat coordinates of the $N$  Darboux-Egoroff metrics $\big\{\eta(\phi_{t^A})\big\}_A$  of Dubrovin.

\begin{Thm} Let $\omega(P)=\int_{\gamma}h(\lambda(Q))W(P,Q)$ be a differential on a surface $C_g$ satisfying (\ref{Diff model}) and (\ref{Assymptions}), $\eta(\omega)$ be the corresponding Darboux-Egoroff metric (\ref{eta-def}) and $E$ be the vector field  $E=\sum_{j=1}^N\lambda_j\partial_{\lambda_j}$. \\
Then, in the flat coordinates $t^{A}(\omega)\in \mathcal{S}(\omega)$ described in (\ref{FBP-W})-(\ref{S-eta}), the entries of the matrix of the intersection form $\mathbf{g}^*(\omega)$
are:
\begin{equation}\label{Entries-g-eta}
\mathbf{g}^*(\omega)\left(dt^A(\omega),dt^B(\omega)\right)=E.t^A(\phi_{t^B})=(d_A+d_B)t^A(\phi_{t^B})+r_{A,B},
\end{equation}
where $r_{A,B}$ is the constant
\begin{equation}\label{r-AB}
r_{A,B}=\sum_{i,j=1}^m\delta_{t^A,s^i}\delta_{t^B,s^j}\Big(\frac{1}{n_0+1}+\delta_{ij}\frac{1}{n_i+1}\Big)=r_{B,A}
\end{equation}
and $d_A$ (resp. $d_B$) is the nonnegative constant appearing in the integral  representation (\ref{Int-rep-e}) of the differential $\phi_{t^A}$ (resp. $\phi_{t^B}$).
\end{Thm}
\emph{Proof:} According to Theorem \ref{Practical F} and formulas (\ref{dual metric-R2})-(\ref{dual metric-R4}), the following two equalities are valid:
$$
\mathbf{g}^*(\omega)\left(dt^A(\omega),dt^B(\omega)\right)=E.t^A(\phi_{t^B})=\sum_{k=1}^N\underset{P_k}{{\rm res}}\frac{\lambda(P)\phi_{t^A}(P)\phi_{t^B}(P)}{d\lambda(P)}.
$$
This shows the first equality in (\ref{Entries-g-eta}).\\
Let $\phi$ be one of the differentials belonging to the set (\ref{Primary-e}). If $\phi\neq \phi_{u^k}$ (i.e. $\phi$ has no jumps), then  the point $\infty^i$ is a pole of the meromorphic function $\frac{\lambda(P)\phi(P)}{d\lambda(P)}$ if and only if $\phi$ is of type $\phi=\phi_{t^{i,\alpha}}$ or $\phi=\phi_{v^{i}}$.
Moreover, according to behaviors (\ref{type1}) and (\ref{type2}) of the Abelian differentials $\phi_{t^{i,\alpha}}$ and $\phi_{v^i}$ near the point $\infty^i$, we have
\begin{align*}
\frac{\lambda(P)\phi(P)}{d\lambda(P)}&\underset{P\sim\infty^i}=-\delta_{\phi,\phi_{t^{i,\alpha}}} \Big(\frac{z_i(P)^{-\alpha}}{\sqrt{n_i+1}}+\text{reg}\Big)-\delta_{\phi,\phi_{v^{i}}}\Big(z_i(P)^{-n_i-1}+\text{reg}\Big)\\
&\underset{P\sim\infty^i}=-\delta_{\phi,\phi_{t^{i,\alpha}}} \Big(\frac{1}{\sqrt{n_i+1}}\lambda(P)^{\frac{\alpha}{n_i+1}}+\text{reg}\Big)-\delta_{\phi,\phi_{v^{i}}}\Big(\lambda(P)+\text{reg}\Big).
\end{align*}
Moreover, when $\phi=\phi_{u^k}$, $k=1,\dots,g$, then the restriction of the function $\frac{\lambda(P)\phi(P)}{d\lambda(P)}$ to the fundamental polygon $F_g$ associated with the surface $C_g$ is holomorphic.\\
Therefore, the following useful identity  holds true
\begin{equation}\label{g-eta-trick}
\underset{\infty^i}{{\rm res}}\bigg(\frac{\lambda(P)\phi(P)}{d\lambda(P)}\phi_{0}(P)\bigg)=-\delta_{\phi,\phi_{t^{i,\alpha}}}\frac{\alpha}{n_i+1}t^{i,\alpha}(\phi_0)-
\delta_{\phi,\phi_{v^i}}v^i(\phi_0)
\end{equation}
holds true whenever the differential $\phi_0$  is holomorphic near $\infty^i$ and $\phi$ is among the differentials listed in (\ref{Primary-e}).\\
Now, in order to prove the second equality in (\ref{Entries-g-eta}),  we distinguish five cases depending on the five types of operations $t^{i,\alpha},\ v^i,\ s^i,\ \rho^k,\ u^k$.\\
\textbf{First case: $t^A=u^k$.} The equality $E.u^k(\phi_{t^B})=(1+d_B)u^k(\phi_{t^B})$ is  a particular case of formula (\ref{u-k-sigma}) established in Appendix 2.\\
\textbf{Second case: $t^A=\rho^k$.} Assume that $\phi\in \big\{\phi_{t^{i,\alpha}}, \phi_{v^i}, \phi_{s^i}, \phi_{\rho^k}=\omega_k\big\}$, i.e. $\phi\neq \phi_{u^k}$  for all $k=1,\dots,g$. Using the residue theorem and applying (\ref{g-eta-trick}) to $\phi_0=\phi_{\rho^k}=\omega_k$ (holomorphic differential), we deduce that
\begin{align*}
E. \rho^k(\phi)
&=-\sum_{i=0}^{m}\sum_{\alpha=1}^{n_i}\delta_{\phi,\phi_{t^{i,\alpha}}}\underset{\infty^i}{{\rm res}}\ \bigg(\frac{\lambda(P)\phi(P)}{d\lambda(P)}\omega_k(P)\bigg)
 -\sum_{i=1}^{m}\delta_{\phi,\phi_{v^{i}}}\underset{\infty^i}{{\rm res}}\ \bigg(\frac{\lambda(P)\phi(P)}{d\lambda(P)}\omega_k(P)\bigg)\\
&=\sum_{i=0}^{m}\sum_{\alpha=1}^{n_i}\delta_{\phi,\phi_{t^{i,\alpha}}}\frac{\alpha}{n_i+1}t^{i,\alpha}\big(\omega_k\big)
+\sum_{i=1}^{m}\delta_{\phi,\phi_{v^{i}}}v^i\big(\omega_k\big)\\
&=\sum_{i=0}^{m}\sum_{\alpha=1}^{n_i}\delta_{\phi,\phi_{t^{i,\alpha}}}\frac{\alpha}{n_i+1}\rho^k\big(\phi_{t^{i,\alpha}}\big)
+\sum_{i=1}^{m}\delta_{\phi,\phi_{v^{i}}}\rho^k\big(\phi_{v^i}\big).
\end{align*}
When $\phi=\phi_{u^k}$, then   the symmetry property (\ref{Fubini-flat}) and the first case imply that
$$
E.\rho^k(\phi_{u^k})=E.u^k(\phi_{\rho^k})=u^k(\phi_{\rho^k})=\rho^k(\phi_{u^k}).
$$
Thus the desired formula (\ref{Entries-g-eta})  is valid in this case. \\
\textbf{Third case: $t^A=v^j$.} By (\ref{Fubini-flat}), we know that $v^j(\phi_{u^k})=u^k(\phi_{v^j})$ and $v^j(\phi_{\rho^k})=\rho^k(\phi_{v^j})$ for all $k=1,\dots,g$.
Hence the action of the vector field $E$ on the functions $v^j(\phi_{u^k})$ and $v^j(\phi_{\rho^k})$ is studied in the first and second cases. Thus, it suffices to consider the cases $E.v^j(\phi)$, where $\phi\in \big\{\phi_{t^{i,\alpha}}, \phi_{v^i}, \phi_{s^i}\big\}$.\\
Given  that the differential $\phi_0(P):=\phi_{v^j}(P)+d\lambda(P)$ is holomorphic near $\infty^j$ (by (\ref{type2}), we see that the left hand side of (\ref{g-eta-trick}) applied to this $\phi_0$ can be rewritten as
$$
\underset{\infty^j}{{\rm res}}\bigg(\frac{\lambda(P)\phi(P)}{d\lambda(P)}\phi_{v^j}(P)\bigg)+v^j(\phi)
$$
while its  right hand side becomes
\begin{align*}
&-\delta_{\phi,\phi_{t^{j,\alpha}}}\frac{\alpha}{n_j+1}\Big(t^{j,\alpha}\big(\phi_{v^j}\big)+t^{j,\alpha}\big(d\lambda(P)\big)\Big)-\delta_{\phi,\phi_{v^j}}
\Big(v^j\big(\phi_{v^j}\big)+v^j\big(d\lambda(P)\big)\Big)\\
&=-\delta_{\phi,\phi_{t^{j,\alpha}}}\frac{\alpha}{n_j+1}t^{j,\alpha}\big(\phi_{v^j}\big)-\delta_{\phi,\phi_{v^j}}v^j\big(\phi_{v^j}\big).
\end{align*}
This amounts to the following equality:
$$
\underset{\infty^j}{{\rm res}}\ \bigg(\frac{\lambda(P)\phi(P)}{d\lambda(P)}\phi_{v^j}(P)\bigg)
=-v^j(\phi)-\delta_{\phi,\phi_{t^{j,\alpha}}}\frac{\alpha}{n_j+1}t^{j,\alpha}\big(\phi_{v^j}\big)-\delta_{\phi,\phi_{v^j}}v^j\big(\phi_{v^j}\big).
$$
This and the key identity (\ref{g-eta-trick}) lead to:
\begin{align*}
&E. v^j(\phi)=\sum_{k=1}^N\underset{P_k}{{\rm res}}\ \bigg(\frac{\lambda(P)\phi(P)}{d\lambda(P)}\phi_{v^j}(P)\bigg)\\
&=-\underset{\infty^j}{{\rm res}}\ \bigg(\frac{\lambda(P)\phi(P)}{d\lambda(P)}\phi_{v^j}(P)\bigg)
-\sum_{i=0, i\neq j}^{m}\sum_{\alpha=1}^{n_i}\delta_{\phi,\phi_{t^{i,\alpha}}}\underset{\infty^i}{{\rm res}}\ \bigg(\frac{\lambda(P)\phi(P)}{d\lambda(P)}\phi_{v^j}(P)\bigg)
 -\sum_{i=1, i\neq j}^{m}\delta_{\phi,\phi_{v^{i}}}\underset{\infty^i}{{\rm res}}\ \bigg(\frac{\lambda(P)\phi(P)}{d\lambda(P)}\phi_{v^j}(P)\bigg)\\
&=v^j(\phi)+\delta_{\phi,\phi_{t^{j,\alpha}}}\frac{\alpha}{n_j+1}t^{j,\alpha}\big(\phi_{v^j}\big)+\delta_{\phi,\phi_{v^j}}v^j\big(\phi_{v^j}\big)+\sum_{i=0,i\neq j}^{m}
\sum_{\alpha=1}^{n_i}\delta_{\phi,\phi_{t^{i,\alpha}}}\frac{\alpha}{n_i+1}t^{i,\alpha}\big(\phi_{v^j}\big)
 +\sum_{i=1, i\neq j}^{m}\delta_{\phi,\phi_{v^{i}}}v^i\big(\phi_{v^j}\big)\\
&=v^j(\phi)+\sum_{i=0}^{m}\sum_{\alpha=1}^{n_i}\delta_{\phi,\phi_{t^{i,\alpha}}}\frac{\alpha}{n_i+1}v^j\big(\phi_{t^{i,\alpha}}\big)
 +\sum_{i=1}^{m}\delta_{\phi,\phi_{v^{i}}}v^j\big(\phi_{v^i}\big),
\end{align*}
with the last equality being a direct consequence of  the following particular cases of (\ref{Fubini-flat}):
$$
t^{i,\alpha}\big(\phi_{v^j}\big)=v^j\big(\phi_{t^{i,\alpha}}\big),\quad \quad v^i\big(\phi_{v^j}\big)=v^j\big(\phi_{v^i}\big).
$$
\noindent \textbf{Forth case: $t^A=t^{j,\beta}$.} Assume that $\phi\in \big\{\phi_{t^{i,\alpha}}, \phi_{s^i}\big\}$. Then by similar arguments we obtain
\begin{align*}
E. t^{j,\beta}(\phi)&=-\underset{\infty^j}{{\rm res}}\ \bigg(\frac{\lambda(P)\phi(P)}{d\lambda(P)}\phi_{t^{j,\beta}}(P)\bigg)
-\sum_{i=0, i\neq j}^{m}\sum_{\alpha=1}^{n_i}\delta_{\phi,\phi_{t^{i,\alpha}}}\underset{\infty^i}{{\rm res}}\ \bigg(\frac{\lambda(P)\phi(P)}{d\lambda(P)}\phi_{t^{j,\beta}}(P)\bigg)\\
&=\frac{\beta}{n_j+1}t^{j,\beta}(\phi)+\sum_{i=0}^{m}\sum_{\alpha=1}^{n_i}\delta_{\phi,\phi_{t^{i,\alpha}}}\frac{\alpha}{n_i+1}t^{j,\beta}\big(\phi_{t^{i,\alpha}}\big).
\end{align*}
\noindent \textbf{Fifth case: $t^A=s^{j}$.} By symmetry, it suffices to treat the case $\phi=\phi_{s^i}=\Omega_{\infty^0\infty^i}$. The action of the vector field $E$ on $s^j(\phi_{s^i})=s_{ij}$ was already described in (\ref{Entrie-g}).

\fd


\section{Hurwitz-Frobenius manifold structures}
\subsection{Frobenius manifold structures on simple Hurwitz spaces}
The goal of this subsection is to build some Frobenius manifold structures  on simple Hurwitz spaces $\widehat{\mathcal{H}}_{g, L}(n_0,\dots,n_m)$ induced by quasi-homogeneous differentials (in the sense of Definition \ref{quasi-homo-diff} below) and construct the corresponding prepotentials via a new approach. As we will see below, Dubrovin's primary differentials in (\ref{Primary-e}) are all quasi-homogeneous differentials. Thus our construction includes semi-simple Hurwitz-Frobenius manifold of Dubrovin as particular cases.  \\
 As pointed out in the introduction, the  prepotentials of the obtained Frobenius manifold structures provide   quasi-homogeneous solutions to the generalized WDVV equations (\ref{WDVV equation}).
\subsubsection{Frobenius algebra structures}
Let $\eta(\omega):=\frac{1}{2}\sum_j\big(\omega(P_j)\big)^2(d\lambda_j)^2$ be the Darboux-Egoroff metric on the manifold
$\mathcal{M}_{\omega}\subset \widehat{\mathcal{H}}_{g, L}(n_0,\dots,n_m)$ determined by
$$
\omega(P_j)\neq 0, \quad \forall\ j=1,\dots,N.
$$
Here, as above, the differential $\omega$ is of the form $\omega(P)=\int_{\gamma}h(\lambda(Q))W(P,Q)$ such that the contour $\gamma$ and the function $h$ satisfy respectively  the conditions \texttt{C1)} and \texttt{C2)} in (\ref{Diff model}).\\
On each tangent space $T_p\mathcal{M}_{\omega}$, we define a commutative and  associative  multiplication by
\begin{equation}\label{multiplication}
\partial_{\lambda_i}\circ\partial_{\lambda_j}=\delta_{ij}\partial_{\lambda_j}.
\end{equation}
Particularly, we easily see that the  vector field $e=\sum_{j}\partial_{\lambda_j}$ is the unit vector field of the multiplication (\ref{multiplication}).\\
For $x=\sum_jx_j\partial_{\lambda_j}$, $y=\sum_jy_j\partial_{\lambda_j}$ and $z=\sum_jz_j\partial_{\lambda_j}\in T_p\mathcal{M}_{\omega}$, consider the $(0,3)$-tensor
\begin{equation}\label{c-tensor}
\mathbf{c}_{\omega}(x,y,z):=\eta(\omega)(x\circ y,z)=\frac{1}{2}\sum_jx_jy_jz_j\omega(P_j)^2.
\end{equation}
Since the tensor $\mathbf{c}_{\omega}(x,y,z)$  is symmetric,   the quadruplet $\big(T_p\mathcal{M}_{\omega},\circ,e, \eta(\omega)\big)$ is a commutative and associative Frobenius algebra with unity (according to Definition \ref{F algebra}).\\
Note that formula (\ref{flat-canonical relation}) allows us to describe the Frobenius algebra structure of $T_p\mathcal{M}_{\omega}$  by means of the new system of flat coordinates $\mathcal{S}(\omega)=\big\{t^{i,\alpha}(\omega), v^i(\omega), s^i(\omega), \rho^k(\omega), u^k(\omega)\big\}$ of the metric $\eta(\omega)$. The upcoming obtained expressions (\ref{mult1-FC})-(\ref{c-FC-eta}) involve the $\eta(\omega)$-duality relations between flat coordinates that were introduced in (\ref{duality}) and (\ref{duality-picture}):
\begin{itemize}
\item[\textbf{a)}] The multiplication ``$\circ$'' in (\ref{multiplication}) takes the following form in flat coordinates:
\begin{align}\label{mult1-FC}
\begin{split}
\partial_{t^A(\omega)}\circ\partial_{t^B(\omega)}
&=\sum_{j=1}^N \frac{\phi_{t^{A^{\prime}}}(P_j)\phi_{t^{B^{\prime}}}(P_j)}{\omega(P_j)^2}\partial_{\lambda_j}\\
&=\sum_C\bigg(\frac{1}{2}\sum_{j=1}^N\frac{\phi_{t^{A^{\prime}}}(P_j)\phi_{t^{B^{\prime}}}(P_j)\phi_{t^{C}}(P_j)}{\omega(P_j)}\bigg)\partial_{t^C(\omega)}.
\end{split}
\end{align}
\item[\textbf{b)}] The unit vector field $e=\sum_{j=1}^N\partial_{\lambda_j}$ is represented by:
\begin{equation}\label{e-FC-eta}
e=\sum_A\bigg(\frac{1}{2}\sum_{j=1}^N\phi_{t^A}(P_j)\omega(P_j)\bigg)\partial_{t^A(\omega)}.
\end{equation}
\item[\textbf{c)}] The symmetric 3-tensor $\mathbf{c}_{\omega}$ defined by (\ref{c-tensor}) has the form:
\begin{equation}\label{c-FC-eta}
\mathbf{c}_{\omega}\big(\partial_{t^A(\omega)},\partial_{t^B(\omega)},\partial_{t^C(\omega)}\big)=\frac{1}{2}
\sum_{j=1}^N\frac{\phi_{t^{A^{\prime}}}(P_j)\phi_{t^{B^{\prime}}}(P_j)\phi_{t^{C^{\prime}}}(P_j)}{\omega(P_j)}.
\end{equation}
This can be rewritten as a residue:
$$
\mathbf{c}_{\omega}\big(\partial_{t^A(\omega)},\partial_{t^B(\omega)},\partial_{t^C(\omega)}\big)=
\sum_{j=1}^N\underset{P_j}{\text{res}}\frac{\phi_{t^{A^{\prime}}}(P)\phi_{t^{B^{\prime}}}(P)\phi_{t^{C^{\prime}}}(P)}{d\lambda(P)\omega(P)}.
$$
\end{itemize}
Note that expressions (\ref{mult1-FC}) and (\ref{c-FC-eta}) imply   that
$$
\partial_{t^A(\omega)}\circ\partial_{t^B(\omega)}
=\textstyle\sum_C\mathbf{c}_{\omega}\big(\partial_{t^A(\omega)},\partial_{t^B(\omega)},\partial_{t^{C^{\prime}}(\omega)}\big)\partial_{t^C(\omega)}
$$
and thus the structure constants of the algebra  multiplication ``$\circ$'' are
$$
\mathbf{c}_{\omega}\big(\partial_{t^A(\omega)},\partial_{t^B(\omega)},\partial_{t^{C^{\prime}}(\omega)}\big)
=\sum_{\alpha=1}^N\eta^{C\alpha}\mathbf{c}_{\omega}\big(\partial_{t^A(\omega)},\partial_{t^B(\omega)},\partial_{t^{\alpha}(\omega)}\big),
$$
with $\eta^{AB}=\delta_{B, A^{\prime}}$ being the constant matrix (\ref{Entries eta}).

\subsubsection{Primary differentials}
\begin{Def} Let $\omega(P)=\int_{\gamma}h(\lambda(Q))W(Q,P)$ be a fixed  differential as in (\ref{Diff model}) and (\ref{Assymptions}) and $\eta(\omega)$ be the corresponding Darboux-Egoroff metric (\ref{eta-def}).\\
We shall call  the differential $\omega$  primary with respect to unit vector field $e=\sum_j\partial_{\lambda_j}$  ($e$-primary)  if
$$
\nabla^{\eta(\omega)}.e=0,
$$
where  $\nabla^{\eta(\omega)}$ is the Levi-Civita connection of the metric $\eta(\omega)$.
\end{Def}
In other words, $e$-primary differentials are exactly those that ensure the flatness of the unit vector field $e$. In this case, the components $e^A(t)$ of $e$ in  flat coordinates $\{t^A(\omega)\}$ of the metric $\eta(\omega)$ are constant functions.

\begin{Prop}\label{ch-e-primary}
For all $j=1,\dots,N$, we have
\begin{equation}\label{e-Levi Civita}
\nabla^{\eta(\omega)}_{\partial_{\lambda_j}}.e=\frac{e.\omega(P_j)}{\omega(P_j)}\partial_{\lambda_j}.
\end{equation}
In particular,  the following statements are equivalent:
\begin{description}
\item[1.] The differential $\omega$ is primary with respect to $e$;
\item[2.] For all $j=1,\dots,N$,
\begin{equation}\label{e-covariantly0}
e.\omega(P_j)=0;
\end{equation}
\item[3.] For all $j=1,\dots,N$,
\begin{equation}\label{e-covariantly}
\int_{\gamma}h'(\lambda(Q))W(Q,P_j)=0.
\end{equation}
\end{description}
\end{Prop}
\emph{Proof:} Bearing in mind the Christoffel coefficients of the diagonal  metric $\eta(\omega)$ given by (\ref{CS}) and the form $e=\sum_j\partial_{\lambda_j}$ of the unit vector field $e$, we obtain
\begin{align*}
\nabla^{\eta(\omega)}_{\partial_{\lambda_i}}e&=\nabla^{\eta(\omega)}_{\partial_{\lambda_i}}\Big(\sum_{j=1}\partial_{\lambda_j}\Big)
=\sum_{j}\sum_k\Gamma_{ij}^k(\eta)\partial_{\lambda_k}
=\sum_{j\neq i}\sum_k\Gamma_{ij}^k(\eta)\partial_{\lambda_k}+\sum_k\Gamma_{ii}^k(\eta)\partial_{\lambda_k}\\
&=\Big(\sum_{j\neq i}\Gamma_{ij}^i(\eta)\Big)\partial_{\lambda_i}+\sum_{j\neq i}\Gamma_{ij}^j(\eta)\partial_{\lambda_j}
+\sum_{k\neq i}\Gamma_{ii}^k(\eta)\partial_{\lambda_k}+\Gamma_{ii}^i(\eta)\partial_{\lambda_i}\\
&=\frac{1}{\omega(P_i)}\Big(\sum_{j\neq i}\partial_{\lambda_j}\omega(P_i)\Big)\partial_{\lambda_i}+\frac{\partial_{\lambda_i}\omega(P_i)}{\omega(P_i)}\partial_{\lambda_i}\\
&=\frac{e.\omega(P_i)}{\omega(P_i)}\partial_{\lambda_i}.
\end{align*}
This proves  (\ref{e-Levi Civita}).\\
The equivalence between the three assertions follows immediately from (\ref{e-Levi Civita}) and the action (\ref{Auxi2}) of the vector field $e$ on the functions $\omega(P_i)$.

\fd

\medskip

As a corollary, the characterization (\ref{e-covariantly}) and the properties of the symmetric bidifferential yield the following.
\begin{Cor}\label{Rmk-primary} Consider the $N$ differentials $\phi_{t^A}\in \Big\{\phi_{t^{i,\alpha}}, \phi_{v^i},\phi_{s^i}, \phi_{\rho^k},\phi_{u^k}\Big\}$  associated with the $\eta(\omega)$-flat coordinates and given by (\ref{Primary-e}), with $N=2g+2m+\textstyle\sum_{i=0}^mn_i$.\\
 Then all the $N$ differentials $\phi_{t^A}$ are  primary with respect to the unit vector field $e=\sum_{j=1}^N\partial_{\lambda_j}$.
\end{Cor}
These primary differentials coincide (up to multiplicative constants) with the Dubrovin  primary differentials \cite{Dubrovin2D, Vasilisa}.
\begin{Prop}
Assume that $\omega_0:=\phi_{t^{A_0}}$ is a fixed primary differential belonging to the family (\ref{Primary-e}). \\
Then the unit vector field $e=\sum_j\partial_{\lambda_j}$ of the Frobenius algebra associated with $\omega_0$ reduces to
\begin{equation}\label{e-tA}
e=\partial_{t^{A_0^{\prime}}(\omega_0)}.
\end{equation}
\end{Prop}
\emph{Proof:} In view of (\ref{e-FC-eta}), (\ref{dual metric-R1}) with $(\beta_1,\beta_2)=(0,1)$ and (\ref{duality}), we obtain
\begin{align*}
e&=\sum_B\bigg(\frac{1}{2}\sum_j\phi_{t^B}(P_j)\phi_{t^{A_0}}(P_j)\bigg)\partial_{t^B(\omega_0)}\\
&=\sum_B\bigg(\eta^*(\omega_0)\big(dt^{A_0}(\omega_0),dt^B(\omega_0)\big)\bigg)\partial_{t^B(\omega_0)}\\
&=\sum_B\delta_{B,A_0^{\prime}}\partial_{t^B(\omega_0)}=\partial_{t^{A_0^{\prime}}(\omega_0)}.
\end{align*}
Alternatively, the result can be also  checked by applying (\ref{flat-canonical relation}) to the partial derivative $\partial_{t^{A_0^{\prime}}(\omega_0)}$ and $\omega_0=\phi_{t^{A_0}}$.

\fd

\subsubsection{Euler vector field and quasi-homogeneous differentials}
\begin{Def}\label{quasi-homo-diff} Let $\omega_0$ be a differential  as in (\ref{Diff model}) and (\ref{Assymptions}).  We shall call  $\omega_0$ quasi-homogenous of degree $d_0\in \R$ if there is a nonzero constant $c_0\in \C$ such that
\begin{equation}\label{homogeneous diff}
\omega_0(P)=c_0\int_{\ell_0}\big(\lambda(Q)\big)^{d_0}W(P,Q).
\end{equation}
More generally, a  differential is called quasi-homogeneous of degree $d_0$ if it can be written as linear combination of differentials of the form  $\int_{\ell_0}\big(\lambda(Q)\big)^{d_0}W(P,Q)$.
\end{Def}

From their integral representation (\ref{Int-rep-e}) with respect to the kernel $W(P,Q)$, primary differentials given in  (\ref{Primary-e}) are quasi-homogeneous and their degrees satisfy  (\ref{degree key relation}). Note that duality relation (\ref{degree key relation})   will be important for our construction  of  prepotentials  associated with a quasi-homogeneous differential (see next subsection).
\begin{Prop}
The $N$ primary differentials listed in (\ref{Primary-e}) are all quasi-homogeneous with degrees
\begin{equation}\label{primary-quasi}
\begin{array}{ccccc}
&{\rm{deg}}\left(\phi_{t^{i,\alpha}}\right)=\frac{\alpha}{n_i+1}, &\hspace{1cm} i=0,\dots,m,  &\hspace{1cm}\alpha=1,\dots,n_i; \\
&{\rm{deg}}\left(\phi_{v^i}\right)=deg\left(\phi_{u^k}\right)=1, &\hspace{1cm} i=1,\dots,m, &\hspace{1cm} k=1,\dots,g; \\
&{\rm{deg}}\left(\phi_{s^i}\right)=deg\left(\phi_{\rho^k}\right)=0, & \hspace{1cm}i=1,\dots,m, &\hspace{1cm} k=1,\dots,g.
\end{array}
\end{equation}
Furthermore, their degrees satisfy  the following duality relation:
\begin{equation}\label{degree key relation}
{\rm{deg}}\big(\phi_{t^A}\big)+{\rm{deg}}\big(\phi_{t^{A^{\prime}}}\big)=1, \quad \text{for all A}.
\end{equation}
Here $t^{A^{\prime}}$ is the $\eta$-dual flat coordinate of $t^A$ in the sense of (\ref{duality}) and (\ref{duality-picture}).
\end{Prop}
\emph{Proof:} Using (\ref{duality-picture}) and (\ref{primary-quasi}), we see that relation (\ref{degree key relation}) holds.

\fd

We are going to consider the following infinite family of quasi-homogenous differentials, none of which is primary with respect the unit vector field $e$. Each quasi-homogenous differential can give a non-normalized (i.e. with non-flat unity) Frobenius manifold structure (see Definition \ref{Frob Man def}).

\medskip

\noindent$\bullet$ For $i=0,\dots,m$ and $\beta\in \N$, $\beta\geq 1$,
$$
\phi_{t^{i,n_i+\beta+1}}(P):=\underset{\infty^i}{\rm res}\ \lambda(Q)^{1+\frac{\beta}{n_i+1}}W(P,Q)
$$
is  an Abelian differential of the second kind with unique pole of order $n_i+\beta+2$ at $\infty^i$. The differential $\phi_{t^{i,n_i+\beta+1}}$ has the following behavior
$$
\phi_{t^{i,n_i+\beta+1}}(P)\underset{P\sim \infty^i}{=}\frac{n_i+\beta+1}{z_i(P)^{n_i+\beta+2}}dz_i(P)+\ \text{holomorphic},\quad\quad  \lambda(P)=z_i(P)^{-n_i-1}.
$$
Let us point out that a formula analogous to (\ref{dual metric-e}) and the residue theorem show that the action of the unit  vector field $e=\sum_{j}\partial_{\lambda_j}$ on the function $t^A\big(\phi_{t^{i,n_i+\beta+1}}\big)$ is determined by
\begin{align*}
e.t^A\big(\phi_{t^{i,n_i+\beta+1}}\big)&=\sum_{r=1}^N\underset{P_r}{{\rm res}}\frac{\phi_{t^A}(P)\phi_{t^{i,n_i+\beta+1}}(P)}{d\lambda(P)}\\
&=-\underset{\infty^i}{{\rm res}}\frac{\phi_{t^A}(P)\phi_{t^{i,n_i+\beta+1}}(P)}{d\lambda(P)}\\
&=\frac{n_i+\beta+1}{n_i+1}\underset{\infty^i}{{\rm res}}\lambda(P)^{\frac{\beta}{n_i+1}}\phi_{t^A}(P).
\end{align*}
Here $t^A$ is one of the operations $\big\{t^{j,\alpha}, v^j, s^j, \rho^k,u^k\big\}$ introduced in (\ref{FBP-W}) and $\phi_{t^A}$ is the corresponding differential.\\
In particular
$$
\begin{array}{ll}
\displaystyle \partial_{\lambda_j}\Big(e.t^A\big(\phi_{t^{i,n_i+\beta+1}}\big)\Big)=\frac{n_i+\beta+1}{2(n_i+1)}\phi_{t^{i,\beta}}(P_j)\phi_{t^A}(P_j),\quad \quad&\hbox{if}\quad j=1,\dots,N;\\
\\
\displaystyle  e.t^A\big(\phi_{t^{i,n_i+\beta+1}}\big)=2v^i(\phi_{t^A}),\quad \quad&\hbox{if}\quad \beta=n_i+1;\\
\\
\displaystyle  e.t^A\big(\phi_{t^{i,n_i+\beta+1}}\big)=\frac{\beta(n_i+\beta+1)}{(n_i+1)^{3/2}}t^{i,\beta}(\phi_{t^A}),\quad \quad &\hbox{if}\quad \beta=1,\dots,n_i.
\end{array}
$$

\medskip

\noindent$\bullet$ Assume that  the covering $(C_g,\lambda)$ belongs to the particular double Hurwitz space $\widehat{\mathcal{H}}_{g,L}(\mathbf{n}; \mathbf{0})$, with $\mathbf{n}=(n_0+1,\dots,n_m+1)$. As indicated before, this is  an open subset of the simple Hurwitz space $\widehat{\mathcal{H}}_{g, L}(n_0,\dots,n_m)$ determined by the conditions $\lambda_j\neq 0$, for all $j=1,\dots,N$.
Let $\mathbf{z}^0,\dots,\mathbf{z}^{L-1}$ be the zeros of the meromorphic function $\lambda$. Then the following (normalized) Abelian differential of the third kind:
$$
\Omega_{\mathbf{z}^0\mathbf{z}^j}(P)=\int_{\mathbf{z}^0}^{\mathbf{z}^j}W(P,Q), \quad  \Omega_{\mathbf{z}^0\infty^0}(P)
=\int_{\mathbf{z}^0}^{\infty^0}W(P,Q), \quad j=1,\dots, L-1,
$$
are quasi-homogeneous differentials of degree 0.\\
In this case, for all $t^A\in \big\{t^{j,\alpha}, v^j, s^j, \rho^k,u^k\big\}$ we have
\begin{align*}
e.t^A\big(\Omega_{\mathbf{z}^0\mathbf{z}^j}\big)&=\sum_{r=1}^N\underset{P_r}{{\rm res}}\frac{\phi_{t^A}(P)\Omega_{\mathbf{z}^0\mathbf{z}^j}(P)}{d\lambda(P)}\\
&=-\underset{\mathbf{z}^j}{{\rm res}}\frac{\phi_{t^A}(P)\Omega_{\mathbf{z}^0\mathbf{z}^j}(P)}{d\lambda(P)}
-\underset{\mathbf{z}^0}{{\rm res}}\frac{\phi_{t^A}(P)\Omega_{\mathbf{z}^0\mathbf{z}^j}(P)}{d\lambda(P)}\\
&=\phi_{t^A}(\mathbf{z}^0)-\phi_{t^A}(\mathbf{z}^j);\\
e.t^A\big(\Omega_{\mathbf{z}^0\infty^0}\big)&=-\underset{\mathbf{z}^0}{{\rm res}}\frac{\phi_{t^A}(P)\Omega_{\mathbf{z}^0\infty^0}(P)}{d\lambda(P)}=\phi_{t^A}(\mathbf{z}^0),
\end{align*}
where  $\phi_{t^A}(\mathbf{z}^j)$ is the evaluation (\ref{notation3}) of the differential $\phi_{t^A}$ at the zero $\mathbf{z}^j$.

\medskip

\noindent$\bullet$ Let $k=1,\dots,g$ and $\sigma, \nu$ be two positive integers. Then the following differentials
$$
\phi_{u^{k,\sigma}}(P):=\oint_{a_k}\big(\lambda(Q)\big)^{\sigma}W(P,Q)\quad\quad  \text{and}\quad\quad  \phi_{\widetilde{u}^{k,\nu}}(P):=\oint_{b_k}\big(\lambda(Q)\big)^{\nu}W(P,Q)
$$
are quasi-homogeneous of degrees $\sigma$ and $\nu$, respectively.\\
The differentials $\phi_{u^{k,\sigma}}$ and $\phi_{\widetilde{u}^{k,\nu}}$  are multivalued and their jumps are studied in (\ref{sigma-nu-jump}) (see Appendix 2 below).  Furthermore, if $t^A$ denotes  one of the operations $t^{i,\alpha}, v^i, s^i, \rho^k,u^k$ listed in (\ref{FBP-W})-(\ref{S-eta}), then the action of the unit vector field $e=\sum_{j}\partial_{\lambda_j}$  on the functions $t^A\big(\phi_{u^{k,\sigma}}\big)$ and $t^A\big(\phi_{\widetilde{u}^{k,\nu}}\big)$  is also described in Appendix 2 (formula (\ref{e-sigma-nu})).

\medskip

Let us introduce a notation for the set of the  quasi-homogeneous differentials mentioned so far.
\begin{equation}\label{QH-1}
\mathrm{QH}(E):=\Big\{\text{$e$-primary differentials (\ref{Primary-e})}\Big\}\bigcup
\Big\{\phi_{t^{i,n_i+\beta+1}}, \Omega_{\mathbf{z}^0\mathbf{z}^j}, \Omega_{\mathbf{z}^0\infty^0},\phi_{u^{k,\sigma}},\phi_{\widetilde{u}^{k,\nu}} \Big\}.
\end{equation}

\medskip

The above quasi-homogeneity terminology is justified by the following property.
\begin{Prop}  Let  $(C_g,\lambda)$ be  a point in the simple Hurwitz space $\widehat{\mathcal{H}}_{g, L}(n_0,\dots,n_m)$.
Let  $D_{E}$ be the vector field defined by:
$$
D_E:=\lambda(P)\partial_{\lambda(P)}+\textstyle \sum_j\lambda_j\partial_{\lambda_j}=\lambda(P)\partial_{\lambda(P)}+E.
$$
If $\omega_0$ is a quasi-homogeneous differential of degree $d_0\geq0$ (\ref{homogeneous diff}), then
\begin{equation}\label{Euler-DE}
D_E.\omega_0(P)=d_0\omega_0(P).
\end{equation}
\end{Prop}
\emph{Proof:}  Given a branched covering $(C_g,\lambda)$ of $\P^1$,
consider the family of branched coverings $\big\{(C_g, \lambda_{\varepsilon})\big\}_{\varepsilon}$  defined by
$$
\lambda_{\varepsilon}(P)=e^{\varepsilon}\lambda(P),
$$
where $\varepsilon\in \C$ and $|\varepsilon|$ is sufficiently small.
This is nothing but  the particular case of (\ref{Dilation-translation}) where $(\beta_1,\beta_2)=(1,0)$.\\
Under this transformation, the differential $\omega_0$ changes as follows:
\begin{align*}
\omega_0^{\varepsilon}(P)=c_0\int_{\ell_0}\big(\lambda_{\varepsilon}(Q)\big)^{d_0}W(P,Q)
=e^{d_0\varepsilon}\omega_0(P).
\end{align*}
Therefore
$$
D_E\omega_0(P)=\frac{d}{d{\varepsilon}}\omega_0^{\varepsilon}(P)\Big|_{\varepsilon=0}=d_0\omega_0(P).
$$

\fd

\begin{Remark}\label{Kernel-E}
Formula (\ref{Euler-DE}) allows us to consider  quasi-homogeneous differentials of degree $d_0\geq0$ as the differentials  belonging to the vector space
$Ker\big(D_E-d_0\big)$.\\
For example, we can view a linear combination of holomorphic differentials as a quasi-homogeneous differential of degree 0.
\end{Remark}

\begin{Prop}\label{Euler Prop1} Let $\omega_{0}$ be a  quasi-homogeneous  differential of degree $d_0\geq0$ belonging to the set $\mathrm{QH}(E)$ (\ref{QH-1})
and $t^A(\omega_0)$ be one of the $\eta(\omega_0)$-flat coordinates (\ref{FBP-W})-(\ref{S-eta}).\\
Then the vector field $E:=\sum_j\lambda_j\partial_{\lambda_j}$ acts on the function $t^A(\omega_0)$ as follows:
\begin{equation}\label{E-tA1}
E.t^A(\omega_0)=(d_A+d_0)t^A(\omega_0)+r_{\omega_0,\phi_{t^A}},
\end{equation}
where $d_A$ is the quasi-homogeneous degree of the primary differential $\phi_{t^A}$ (see relations (\ref{homogeneous diff}) and (\ref{primary-quasi})) and $r_{\omega_0,\phi_{t^A}}$ is the constant defined by
\begin{equation}\label{r0A}
r_{\omega_0,\phi_{t^A}}=\left\{
\begin{array}{ll}
\displaystyle r_{B,A}=r_{A,B} & \hbox{if} \quad \text{$\omega_0=\phi_{t^B}$ is one of the primary differentials (\ref{Primary-e})};\\
\\
\displaystyle -\frac{1}{n_0+1}\sum_{i=1}^m\delta_{\phi_{t^A},\phi_{s^i}},& \hbox{if} \quad \omega_0=\Omega_{\mathbf{z}^0\infty^0};\\
\\
0,& \hbox{} \quad\text{otherwise}.
\end{array}
\right.
\end{equation}
Here $r_{A,B}$ are given by (\ref{r-AB}). \\
Moreover, in flat coordinates (\ref{FBP-W})-(\ref{S-eta}) of the metric $\eta(\omega_0)$ (\ref{eta-def}), the vector field $E$ admits the following representation
\begin{equation}\label{E-tA2}
E=\sum_A\Big((d_A+d_0)t^A(\omega_0)+r_{\omega_0,\phi_{t^A}}\Big)\partial_{t^A(\omega_0)}.
\end{equation}
\end{Prop}
\emph{Proof:} As in Section  3.4, in order  to calculate the action of the vector field $E$ on the functions $t^A(\omega_0)$, we shall use the following identity:
$$
E.t^A(\omega_0):=\sum_j\lambda_j\partial_{\lambda_j}t^A(\omega_0)=\sum_j\underset{P_j}{\text{res}}\frac{\lambda(P)\phi_{t^A}(P)\omega_0(P)}{d\lambda(P)}.
$$
Here the second equality follows directly from (\ref{dual metric-R2}) and (\ref{dual metric-R4}) where we take $(\beta_1,\beta_2)=(1,0)$.\\
If $\omega_0=\phi_{t^B}$ is among the $e$-primary differentials (\ref{Primary-e}), then (\ref{E-tA1}) is nothing but the second equality in (\ref{Entries-g-eta}).\\
In the case where $\omega_0$ is one of the multivalued differentials $\big\{\phi_{u^{k,\sigma}},\phi_{\widetilde{u}^{k,\nu}}\big\}$, we  give a detailed calculation that proves (\ref{E-tA1}) in   Appendix 2 (formulas (\ref{u-k-sigma}) and (\ref{v-k-nu})). With the help of the residue theorem and identity (\ref{g-eta-trick}), we can straightforwardly obtain the remaining cases. \\
Finally,  we have
$$
E=\sum_j\lambda_j\partial_{\lambda_j}=\sum_j\lambda_j\bigg(\sum_A\Big(\partial_{\lambda_j}t^A(\omega_0)\Big)\partial_{t^A(\omega_0)}\bigg)
=\sum_A\Big(E.t^A(\omega_0)\Big)\partial_{t^A(\omega_0)}.
$$
Thus representation (\ref{E-tA2}) of the vector field $E$ is an immediate consequence of (\ref{E-tA1}).

\fd

\begin{Prop}\label{Euler Prop2}
Let $\omega_{0}(P):=c_0\int_{\ell_{0}}\left(\lambda(Q)\right)^{d_{0}}W(P,Q)$ be a fixed quasi-homogeneous  differential of degree $d_0\geq0$. \\
Then the vector field $E=\sum_j\lambda_j\partial_{\lambda_j}$ satisfies:
\begin{align*}
\begin{array}{llllll}
&[E,e]=-e;\\
\\
&[E,x\circ y]-[E,x]\circ y-x\circ[E,y]=x\circ y;\\
\\
&E.\omega_0(P_j)=(d_0-1/2)\omega_{0}(P_j),&\hbox{}\quad \quad j=1,\dots,N;\\
\\
&\big(Lie_{E}.\eta(\omega_{0})\big)\big(\partial_{\lambda_j},\partial_{\lambda_j}\big)
=(2d_{0}-3)\eta(\omega_{0})\big(\partial_{\lambda_j},\partial_{\lambda_j}\big),&\hbox{}\quad \quad j=1,\dots,N.
\end{array}
\end{align*}

\end{Prop}
\noindent\emph{Proof:}  By simple calculation, we can check that the first two relations hold true. The third one is a particular case of (\ref{Auxi2}) where we take  $h(\lambda)=c_0\lambda^{d_0}$   and $(\beta_1,\beta_2)=(1,0)$. \\
Finally, the action of the Lie derivative along the vector field $E$ on the diagonal metric $\eta(\omega_{0})=\frac{1}{2}\sum_j\omega_0(P_j)^2(d\lambda_j)^2$ is as follows
\begin{align*}
\big(Lie_{E}.\eta(\omega_0)\big)\big(\partial_{\lambda_j},\partial_{\lambda_j}\big)
&:=E.\Big(\eta(\omega_0)\big(\partial_{\lambda_j},\partial_{\lambda_j}\big)\Big)-2\eta(\omega_0)\big([E,\partial_{\lambda_j}],\partial_{\lambda_j}\big)\\
&=\big(E.\omega_0(P_j)\big)\omega_0(P_j)+\omega_0(P_j)^2\\
&=(2d_0-3)\frac{\omega_0(P_j)^2}{2}=(2d_0-3)\eta(\omega_0)\big(\partial_{\lambda_j},\partial_{\lambda_j}\big).
\end{align*}

\fd

\bigskip

Summarizing the above results,  Propositions \ref{Euler Prop1} and \ref{Euler Prop2} tell us that the vector field $E=\sum_{j=1}^N\lambda_j\partial_{\lambda_j}$ is covariantly  linear and obeys the conditions   (\ref{Euler vector def}) and thus it is an Euler vector field.

\subsubsection{A new formula for the prepotential}
A prepotential of a Frobenius manifold structure $(\mathcal{M},\circ, \prs{\cdot}{\cdot})$  is a function $\mathrm{F}$ of flat coordinates $\{t^A\}_A$ of the corresponding flat metric $\prs{\cdot}{\cdot}$  such that its third derivatives are given by the symmetric 3-tensor $\mathrm{c}$ from the Definition \ref{Frob Man def} of a Frobenius manifold:
\begin{align*}
\partial_{t^A}\partial_{t^B}\partial_{t^C}\mathrm{F}=\mathrm{c}\left(\partial_{t^A}, \partial_{t^B}, \partial_{t^C}\right)
=\prs{\partial_{t^A}\circ\partial_{t^B}}{\partial_{t^C}}.
\end{align*}

Let us start with the following  crucial observation concerning  the symmetric 3-tensor (\ref{c-FC-eta}):
\begin{Prop} Let $\omega_0$ be a differential satisfying  (\ref{Diff model}) and (\ref{Assymptions}) and $\eta(\omega_0)$ be the corresponding Darboux-Egoroff metric (\ref{eta-def}). Then  the  symmetric 3-tensor $\mathbf{c}_{\omega_0}$ (\ref{c-FC-eta}) associated with $\omega_0$ enjoys the following formula:
\begin{equation}\label{c-R1}
\mathbf{c}_{\omega_0}\big(\partial_{t^A(\omega_0)},\partial_{t^B(\omega_0)},\partial_{t^C(\omega_0)}\big)
=\partial_{t^C(\omega_0)}t^{A^{\prime}}\left(\phi_{t^{B^{\prime}}}\right)=\partial_{t^C(\omega_0)}t^{B^{\prime}}\left(\phi_{t^{A^{\prime}}}\right).
\end{equation}
\end{Prop}
\emph{Proof:} From  (\ref{flat-canonical relation}) we know that
$$
\partial_{t^C(\omega_0)}=\sum_j\frac{\phi_{t^{C^{\prime}}}(P_j)}{\omega_0(P_j)}\partial_{\lambda_j}.
$$
This relation, Rauch's formula (\ref{partial-t-W}) and expression (\ref{c-FC-eta}) for the tensor $\mathbf{c}_{\omega_0}$ imply that  the desired result
(\ref{c-R1}) holds true.

\fd

\medskip

In view of (\ref{c-R1}), given that one can permute the order of differentiation, we see that the 4-tensor
$$
\partial_{t^D(\omega_0)}\mathbf{c}_{\omega_0}\big(\partial_{t^A(\omega_0)},\partial_{t^B(\omega_0)},\partial_{t^C(\omega_0)}\big)
$$
is totally symmetric in the four components. As a consequence, according to the above results and Definitions \ref{Frob Man def} and \ref{Flatness-e},  we reach  the following conclusions.
\begin{Cor}\label{Fro-C} Assume that  $\omega_0$ is a quasi-homogeneous differential (\ref{homogeneous diff}) of degree $d_0\geq 0$. Let  $\mathcal{M}_{\omega_0}$ be the open subset of the simple Hurwitz space  $\widehat{\mathcal{H}}_{g, L}(n_0,\dots,n_m)$ determined by the conditions:
$$
\omega_0(P_j)\neq0,\quad \forall\ j=1,\dots,N.
$$
Then the open set $\mathcal{M}_{\omega_0}$ carries  a Frobenius manifold structure given by the flat metric $\eta(\omega_0)$ (\ref{eta-def}),  multiplication law (\ref{multiplication}) and Euler vector field (\ref{E-tA2}).\\
If the differential $\omega_0$ is additionally chosen among primary differentials (\ref{Primary-e}) with respect to the unit vector field $e$, then the above Frobenius manifold structure becomes  normalized in the sense of Definition \ref{Flatness-e}.
\end{Cor}

The following theorem is the central result of this work. It establishes a new formula for prepotentials associated with the obtained Frobenius manifold structures that are induced by  quasi-homogenous differentials and described in the above corollary.
\begin{Thm}\label{Prep-eta-thm}
Let  $\omega_0$ be a given  quasi-homogeneous  differential (\ref{homogeneous diff})  of degree $d_0\geq0$  satisfying  (\ref{Assymptions}). Let $\big(\mathcal{M}_{\omega_0}, \eta(\omega_0), \circ, e, E\big)$ be the  Frobenius manifold structure from Corollary \ref{Fro-C}. Let also
\begin{description}
\item[i)] $E.t^A(\omega_0)$ be the affine function $E.t^A(\omega_0):=(d_0+d_A)t^A(\omega_0)+r_{\omega_0,\phi_{t^A}}$, for all $A$ (see (\ref{E-tA1}));
\item[ii)]  $t^{A^{\prime}}$ denote the $\eta$-dual flat coordinate of $t^A$ introduced in (\ref{duality})-(\ref{duality-picture});
\item[iii)] $\big\{\phi_{t^A}\big\}_{A}$ be  the family of the $N$ primary differentials (\ref{Primary-e}) of Dubrovin;
\item[iv)] $r_{A,B}$ be  the constants given by (\ref{r-AB}).
\end{description}
Then the following function is a  prepotential of the  Frobenius manifold   $\big(\mathcal{M}_{\omega_0}, \eta(\omega_0), \circ, e, E\big)$:
\begin{equation}\label{Prep-eta}
\begin{split}
\mathbf{F}_{\omega_0}&=\frac{1}{4(1+d_0)}\sum_{A,B}\frac{\left(E.t^A(\omega_0)\right)\left(E.t^B(\omega_0)\right)}{1+d_0+d_{A^{\prime}}}
\Big(t^{A^{\prime}}(\phi_{t^{B^{\prime}}})+t^{B^{\prime}}(\phi_{t^{A^{\prime}}})\Big)\\
&\quad-\frac{1}{4(1+d_0)}\sum_{A,B}\bigg(2\frac{E.t^A(\omega_0)}{1+d_0+d_{A^{\prime}}}+t^A(\omega_0)\bigg)r_{A^{\prime},B^{\prime}}t^B(\omega_0).
\end{split}
\end{equation}
Moreover, the Hessian matrix of the function $\mathbf{F}_{\omega_0}$ does not depend on the chosen  quasi-homogeneous  differential $\omega_0$ and its entries are  given by
\begin{equation}\label{Prep-eta-H}
\partial_{t^A(\omega_0)}\partial_{t^B(\omega_0)}\mathbf{F}_{\omega_0}
=\frac{1}{2}\left(t^{A^{\prime}}\left(\phi_{t^{B^{\prime}}}\right)+t^{B^{\prime}}\left(\phi_{t^{A^{\prime}}}\right)\right).
\end{equation}
\end{Thm}

\medskip

The proof of this theorem is a direct consequence of  relation (\ref{c-R1}) combined  with  upcoming results
(\ref{FA-derivation}) and (\ref{FA-F2}).
\begin{Prop}
Consider the $N$ functions $\big\{\mathbf{F}_{A,\omega_0}\big\}_{A}$, with
\begin{equation}\label{FA}
\mathbf{F}_{A,\omega_0}=\frac{1}{2\big(1+d_0+d_{A^{\prime}}\big)}\sum_B\bigg(\left(E.t^{B}(\omega_0)\right)\Big(t^{A^{\prime}}\left(\phi_{t^{B^{\prime}}}\right)
+t^{B^{\prime}}\left(\phi_{t^{A^{\prime}}}\right)\Big)-2r_{A^{\prime},B^{\prime}}t^B(\omega_0)\bigg).
\end{equation}
Then the following statements  hold:
\begin{description}
\item[1.] For any $B$, we have
\begin{equation}\label{FA-derivation}
\partial_{t^B(\omega_0)}\mathbf{F}_{A,\omega_0}=\frac{1}{2}\Big(t^{A^{\prime}}\left(\phi_{t^{B^{\prime}}}\right)+t^{B^{\prime}}\left(\phi_{t^{A^{\prime}}}\right)\Big)
=\partial_{t^A(\omega_0)}\mathbf{F}_{B,\omega_0}.
\end{equation}
\item[2.] The Euler vector field (\ref{E-tA2}) acts on the function $\mathbf{F}_{A,\omega_0}$ as follows:
\begin{equation}\label{FA-Euler}
E.\mathbf{F}_{A,\omega_0}=\left(1+d_0+d_{A^{\prime}}\right)\mathbf{F}_{A,\omega_0}+\sum_Br_{A^{\prime},B^{\prime}}t^B(\omega_0).
\end{equation}
\item[3.] The function $\mathbf{F}_{\omega_0}$ defined by  (\ref{Prep-eta}) admits the following representation involving the functions  $\mathbf{F}_{A,\omega_0}$:
\begin{equation}\label{FA-F1}
\mathbf{F}_{\omega_0}=\frac{1}{2(1+d_0)}\sum_{A}\big(E.t^A(\omega_0)\big)\mathbf{F}_{A,\omega_0}
-\frac{1}{4(1+d_0)}\sum_{A,B}r_{A^{\prime},B^{\prime}}t^A(\omega_0)t^B(\omega_0).
\end{equation}
\item[4.] The partial derivatives  of $\mathbf{F}_{\omega_0}$ are:
\begin{equation}\label{FA-F2}
\partial_{t^A(\omega_0)}\mathbf{F}_{\omega_0}=\mathbf{F}_{A,\omega_0}.
\end{equation}
\end{description}
\end{Prop}
\emph{Proof:} Using  the symmetry property  of the tensor $\mathbf{c}_{\omega_0}$ following from expression
(\ref{c-FC-eta}) together  with  the quasi-homogeneity property (\ref{Entries-g-eta})  of the functions $t^{A}(\phi_{t^B})$, i.e. $E.t^{A}(\phi_{t^B})=(d_A+d_B) t^A(\phi_{t^B})+r_{A,B}$ for the Euler vector field $E$, we deduce that
\begin{align*}
&2\big(1+d_0+d_{A^{\prime}}\big)\partial_{t^B(\omega_0)}\mathbf{F}_{A,\omega_0}\\
&=(d_B+d_0)\Big(t^{A^{\prime}}\left(\phi_{t^{B^{\prime}}}\right)+t^{B^{\prime}}\left(\phi_{t^{A^{\prime}}}\right)\Big)-2r_{A^{\prime},B^{\prime}}
+\sum_D\left(E.t^{D}(\omega_0)\right)\partial_{t^B(\omega_0)}\Big(t^{A^{\prime}}\left(\phi_{t^{D^{\prime}}}\right)+t^{D^{\prime}}\left(\phi_{t^{A^{\prime}}}\right)\Big)\\
&=(d_B+d_0)\Big(t^{A^{\prime}}\left(\phi_{t^{B^{\prime}}}\right)+t^{B^{\prime}}\left(\phi_{t^{A^{\prime}}}\right)\Big)-2r_{A^{\prime},B^{\prime}}
+\sum_D\left(E.t^{D}(\omega_0)\right)\partial_{t^D(\omega_0)}\Big(t^{A^{\prime}}\left(\phi_{t^{B^{\prime}}}\right)+t^{B^{\prime}}\left(\phi_{t^{A^{\prime}}}\right)\Big)\\
&=(d_B+d_0)\Big(t^{A^{\prime}}\left(\phi_{t^{B^{\prime}}}\right)+t^{B^{\prime}}\left(\phi_{t^{A^{\prime}}}\right)\Big)-2r_{A^{\prime},B^{\prime}}
+E.t^{A^{\prime}}\left(\phi_{t^{B^{\prime}}}\right)+E.t^{B^{\prime}}\left(\phi_{t^{A^{\prime}}}\right)\\
&=\big(d_B+d_0+d_{B^{\prime}}+ d_{A^{\prime}}\big)\Big(t^{A^{\prime}}\left(\phi_{t^{B^{\prime}}}\right)+t^{B^{\prime}}\left(\phi_{t^{A^{\prime}}}\right)\Big)\\
&=\big(1+d_0+ d_{A^{\prime}}\big)\Big(t^{A^{\prime}}\left(\phi_{t^{B^{\prime}}}\right)+t^{B^{\prime}}\left(\phi_{t^{A^{\prime}}}\right)\Big),
\end{align*}
where we have used the key duality formula $d_B+d_{B^{\prime}}=1$ in the last line.\\
Now, by employing (\ref{FA}) and (\ref{FA-derivation}) and  expression (\ref{E-tA2}) for the Euler vector field, we directly arrive at (\ref{FA-Euler}).  Furthermore, replacing the function $\mathbf{F}_{A,\omega_0}$ by its expression (\ref{FA}) in the left hand side of formula (\ref{FA-F1}) and comparing the resulting with  (\ref{Prep-eta}), we conclude that (\ref{FA-F1}) holds.\\
Lastly, we are going to look at (\ref{FA-F2}). Differentiating in (\ref{FA-F1}) with respect to $t^A(\omega_0)$, we obtain
\begin{align*}
\partial_{t^A(\omega_0)}\mathbf{F}_{\omega_0}
&=\frac{d_A+d_0}{2(1+d_0)}\mathbf{F}_{A,\omega_0}+\frac{1}{2(1+d_0)}\sum_{B}\big(E.t^B(\omega_0)\big)\partial_{t^A(\omega_0)}\mathbf{F}_{B,\omega_0}
-\frac{1}{2(1+d_0)}\sum_{B}r_{A^{\prime},B^{\prime}}t^B(\omega_0)\\
&=\frac{d_A+d_0}{2(1+d_0)}\mathbf{F}_{A,\omega_0}+\frac{1}{2(1+d_0)}\sum_{B}\big(E.t^B(\omega_0)\big)\partial_{t^B(\omega_0)}\mathbf{F}_{A,\omega_0}
-\frac{1}{2(1+d_0)}\sum_{B}r_{A^{\prime},B^{\prime}}t^B(\omega_0)\\
&=\frac{d_A+d_0}{2(1+d_0)}\mathbf{F}_{A,\omega_0}+\frac{1}{2(1+d_0)}E.\mathbf{F}_{A,\omega_0}-\frac{1}{2(1+d_0)}\sum_{B}r_{A^{\prime},B^{\prime}}t^B(\omega_0)\\
&=\frac{d_A+d_{A^{\prime}}+2d_0+1}{2(1+d_0)}\mathbf{F}_{A,\omega_0}\\
&=\mathbf{F}_{A,\omega_0};
\end{align*}
where we have used  relation (\ref{FA-derivation})  in the second equality, (\ref{E-tA2}) in the third equality, (\ref{FA-Euler}) in forth one and once again the duality relation $d_A+d_{A^{\prime}}=1$ in the last equality. The proof of the  proposition is now complete.

\fd

\noindent\emph{Proof of Theorem \ref{Prep-eta-thm}:} Differentiating three times the function $\mathbf{F}_{\omega_0}$ given by (\ref{Prep-eta}) and using successively (\ref{FA-F2}), (\ref{FA-derivation}) and (\ref{c-R1}), we obtain
\begin{align*}
\partial_{t^C(\omega_0)}\partial_{t^B(\omega_0)}\partial_{t^A(\omega_0)}\mathbf{F}_{\omega_0}
&=\partial_{t^C(\omega_0)}\partial_{t^B(\omega_0)}\mathbf{F}_{A, \omega_0}
=\frac{1}{2}\partial_{t^C(\omega_0)}\Big(t^{A^{\prime}}\left(\phi_{t^{B^{\prime}}}\right)+t^{B^{\prime}}\left(\phi_{t^{A^{\prime}}}\right)\Big)\\
&=\mathbf{c}_{\omega_0}\big(\partial_{t^A(\omega_0)},\partial_{t^B(\omega_0)},\partial_{t^C(\omega_0)}\big).
\end{align*}

\fd

\begin{Cor}
The prepotential $\mathbf{F}_{\omega_0}$ in (\ref{Prep-eta}) as well as its first,  second and third partial derivatives are  quasi-homogeneous functions, with respect to the Euler vector field $E$ given by  (\ref{E-tA2}), in the sense of (\ref{quasihomog-def}).\\
More precisely, for the quantities
$$
\mathbf{F}_{A, \omega_0}:=\partial_{t^A(\omega_0)}\mathbf{F}_{\omega_0},\quad\quad \mathbf{F}_{AB, \omega_0}
:=\partial_{t^A(\omega_0)}\partial_{t^B(\omega_0)}\mathbf{F}_{\omega_0},
\quad \quad \mathbf{F}_{ABC, \omega_0}:=\partial_{t^A(\omega_0)}\partial_{t^B(\omega_0)}\partial_{t^C(\omega_0)}\mathbf{F}_{\omega_0},
$$
we have
\begin{equation}\label{E-F-FA}
\begin{split}
&E.\mathbf{F}_{\omega_0}=\textstyle 2(1+d_0)\mathbf{F}_{\omega_0}+\frac{1}{2}\sum_{A,B}r_{A^{\prime},B^{\prime}}t^A(\omega_0)t^B(\omega_0);\\
&E.\mathbf{F}_{A, \omega_0}=\textstyle \left(1+d_0+d_{A^{\prime}}\right)\mathbf{F}_{A, \omega_0}+\sum_Br_{A^{\prime},B^{\prime}}t^B(\omega_0);\\
&E.\mathbf{F}_{AB, \omega_0}=\left(d_{A^{\prime}}+d_{B^{\prime}}\right)\mathbf{F}_{AB, \omega_0}+r_{A^{\prime},B^{\prime}};\\
&E.\mathbf{F}_{ABC, \omega_0}=\big(2-d_0-d_A-d_B-d_C\big)\mathbf{F}_{ABC, \omega_0}.
\end{split}
\end{equation}
Here $r_{A,B}$ are the constants given by (\ref{r-AB}).
\end{Cor}
\emph{Proof:} In view of (\ref{FA-F1}) and (\ref{FA-F2}), we obtain
\begin{align*}
E.\mathbf{F}_{\omega_0}&=\sum_{A}\big(E.t^A(\omega_0)\big)\partial_{t^A(\omega_0)}\mathbf{F}_{\omega_0}
=\sum_{A}\big(E.t^A(\omega_0)\big)\mathbf{F}_{A,\omega_0}\\
&=2(1+d_0)\mathbf{F}_{\omega_0}+\frac{1}{2}\sum_{A,B}r_{A^{\prime},B^{\prime}}t^A(\omega_0)t^B(\omega_0).
\end{align*}
The second relation in (\ref{E-F-FA}) is nothing but (\ref{FA-Euler}).
In addition, using (\ref{Prep-eta-H})  and (\ref{Entries-g-eta}), we see that the second partial derivatives $\mathbf{F}_{AB, \omega_0}$   of $\mathbf{F}_{\omega_0}$
 are also quasi-homogeneous functions.\\
Now let us deal with the last equality in (\ref{E-F-FA}). By employing Theorem \ref{Prep-eta-thm}, relation (\ref{c-R1}), the duality formula $d_A+d_{A^{\prime}}=1$ as well as expression (\ref{E-tA2}) for the Euler vector field, we can write
\begin{align*}
E.\mathbf{F}_{ABC, \omega_0}&=E.\mathbf{c}_{\omega_0}\big(\partial_{t^A(\omega_0)},\partial_{t^B(\omega_0)},\partial_{t^C(\omega_0)}\big)
=E. \partial_{t^C(\omega_0)}t^{A^{\prime}}\left(\phi_{t^{B^{\prime}}}\right)\\
&= \partial_{t^C(\omega_0)}\Big(E.t^{A^{\prime}}\left(\phi_{t^{B^{\prime}}}\right)\Big)+[E, \partial_{t^C(\omega_0)}].t^{A^{\prime}}\left(\phi_{t^{B^{\prime}}}\right)\\
&=(d_{A^{\prime}}+d_{B^{\prime}})\partial_{t^C(\omega_0)}t^{A^{\prime}}\left(\phi_{t^{B^{\prime}}}\right)
-(d_0+d_C)\partial_{t^C(\omega_0)}t^{A^{\prime}}\left(\phi_{t^{B^{\prime}}}\right)\\
&=\big(2-d_0-d_A-d_B-d_C\big)\mathbf{F}_{ABC, \omega_0}.
\end{align*}

\fd

\medskip

The aim of the upcoming  result is to improve formula (\ref{Prep-eta}) by providing further precisions for some terms.
\begin{Prop} The prepotential $\mathbf{F}_{\omega_0}$ in  (\ref{Prep-eta}) can be written in the following form:
\begin{equation}\label{Prep-eta2}
\begin{split}
&\mathbf{F}_{\omega_0}=\frac{1}{2(1+d_0)}\sum_{A,B}\frac{\big((d_0+d_A)t^A(\omega_0)+r_{\omega_0,\phi_{t^A}}\big)
\big((d_0+d_B)t^B(\omega_0)+r_{\omega_0,\phi_{t^B}}\big)}{1+d_0+d_{A^{\prime}}}t^{A^{\prime}}(\phi_{t^{B^{\prime}}})\\
&\quad +\frac{\log(-1)}{4}\sum_{i,j=1, i\neq j}^mv^i(\omega_0)v^j(\omega_0)
-\frac{3}{4(1+d_0)}\sum_{i,j=1}^m\bigg(\frac{1}{n_0+1}+\delta_{ij}\frac{1}{n_i+1}\bigg)v^i(\omega_0)v^j(\omega_0),
\end{split}
\end{equation}
where $v^i(\omega_0)$ is the function $v^i(\omega_0):= \underset{\infty^i}{{\rm res}\ }\lambda(P)\omega_0(P)$.
Moreover, the first relation in (\ref{E-F-FA}) becomes
\begin{equation}\label{Prep-eta-homog}
E.\mathbf{F}_{\omega_0}=2(1+d_0)\mathbf{F}_{\omega_0}+\frac{1}{2}\sum_{i,j=1}^m\bigg(\frac{1}{n_0+1}+\delta_{ij}\frac{1}{n_i+1}\bigg)v^i(\omega_0)v^j(\omega_0).
\end{equation}
\end{Prop}
\emph{Proof:} The proof of formula (\ref{Prep-eta2}) is  based on the symmetry  property (\ref{Fubini-flat}) of the functions $t^A(\phi_{t^B})$ and the fact that the constants $r_{A,B}$, given by  (\ref{r-AB}), are such that $r_{A,B}=0$ whenever at least one of the operations $t^A$ and $t^B$ is not of type $s^i$. Here $t^A$, $t^B$ and $s^i$  are among the five types of operations given by (\ref{FBP-W}) and (\ref{S-eta}) with  $s^i:=p.v.\int_{\infty^0}^{\infty^i}$. \\
Denote by $\widetilde{F}_{\omega_0}$ (resp. $\widehat{F}_{\omega_0}$ ) the first (resp. second)  sum in expression (\ref{Prep-eta}) for the prepotential $\mathbf{F}_{\omega_0}$. That is we can write $\mathbf{F}_{\omega_0}=\widetilde{F}_{\omega_0}-\widehat{F}_{\omega_0}$.\\
Let us first mention the following remarks:
\begin{enumerate}
\item By (\ref{duality-picture}) we have $\delta_{t^{A^{\prime}},s^i}=\delta_{t^{A},v^i}$ (the Kronecker delta symbol).
\item Let $d_A:={\rm{deg}}\big(\phi_{t^A}\big)$ be the degree of the quasi-homogeneous (and primary) differential $\phi_{t^A}$ defined by (\ref{homogeneous diff}). According to (\ref{primary-quasi}), we have
$$
\delta_{t^A,v^i}d_{A^{\prime}}={\rm{deg}}\big(\phi_{{v^i}^{\prime}}\big)={\rm{deg}}\big(\phi_{s^i}\big)=0.
$$
\item $v^i(\omega_0)$ is an eigenfunction of the Euler vector field $E$ with $E.v^{i}(\omega_0)=(d_0+1)v^i(\omega_0)$.
\end{enumerate}
Now, using these remarks as well  as the symmetry relation (\ref{Fubini-flat}), we deduce that
\begin{align*}
&\widetilde{F}_{\omega_0}-\frac{1}{2(1+d_0)}\sum_{A,B}\frac{\left(E.t^A(\omega_0)\right)\left(E.t^B(\omega_0)\right)}{1+d_0+d_{A^{\prime}}}
t^{A^{\prime}}(\phi_{t^{B^{\prime}}})\\
&=\frac{1}{4(1+d_0)}\sum_{A,B}\frac{\left(E.t^A(\omega_0)\right)\left(E.t^B(\omega_0)\right)}{1+d_0+d_{A^{\prime}}}
\bigg(\delta_{t^{A^{\prime}},s^i}\delta_{t^{B^{\prime}},s^j}(1-\delta_{ij})\log(-1)\bigg)\\
&=\frac{\log(-1)}{4(1+d_0)}\sum_{i,j=1}^m\big(1-\delta_{ij}\big)\frac{\left(E.v^i(\omega_0)\right)\left(E.v^j(\omega_0)\right)}{1+d_0+deg(\phi_{{v^i}^{\prime}})}\\
&=\frac{\log(-1)}{4}\sum_{i,j=1}^m\big(1-\delta_{ij}\big)v^i(\omega_0)v^j(\omega_0).
\end{align*}
On the other hand, by employing once again the aforementioned remarks and the precise values of the constants $r_{A^{\prime},B^{\prime}}$ (\ref{r-AB}):
$$
r_{A^{\prime},B^{\prime}}=\sum_{i,j=1}^m\delta_{t^{A^{\prime}},s^i}\delta_{t^{B^{\prime}},s^j}\Big(\frac{1}{n_0+1}+\delta_{ij}\frac{1}{n_i+1}\Big)
=\sum_{i,j=1}^m\delta_{t^A,v^i}\delta_{t^B,v^j}\Big(\frac{1}{n_0+1}+\delta_{ij}\frac{1}{n_i+1}\Big),
$$
we arrive at
\begin{align*}
\widehat{F}_{\omega_0}&:=\frac{1}{4(1+d_0)}\sum_{A,B}\bigg(2\frac{E.t^A(\omega_0)}{1+d_0+d_{A^{\prime}}}+t^A(\omega_0)\bigg)r_{A^{\prime},B^{\prime}}t^B(\omega_0)\\
&=\frac{3}{4(1+d_0)}\sum_{i,j=1}^m\bigg(\frac{1}{n_0+1}+\delta_{ij}\frac{1}{n_i+1}\bigg)v^i(\omega_0)v^j(\omega_0).
\end{align*}
Thus, we obtain (\ref{Prep-eta2}). On the other hand, by using the first relation in (\ref{E-F-FA}) and the mentioned values of the constants $r_{A^{\prime},B^{\prime}}$, we conclude that the function $\mathbf{F}_{\omega_0}$ satisfies the quasi-homogeneity property (\ref{Prep-eta-homog}).

\fd

\begin{Remark}
\end{Remark}
\noindent $\bullet$  According to Theorem \ref{Prep-eta-thm} and relation (\ref{Prep-eta-homog}),  for each quasi-homogeneous differential $\omega_0$ of degree $d_0\geq0$,  the function $\mathbf{F}_{\omega_0}$ (\ref{Prep-eta}) is a $(2+2d_0)$-quasi-homogeneous solution to the  generalized WDVV equations (\ref{WDVV equation}).
\medskip

\noindent $\bullet$ By using (\ref{Prep-eta2}) and omitting the quadratic terms, we deduce that
\begin{equation}\label{Prep-eta3}
\mathcal{F}_{\omega_0}=\frac{1}{2(1+d_0)}\sum_{A,B}\frac{\big((d_0+d_A)t^A(\omega_0)+r_{\omega_0,\phi_{t^A}}\big)
\big((d_0+d_B)t^B(\omega_0)+r_{\omega_0,\phi_{t^B}}\big)}{1+d_0+d_{A^{\prime}}}t^{A^{\prime}}(\phi_{t^{B^{\prime}}})
\end{equation}
is also a prepotential of the described Frobenius manifold determined by the quasi-homogeneous differential $\omega_0$.

\medskip

\noindent $\bullet$ In formulas (\ref{Prep-eta}) and (\ref{Prep-eta2}), we have chosen to add quadratic terms  in order to construct  prepotentials that share the Hessian matrix (\ref{Prep-eta-H}) when they correspond to Frobenius manifold structures of the same family.\\
This property could be useful in various situation. We refer to Section 7 for some examples of applications.
\medskip

\noindent $\bullet$ According to the symmetry relations (\ref{Fubini-flat}) and the relation $\delta_{t^{A^{\prime}},s^i}=\delta_{t^{A},v^i}$, formula (\ref{Prep-eta-H}) can be rewritten as follows:
\begin{equation}\label{Hessian}
t^{A^{\prime}}\left(\phi_{t^{B^{\prime}}}\right)=\partial_{t^A(\omega_0)}\partial_{t^B(\omega_0)}\mathbf{F}_{\omega_0}-
\frac{\log(-1)}{2}\sum_{i,j=1}^m(1-\delta_{ij})\delta_{t^A,v^i}\delta_{t^B,v^j}.
\end{equation}

\medskip

\noindent $\bullet$ Assume that the simple Hurwitz space is of type $\mathcal{H}_{g,L}(n_0)$, i.e. all the considered  coverings $\lambda$ have a unique pole $\infty^0$
of order $n_0+1=L$ ($L$ is the number of sheets).
\begin{itemize}
\item[i)] The list of primary differentials (\ref{Primary-e}) becomes $\big\{\phi_{t^{0,\alpha}}, \phi_{\rho^k}=\omega_k, \phi_{u^k}\big\}$.
\item[ii)] We only have three types of flat coordinates among those in (\ref{FBP-W})-(\ref{S-eta}) determined   by the operations:
$$
t^{0,\alpha},\quad \rho^k,\quad u^k,\quad \quad \alpha=1,\dots,n_0,\  \  k=1,\dots,g.
$$
\item[iii)] The set $\mathrm{QH}(E)$ (\ref{QH-1}) consisting of quasi-homogeneous differential reduces to
$$
\Big\{\phi_{t^{0,\alpha}},\quad \phi_{\rho^k}=\omega_k,\quad \phi_{u^k},\quad  \phi_{t^{0,n_0+\beta+1}},\quad \Omega_{\mathbf{z}^0\mathbf{z}^j},\quad  \phi_{u^{k,\sigma}},\quad \phi_{\widetilde{u}^{k,\nu}}\Big\}.
$$
\item[iv)] When $\omega_0$ is among these quasi-homogeneous  differentials, then, by (\ref{E-tA1}) and (\ref{r0A}), the flat functions $t^A(\omega_0)$ are eigenfunctions of the Euler vector field $E$ with eigenvalues $d_0+d_A$.\\
 As a consequence, formulas (\ref{Prep-eta2}) and (\ref{Prep-eta3}) coincide and reduce to the following
\begin{equation}\label{Prep-eta4}
\mathbf{F}_{\omega_0}=\mathcal{F}_{\omega_0}
=\frac{1}{2(1+d_0)}\sum_{A,B}\frac{(d_0+d_A)(d_0+d_B)}{1+d_0+d_{A^{\prime}}}t^A(\omega_0)t^B(\omega_0)t^{A^{\prime}}(\phi_{t^{B^{\prime}}}).
\end{equation}
\end{itemize}
Note that the described situation in the last point is realized by the genus zero Hurwitz space $\mathcal{H}_{0,n+1}(n)$ of polynomial functions
$$
\lambda(P)=z_P^{n+1}+a_nz_P^{n-1}+\dots+a_1,
$$
where $a_1,\dots,a_n$, $n\geq 3$, are complex numbers chosen in such a way the $n$ zeros $\lambda'(P)$ are distinct.\\

\medskip

\noindent $\bullet$ Assume that $\omega_0$ is a linear combination of quasi-homogeneous differentials of the same degree $d_0\geq0$ such that the differential $\omega_0$ fulfills the conditions in (\ref{Assymptions}). According  to Remark \ref{Kernel-E}, $\omega_0$ is again a quasi-homogeneous differential of degree $d_0$. Let  $t^A(\omega_0)$ be function from the set $\Big\{t^{i,\alpha}(\omega_0), v^i(\omega_0), s^i(\omega_0), \rho^k(\omega_0), u^k(\omega_0)\Big\}$ defined by (\ref{FBP-W})-(\ref{S-eta}).  Since  the Euler vector field $E=\sum_j\lambda_j\partial_{\lambda_j}$ acts on
$t^A(\omega_0)$ in analogous way as in (\ref{E-tA1}), it follows that the quasi-homogeneous differential $\omega_0$ gives  a Frobenius manifold structure.\\
As example, we can produce a quasi-homogeneous differential $\omega_0$ of \\
- degree 0 by taking $\omega_0$ as a linear combination of holomorphic differentials $\omega_k=\phi_{\rho^k}$ and Abelian differentials of the third kind $\phi_{s^i}=\Omega_{\infty^0\infty^i}$;\\
-degree 1 by taking $\omega_0$ in the  vector space generating by the differentials $\phi_{v^i}, \phi_{u^k}$ and $\phi_{\widetilde{u}^k}(P):=\oint_{b_k}\lambda(Q)W(P,Q)$.

\subsubsection{Prepotentials associated with holomorphic differentials: examples in high dimension}
In this subsection we  further investigate the quasi-homogeneous solutions (\ref{Prep-eta})-(\ref{Prep-eta2}) to the WDVV equations induced by normalized holomorphic differentials $\omega_1,\dots,\omega_g$, $g\geq 1$. \\
Note that, by Corollary \ref{Rmk-primary} and Proposition \ref{primary-quasi},   normalized holomorphic differentials  are all primary with respect to the unit vector field $e=\sum_{j=1}^N\partial_{\lambda_j}$ and quasi-homogeneous of degree 0.\\
Let us begin with the following result which gives the general explicit form of the prepotential $\mathbf{F}_{\omega_0}$ (\ref{Prep-eta})-(\ref{Prep-eta2}) corresponding to the differential $\omega_0\in \big\{\omega_1,\dots,\omega_g\big\}$.
\begin{Thm} Assume that $g\geq 1$ and  $\omega_0$ is one of the normalized holomorphic differentials. Then the prepotential associated with $\omega_0$ is given by:
\begin{equation}\label{Prep-holom}
\begin{split}
\mathbf{F}_{\omega_0}
&=\frac{1}{2}\sum_{i,j=0}^m\sum_{\alpha=1}^{n_i}\sum_{\beta=1}^{n_j}\frac{\alpha\beta}{(2n_i+2-\alpha)(n_j+1)}t^{i,\alpha}(\omega_0)
t^{j,\beta}(\omega_0)t^{i,n_i+1-\alpha}(\phi_{t^{j,n_j+1-\beta}})\\
&\quad +\frac{1}{2}\sum_{i=0}^m\sum_{\alpha=1}^{n_i}\sum_{j=1}^m\Big(\frac{\alpha}{2n_i+2-\alpha}+\frac{\alpha}{n_i+1}\Big)
t^{i,\alpha}(\omega_0)v^j(\omega_0)t^{i,n_i+1-\alpha}(\phi_{s^j})\\
&\quad +\frac{1}{2}\sum_{i=0}^m\sum_{\alpha=1}^{n_i}\sum_{k=1}^g\Big(\frac{\alpha}{2n_i+2-\alpha}+\frac{\alpha}{n_i+1}\Big)
t^{i,\alpha}(\omega_0)u^k(\omega_0)t^{i,n_i+1-\alpha}(\phi_{\rho^k})\\
&\quad+\frac{1}{2}\sum_{i,j=1}^mv^i(\omega_0)v^j(\omega_0)s^{i}(\phi_{s^j})+\sum_{i=1}^m\sum_{k=1}^gv^i(\omega_0)u^k(\omega_0)s^{i}(\phi_{\rho^k})
+\frac{1}{2}\sum_{k,l=1}^gu^k(\omega_0)u^l(\omega_0)\rho^k(\phi_{\rho^l})\\
&\quad +\frac{\log(-1)}{4}\sum_{i,j=1, i\neq j}^mv^i(\omega_0)v^j(\omega_0)
-\frac{3}{4}\sum_{i,j=1}^m\bigg(\frac{1}{n_0+1}+\delta_{ij}\frac{1}{n_i+1}\bigg)v^i(\omega_0)v^j(\omega_0).
\end{split}
\end{equation}
In addition, this prepotential  satisfies
$$
E_{\omega_0}.\mathbf{F}_{\omega_0}=2\mathbf{F}_{\omega_0}
+\frac{1}{2}\sum_{i,j=1}^m\Big(\frac{1}{n_0+1}+\delta_{ij}\frac{1}{n_i+1}\Big)v^i(\omega_0)v^j(\omega_0),
$$
where $E_{\omega_0}$ is the following Euler vector field:
\begin{equation}\label{E-holom}
E_{\omega_0}=\sum_{i=0}^m\sum_{\alpha=1}^{n_i}\frac{\alpha}{n_i+1}t^{i,\alpha}(\omega_0)\partial_{t^{i,\alpha}(\omega_0)}+
\sum_{i=1}^mv^i(\omega_0)\partial_{v^i(\omega_0)}+ \sum_{k=1}^gu^k(\omega_0)\partial_{u^k(\omega_0)}.
\end{equation}

\end{Thm}
\emph{Proof:}
Since the holomorphic differential $\omega_0$ is of degree $d_0=0$, formulas (\ref{Entries-g-eta}) and (\ref{E-tA1}) tell us that
$$
E.t^A(\omega_0)=d_At^A(\omega_0),\quad\quad \text{for all $A$}.
$$
Accordingly, formula  (\ref{Prep-eta2}) reads as:
\begin{align*}
&\mathbf{F}_{\omega_0}-\frac{\log(-1)}{4}\sum_{i,j=1, i\neq j}^mv^i(\omega_0)v^j(\omega_0)
+\frac{3}{4}\sum_{i,j=1}^m\bigg(\frac{1}{n_0+1}+\delta_{ij}\frac{1}{n_i+1}\bigg)v^i(\omega_0)v^j(\omega_0)\\
&=\frac{1}{2}\sum_{t^A,t^B}\frac{d_Ad_B}{2-d_A}t^A(\omega_0)t^B(\omega_0)t^{A^{\prime}}(\phi_{t^{B^{\prime}}})\\
&=\frac{1}{2}\sum_{t^A,t^B\in \{t^{i,\alpha}, v^i,u^k\}}\frac{d_Ad_B}{2-d_A}t^A(\omega_0)t^B(\omega_0)t^{A^{\prime}}(\phi_{t^{B^{\prime}}}),
\end{align*}
where we have used the relation $d_{A^{\prime}}=1-d_A$ in the first equality and in the second equality we used the fact that $d_A\neq 0$ only when $t^A$ is one of the three  operations $t^{i,\alpha}, v^i,u^k$ belonging to the list (\ref{FBP-W})-(\ref{S-eta}). \\
Thus we arrive at (\ref{Prep-holom}) by considering the degrees of the differentials $\phi_{t^{i,\alpha}}$, $\phi_{v^i}$ and $\phi_{u^k}$ in (\ref{primary-quasi})  and duality relations (\ref{duality-picture}).\\
On the other hand, expression (\ref{E-holom}) of the Euler vector field   is nothing but the particular case of (\ref{E-tA2}) where $d_0=0$ and $r_{\omega_0,\phi_{t^A}}=0$ for all $A$. In addition, the stated quasi-homogeneity  property of the prepotential (\ref{Prep-holom}) is an immediate consequence of (\ref{Prep-eta-homog}).

\fd

\begin{Remark}
For the forth line of the prepotential (\ref{Prep-holom}), we mention the following:
\begin{equation}\label{B-S}
\begin{split}
&\rho^k(\phi_{\rho^l}):=\frac{1}{2{\rm i}\pi}\oint_{b_k}\omega_l=\frac{\mathbb{B}_{kl}}{2{\rm{i}}\pi };\\
&s^{i}(\phi_{\rho^k})=s^i(\omega_k):=\int_{\infty^0}^{\infty^i}\omega_k;\\
&s^{i}(\phi_{s^j})=s_{ji}=\text{the functions given by (\ref{s-ij Formula})}.
\end{split}
\end{equation}
\end{Remark}

\medskip

Now, by considering some specific cases of the combinatorial parameters $(g,L,n_0,\dots,n_m)$ of the simple Hurwitz space $\widehat{\mathcal{H}}_{g, L}(n_0,\dots,n_m)$, we shall discuss  some consequences of formula  (\ref{Prep-holom}).
\begin{Prop} Assume that the Hurwitz space is $\widehat{\mathcal{H}}_{g, n+1}(n)$, i.e. the considered meromorphic functions have a unique pole $\infty^0$ of order $n+1\geq 2$. Then the prepotential induced by a normalized holomorphic differential $\omega_0$ is of the following  form:
\begin{equation}\label{Prep-holom1}
\begin{split}
\mathbf{F}_{\omega_0}
&=\frac{1}{2(n+1)}\sum_{\alpha,\beta=1}^{n}\frac{\alpha\beta}{(2n+2-\alpha)}t^{\alpha}(\omega_0)
t^{\beta}(\omega_0)t^{n+1-\alpha}(\phi_{t^{n+1-\beta}})\\
&\quad +\frac{1}{2}\sum_{\alpha=1}^{n}\sum_{k=1}^g\Big(\frac{\alpha}{2n+2-\alpha}+\frac{\alpha}{n+1}\Big)
t^{\alpha}(\omega_0)u^k(\omega_0)t^{n+1-\alpha}(\phi_{\rho^k})\\
&\quad+\frac{1}{4{\rm{i}}\pi}\sum_{k,l=1}^gu^k(\omega_0)u^l(\omega_0)\mathbb{B}_{kl}.
\end{split}
\end{equation}
\end{Prop}
\emph{Proof:} Formula (\ref{Prep-holom1}) follows from (\ref{Prep-holom}) and (\ref{B-S}).

\fd

\medskip

In the following corollary, we are going to rewrite   (\ref{Prep-holom1}) when $n=1$, i.e., the considered  Hurwitz space is that of hyperelliptic curves
\begin{equation}\label{hyper1}
\Big\{(\mu,\lambda)\in\C^2:\quad \mu^2=\prod_{k=1}^{2g+1}(\lambda-\lambda_i),\quad \text{$\lambda_i\neq \lambda_j$ for $i\neq j$} \Big\}.
\end{equation}
\begin{Cor} Assume that the Hurwitz space is $\mathcal{H}_{g, 2}(1)$. Then the prepotential in (\ref{Prep-holom1}) reduces to:
\begin{equation}\label{Prep-holom11}
\begin{split}
\mathbf{F}_{\omega_0}
&=\frac{1}{12}\big(t(\omega_0)\big)^2t(\phi_t)+\frac{5}{12}t(\omega_0)\sum_{k=1}^gu^k(\omega_0)t(\omega_k)
+\frac{1}{4{\rm{i}}\pi}\sum_{k,l=1}^gu^k(\omega_0)u^l(\omega_0)\mathbb{B}_{kl}.
\end{split}
\end{equation}
Here $\phi_t(P)=\sqrt{2}W(P,\infty^0)$ is the Abelian differential of the second kind
and $t(\phi)$ is the flat function  $t(\phi):=\sqrt{2}\underset{\infty^0}{{\rm res}\ }\sqrt{\lambda(P)}\phi(P)=\sqrt{2}\phi(\infty^0)$, with $\phi\in \big\{\phi_t,\omega_1,\dots, \omega_g\big\}$.
\end{Cor}

As another  corollary, we shall analyse formula (\ref{Prep-holom1}) in the genus one case. For this, let us denote the chosen  flat coordinates (\ref{FBP-W})-(\ref{S-eta}) associated with  the unique normalized holomorphic differential $\omega_0$ as follows:
\begin{align*}
&t_0=\tau/(2{\rm{i}}\pi), \quad \text{with}\quad \Im(\tau)>0;\\
&t_{\alpha}=t^{\alpha}(\omega_0):=\underset{\infty^0}{\rm res}\ \lambda(P)^{\frac{\alpha}{n+1}}\omega_0(P),\quad \alpha=1,\dots,n;\\
&t_{n+1}:=u(\omega_0):=\oint_a\lambda(P)\omega_0(P).
\end{align*}
With respect to this system of flat coordinates, the Dubrovin flat metric $\eta(\omega_0)$ (\ref{eta-def}) becomes anti-diagonal. Moreover, (\ref{e-tA}) yields that the unit vector $e=\sum_{j=1}^{n+2}\partial_{\lambda_j}$ is represented by $e=\partial_{t_{n+1}}$.
\begin{Cor} With the above notation, the genus one case of the prepotential (\ref{Prep-holom1}) is as follows:
\begin{equation}\label{Prep-holom2}
\begin{split}
\mathbf{F}_{\omega_0}(t_0,t_1,\dots,t_{n+1})
&=\frac{t_0}{2}(t_{n+1})^2+\frac{t_{n+1}}{2}\sum_{\alpha=1}^{n}\Big(\frac{\alpha}{2n+2-\alpha}+\frac{\alpha}{n+1}\Big)t_{\alpha}t_{n+1-\alpha}\\
&\quad+\frac{1}{2(n+1)}\sum_{\alpha,\beta=1}^{n}\frac{\alpha\beta}{(2n+2-\alpha)}t_{\alpha}t_{\beta}t^{n+1-\alpha}(\phi_{t^{n+1-\beta}}).
\end{split}
\end{equation}
\end{Cor}

Note that the unknown functions  $\big\{t^{\alpha}(\phi_{t^{\beta}}):\quad 1\leq\alpha,\beta\leq n\big\}$ in (\ref{Prep-holom2}) are
functions of $t_0,t_1,\dots,t_n, t_{n+1}$ such that
$$
\partial_{t_{n+1}}t^{\alpha}(\phi_{t^{\beta}})=e.t^{\alpha}(\phi_{t^{\beta}})=\delta_{n+1,\alpha+\beta}.
$$
Moreover, since the Euler vector field (\ref{E-holom}) is represented by $E=\sum_{k=1}^{n+1}\frac{k}{n+1}\partial_{t_k}$, it follows that the function $t^{\alpha}(\phi_{t^{\beta}})$ satisfies the following partial differential equation:
$$
\sum_{k=1}^nk\frac{\partial}{\partial{t_k}}t^{\alpha}(\phi_{t^{\beta}})=(\alpha+\beta) t^{\alpha}(\phi_{t^{\beta}})-(n+1)\delta_{n+1,\alpha+\beta}.
$$
On the other hand, according to (\ref{Fubini-flat}), the functions $\big\{t^{\alpha}(\phi_{t^{\beta}})\big\}$ satisfy: $t^{\alpha}(\phi_{t^{\beta}})=t^{\beta}(\phi_{t^{\alpha}})$.

\begin{Prop} Assume that the simple Hurwitz space is of type $\widehat{\mathcal{H}}_{g, m+1}(0,\dots,0)$, that is the meromorphic functions have $m+1$ simples poles $\infty^0,\dots,\infty^m$, with $m\geq 1$.\\
Then the  prepotential (\ref{Prep-holom}) associated with the holomorphic differential $\omega_0\in \big\{\omega_1,\dots,\omega_g\big\}$ takes the following form:
\begin{equation}\label{Prep-holom3}
\begin{split}
&\mathbf{F}_{\omega_0}
=\frac{1}{2}\sum_{i=1}^m\big(v^i(\omega_0)\big)^2\log\bigg(\sum_{k=1}^gv^i(\omega_k)\partial_{z_k}\Theta_{\Delta}\big(0\big|\mathbb{B}\big)\bigg)\\
&\quad -\bigg(\sum_{i=1}^m\big(v^i(\omega_0)\big)^2\bigg)\log\Big(\Theta_{\Delta}\Big(s^i(\omega_1),\dots,s^i(\omega_g)\big|\mathbb{B}\Big)\Big)\\
&\quad+\frac{1}{2}\bigg(\sum_{i,j=1}^mv^i(\omega_0)v^j(\omega_0)\bigg)
\log\left(\sum_{r=1}^m\sum_{k=1}^gv^r(\omega_k)\partial_{z_k}\Theta_{\Delta}\big(0\big|\mathbb{B}\big)\right)\\
&\quad +\frac{1}{2}\sum_{i,j=1, i\neq j}^mv^i(\omega_0)v^j(\omega_0)
\log\left(\frac{\Theta_{\Delta}\Big(s^i(\omega_1)-s^j(\omega_1),\dots,s^i(\omega_g)-s^j(\omega_g)\big|\mathbb{B}\Big)}
{\Theta_{\Delta}\Big(s^i(\omega_1),\dots,s^i(\omega_g)\big|\mathbb{B}\Big)\Theta_{\Delta}\Big(s^j(\omega_1),\dots,s^j(\omega_g)\big|\mathbb{B}\Big)}\right)\\
&\quad +\sum_{i=1}^m\sum_{k=1}^gv^i(\omega_0)u^k(\omega_0)s^{i}(\omega_k)
+\frac{1}{2}\sum_{k,l=1}^gu^k(\omega_0)u^l(\omega_0)\Big(\mathbb{B}_{kl}/(2{\rm{i}}\pi)\Big)\\
&\quad +\frac{\log(-1)}{4}\sum_{i,j=1, i\neq j}^mv^i(\omega_0)v^j(\omega_0)
-\frac{3}{4}\sum_{i,j=1}^m\big(1+\delta_{ij}\big)v^i(\omega_0)v^j(\omega_0).
\end{split}
\end{equation}
Here, as before, $\Theta_{\Delta}$ is the Riemann theta function (\ref{Theta}) corresponding to an odd and non-singular  half integer characteristic $\Delta$.
\end{Prop}
\emph{Proof:} Due to  (\ref{Prep-holom}), (\ref{B-S}) and the fact that the poles of $\lambda$ are all simple, we have
\begin{align*}
\mathbf{F}_{\omega_0}
&=\frac{1}{2}\sum_{i,j=1}^mv^i(\omega_0)v^j(\omega_0)s^{i}(\phi_{s^j})+\sum_{i=1}^m\sum_{k=1}^gv^i(\omega_0)u^k(\omega_0)s^{i}(\omega_k)
+\frac{1}{2}\sum_{k,l=1}^gu^k(\omega_0)u^l(\omega_0)\Big(\mathbb{B}_{kl}/(2{\rm{i}}\pi)\Big)\\
&\quad +\frac{\log(-1)}{4}\sum_{i,j=1, i\neq j}^mv^i(\omega_0)v^j(\omega_0)
-\frac{3}{4}\sum_{i,j=1}^m\big(1+\delta_{ij}\big)v^i(\omega_0)v^j(\omega_0).
\end{align*}
In order to establish the desired formula, it suffices to express the functions $s^{i}(\phi_{s^j})=s_{ji}$, obtained in (\ref{s-ij Formula}), by means of flat coordinates $v^i(\omega_k),s^i(\omega_k), \mathbb{B}_{kl}$ corresponding to the normalized holomorphic differentials $\omega_1,\dots,\omega_g$. \\
Let us first mention that by notation (\ref{notation3}) and the residue theorem, we can write
$$
\omega_k(\infty^i)=\underset{\infty^i}{{\rm res}}\ \lambda(P)\omega_k(P)=v^i(\omega_k)
$$
and
$$
\omega_k(\infty^0)=\underset{\infty^0}{{\rm res}}\ \lambda(P)\omega_k(P)=-\sum_{i=1}^m\underset{\infty^i}{{\rm res}}\ \lambda(P)\omega_k(P)=-\sum_{i=1}^mv^i(\omega_k).
$$
Therefore, by substituting this into formula (\ref{s-ij Formula}), we get
\begin{align*}
s_{ii}&=\log\left(\sum_{k=1}^g\omega_k(\infty^i)\partial_{z_k}\Theta_{\Delta}\big(0\big)\right)
+\log\left(\sum_{k=1}^g\omega_k(\infty^0)\partial_{z_k}\Theta_{\Delta}\big(0\big)\right)
-2\log\left(\Theta_{\Delta}\big(\mathcal{A}(\infty^i)-\mathcal{A}(\infty^0)\big)\right)-\log(-1)\\
&=\log\left(\sum_{k=1}^gv^i(\omega_k)\partial_{z_k}\Theta_{\Delta}\big(0\big|\mathbb{B}\big)\right)
+\log\left(\sum_{r=1}^m\sum_{k=1}^gv^r(\omega_k)\partial_{z_k}\Theta_{\Delta}\big(0\big|\mathbb{B}\big)\right)
-2\log\Big(\Theta_{\Delta}\Big(s^i(\omega_1),\dots,s^i(\omega_g)\big|\mathbb{B}\Big)\Big)\\
\end{align*}
and when $i\neq j$,
\begin{align*}
s^{i}(\phi_{s^j})&=s_{ji}=\log\left(\frac{\Theta_{\Delta}\big(\mathcal{A}(\infty^i)-\mathcal{A}(\infty^j)\big)}
{\Theta_{\Delta}\big(\mathcal{A}(\infty^i)-\mathcal{A}(\infty^0)\big)\Theta_{\Delta}\big(\mathcal{A}(\infty^j)-\mathcal{A}(\infty^0)\big)}\right)
+\log\left(\sum_{k=1}^g\omega_k(\infty^0)\partial_{z_k}\Theta_{\Delta}\big(0\big)\right)-\log(-1)\\
&=\log\left(\frac{\Theta_{\Delta}\Big(s^i(\omega_1)-s^j(\omega_1),\dots,s^i(\omega_g)-s^j(\omega_g)\big|\mathbb{B}\Big)}
{\Theta_{\Delta}\Big(s^i(\omega_1),\dots,s^i(\omega_g)\big|\mathbb{B}\Big)\Theta_{\Delta}\Big(s^j(\omega_1),\dots,s^j(\omega_g)\big|\mathbb{B}\Big)}\right)
+\log\left(\sum_{r=1}^m\sum_{k=1}^gv^r(\omega_k)\partial_{z_k}\Theta_{\Delta}\big(0\big|\mathbb{B}\big)\right).
\end{align*}
This completes the proof of (\ref{Prep-holom3}).

\fd

\medskip

Note that we have obtained the prepotential (\ref{Prep-holom3}) by simple specialization of the general formula (\ref{Prep-eta}) (see also (\ref{Prep-eta2})) to the case of the Hurwitz space $\widehat{\mathcal{H}}_{g, m+1}(0,\dots,0)$ and the holomorphic primary differential $\omega_0$. Now we show how formula (\ref{Prep-holom3}) leads directly to an explicit solution of the (non-generalized) WDVV equations if we specialize it to the genus one case. In other words, we obtain an explicit solution depending on $2m+2$ variables without much calculation but simply by specializing the general formula to the case of the Hurwitz space $\widehat{\mathcal{H}}_{1, m+1}(0,\dots,0)$ and
a holomorphic differential $\omega_0$.\\
Consider the lattice $\mathbb{L}:=\Z+\tau\Z$, with $\Im(\tau)>0$.   Then the Riemann theta function $\Theta_{\Delta}(\cdot|\tau)$ (\ref{Theta}), corresponding to the odd and non-singular  half integer characteristic $\Delta=[1/2,1/2]$, is closely related to  the  $\theta_1$-Jacobi function
\begin{equation}\label{theta1}
\Theta_{[\frac{1}{2},\frac{1}{2}]}(u|\tau):=i\sum_{n\in \Z}(-1)^ne^{{\rm{i}}\pi\tau(n+1/2)^2+2{\rm{i}}\pi(n+1/2)u}=-\theta_1(u|\theta), \quad \quad \Im(\tau)>0.
\end{equation}
We refer to  \cite{Erdelyi} (Section 13.20) and \cite{Akhiezer, Chandra, Watson} for details on the  Jacobi theta functions. \\
The dimension of the Frobenius manifold associated with the normalized holomorphic differential $\omega_0$ is  $N=2m+2$, with $m+1$ being the common degree of the coverings belonging to the Hurwitz space $\mathcal{H}_{1,m+1}(0,\dots,0)$.\\
Moreover, according to Theorem \ref{FC-thm-eta}, the following functions provide a system of flat coordinates of the flat metric $\eta(\omega_0)$ (\ref{eta-def}):
\begin{equation}\label{FC-holo-g1}
\begin{split}
&t_0:=\tau/{2{\rm i}\pi}=\frac{1}{2{\rm i}\pi}\oint_b\omega_0;\\
&t_k:=s^k(\omega_0)=\int_{\infty^0}^{\infty^k}\omega_0,\quad k=1,\dots,m;\\
&t_{2m+1-k}:=v^k(\omega_0)=\underset{\infty^k}{{\rm res}}\ \lambda(P)\omega_0(P),\quad k=1,\dots,m;\\
&t_{2m+1}:=u(\omega_0)=\oint_a\lambda(P)\omega_0(P).
\end{split}
\end{equation}
According to (\ref{Entries eta}) and (\ref{eta-tA}), the constant matrix of the metric $\eta(\omega_0)$, with respect to flat coordinates (\ref{FC-holo-g1}),  is anti-diagonal :
$$
\eta_{\alpha\beta}=\delta_{2m+1,\alpha+\beta}=\eta^{\alpha\beta},\quad \alpha,\beta=0,\dots,2m+1.
$$
\begin{Cor} Assume that the Hurwitz space is $\mathcal{H}_{1,m+1}(0,\dots,0)$.
With the preceding notation, the  prepotential $\mathbf{F}_{\omega_0}=\mathbf{F}_{\omega_0}(t_0,t_1,\dots,t_{2m+1})$  induced by the normalized holomorphic
differential $\omega_0$ is given by:
\begin{equation}\label{F-holo-g1}
\begin{split}
\mathbf{F}_{\omega_0}
&=\frac{(t_{2m+1})^2}{2}t_0+t_{2m+1}\sum_{k=1}^mt_kt_{2m+1-k}+\frac{1}{2}\sum_{k=1}^m\big(t_{2m+1-k}\big)^2\log(t_{2m+1-k})\\
&\quad +\frac{1}{2}\bigg(\sum_{k=1}^mt_{2m+1-k}\bigg)^2\log\bigg(\sum_{j=1}^mt_{2m+1-j}\bigg)
-\sum_{k=1}^m\big(t_{2m+1-k}\big)^2\log\left(\frac{\theta_{1}\big(t_k\big|{2{\rm i}\pi}t_0\big)}{\theta_1'\big(0\big|2{\rm i}\pi t_0\big)}\right)\\
&\quad +\frac{1}{2}\sum_{j,k=1, j\neq k}^mt_{2m+1-j}t_{2m+1-k}
\log\bigg(\frac{\theta_1'\big(0\big|{2{\rm i}\pi}t_0\big)\theta_1\big(t_j-t_k\big|{2{\rm i}\pi}t_0\big)}{\theta_1\big(t_j\big|{2{\rm i}\pi}t_0\big)
\theta_1\big(t_k\big|{2{\rm i}\pi}t_0\big)}\bigg)\\
&\quad +\frac{\log(-1)}{4}\sum_{j,k=1, j\neq k}^mt_{2m+1-j}t_{2m+1-k}-\frac{3}{4}\sum_{j,k=1}^m(1+\delta_{jk})t_{2m+1-j}t_{2m+1-k}.
\end{split}
\end{equation}
In addition, the function $\mathbf{F}_{\omega_0}$ is quasi-homogeneous of degree $2$ with respect to the Euler vector field $E_{\omega_0}=\sum_{k=0}^mt_{2n+1-k}\partial_{t_{2n+1-k}}$ and
$$
E_{\omega_0}\mathbf{F}_{\omega_0}=2\mathbf{F}_{\omega_0}+\frac{1}{2}\sum_{k=1}^{n}(t_{2n+1-k})^2+\frac{1}{2}\left(\sum_{k=1}^{n}t_{2n+1-k}\right)^2.
$$
\end{Cor}
\emph{Proof:} Formula (\ref{Prep-holom3}) implies that
\begin{align*}
&\mathbf{F}_{\omega_0}
=\frac{1}{2}\sum_{i=1}^m(v^i)^2\log\Big(v^i\theta_{[\frac{1}{2},\frac{1}{2}]}'\big(0\big|\tau\big)\Big)-\bigg(\sum_{i=1}^m(v^i)^2\bigg)
\log\Big(\theta_{1}\big(s^i\big|\tau\big)\Big)\\
&\quad+\frac{1}{2}\bigg(\sum_{i,j=1}^mv^iv^j\bigg)
\log\bigg(\sum_{r=1}^mv^r\theta_{[\frac{1}{2},\frac{1}{2}]}'\big(0\big|\tau\big)\bigg) +\frac{1}{2}\sum_{i,j=1, i\neq j}^mv^iv^j
\log\bigg(\frac{\theta_{[\frac{1}{2},\frac{1}{2}]}\big(s^i-s^j\big|\tau\big)}{\theta_{[\frac{1}{2},\frac{1}{2}]}\big(s^i\big|\tau\big)
\theta_{[\frac{1}{2},\frac{1}{2}]}\big(s^j\big|\tau\big)}\bigg)\\
&\quad +u\sum_{i=1}^mv^is^{i}+\frac{(u)^2}{4{\rm i}\pi}\tau+\frac{\log(-1)}{4}\sum_{i,j=1, i\neq j}^mv^iv^j
-\frac{3}{4}\sum_{i,j=1}^m\big(1+\delta_{ij}\big)v^iv^j\\
&=\frac{1}{2}\sum_{i=1}^m(v^i)^2\log(v^i)+\bigg(\sum_{i=1}^m(v^i)^2\bigg)\log\Big(\theta_{[\frac{1}{2},\frac{1}{2}]}'\big(0\big|2{\rm i}\pi t_0\big)\Big)-\bigg(\sum_{i=1}^m(v^i)^2\bigg)
\log\Big(\theta_{1}\big(s^i\big|\tau\big)\Big)\\
&\quad+\frac{1}{2}\bigg(\sum_{i,j=1}^mv^iv^j\bigg)\log\bigg(\sum_{r=1}^mv^r\bigg)+\frac{1}{2}\sum_{i,j=1, i\neq j}^mv^iv^j
\log\bigg(\frac{\theta_{[\frac{1}{2},\frac{1}{2}]}'\big(0\big|\tau\big)\theta_{[\frac{1}{2},\frac{1}{2}]}\big(s^i-s^j\big|\tau\big)}
{\theta_{[\frac{1}{2},\frac{1}{2}]}\big(s^i\big|\tau\big)\theta_{[\frac{1}{2},\frac{1}{2}]}\big(s^j\big|\tau\big)}\bigg)\\
&\quad +u\sum_{i=1}^mv^is^{i}+\frac{(u)^2}{4{\rm i}\pi}\tau+\frac{\log(-1)}{4}\sum_{i,j=1, i\neq j}^mv^iv^j-\frac{3}{4}\sum_{i,j=1}^m\big(1+\delta_{ij}\big)v^iv^j.
\end{align*}
Thus we obtain (\ref{F-holo-g1}) by using (\ref{FC-holo-g1}) and (\ref{theta1}).

\fd

\begin{Remark}  When $m=1$, (\ref{F-holo-g1}) becomes
\begin{equation}\label{F-holo-g11}
\mathbf{F}_{\omega_0}=\frac{(t_3)^2}{2}t_0+t_1t_2t_3+(t_2)^2\log(t_2)-
(t_2)^2\log\bigg(\frac{\theta_1(t_1|2\rm{i}\pi t_0)}{\theta'_1\big(0|2{\rm{i}}\pi t_0\big)}\bigg)-\frac{3}{2}(t_2)^2.
\end{equation}
Note that, up to the change of variables  $(t_0,t_1,t_2,t_3)\longleftrightarrow (u_1,u_2,u_3,u_4)=(-t_3,t_2,t_1,t_0)$ and up to the quadratic term $-\frac{3}{2}t_2^2$, our prepotential (\ref{F-holo-g11}) coincides with the one appearing in \cite{Dubrovin2009, Miguel} and obtained  using a long  computation based on Dubrovin's bilinear pairing method. The mentioned change of variables does not affect the WDVV equations.
\end{Remark}

\subsection{Almost Frobenius manifold structures on double Hurwitz spaces}
In order to simplify formulas, summation over repeated \emph{Greek indices} will be assumed throughout  this subsection.\\
Let us  start by the following definition of an almost Frobenius structure which is motivated by Definition 9 in \cite{Dubrovin2004}.
\begin{Def}\label{Almost-def} Let $\mathcal{M}$ be a manifold of dimension N. An almost Frobenius manifold structure on $\mathcal{M}$ of charge $D\neq 1$  is the data of $\big(\mathcal{M},\ast,\mathbf{g},E\big)$ such that each tangent space $T_p\mathcal{M}$ is a Frobenius algebra varying smoothly or analytically over $\mathcal{M}$ with the additional properties:
\begin{itemize}
\item[\textbf{\emph{1.}}] the inner product $\mathbf{g}$ is a  flat metric on $\mathcal{M}$;
\item[\textbf{\emph{2.}}] in flat coordinates $\xi=(\xi^1,\dots,\xi^N)$ for the metric $\mathbf{g}$,
\begin{itemize}
\item[i)] the structure constants of the  multiplication $\partial_{\xi^i}\ast\partial_{\xi^j}=\mathbf{c}_{\ast, ij}^{\alpha}(\xi)\partial_{\xi^{\alpha}}$
can be locally represented in the form
$$
\mathbf{c}_{\ast, ij}^{k}(\xi)=G^{k\alpha}\partial_{\xi^{\alpha}}\partial_{\xi^i}\partial_{\xi^j}F_{\ast}(\xi)
$$
for some function $F_{\ast}$,  with $G_{\alpha\beta}=\mathbf{g}\big(\partial_{\xi^{\alpha}},\partial_{\xi^{\beta}}\big)$ and $\big(G^{\alpha\beta}\big)$ being the inverse matrix of $\big(G_{\alpha\beta}\big)$.
\item[ii)] the Euler vector field $E=\frac{(1-D)}{2}\sum_{j}\xi^j\partial_{\xi^j}$ is the unity of the Frobenius algebra;
\item[iii)] the function $F_{\ast}$ satisfies the homogeneity property:
$$
E.F_{\ast}=(1-D)F_{\ast}+2G_{\alpha\beta}\xi^{\alpha}\xi^{\beta}.
$$

\end{itemize}
\end{itemize}
\end{Def}
\subsubsection{Frobenius algebras associated with intersection forms}
Consider the intersection form $\mathbf{g}(\omega_0):=\frac{1}{2}\sum_j\frac{\omega_0(P_j)^2}{\lambda_j}(d\lambda_j)^2$ (\ref{Inter-form}) induced by the differential $\omega_0$ satisfying (\ref{Diff model}) and (\ref{Assymptions}). From Appendix 1, we know that $\mathbf{g}(\omega_0)$ is a flat metric  defined on the open subset $\mathcal{M}_{\omega_0}$ of the double Hurwitz space $\widehat{\mathcal{H}}_{g,L}(\mathbf{n}; \mathbf{m})$ determined by the conditions
$$
\omega_0(P_j)\neq 0, \quad \forall\ j=1,\dots,N.
$$
We have a commutative and associative algebra structure on the tangent spaces of the complex manifold $\mathcal{M}_{\omega_0}$ defined by the  multiplication $``\ast"$:
\begin{equation}\label{multiplication-g}
\partial_{\lambda_i}\ast\partial_{\lambda_j}=\frac{\delta_{ij}}{\lambda_j}\partial_{\lambda_j}.
\end{equation}
Moreover, the multiplication (\ref{multiplication-g}) admits the  vector field $E=\sum_{j}\lambda_j\partial_{\lambda_j}$ as the unit vector field.\\
For $x=\sum_jx_j\partial_{\lambda_j}$, $y=\sum_jy_j\partial_{\lambda_j}$ and $z=\sum_jz_j\partial_{\lambda_j}\in T_p\mathcal{M}_{\omega_0}$, consider the $(0,3)$-tensor
\begin{equation}\label{c-tensor-g}
\mathbf{c}_{\ast,\omega_0}(x,y,z):=\mathbf{g}(\omega_0)(x\ast y,z)=\frac{1}{2}\sum_jx_jy_jz_j\frac{\omega_0(P_j)^2}{\lambda_j^2}.
\end{equation}
In particular, the multiplication $``\ast"$ is compatible with the metric $\mathbf{g}(\omega_0)$ and thus, according to Definition \ref{F algebra},
the data $\big(\ast,E, \mathbf{g}(\omega_0)\big)$ gives rise to a Frobenius algebra structure on tangent spaces of $\mathcal{M}_{\omega_0}$.\\

In the following result, we  redescribe such Frobenius algebras structure on tangent spaces by using   the already chosen  system (\ref{flat basis-g}) of flat coordinates $\{u^A(\omega_0)\}$ of the metric $\mathbf{g}(\omega_0)$. The proof  is an immediate consequence of  formulas (\ref{g-partial der-FC}), (\ref{multiplication-g}) and (\ref{c-tensor-g}).
\begin{Prop} Let $\big(G_{AB}\big)$ be the constant matrix (\ref{g-entries}) of the metric $\mathbf{g}(\omega_0)$ (\ref{Inter-form}) with respect to flat  coordinates (\ref{flat basis-g}).  The following assertions hold:
\begin{description}
\item[a)] In $\mathbf{g}(\omega_0)$-flat coordinates (\ref{flat basis-g}), the algebra multiplication ``$\ast$'' (\ref{multiplication-g}) is given by:
\begin{align}\label{mult-flat coord-g}
\begin{split}
\partial_{u^A(\omega_0)}\ast\partial_{u^B(\omega_0)}
&=\sum_{j}\bigg(\frac{\lambda_j}{\omega_0(P_j)^2}\textsf{G}_{A\alpha}\phi_{u^{\alpha}}(P_j)\textsf{G}_{B\beta}\phi_{u^{\beta}}(P_j)\bigg)\partial_{\lambda_j}\\
&=\sum_C\bigg(\frac{1}{2}\sum_{j}\bigg(\frac{\lambda_j\phi_{u^C}(P_j)}{\omega_0(P_j)}\textsf{G}_{A\alpha}\phi_{u^{\alpha}}(P_j)\textsf{G}_{B\beta}\phi_{u^{\beta}}(P_j)
\bigg)\bigg)\partial_{u^C(\omega_0)}.
\end{split}
\end{align}
\item[b)] In flat coordinates $\{u^A(\omega_0)\}_A$ (\ref{flat basis-g}) of the metric $\mathbf{g}(\omega_0)$, the unit vector field $E$ takes the following form:
\begin{equation}\label{E-flat coord-g}
E=\sum_{A}\bigg(\frac{1}{2}\sum_{j=1}^N\lambda_j\phi_{u^A}(P_j)\omega_0(P_j)\bigg)\partial_{u^{A}(\omega_0)};
\end{equation}
\item[c)] The symmetric 3-tensor $\mathbf{c}_{\ast,\omega}$ (\ref{c-tensor-g}):
\begin{equation}\label{c-flat coord-g}
\mathbf{c}_{\ast,\omega_0}\big(\partial_{u^A(\omega_0)},\partial_{u^B(\omega_0)},\partial_{u^C(\omega_0)}\big)
=\frac{1}{2}\sum_{j}\frac{\lambda_j}{\omega_0(P_j)} \textsf{G}_{A\alpha}\phi_{u^{\alpha}}(P_j)\textsf{G}_{B\beta}\phi_{u^{\beta}}(P_j)\textsf{G}_{C\sigma}\phi_{u^{\sigma}}(P_j).
\end{equation}
\item[d)] Structure constants of the multiplication ``$\ast$'' are such that:
\begin{equation}\label{c-structure constants}
\partial_{u^A(\omega_0)}\ast\partial_{u^B(\omega_0)}=\sum_{C} \Big( \textsf{G}^{C\theta}\mathbf{c}_{\ast,\omega_0}\big(\partial_{u^A(\omega_0)},\partial_{u^B(\omega_0)},\partial_{u^{\theta}(\omega_0)}\big)\Big)\partial_{u^C(\omega_0)}.
\end{equation}
\end{description}
\end{Prop}

\begin{Def} Let $\nabla^{\mathbf{g}(\omega_0)}$ be the Levi-Civita connection of the metric $\mathbf{g}(\omega_0)$. The differential $\omega_0$ is called primary with respect to unit vector field $E=\sum_j\lambda_j\partial_{\lambda_j}$  if
$$
\nabla^{\mathbf{g}(\omega_0)}E=0.
$$

\end{Def}

\begin{Prop}\label{ch-E-primary} Let $\omega_0(P):=\int_{\gamma}h(\lambda(Q))W(P,Q)$.
\begin{description}
\item[1.] The following statements are equivalent:
\begin{description}
\item[i)] The differential $\omega_0$ is primary with respect to $E=\sum_j\lambda_j\partial_{\lambda_j}$;
\item[ii)] For all $j=1,\dots,N$,
$$
E.\omega_0(P_j)=-\frac{\omega_0(P_j)}{2};
$$
\item[iii)] For all $j=1,\dots,N$,
\begin{equation}\label{E-covariantly}
\int_{\gamma}\lambda(Q)h'(\lambda(Q))W(Q,P_j)=0.
\end{equation}
\end{description}
\item[2.] The $N$ differentials $\psi_{u^A}$ in (\ref{Primary-E}) are all primary with respect to the  vector field $E$.
\item[3.] Assume that  $\omega_0=\psi_{u^{A}}$ is one of primary differentials (\ref{Primary-E}). Then, with respect to flat coordinates of the metric $\mathbf{g}(\omega_0)$, the unit vector field  $E$ is represented by (sum over $\alpha$ is assumed)
$$
E=\textsf{G}^{A\alpha}\partial_{u^{\alpha}(\omega_0)},
$$
\end{description}
with $\textsf{G}^{A\alpha}:=\mathbf{g}^*(\omega_0)\big(du^A(\omega_0),du^{\alpha}(\omega_0)\big)$ being
the entries (\ref{Entrie-g}) of the intersection form $\mathbf{g}^*(\omega_0)$.
\end{Prop}
\emph{Proof:} \textbf{1.}  By using  expressions (\ref{CS}) for Christoffel coefficients of the
intersection form $\mathbf{g}(\omega_0):=\mathbf{ds^2_{1,0}}(\omega_0)$, we obtain
\begin{align*}
\nabla^{\mathbf{g}(\omega_0)}_{\partial_{\lambda_i}}E&=\partial_{\lambda_i}+\sum_{j}\lambda_j\sum_k\Gamma_{ij}^k(\mathbf{g})\partial_{\lambda_k}
=\partial_{\lambda_i}+\sum_{j\neq i}\lambda_j\sum_k\Gamma_{ij}^k(\mathbf{g})\partial_{\lambda_k}
+\lambda_i\sum_k\Gamma_{ii}^k(\mathbf{g})\partial_{\lambda_k}\\
&=\partial_{\lambda_i}+\Big(\sum_{j\neq i}\lambda_j\Gamma_{ij}^i(\mathbf{g})\Big)\partial_{\lambda_i}
+\sum_{j\neq i}\lambda_j\Gamma_{ij}^j(\mathbf{g})\partial_{\lambda_j}
+\lambda_i\sum_{k\neq i}\Gamma_{ii}^k(\mathbf{g})\partial_{\lambda_k}+\lambda_i\Gamma_{ii}^i(\mathbf{g})\partial_{\lambda_i}\\
&=\partial_{\lambda_i}+\frac{1}{\omega(P_i)}\Big(\sum_{j\neq i}\lambda_j\partial_{\lambda_j}\omega(P_i)\Big)\partial_{\lambda_i}
+\lambda_i\Big(-\frac{1}{2\lambda_i}+\frac{\partial_{\lambda_i}\omega(P_i)}{\omega(P_i)}\Big)\partial_{\lambda_i}\\
&=\Big(\frac{E.\omega(P_i)}{\omega(P_i)}+\frac{1}{2}\Big)\partial_{\lambda_i}.
\end{align*}
Therefore, this and the action (\ref{Auxi2}) of the vector field $E$ on the functions $\omega(P_i)$ show
the equivalence between the three assertions. \\
\textbf{2.} Since the functions $h(\lambda)$ defining the Abelian differentials $\psi_{s^i}=\Omega_{\infty^0\infty^i}$, $\psi_{y^j}=\Omega_{\mathbf{z}^0\mathbf{z}^j}$ and $\psi_{\rho^k}=\omega_k$ are  constants,   they satisfy  conditions (\ref{E-covariantly}). Furthermore, because of (\ref{W-periods}), these conditions are also fulfilled by the multivalued differentials $\psi_{q^k}$.\\
\textbf{3.} Due to formula (\ref{dual metric-E}) we have
$$
\textstyle E=\sum_j\lambda_j\partial_{\lambda_j}=\sum_B\Big(E.u^B(\omega_0)\Big)\partial_{u^{B}(\omega_0)}
=\sum_B\Big(\mathbf{g}^*(\omega_0)\big(du^A(\omega_0),du^B(\omega_0)\big)\Big)\partial_{u^{B}(\omega_0)}.
$$

\fd

\begin{Remark}
The Abelian differential of the third kind $\omega_0(P)=\Omega_{\mathbf{z}^0\infty^0}(P)=\int_{\mathbf{z}^0}^{\infty^0}W(P,Q)$ is  primary with respect to the vector field $E$, but it does not belong to the vector space generated by the differentials (\ref{Primary-E}).\\
Moreover, in flat coordinates (\ref{flat basis-g}) of the metric $\mathbf{g}(\omega_0)$, $E$ has the following form:
\begin{align*}
E=\sum_{A}\Big(\sum_{k=1}^N\underset{P_k}{\text{res}}\frac{\lambda(P)\psi_{u^A}(P)\omega_0(P)}{d\lambda(P)}\Big)\partial_{u^A(\omega_0)}
=\frac{1}{n_0+1}\sum_{i=1}^m\partial_{s^i(\omega_0)}-\frac{1}{m_0+1}\sum_{i=1}^n\partial_{y^i(\omega_0)}.
\end{align*}
\end{Remark}

\subsubsection{Prepotentials of almost Frobenius manifolds}
Let $\omega_0$ be a fixed quasi-homogeneous differential  of degree $d_0\neq 0$ with vanishing $a$-periods. Essentially, $\omega_0$ will belong to the following family of Abelian differentials of the second kind:
\begin{align*}
&\psi_{t^{i,\alpha}}(P):=\underset{\infty^i}{\rm res}\lambda(Q)^{\frac{\alpha}{n_i+1}}W(P,Q),\quad\quad\quad \text{$i=1,\dots,m$\quad  and\quad  $\alpha\in\N$, $\alpha\geq1$};\\
&\widetilde{\psi}_{t^{j,\beta}}(P):=\underset{\mathbf{z}^j}{\rm res}\lambda(Q)^{-\frac{\beta}{m_j+1}}W(P,Q),\quad \quad \text{$j=1,\dots,n$\quad  and\quad $\beta\in \N$, $\beta\geq1$};
\end{align*}
or  the following family of  multivalued differentials (studied in Appendix 2):
\begin{align*}
\phi_{u^{k,\sigma}}(P):=\oint_{a_k}\big(\lambda(Q)\big)^{\sigma}W(P,Q),\quad \quad\quad \quad \quad \text{$k=1,\dots,g$\quad  and\quad  $\sigma\in \N$, $\sigma\geq1$}.
\end{align*}
Note that the degrees of these differentials are:
$$
deg\big(\psi_{t^{i,\alpha}}\big)=\frac{\alpha}{n_i+1},\quad \quad deg\big(\widetilde{\psi}_{t^{j,\beta}}\big)=-\frac{\beta}{m_j+1},
\quad \quad deg\big(\phi_{u^{k,\sigma}}\big)=\sigma.
$$
\begin{Prop} Let $\omega_0\in \big\{\psi_{t^{i,\alpha}}, \widetilde{\psi}_{t^{j,\beta}}, \phi_{u^{k,\sigma}}\big\}$ and $\big\{u^A(\omega_0)\}_A$ be the corresponding  system of flat coordinates (\ref{flat basis-g}) and $E=\sum_j\lambda_j\partial_{\lambda_j}$ be the unit vector field. Then
\begin{equation}\label{E-uA}
\begin{split}
&E.u^A(\omega_0)=d_0u^A(\omega_0), \quad \text{for all $A$};\\
&E=d_0\sum_Au^A(\omega_0)\partial_{u^A(\omega_0)}.
\end{split}
\end{equation}
\end{Prop}
\emph{Proof:} The second formula in (\ref{E-uA}) follows from the first one. In order to prove that $u^A(\omega_0)$ is an eigenfunction of the vector field $E$, we shall use formula (\ref{dual metric-E}) where we replace $\psi_{u^B}$ by $\omega_0$.
Assume first that $\omega_0=\psi_{t^{i_0,\alpha}}(P)$. Similarly to
(\ref{type1}), we have
$$
\psi_{t^{i_0,\alpha}}(P)\underset{P\sim \infty^{i_0}}=\frac{\alpha}{z_{i_0}(P)^{\alpha+1}}dz_{i_0}(P)+\text{holomorphic},\quad \quad \lambda(P)=z_{i_0}(P)^{-n_{i_0}-1}.
$$
When $\psi_{u^A}$ is among the $E$-primary differentials $\psi_{s^i},\psi_{y^j},\psi_{\rho^k}$ (see (\ref{Primary-E})), the function $\frac{\lambda(P)\psi_{u^A}(P)}{d\lambda(P)}$ is holomorphic near the point $\infty^{i_0}$.  Therefore by writing its local behavior near $\infty^{i_0}$ and using notation (\ref{notation3}) we deduce that
\begin{align*}
E.u^A(\omega_0)&=\sum_{k=1}^N\underset{P_k}{{\rm res}}\frac{\lambda(P)\psi_{u^A}(P)\omega_0(P)}{d\lambda(P)}
=-\underset{\infty^{i_0}}{{\rm res}}\frac{\lambda(P)\psi_{u^A}(P)\omega_0(P)}{d\lambda(P)}\\
&=\frac{\alpha}{(n_{i_0}+1)(\alpha-1)!}\psi_{u^A}^{(\alpha-1)}(\infty^{i_0})
=\frac{\alpha}{n_{i_0}+1}\underset{\infty^{i_0}}{{\rm res}}\big(\lambda(P)\big)^{\frac{\alpha}{n_{i_0}+1}}\psi_{u^A}(P)\\
&=\frac{\alpha}{n_{i_0}+1}u^A(\omega_0)\quad \quad (\text{by changing the order of integration}).
\end{align*}
When $\psi_{u^A}$ is among the multivalued differentials $\psi_{q^k}(P):=\oint_{a_k}\log(\lambda(Q))W(P,Q)$, the restriction of function $\frac{\lambda(P)\psi_{q^k}(P)}{d\lambda(P)}$ to a chosen fundamental polygon $F_g$ is also holomorphic near the point $\infty^{i_0}$. Accordingly,  in view of  the jumps (\ref{jump-E}) of $\psi_{q^k}$ and the residue theorem, we obtain
\begin{align*}
E.q^k(\omega_0)&=\sum_{j=1}^N\underset{P_j}{{\rm res}}\frac{\lambda(P)\psi_{q^k}(P)\omega_0(P)}{d\lambda(P)}
=\frac{1}{2{\rm{i}}\pi}\oint_{\partial{F_g}}\bigg(\frac{\lambda(P)\omega_0(P)}{d\lambda(P)}\psi_{q^k}(P)\bigg)
-\underset{\infty^{i_0}}{{\rm res}}\frac{\lambda(P)\psi_{q^k}(P)\omega_0(P)}{d\lambda(P)}\\
&=\frac{1}{2{\rm{i}}\pi}\oint_{a_k}\bigg(\frac{\lambda(P)\omega_0(P)}{d\lambda(P)}\Big(\psi_{q^k}(P)-\psi_{q^k}(P+b_k)\Big)\bigg)
+\frac{\alpha}{(n_{i_0}+1)(\alpha-1)!}\psi_{q^k}^{(\alpha-1)}(\infty^{i_0})\\
&=\oint_{a_k}\omega_0(P)+\frac{\alpha}{n_{i_0}+1}\underset{\infty^{i_0}}{{\rm res}}\big(\lambda(P)\big)^{\frac{\alpha}{n_{i_0}+1}}\psi_{q^k}(P)\\
&=\frac{\alpha}{n_{i_0}+1}q^k(\omega_0)\quad \quad (\text{by changing the order of integration}).
\end{align*}
The case where $\omega_0=\widetilde{\psi}_{t^{j,\beta}}$ can be done similarly. Let us now take $\omega_0=\phi_{u^{k_0,\sigma}}$. Then the action of $E$ on the functions
$s^i(\omega_0),\rho^k(\omega_0), q^j(\omega_0)$ is included in formulas (\ref{u-k-sigma}) and (\ref{E-q-sigma}). Moreover, the  arguments similar to those used in the preceeding case $E.q^k(\omega_0)$ allow us to arrive at the relation $E.y^j(\omega_0)=\sigma y^j(\omega_0)$.

\fd

\begin{Thm}\label{Prep-almost}
Let $\omega_0\in \big\{\psi_{t^{i,\alpha}}, \widetilde{\psi}_{t^{j,\beta}}, \phi_{u^{k,\sigma}}\big\}$ be a quasi-homogeneous  differential of degree $d_0\neq0$.
\begin{description}
\item[1)] Consider the symmetric matrix $(H_{AB})$ defined by
\begin{equation}\label{H-g}
H_{AB}:=\frac{1}{2}\textsf{G}_{A\alpha}\textsf{G}_{B\beta}\Big(u^{\alpha}\big(\psi_{u^{\beta}}\big)+u^{\beta}\big(\psi_{u^{\alpha}}\big)\Big).
\end{equation}
Then the 3-tensor (\ref{c-flat coord-g}) can be expressed as:
\begin{equation}\label{c-flat coord-g1}
\begin{split}
\partial_{u^{C}(\omega_0)}H_{AB}=\mathbf{c}_{\ast,\omega_0}\big(\partial_{u^A(\omega_0)},\partial_{u^B(\omega_0)},\partial_{u^C(\omega_0)}\big)
=\partial_{u^{B}(\omega_0)}H_{AC}.
\end{split}
\end{equation}
\item[2)] Consider the  function
\begin{equation}\label{prepotential-NP-W}
\mathbf{F}_{\ast,\omega_0}=\frac{1}{2}\sum_{A,B}u^A(\omega_0)u^B(\omega_0)\Big(H_{AB}-\frac{3}{2d_0}\textsf{G}_{AB}\Big).
\end{equation}
Then $\mathbf{F}_{\ast,\omega_0}$ satisfies the following properties:
\begin{description}
\item[i.] The third order partial derivatives of $\mathbf{F}_{\ast,\omega_0}$ are such that
$$
\partial_{u^A(\omega_0)}\partial_{u^B(\omega_0)}\partial_{u^C(\omega_0)}\mathbf{F}_{\ast,\omega_0}=
\mathbf{c}_{\ast,\omega_0}\big(\partial_{u^A(\omega_0)},\partial_{u^B(\omega_0)},\partial_{u^C(\omega_0)}\big).
$$
\item[ii.] the function $\mathbf{F}_{\ast,\omega_0}$ is quasi-homogeneous w.r.t. the vector field $E$ in (\ref{E-uA}) with
$$
E.\mathbf{F}_{\ast,\omega_0}=2d_0\mathbf{F}_{\ast,\omega_0}+\frac{1}{2}\sum_{A,B}\textsf{G}_{AB}u^A(\omega_0)u^B(\omega_0);
$$
\item[iii.] the matrix $(H_{AB})$ in (\ref{H-g}), which does not depend on $\omega_0$, is the Hessian matrix of $\mathbf{F}_{\ast,\omega_0}$.
\end{description}
\end{description}
\end{Thm}
\emph{Proof:} \textbf{1)}  From (\ref{g-partial der-FC}) we know that:
$$
\partial_{u^{C}(\omega_0)}=\textstyle\sum_{j}\bigg(\frac{\lambda_j}{\omega_0(P_j)}\textsf{G}_{C\sigma}\psi_{u^{\sigma}}(P_j)\bigg)\partial_{\lambda_j}.
$$
Using this,  the Rauch formula  and (\ref{c-flat coord-g}) we deduce that
\begin{align*}
\partial_{u^{C}(\omega_0)}H_{AB}
&=\frac{1}{2}\sum_{j}\frac{\lambda_j}{\omega_0(P_j)}\textsf{G}_{C\sigma}\psi_{u^{\sigma}}(P_j)
\textsf{G}_{A\alpha}\textsf{G}_{B\beta}\psi_{u^{\alpha}}(P_j)\psi_{u^{\beta}}(P_j)\\
&=\mathbf{c}_{\ast,\omega_0}\big(\partial_{u^A(\omega_0)},\partial_{u^B(\omega_0)},\partial_{u^C(\omega_0)}\big),
\end{align*}
which gives (\ref{c-flat coord-g1}).\\
\textbf{2)} Using the first equality in  (\ref{dual metric-E}), we deduce that the action of the vector field $E$  on the functions $H_{AB}$ is determined by
\begin{align*}
E.H_{AB}=\frac{1}{2}\textsf{G}_{A\alpha}\textsf{G}_{B\beta}\Big(E.u^{\alpha}\big(\psi_{u^{\beta}}\big)+E.u^{\alpha}\big(\psi_{u^{\beta}}\big)\Big)
&=\textsf{G}_{A\alpha}\textsf{G}_{B\beta}G^{\alpha\beta}=G_{AB}.
\end{align*}
This and (\ref{E-uA}) lead to:
\begin{align*}
\partial_{u^A(\omega_0)}\mathbf{F}_{\ast,\omega_0}
&=\sum_{C}u^C(\omega_0)\Big(H_{AC}-\frac{3}{2d_0}\textsf{G}_{AC}\Big)+\frac{1}{2}\sum_{C}u^C(\omega_0)\left(\sum_Du^D(\omega_0)\partial_{u^A(\omega_0)}H_{CD}\right)\\
&=\sum_{C}u^C(\omega_0)\Big(H_{AC}-\frac{3}{2d_0}\textsf{G}_{AC}\Big)+\frac{1}{2}\sum_{C}u^C(\omega_0)\Big(\sum_Du^D(\omega_0)\partial_{u^D(\omega_0)}H_{AC}\Big)\\
&=\sum_{C}u^C(\omega_0)\Big(H_{AC}-\frac{3}{2d_0}\textsf{G}_{AC}\Big)+\frac{1}{2d_0}\sum_{C}u^C(\omega_0)E.H_{AC}\\
&=\sum_{C}u^C(\omega_0)H_{AC}-\frac{1}{d_0}\sum_{C}\textsf{G}_{AC}u^C(\omega_0).
\end{align*}
In particular
$$
E.\mathbf{F}_{\ast,\omega_0}=d_0\sum_Au^A(\omega_0)\partial_{u^A(\omega_0)}\mathbf{F}_{\ast,\omega_0}
=2d_0\mathbf{F}_{\ast,\omega_0}+\frac{1}{2}\sum_{A,B}\textsf{G}_{AB}u^A(\omega_0)u^B(\omega_0).
$$
Moreover, like before we have
\begin{align*}
\partial_{u^B(\omega_0)}\partial_{u^A(\omega_0)}\mathbf{F}_{\ast,\omega_0}
&=H_{AB}+\sum_{C}u^C(\omega_0)\partial_{u^B(\omega_0)}H_{AC}-\frac{1}{d_0}\textsf{G}_{AB}\\
&=H_{AB}+\frac{1}{d_0}E.H_{AB}-\frac{1}{d_0}\textsf{G}_{AB}\\
&=H_{AB}.
\end{align*}
Finally, by relation (\ref{c-flat coord-g1}) we conclude that
$$
\partial_{u^C(\omega_0)}\partial_{u^B(\omega_0)}\partial_{u^A(\omega_0)}\mathbf{F}_{\ast,\omega_0}=
\partial_{u^{C}(\omega_0)}H_{AB}=\mathbf{c}_{\ast,\omega_0}\big(\partial_{u^A(\omega_0)},\partial_{u^B(\omega_0)},\partial_{u^C(\omega_0)}\big).
$$

\fd

\medskip

Theorem \ref{Prep-almost} and Definition \ref{Almost-def} enable us to conclude the following.
\begin{Cor}Let $\omega_0\in \big\{\psi_{t^{i,\alpha}}, \widetilde{\psi}_{t^{j,\beta}}, \phi_{u^{k,\sigma}}\big\}$ be a quasi-homogeneous  differential of degree $d_0\neq0$.
Then the open set $\mathcal{M}_{\omega_0}\subset \widehat{\mathcal{H}}_{g,L}(\mathbf{n}; \mathbf{m})$ carries an almost Frobenius structure determined by the intersection form $\mathbf{g}(\omega_0)$ (\ref{Inter-form}), the multiplication $``\ast"$ (\ref{mult-flat coord-g}), the unit=Euler vector field $E$ (\ref{E-uA}) and the prepotential
$\mathbf{F}_{\ast,\omega_0}$ (\ref{prepotential-NP-W}).
The charge is given by $D=1-2d_0\neq 1$.
\end{Cor}


\section{Prepotentials of deformed Hurwitz-Frobenius manifold structures}
In this section, our goal is to extend formula (\ref{Prep-eta})-(\ref{Prep-eta2}) to quasi-homogeneous solutions  associated  with  deformations of Frobenius manifold structures on the simple Hurwitz spaces $\widehat{\mathcal{H}}_{g, L}(n_0,\dots,n_m)$, with $g\geq 1$.
As already indicated in the introduction, the deformed Frobenius structures on  $\widehat{\mathcal{H}}_{g, L}(n_0,\dots,n_m)$ depending $g(g+1)/2$ complex parameters were studied by  Shramchenko in \cite{Vasilisa2}.\\
In order to achieve our goal, we  follow the pattern  of the Sections 3.3 and 4.1 and use the results obtained there.
\subsection{Deformations of Frobenius manifold structures on simple Hurwitz spaces}
Let  $\left\{(C_g,\lambda), \{a_k,b_k\}\right\}$ be a point in $\widehat{\mathcal{H}}_{g, L}(n_0,\dots,n_m)$ and  $\omega_1,\dots,\omega_g$ be the basis of  normalized holomorphic differential on the surface $C_g$, with $\oint_{a_l}\omega_k=\delta_{kl}$.  Here, the cycles $a_k,b_k$ are assumed to start at a chosen marked point $P_0\in C_g$ so that $\lambda(P_0)=0$.\\
We will use repeatedly the following preliminary result about the Riemann matrix of $b$-periods $\mathbb{B}=(\mathbb{B}_{kl})$.
\begin{Lemma} We have
\begin{equation}\label{B-eE}
e.\mathbb{B}_{kl}=0=E.\mathbb{B}_{kl}, \quad\quad  \forall\ k,l=1,\dots,g,
\end{equation}
where $e=\sum_{j}\partial_{\lambda_j}$ and $E=\sum_{j}\lambda_j\partial_{\lambda_j}$.
\end{Lemma}
\emph{Proof:} Using notation (\ref{FBP-W}), we have $\mathbb{B}_{kl}={2{\rm i}\pi}\rho^k(\phi_{\rho^l})$. Thus (\ref{B-eE}) follows from (\ref{dual metric-e}) and (\ref{Entries-g-eta}) and the fact that normalized holomorphic differentials are quasi-homogeneous differentials of degree 0 (as we know from  (\ref{primary-quasi})).

\fd

\begin{description}
\item[$\bullet$]\textbf{Darboux-Egoroff metrics in terms of the $\mathbf{q}$-bidifferentials}
\end{description}
 Let $\mathfrak{p}=\big(\mathfrak{p}_{kl}\big)$ and $\mathfrak{q}=\big(\mathfrak{q}_{kl}\big)$ be  two  $g\times g$ matrices such that
\begin{description}
\item[i)] the matrix $\mathfrak{q}$ is nondegenerate and the matrix $\mathbf{q}:=\mathfrak{q}^{-1}\mathfrak{p}=\big(\mathbf{q}_{kl}\big)$ is symmetric;
\item[ii)] the matrix $I+\mathbf{q}\mathbb{B}$ is nondegenerate (here $\mathbb{B}$ is the Riemann matrix and $I$ is the identity matrix);
\item[iii)] the matrix $\big(I+\mathbf{q}\mathbb{B}\big)^{-1}\mathbf{q}$ is symmetric;
\item[iv)] the parameters $\mathbf{q}_{kl}$  are constants with respect to the branch points $\{\lambda_j\}$ of $(C_g,\lambda)$ and $\lambda$.
\end{description}
Note that  item \textbf{i)} tells us that  the following $2g\times 2g$  matrix is symplectic:
\begin{equation}\label{Symp}
\mathcal{S}:=
\begin{pmatrix}
\left(\mathfrak{q}^{-1}\right)^T & 0 \\
\mathfrak{p} & \mathfrak{q}
\end{pmatrix},
\end{equation}
that is $\mathcal{S}^T\mathcal{J}\mathcal{S}=\mathcal{J}$, with $\mathcal{J}=
\begin{pmatrix}
0 & -I \\
I & 0
\end{pmatrix}$.\\
Moreover, when the matrix $\mathbf{q}$ is nondegenerate (and symmetric), $\big(I+\mathbf{q}\mathbb{B}\big)^{-1}\mathbf{q}=\big(\mathbb{B}+\mathbf{q}^{-1}\big)^{-1}$ is symmetric. \\
By following \cite{Vasilisa2}, let us consider the following $\mathbf{q}$-deformation of the bidifferential $W(P,Q)$ (\ref{W-def}):
\begin{equation}\label{q-W-def}
\begin{split}
W_{\mathbf{q}}(P,Q)&:=W(P,Q)-{2{\rm i}\pi}\Big(\big(\mathfrak{q}+\mathfrak{p}\mathbb{B}\big)^{-1}\mathfrak{p}\Big)_{\alpha\beta}\omega_{\alpha}(P)\omega_{\beta}(Q)\\
&=W(P,Q)-{2{\rm i}\pi}\Big(\big(I+\mathbf{q}\mathbb{B}\big)^{-1}\mathbf{q}\Big)_{\alpha\beta}\omega_{\alpha}(P)\omega_{\beta}(Q),
\end{split}
\end{equation}
where $\omega_1,\dots,\omega_g$ are the normalized holomorphic differentials.\\
\emph{In formula (\ref{q-W-def}) as well as during this  subsection and the next one, summation over  the repeated Greek indices $\alpha,\beta=1,\dots,g$  is assumed}.\\
We shall call $W_{\mathbf{q}}(P,Q)$ the $\mathbf{q}$-bidifferential. Expression (\ref{q-W-def}) for the $\mathbf{q}$-bidifferential  differs slightly from the one  used in \cite{Vasilisa2} where the symmetric matrix $\mathbf{q}$  can be  degenerate.   From the properties of the symmetric bidifferential $W(P,Q)$ (\ref{W-def}),  the $\mathbf{q}$-bidifferential $W_{\mathbf{q}}(P,Q)$ (\ref{q-W-def}) is also symmetric and has a pole of second order on the diagonal with biresidue 1. Moreover, for all $k=1,\dots,g$ we have
\begin{align*}
\oint_{Q\in a_k}W_{\mathbf{q}}(P,Q)&=-{2{\rm i}\pi}\Big(\big(I+\mathbf{q}\mathbb{B}\big)^{-1}\mathbf{q}\Big)_{k\alpha}\omega_{\alpha}(P);\\
\oint_{Q\in b_k}W_{\mathbf{q}}(P,Q)
&={2{\rm i}\pi}\omega_k(P)-{2{\rm i}\pi}\Big(\big(I+\mathbf{q}\mathbb{B}\big)^{-1}\mathbf{q}\Big)_{\alpha\beta}\mathbb{B}_{\beta{k}}\omega_{\alpha}(P)\\
&={2{\rm i}\pi}\Big(\big(I+\mathbf{q}\mathbb{B}\big)^{-1}\Big)_{\alpha{k}}\omega_{\alpha}(P).
\end{align*}
Thus the $\mathbf{q}$-bidifferential $W_{\mathbf{q}}(P,Q)$ fulfills the normalization condition
\begin{equation}\label{q-W-norm}
\oint_{Q\in a_k}W_{\mathbf{q}}(P,Q)+\mathbf{q}_{k\alpha}\oint_{Q\in b_{\alpha}}W_{\mathbf{q}}(P,Q)=0,\quad \quad \forall\ k=1,\dots,g,
\end{equation}
with $\mathbf{q}_{\alpha\beta}$ being the coefficient of the matrix $\mathbf{q}$.\\
From its expression (\ref{q-W-def}), we see that the $\mathbf{q}$-bidifferential $W_{\mathbf{q}}(P,Q)$ turns into the bidifferential $W(P,Q)$ provided that  $\mathbf{q}$ is  the zero matrix.\\
Before delving more into the properties of the  $\mathbf{q}$-bidifferential,  let us emphasize that the deformation (\ref{q-W-def}) is motivated by the following viewpoint.  Assume  that the symplectic matrix (\ref{Symp})  belongs to $Sp(2g,\Z)$. Then the cycles $\widetilde{\mathbf{a}}:=\big(\widetilde{a}_1,\dots,\widetilde{a}_g\big)^T$ and $\widetilde{\mathbf{b}}:=\big(\widetilde{b}_1,\dots,\widetilde{b}_g\big)^T$ given by
$$
\begin{pmatrix}
  \widetilde{\mathbf{b}} \\
  \widetilde{\mathbf{a}}
\end{pmatrix}
=
\begin{pmatrix}
\big(\mathfrak{q}^{-1}\big)^T & 0 \\
\mathfrak{p} & \mathfrak{q}
\end{pmatrix}
\begin{pmatrix}
  \mathbf{b} \\
  \mathbf{a}
\end{pmatrix},
\quad \quad \text{with}\quad \mathbf{a}:=\big(a_1,\dots,a_g\big)^T,\quad \mathbf{b}:=\big(b_1,\dots,b_g\big)^T
$$
define a new canonical homology basis on the surface $C_g$.
Thus the  $\mathbf{q}$-bidifferential (\ref{q-W-def}) becomes normalized by
\begin{align*}
\oint_{Q\in \widetilde{a}_k}W_{\mathbf{q}}(P,Q)&=\mathfrak{p}_{k\alpha}\oint_{Q\in b_{\alpha}}W_{\mathbf{q}}(P,Q)+\mathfrak{q}_{k\beta}\oint_{Q\in a_{\beta}}W_{\mathbf{q}}(P,Q)\\
&=\mathfrak{q}_{k\beta}\left(\mathbf{q}_{\beta\alpha}\oint_{Q\in b_{\alpha}}W_{\mathbf{q}}(P,Q)+\oint_{Q\in a_{\beta}}W_{\mathbf{q}}(P,Q)\right)=0,
\end{align*}
where we used the equality $\mathfrak{p}=\mathfrak{q}\mathbf{q}$ and (\ref{q-W-norm}) in the second line.\\
As consequence, the characterization  (\ref{W-asymptotic})-(\ref{W-periods}) allows us to view  $W_{\mathbf{q}}(P,Q)$ (\ref{q-W-def}) as the canonical symmetric bidifferential with respect to the new  canonical homology basis $\{\widetilde{a}_k, \widetilde{b}_k\}$ on $C_g$ and the corresponding basis of normalized holomorphic differentials
$\widetilde{\omega}_1,\dots,\widetilde{\omega}_g$, with
$$
\big(\widetilde{\omega}_1,\dots,\widetilde{\omega}_g\big)^T=\big(\mathbb{B}\mathfrak{p}^T+\mathfrak{q}^T\big)^{-1}\big(\omega_1,\dots,\omega_g\big)^T.
$$

The following result was proven in \cite{Vasilisa2} when the $\mathfrak{p}=I$ and $\mathbf{q}=\mathfrak{q}^{-1}$ is symmetric and nondegenerate.  We  include a proof for the reader’s convenience.
\begin{Prop} The $\mathbf{q}$-bidifferential $W_{\mathbf{q}}(P,Q)$ satisfies the following variational Rauch formulas:
\begin{equation}\label{q-W-Rauch}
\partial_{\lambda_j}W_{\mathbf{q}}(P,Q)=\frac{1}{2}W_{\mathbf{q}}(P,P_j)W_{\mathbf{q}}(P_j,Q),
\end{equation}
where $W_{\mathbf{q}}(P,P_j)$ is the evaluation (\ref{notation1}) of the $\mathbf{q}$-bidifferential $W_{\mathbf{q}}(P,Q)$ at $Q=P_j$.\\
In particular, the functions
$$
\kappa_{ij}(\mathbf{q}):=\frac{1}{2}W_{\mathbf{q}}(P_i,P_j):=\frac{1}{2}\frac{W_{\mathbf{q}}(P,Q)}{dx_i(P)dx_j(Q)}, \quad \quad i\neq j,
$$
solve the Darboux-Egoroff equations:
\begin{equation}\label{Darboux}
\begin{split}
&\partial_{\lambda_k}\kappa_{ij}(\mathbf{q})=\kappa_{ik}(\mathbf{q})\kappa_{kj}(\mathbf{q}),\quad \quad \text{$i,j,k$ are distinct};\\
&e.\kappa_{ij}(\mathbf{q}):=\sum_{j}\partial_{\lambda_j}\kappa_{ij}(\mathbf{q})=0.
\end{split}
\end{equation}
\end{Prop}
\emph{Proof:} Since $\partial_{\lambda_j}\mathbb{B}_{kl}={{\rm i}\pi}\omega_k(P_j)\omega_l(P_j)$ (by Rauch's formula (\ref{Rauch})), it follows that
\begin{align*}
\partial_{\lambda_j}\Big(\big(I+\mathbf{q}\mathbb{B}\big)^{-1}\mathbf{q}\Big)_{\alpha\beta}
&=\partial_{\lambda_j}\Big(\big(I+\mathbf{q}\mathbb{B}\big)^{\alpha\sigma}\mathbf{q}_{\sigma\beta}\Big)\\
&=\big(I+\mathbf{q}\mathbb{B}\big)^{\alpha\mu}\big(I+\mathbf{q}\mathbb{B}\big)^{\nu\sigma}\mathbf{q}_{\sigma\beta}
\partial_{\lambda_j}\Big(\mathbf{q}\mathbb{B}\Big)_{\mu\nu}\\
&={{\rm i}\pi}\big(I+\mathbf{q}\mathbb{B}\big)^{\alpha\mu}\big(I+\mathbf{q}\mathbb{B}\big)^{\nu\sigma}\mathbf{q}_{\sigma\beta}
\mathbf{q}_{\mu\epsilon}\omega_{\epsilon}(P_j)\omega_{\nu}(P_j)\\
&={{\rm i}\pi}\Big(\big(I+\mathbf{q}\mathbb{B}\big)^{-1}\mathbf{q}\Big)_{\alpha\epsilon}
\Big(\big(I+\mathbf{q}\mathbb{B}\big)^{-1}\mathbf{q}\Big)_{\nu\beta}\omega_{\epsilon}(P_j)\omega_{\nu}(P_j)
\end{align*}
where we have taken the sum over all the  repeated Greek indices and denoted by $\Lambda^{ij}$ the coefficient of a nondegenerate matrix $(\Lambda_{ij})$.
This  relation, Rauch's formulas (\ref{Rauch}) and expression (\ref{q-W-def}) for the $\mathbf{q}$-bidifferential imply that
\begin{align*}
\partial_{\lambda_j}W_{\mathbf{q}}(P,Q)
&=\frac{1}{2}W(P,P_j)W_{\mathbf{q}}(P_j,Q)-{{\rm i}\pi}W(P_j,Q)\Big(\big(I+\mathbf{q}\mathbb{B}\big)^{-1}\mathbf{q}\Big)_{\alpha\beta}
\omega_{\alpha}(P)\omega_{\beta}(P_j)\\
&\quad-2{{\rm i}\pi}\bigg({{\rm i}\pi}\Big(\big(I+\mathbf{q}\mathbb{B}\big)^{-1}\mathbf{q}\Big)_{\alpha\epsilon}
\Big(\big(I+\mathbf{q}\mathbb{B}\big)^{-1}\mathbf{q}\Big)_{\nu\beta}\omega_{\epsilon}(P_j)\omega_{\nu}(P_j)\bigg)
\omega_{\alpha}(P)\omega_{\beta}(Q)\\
&=\frac{1}{2}W_{\mathbf{q}}(P,P_j)W_{\mathbf{q}}(P_j,Q).
\end{align*}
The first equation in (\ref{Darboux}) is a direct consequence of Rauch's variational formulas (\ref{q-W-Rauch}). On the other hand, (\ref{q-W-def}) and (\ref{B-eE}) imply that
$$
e.W_{\mathbf{q}}(P_i,P_j)
=e.W(P_i,P_j)-2{{\rm i}\pi}\Big(\big(I+\mathbf{q}\mathbb{B}\big)^{-1}\mathbf{q}\Big)_{\alpha\beta}\Big(e.\big(\omega_{\alpha}(P_i)\omega_{\beta}(P_j)\big)\Big).
$$
Thus the second equation in (\ref{Darboux}) follows from  $e.W(P_i,P_j)=0=e.\omega_k(P_j)$, by (\ref{Auxi1}) and (\ref{Auxi2}).

\fd

\medskip

The above result leads to a family of Darboux-Egoroff metrics with respect to the introduced  $\mathbf{q}$-bidifferential $W_{\mathbf{q}}(P,Q)$ (\ref{q-W-def}), similarly to (\ref{eta-def}).
\begin{Cor} Let $\omega_{\mathbf{q}}$ be the $\mathbf{q}$-differential on $C_g$ defined by
\begin{equation}\label{q-Diff model}
\omega_{\mathbf{q}}(P)=\int_{\gamma}h\big(\lambda(Q))W_{\mathbf{q}}(P,Q),
\end{equation}
where the contour $\gamma$ and the  function $h$ are as in (\ref{Diff model}). Then the following formula
\begin{equation}\label{q-eta}
\eta\big(\omega_{\mathbf{q}}\big):=\frac{1}{2}\sum_{j=1}^N\big(\omega_{\mathbf{q}}(P_j)\big)^2(d\lambda_j)^2
=\sum_{j=1}^N\bigg(\underset{P_j}{{\rm res}}\frac{\big(\omega_{\mathbf{q}}(P)\big)^2}{d\lambda(P)}\bigg)(d\lambda_j)^2.
\end{equation}
defines a Darboux-Egoroff metric on the open subset $\mathcal{M}_{\omega_{\mathbf{q}}}$ of the Hurwitz space $\widehat{\mathcal{H}}_{g, L}(n_0,\dots,n_m)$ determined by the conditions
\begin{equation}\label{q-assumptions}
det(I+\mathbf{q}\mathbb{B})\neq0\quad \quad\text{and}\quad \quad \omega_{\mathbf{q}}(P_j)\neq 0,\quad \quad j=1,\dots,N.
\end{equation}
The rotation coefficients of the metric $\eta\big(\omega_{\mathbf{q}}\big)$ are given by $\kappa_{ij}(\mathbf{q}):=W_{\mathbf{q}}(P_i,P_j)/2$.
\end{Cor}

\begin{description}
\item[$\bullet$]\textbf{Normalized holomorphic $\mathbf{q}$-differentials}
\end{description}

\noindent For $k=1,\dots,g$, let us denote by $\omega_{\mathbf{q},k}$  the holomorphic $\mathbf{q}$-differentials defined by
\begin{align*}
\omega_{\mathbf{q},k}(P)&:=\frac{1}{2{\rm i}\pi}\oint_{b_k}W_{\mathbf{q}}(P,Q)
=\omega_k(P)-\Big(\big(I+\mathbf{q}\mathbb{B}\big)^{-1}\mathbf{q}\Big)_{\alpha\beta}\mathbb{B}_{k\beta}\omega_{\alpha}(P).
\end{align*}
Therefore
\begin{equation}\label{holomorphic-q}
\omega_{\mathbf{q},k}(P)=\big(I+\mathbb{B}\mathbf{q}\big)^{k\alpha}\omega_{\alpha}(P).
\end{equation}
The holomorphic $\mathbf{q}$-differentials $\omega_{\mathbf{q},k}$ enjoy the following normalization property:
\begin{equation}\label{holom-q-norm}
\oint_{a_j}\omega_{\mathbf{q},k}+\mathbf{q}_{j\alpha}\oint_{b_{\alpha}}\omega_{\mathbf{q},k}=\delta_{jk},\quad \quad j=1,\dots,g.
\end{equation}
Let us now introduce the $\mathbf{q}$-Riemann matrix of $b$-periods of normalized holomorphic $\mathbf{q}$-differentials (\ref{holomorphic-q}) by
\begin{equation}\label{B-q}
(\mathbb{B}_{\mathbf{q}})_{kl}:=\oint_{b_l}\omega_{\mathbf{q},k}=\big(I+\mathbb{B}\mathbf{q}\big)^{k\alpha}\mathbb{B}_{\alpha{l}}.
\end{equation}
Note that the normalization property (\ref{holom-q-norm}) specifies the holomorphic  $\mathbf{q}$-differentials $\omega_{\mathbf{q},k}$ uniquely. This follows from the fact that a holomorphic differential $\phi$ on $C_g$ satisfying
\begin{equation}\label{condition norm}
\oint_{a_j}\phi+\mathbf{q}_{j\alpha}\oint_{b_{\alpha}}\phi=0,\quad \quad \forall\ j=1,\dots,g,
\end{equation}
must vanish identically. Indeed, if we write $\phi=c_1\omega_1+\dots+c_g\omega_g$, then (\ref{condition norm}) implies that
$$
\big(I+\mathbf{q}\mathbb{B}\big)(c_1,\dots,c_g)^T=0
$$
and thus we get $c_1=\dots=c_g=0$.

\begin{description}
\item[$\bullet$] \textbf{Quasi-homogeneous $\mathbf{q}$-differentials and primary $\mathbf{q}$-differentials}
\end{description}
We start with the following definition of  quasi-homogeneous differentials with respect to the $\mathbf{q}$-bidifferential (\ref{q-W-def}). It is similar to Definition  \ref{quasi-homo-diff}.
\begin{Def}\label{q-quasih}
Let $d_0$ be a nonnegative real number. A  $\mathbf{q}$-differential $\omega_{\mathbf{q}}$ is called quasi-homogeneous of degree $d_0$  if $\omega_{\mathbf{q}}$ can be written as a linear combination of $\mathbf{q}$-differentials of the following form:
$$
\int_{\ell_0}\big(\lambda(Q)\big)^{d_0}W_{\mathbf{q}}(P,Q)
$$
with $\ell_0$ being a contour satisfying  condition \texttt{C1}) in (\ref{Diff model}).
\end{Def}

 We are going to introduce the following family of $N$ typical  examples of quasi-homogeneous $\mathbf{q}$-differentials.
They can be regarded as the $\mathbf{q}$-analogue of the Dubrovin primary differentials (\ref{Primary-e}).  These $\mathbf{q}$-differentials and the corresponding Darboux-Egoroff metrics (\ref{q-eta})  turn out to the so-called $\mathbf{q}$-deformed semi-simple Frobenius manifold structures, studied in \cite{Vasilisa2}.
\medskip

\noindent\textbf{i)} For $t^A\in \big\{t^{i,\epsilon},v^i,s^i,\rho^k\big\}$, with $t^{i,\epsilon}$, $v^i$, $s^i$, $\rho^k$ being the operations   defined by (\ref{FBP-W}), consider the $\mathbf{q}$-differential
\begin{equation}\label{q-phi-tA}
\begin{split}
\phi_{\mathbf{q},t^A}(P)&:=c_A\int_{\ell_A}\big(\lambda(Q)\big)^{d_A}W_{\mathbf{q}}(P,Q)
=\phi_{t^A}(P)-{2{\rm i}\pi}\Big(\big(I+\mathbf{q}\mathbb{B}\big)^{-1}\mathbf{q}\Big)_{\alpha\beta}t^A(\omega_{\beta})\omega_{\alpha}(P),
\end{split}
\end{equation}
where the triplet $(d_A,c_A,\ell_A)$ is determined by  integral representation formulas (\ref{Primary-e}) and (\ref{Int-rep-e}) for the Abelian differential $\phi_{t^A}$.\\
In particular, the $\mathbf{q}$-differential $\phi_{\mathbf{q},t^A}$ is a quasi-homogeneous $\mathbf{q}$-differential of degree $d_A$ and  has the same singularity structure as the primary differential $\phi_{t^A}$ (\ref{Primary-e})  described in Proposition \ref{Primary-e-prop}. Furthermore, because of the symmetry property $\oint_{b_r}\phi_{t^A}={2{\rm i}\pi}\rho^r(\phi_{t^A})=t^A(\omega_r)$ (\ref{Fubini-flat}),  differential  $\phi_{\mathbf{q},t^A}$ has the following property:
\begin{equation}\label{q-phi-tA-period}
\oint_{a_j}\phi_{\mathbf{q},t^A}+\mathbf{q}_{j\alpha}\oint_{b_{\alpha}}\phi_{\mathbf{q},t^A}=\delta_{t^A,\rho^j},\quad\quad  j=1,\dots,g.
\end{equation}
On the other hand, (\ref{q-phi-tA}) implies that $\phi_{\mathbf{q},t^A}$ coincides with the differential $\phi_{t^A}$ provided that the matrix $\mathbf{q}$ is zero.

\medskip

\noindent\textbf{ii)} For $k=1,\dots,g$, let $\phi_{\mathbf{q},\xi^k}$ be the multivalued $\mathbf{q}$-differential defined by
\begin{equation}\label{q-phi-xik}
\begin{split}
\phi_{\mathbf{q},\xi^k}(P)&:=\oint_{a_k}\lambda(Q)W_{\mathbf{q}}(P,Q)+\mathbf{q}_{k\sigma}\oint_{Q\in b_{\sigma}}\lambda(Q)W_{\mathbf{q}}(P,Q)\\
&=\phi_{u^k}(P)-{2{\rm i}\pi}\Big(\big(I+\mathbf{q}\mathbb{B}\big)^{-1}\mathbf{q}\Big)_{\alpha\beta}u^k(\omega_{\beta})\omega_{\alpha}(P)\\
&\quad +\mathbf{q}_{k\sigma}\Big(\phi_{{\widetilde{u}}^{\sigma}}(P)-{2{\rm i}\pi}
\Big(\big(I+\mathbf{q}\mathbb{B}\big)^{-1}\mathbf{q}\Big)_{\alpha\beta}{\widetilde{u}}^{\sigma}(\omega_{\beta})\omega_{\alpha}(P)\Big),
\end{split}
\end{equation}
with $\phi_{u^k}$ being the primary differential in (\ref{Primary-e}), $u^k(\omega)$ the function $u^k(\omega)=\oint_{a_k}\lambda(P)\omega(P)$ and  $\phi_{\widetilde{u}^r}$ and $\widetilde{u}^r$ being respectively the differential and the corresponding operation defined by
\begin{equation}\label{phi-u-b}
\begin{split}
&\phi_{{\widetilde{u}}^r}(P):=\oint_{b_r}\lambda(Q)W(P,Q);\\
&{\widetilde{u}}^r(\omega):=\oint_{b_r}\lambda(P)\omega(P),
\end{split}
\quad\quad  r=1,\dots,g.
\end{equation}
In expression (\ref{q-phi-xik}), we have assumed the sum over the repeated indices $\alpha,\beta,\sigma$.\\
The multivalued  differential $\phi_{{\widetilde{u}}^r}$ is  the particular case $\nu=1$ of the differentials $\big\{\phi_{{\widetilde{u}}^{k,\nu}}\big\}_{k,\nu}$ studied in  Appendix 2.\\
According to the described  jumps  (\ref{type5}) and  (\ref{sigma-nu-jump}) of the differentials $\phi_{u^k}$ and $\phi_{{\widetilde{u}}^r}$, we get
\begin{equation}\label{q-phi-xik-jump}
\begin{split}
&\phi_{\mathbf{q},\xi^k}(P+a_j)-\phi_{\mathbf{q},\xi^k}(P)={2{\rm i}\pi}\mathbf{q}_{jk}d\lambda(P);\\
&\phi_{\mathbf{q},\xi^k}(P+b_j)-\phi_{\mathbf{q},\xi^k}(P)=-{2{\rm i}\pi}\delta_{jk}d\lambda(P).
\end{split}
\end{equation}
Moreover, using (\ref{sigma-nu-periods}) and taking into account the fact that the cycles $a_k, b_k$ start at a marked point $P_0$ such  that $\lambda(P_0)=0$, we deduce that  the multivalued differential $\phi_{\mathbf{q},\xi^k}$ (\ref{q-phi-xik}) satisfies the normalization property  (\ref{q-phi-tA-period}). Note that, the first equality in (\ref{q-phi-xik}) and Definition \ref{q-quasih} tell us that $\phi_{\mathbf{q},\xi^k}$ is a quasi-homogeneous $\mathbf{q}$-differential of degree 1. In addition, $\phi_{\mathbf{q},\xi^k}$ turns into the multivalued  differential $\phi_{u^k}$ defined in (\ref{Primary-e}) when $\mathbf{q}=0$.
\begin{Prop}\label{primary-prop} Assume that $\phi_{\mathbf{q},t^A}$ is among the list of  $\mathbf{q}$-differentials
\begin{equation}\label{Primary-e-q}
\Big\{\phi_{\mathbf{q},t^{i,\epsilon}}, \phi_{\mathbf{q},v^i}, \phi_{\mathbf{q},s^i}, \phi_{\mathbf{q},\rho^k}, \phi_{\mathbf{q},\xi^k}\Big\}
\end{equation}
 defined by (\ref{q-phi-tA}) and (\ref{q-phi-xik}). \\
Then the $\mathbf{q}$-differential $\phi_{\mathbf{q},t^A}$ is  primary with respect to vector field $e=\sum_{j=1}^N\partial_{\lambda_j}$, where by this we mean that  $e$ is covariantly constant with respect to the Levi-Civita connection  of the metric $\eta\big(\phi_{{\mathbf{q}},t^A}\big)$ defined by  (\ref{q-eta}).
\end{Prop}
\emph{Proof:} Let $\nabla$ be the Levi-Civita connection of the metric $\eta\big(\phi_{{\mathbf{q}},t^A}\big)$. Analogously  to (\ref{e-covariantly0}), we can see that the flatness  property  $\nabla{e}=0$ of the vector field $e$ can be characterized by the N conditions
$$
e.\phi_{\mathbf{q},t^A}(P_j)=0,\quad  \quad \forall\ j=1,\dots,N.
$$
Suppose first that $t^A\in \big\{t^{i,\epsilon},v^i,s^i,\rho^k\big\}$.  We have
\begin{align*}
e.\phi_{\mathbf{q},t^A}(P_j)
&=e.\phi_{t^A}(P_j)-{2{\rm i}\pi}\Big(e.\Big(\big(I+\mathbf{q}\mathbb{B}\big)^{-1}\mathbf{q}\Big)_{\alpha\beta}\Big)t^A(\omega_{\beta})\omega_{\alpha}(P_j)\\
&\quad -{2{\rm i}\pi}\Big(\big(I+\mathbf{q}\mathbb{B}\big)^{-1}\mathbf{q}\Big)_{\alpha\beta}\Big(e.\big[t^A(\omega_{\beta})\omega_{\alpha}(P_j)\big]\Big)\\
&=0,
\end{align*}
where we used  expression  (\ref{q-phi-tA}) for $\phi_{\mathbf{q},t^A}$ in the first equality and in the second equality, we used the following arguments:\\
1) relations $e.\mathbb{B}=0=e. t^A(\omega_k)$: the first one was observed in  (\ref{B-eE}) and the second follows from  (\ref{Entries eta}) and (\ref{dual metric-e});\\
2) the differentials $\phi_{t^{i,\epsilon}}$, $\phi_{v^i}$, $\phi_{s^i}$ and  $\phi_{\rho^k}=\omega_k$ are  $e$-primary (by Corollary \ref{Rmk-primary}) and thus $e.\phi_{t^A}(P_j)=0$.\\
Let us now consider the case $t^A=\xi^k$. We first observe that formula (\ref{Auxi2}) applied to the differential $\omega=\phi_{{\widetilde{u}}^r}$ given by (\ref{phi-u-b})
and  relation (\ref{e-sigma-nu}) imply respectively the following:
$$
e.\phi_{{\widetilde{u}}^{\sigma}}(P_j)={2{\rm i}\pi}\omega_{\sigma}(P_j)\quad \text{and}\quad e.{\widetilde{u}}^{\sigma}(\omega_{\beta})=\mathbb{B}_{\beta\sigma}.
$$
These relations, formula (\ref{q-phi-xik}) and the  arguments used in the previous case enable us to arrive at the following:
\begin{align*}
e.\phi_{\mathbf{q},\xi^k}(P_j)
&=e.\phi_{u^k}(P_j)-{2{\rm i}\pi}\Big(\big(I+\mathbf{q}\mathbb{B}\big)^{-1}\mathbf{q}\Big)_{\alpha\beta}\big(e.u^k(\omega_{\beta})\big)\omega_{\alpha}(P_j)\\
&\quad +\mathbf{q}_{k\sigma}\bigg(e.\phi_{{\widetilde{u}}^{\sigma}}(P_j)-{2{\rm i}\pi}
\Big(\big(I+\mathbf{q}\mathbb{B}\big)^{-1}\mathbf{q}\Big)_{\alpha\beta}\big(e.{\widetilde{u}}^{\sigma}(\omega_{\beta})\big)\omega_{\alpha}(P_j)\bigg)\\
&=-{2{\rm i}\pi}\Big(\big(I+\mathbf{q}\mathbb{B}\big)^{-1}\mathbf{q}\Big)_{k\alpha}\omega_{\alpha}(P_j)
 +{2{\rm i}\pi}\mathbf{q}_{k\sigma}\bigg(\omega_{\sigma}(P_j)-\Big(\big(I+\mathbf{q}\mathbb{B}\big)^{-1}\mathbf{q}\Big)_{\alpha\beta}
 \mathbb{B}_{\beta\sigma}\omega_{\alpha}(P_j)\bigg)\\
&=0.
\end{align*}

\fd

\begin{description}
\item[$\bullet$]\textbf{Flat coordinates of the flat metric $\eta\big(\omega_{\mathbf{q}}\big)$}
\end{description}
Similarly to the link between (\ref{FF-model}) and (\ref{phi-t}), we consider the list of $N$  operations
$\big\{t^{i,\epsilon},v^i,s^i,\rho^k,\xi^k\big\}$ corresponding to the primary  $\mathbf{q}$-differentials (\ref{q-phi-tA}) and (\ref{q-phi-xik}) where  $t^{i,\epsilon},v^i,s^i,\rho^k$ are the same as in (\ref{FBP-W}) and $\xi^k$ is the following operation:
\begin{equation}\label{xik}
\xi^k(\omega):=\oint_{a_k}\lambda(P)\omega(P)+\mathbf{q}_{k\alpha}\oint_{b_{\alpha}}\lambda(P)\omega(P)
=u^k(\omega)+\mathbf{q}_{k\alpha}\widetilde{u}^{\alpha}(\omega).
\end{equation}
Here $\omega$ is a differential as in (\ref{Diff model}) or a $\mathbf{q}$-differential as in (\ref{q-Diff model}) and $u^k$ and $\widetilde{u}^{\alpha}$ are respectively the operations defined in (\ref{S-eta}) and (\ref{phi-u-b}).

\medskip

\noindent\textbf{i)} Assume that  $t^A, t^B\in \big\{t^{i,\epsilon},v^i,s^i,\rho^k\big\}$. Then,  due to (\ref{q-phi-tA}) and the symmetry property (\ref{Fubini-flat}), we have
\begin{equation}\label{q-tAB-1}
\begin{split}
t^A\big(\phi_{\mathbf{q},t^B}\big)
&=t^A\big(\phi_{t^B}\big)-{2{\rm i}\pi}\Big(\big(I+\mathbf{q}\mathbb{B}\big)^{-1}\mathbf{q}\Big)_{\alpha\beta}t^A(\omega_{\alpha})t^B(\omega_{\beta})\\
&=t^B\big(\phi_{\mathbf{q},t^A}\big)+\log(-1)\sum_{i,j=1}^m(1-\delta_{ij})\delta_{t^A,s^i}\delta_{t^B,s^j}.
\end{split}
\end{equation}
\noindent\textbf{ii)} When $t^A\in \big\{t^{i,\epsilon},v^i,s^i,\rho^k\big\}$ and $t^B=\xi^k$,  then we have
\begin{equation}\label{q-tAB-2}
\begin{split}
t^A\big(\phi_{\mathbf{q},\xi^k}\big)
&=t^A\big(\phi_{u^k}\big)-{2{\rm i}\pi}\Big(\big(I+\mathbf{q}\mathbb{B}\big)^{-1}\mathbf{q}\Big)_{\alpha\beta}t^A(\omega_{\alpha})u^k(\omega_{\beta})\\
&\quad +\mathbf{q}_{k\sigma}\bigg(t^A\big(\phi_{\widetilde{u}^{\sigma}}\big)-{2{\rm i}\pi}
\Big(\big(I+\mathbf{q}\mathbb{B}\big)^{-1}\mathbf{q}\Big)_{\alpha\beta}t^A(\omega_{\alpha}){\widetilde{u}}^{\sigma}(\omega_{\beta})\bigg)\\
&=\xi^k\big(\phi_{\mathbf{q},t^A}\big),
\end{split}
\end{equation}
where we have used expression (\ref{q-phi-xik}) for the multivalued  differential $\phi_{\mathbf{q},\xi^k}$ in the first equality and the symmetry properties
$$
t^A\big(\phi_{u^r}\big)=u^r\big(\phi_{t^A}\big)\quad \quad \text{and}\quad \quad t^A\big(\phi_{\widetilde{u}^r}\big)=\widetilde{u}^r\big(\phi_{t^A}\big), \quad\quad  \forall\ t^A=t^{i,\epsilon},v^i,s^i,\rho^k
$$
in the second equality. Here, as in (\ref{Fubini1}) and (\ref{Fubini-flat}),  we have taken into account the fact that the cycles $a_k,b_k$  start at a marked point $P_0\in C_g$ so that $\lambda(P_0)=0$.

\medskip

\noindent\textbf{iii)} If $t^A=\xi^k$ and $t^B=\xi^l,$ with $k,l=1,\dots,g$,  then by assuming the sum over the indices $\alpha,\beta,\mu,\nu$, we have
\begin{equation}\label{q-tAB-3}
\begin{split}
\xi^k\big(\phi_{\mathbf{q},\xi^l}\big)
&=u^k\big(\phi_{u^l}\big)-{2{\rm i}\pi}\Big(\big(I+\mathbf{q}\mathbb{B}\big)^{-1}\mathbf{q}\Big)_{\alpha\beta}u^k(\omega_{\alpha})u^l(\omega_{\beta})
+\mathbf{q}_{k\mu}\widetilde{u}^{\mu}\big(\phi_{u^l}\big)+\mathbf{q}_{l\nu}u^k\big(\phi_{\widetilde{u}^{\nu}}\big)\\
&\quad-{2{\rm i}\pi}\mathbf{q}_{k\mu}\Big(\big(I+\mathbf{q}\mathbb{B}\big)^{-1}\mathbf{q}\Big)_{\alpha\beta}\widetilde{u}^{\mu}(\omega_{\alpha})u^l(\omega_{\beta})
-{2{\rm i}\pi}\mathbf{q}_{l\nu}\Big(\big(I+\mathbf{q}\mathbb{B}\big)^{-1}\mathbf{q}\Big)_{\alpha\beta}u^k(\omega_{\alpha})\widetilde{u}^{\nu}(\omega_{\beta})\\
&\quad +\mathbf{q}_{k\mu}\mathbf{q}_{l\nu}\bigg(\widetilde{u}^{\mu}\big(\phi_{\widetilde{u}^{\nu}}\big)-{2{\rm i}\pi}
\Big(\big(I+\mathbf{q}\mathbb{B}\big)^{-1}\mathbf{q}\Big)_{\alpha\beta}\widetilde{u}^{\mu}(\omega_{\alpha})\widetilde{u}^{\nu}(\omega_{\beta})\bigg)\\
&=\xi^l\big(\phi_{\mathbf{q},\xi^k}\big).
\end{split}
\end{equation}
Therefore,  equalities (\ref{q-tAB-1}),  (\ref{q-tAB-2}) and  (\ref{q-tAB-3}) yield that
$$
t^A\big(\phi_{\mathbf{q},t^B}\big)=t^B\big(\phi_{\mathbf{q},t^A}\big)+\log(-1)\sum_{i,j=1}^m(1-\delta_{ij})\delta_{t^A,s^i}\delta_{t^B,s^j},
\quad \quad \forall\  t^A, t^B\in \big\{t^{i,\epsilon},v^i,s^i,\rho^k, \xi^k\big\}
$$
provided that the cycles of the canonical homology basis $\{a_k,b_k\}$ start at a marked point $P_0$ with $\lambda(P_0)=0$. Furthermore, using the same relations we see  that the functions $\big\{t^A\big(\phi_{\mathbf{q},t^B}\big)\big\}$ satisfy the following limiting behavior as the matrix $\mathbf{q}$ is the zero matrix:
\begin{equation}\label{q-limit-FC}
t^A\big(\phi_{\mathbf{q},t^B}\big)\Big|_{\mathbf{q}=0}=
\left\{
\begin{array}{lll}
t^A\big(\phi_{t^B}\big),\quad  & \hbox{if}\quad  t^A, t^B\in \big\{t^{i,\epsilon},v^i,s^i,\rho^k\big\};\\
\\
t^A\big(\phi_{u^k}\big), \quad  & \hbox{if}\quad t^A\in \big\{t^{i,\epsilon},v^i,s^i,\rho^k\big\}\quad \text{and}\quad t^B=\xi^k;\\
\\
u^k\big(\phi_{u^l}\big),\quad  & \hbox{if}\quad  t^A=\xi^k\quad \text{and}\quad t^B=\xi^l.
\end{array}
\right.
\end{equation}

\begin{Prop} Let $\omega_{\mathbf{q}}$ be the $\mathbf{q}$-differential defined by (\ref{q-Diff model}) and $\eta(\omega_{\mathbf{q}}\big)$ be the corresponding Darboux-Egoroff metric (\ref{q-eta}) defined on the open set $\mathcal{M}_{\omega_{\mathbf{q}}}$ determined by  conditions (\ref{q-assumptions}).\\
Then the following set
\begin{equation}\label{q-S-eta}
\mathcal{S}(\omega_{\mathbf{q}}):=\Big\{t^{i,\epsilon}\big(\omega_{\mathbf{q}}\big),v^i\big(\omega_{\mathbf{q}}\big),s^i\big(\omega_{\mathbf{q}}\big),
\rho^k\big(\omega_{\mathbf{q}}\big),\xi^k\big(\omega_{\mathbf{q}}\big)\Big\},
\end{equation}
where $t^{i,\epsilon},v^i,s^i,\rho^k$ are the operations defined  in (\ref{FBP-W}) and $\xi^k$ is given by (\ref{xik}), provides a system of flat coordinates of the metric
$\eta(\omega_{\mathbf{q}}\big)$. \\
In  coordinates (\ref{q-S-eta}), the nonzero entries of the constant  matrix of the contravariant metric $\eta^*\big(\omega_{\mathbf{q}}\big)$ are as follows:
\begin{equation}\label{Entries eta-q}
\begin{split}
&\eta^*\big(\omega_{\mathbf{q}}\big)(dt^{i,\epsilon_1}\big(\omega_{\mathbf{q}}\big),dt^{j,\epsilon_2}\big(\omega_{\mathbf{q}}\big))
=\delta_{ij}\delta_{\epsilon_1+\epsilon_2,n_j+1};\\
&\eta^*\big(\omega_{\mathbf{q}}\big)(dv^{i}\big(\omega_{\mathbf{q}}\big),ds^{j}\big(\omega_{\mathbf{q}}\big))=\delta_{ij};\\
&\eta^*\big(\omega_{\mathbf{q}}\big)(d\rho^{k}\big(\omega_{\mathbf{q}}\big),d\xi^{l}\big(\omega_{\mathbf{q}}\big))=\delta_{kl}.
\end{split}
\end{equation}
\end{Prop}
\emph{Proof:} Due to  Rauch variational formula (\ref{q-W-Rauch}), we have
\begin{align*}
\eta^*\big(\omega_{\mathbf{q}}\big)\big(dt^A\big(\omega_{\mathbf{q}}\big),dt^B\big(\omega_{\mathbf{q}}\big)\big)
&=2\sum_{j=1}^N \frac{\partial_{\lambda_j}t^A\big(\omega_{\mathbf{q}}\big)\partial_{\lambda_j}t^B\big(\omega_{\mathbf{q}}\big)}{\big(\omega_{\mathbf{q}}(P_j)\big)^2}
=\frac{1}{2}\sum_{j=1}^N\phi_{\mathbf{q},t^A}(P_j)\phi_{\mathbf{q},t^B}(P_j)
=e.t^A\big(\phi_{\mathbf{q},t^B}\big).
\end{align*}
We consider  two cases. \\
\emph{First case:} $t^A, t^B\in \big\{t^{i,\epsilon},v^i,s^i,\rho^k\big\}$. Using (\ref{q-tAB-1}) and  properties  (\ref{Entries eta}), (\ref{dual metric-e}) and (\ref{duality}) we deduce that
\begin{align*}
e.t^A\big(\phi_{\mathbf{q},t^B}\big)=e.t^A\big(\phi_{t^B}\big)=\delta_{t^B,t^{A^{\prime}}},
\end{align*}
with $t^{A^{\prime}}$ being the dual flat  coordinate of $t^A$ in the sense of (\ref{duality}) and (\ref{duality-picture}).\\
\emph{Second case:}  $t^A\in \big\{t^{i,\epsilon},v^i,s^i,\rho^k,\xi^k\big\}$ and $t^B=\xi^l$. We proceed as in the proof of Theorem \ref{FC-thm-eta}. Consider the multivalued differential
\begin{equation*}
\Phi_{\mathbf{q},A,l}(P):=\frac{\phi_{\mathbf{q},t^A}(P)\phi_{\mathbf{q},\xi^l}(P)}{d\lambda(P)}.
\end{equation*}
Relation (\ref{q-phi-xik-jump}) implies that the jumps of $\Phi_{\mathbf{q},A,l}$ are as follows:
\begin{align*}
\Phi_{\mathbf{q},A,l}(P+a_j)-\Phi_{\mathbf{q},A,l}(P)&={2{\rm i}\pi}
\delta_{t^A,\xi^k}\Big(\mathbf{q}_{lj}\phi_{\mathbf{q},\xi^k}(P)+\mathbf{q}_{kj}\phi_{\mathbf{q},\xi^l}(P)+{2{\rm i}\pi}
\mathbf{q}_{kj}\mathbf{q}_{lj}d\lambda(P)\Big)\\
&\quad +{2{\rm i}\pi}(1-\delta_{t^A,\xi^k})\mathbf{q}_{lj}\phi_{\mathbf{q},t^A}(P);\\
\Phi_{\mathbf{q},A,l}(P+b_j)-\Phi_{\mathbf{q},A,l}(P)&=-{2{\rm i}\pi}\delta_{t^A,\xi^k}\Big(\delta_{jl}\phi_{\mathbf{q},\xi^k}(P)
+\delta_{jk}\phi_{\mathbf{q},\xi^l}(P)-{2{\rm i}\pi}\delta_{jk}\delta_{jl}d\lambda(P)\Big)\\
&\quad -{2{\rm i}\pi}(1-\delta_{t^A,\xi^k})\delta_{jl}\phi_{\mathbf{q},t^A}(P).
\end{align*}
Now by using the residue theorem and the normalization property (\ref{q-phi-tA-period}) of the primary $\mathbf{q}$-differentials $\big\{\phi_{\mathbf{q},t^A}\big\}$, we conclude that
\begin{align*}
e.t^A\big(\phi_{\mathbf{q},\xi^l}\big)&=\frac{1}{2}\sum_{r=1}^N\phi_{\mathbf{q},t^A}(P_r)\phi_{\mathbf{q},\xi^l}(P_r)
=\sum_{r=1}^N\underset{P_r}{{\rm res}}\Phi_{\mathbf{q},A,l}(P)\\
&=\frac{1}{{2{\rm i}\pi}}\sum_{j=1}^g\oint_{a_j}\Big(\Phi_{\mathbf{q},A,l}(P)-\Phi_{\mathbf{q},A,l}(P+b_j)\Big)
+\frac{1}{{2{\rm i}\pi}}\sum_{j=1}^g\oint_{b_j}\Big(\Phi_{\mathbf{q},A,l}(P)-\Phi_{\mathbf{q},A,l}(P+a_j^{-1})\Big)\\
&=(1-\delta_{t^A,\xi^k})\bigg(\oint_{a_l}\phi_{\mathbf{q},t^A}+\sum_{j=1}^g\mathbf{q}_{lj}\oint_{b_j}\phi_{\mathbf{q},t^A}\bigg)
+\delta_{t^A,\xi^k}\bigg(\oint_{a_l}\phi_{\mathbf{q},\xi^k}+\sum_{j=1}^g\mathbf{q}_{lj}\oint_{b_j}\phi_{\mathbf{q},\xi^k}\bigg)\\
&\quad +\delta_{t^A,\xi^k}\bigg( \oint_{a_k}\phi_{\mathbf{q},\xi^l}+\sum_{j=1}^g\mathbf{q}_{kj}\oint_{b_j}\phi_{\mathbf{q},\xi^l}\bigg)\\
&=(1-\delta_{t^A,\xi^k})\delta_{t^A,\rho^l}=\delta_{t^A,\rho^l}.
\end{align*}
\fd

\begin{Remark}
As in (\ref{duality}) and (\ref{duality-picture}),  the constant Gram matrix (\ref{Entries eta-q}), which does not depend on the choice of the $\mathbf{q}$-differential $\omega_{\mathbf{q}}$, permits us to the following duality relations that connect  the five families of operations  $\big\{t^{i,\epsilon}, v^i, s^i, \rho^k, \xi^k\big\}$  as well as the corresponding primary $\mathbf{q}$-differentials (\ref{Primary-e-q}):
\begin{equation}\label{q-duality-picture}
\begin{array}{cccc}
t^{({i,\epsilon})^{\prime}}=t^{i,n_i+1-\epsilon},&\quad \quad \quad v^{{i}^{\prime}}=s^i,&\quad \quad  \quad \rho^{k^{\prime}}=\xi^k;\\
\\
\phi_{\mathbf{q}, t^{({i,\epsilon})^{\prime}}}=\phi_{\mathbf{q}, t^{i,n_i+1-\epsilon}},&\quad \quad \quad \phi_{\mathbf{q}, v^{i^{\prime}}}=\phi_{\mathbf{q}, s^i},&\quad \quad \quad \phi_{\mathbf{q}, \rho^{k^{\prime}}}=\phi_{\mathbf{q}, \xi^k}.
\end{array}
\end{equation}
In particular, we observe that the degrees of the quasi-homogeneous and primary $\mathbf{q}$-differentials (\ref{Primary-e-q})  enjoy   the following duality relation, identical to (\ref{degree key relation}):
\begin{equation}\label{degree-q}
{\rm deg}\big(\phi_{\mathbf{q},t^A}\big)+{\rm deg}\big(\phi_{\mathbf{q},t^{A^{\prime}}}\big)=1, \quad \quad \text{for all\ A}.
\end{equation}
\end{Remark}

\medskip

As  two direct consequences, we can use  (\ref{Entries eta-q}) together with the already mentioned duality relations (\ref{q-duality-picture}) to get  the following results analogous to (\ref{flat-canonical relation})  and (\ref{e-tA}), respectively.
\begin{Cor}
The partial derivative with respect to $t^A\big(\omega_{\mathbf{q}}\big)\in \mathcal{S}(\omega_{\mathbf{q}})$ (\ref{q-S-eta}) is determined by
\begin{equation}\label{tA-q-R1}
\partial_{t^A(\omega_{\mathbf{q}})}= \sum_j\frac{\phi_{\mathbf{q}, t^{A^{\prime}}}(P_j)}{\omega_{\mathbf{q}}(P_j)}\partial_{\lambda_j}.
\end{equation}
\end{Cor}
\begin{Cor}
Let $\phi_{\mathbf{q}, t^A}$ be one of the primary $\mathbf{q}$-differentials listed in (\ref{Primary-e-q}) and defined by (\ref{q-phi-tA}) and (\ref{q-phi-xik}). Then, in the system of flat coordinates $\mathcal{S}\big(\phi_{\mathbf{q}, t^A}\big)$ (\ref{q-S-eta}) of the Darboux-Egoroff metric $\eta\big(\phi_{\mathbf{q}, t^A}\big)$ (\ref{q-eta}), the vector field $e=\sum_{j=1}^N\partial_{\lambda_j}$
is represented by
\begin{equation}\label{e-tA-q}
e=\frac{\partial}{\partial{t^{A^{\prime}}\big(\phi_{\mathbf{q}, t^A}\big)}}.
\end{equation}
\end{Cor}

\medskip

In the following result we discuss  the quasi-homogeneity property of the flat functions $\big\{t^A\big(\phi_{\mathbf{q}, t^B}\big)\big\}_{A,B}$. This is also similar  to
(\ref{Entries-g-eta}).
\begin{Prop} Let $E$ be the vector field $E=\sum_j\lambda_j\partial_{\lambda_j}$ and $t^A,t^B$ be two operations from the set
$\big\{t^{i,\alpha},v^i,s^i,\rho^k,\xi^k\big\}$. Then
\begin{equation}\label{E-q-tAtB}
E.t^A\big(\phi_{\mathbf{q}, t^B}\big)=(d_A+d_B)t^A\big(\phi_{\mathbf{q}, t^B}\big)
+\sum_{i,j=1}^m\delta_{t^A,s^i}\delta_{t^B,s^j}\Big(\frac{1}{n_0+1}+\delta_{ij}\frac{1}{n_i+1}\Big).
\end{equation}
\end{Prop}
\emph{Proof:}
Assume that $t^A, t^B\in \big\{t^{i,\epsilon},v^i,s^i,\rho^k\big\}$. By using (\ref{q-tAB-1}), (\ref{B-eE}) and applying twice (\ref{Entries-g-eta}), we see that
\begin{align*}
E.t^A\big(\phi_{\mathbf{q},t^B}\big)&=E.t^A\big(\phi_{t^B}\big)-{2{\rm i}\pi}\Big(\big(I+\mathbf{q}\mathbb{B}\big)^{-1}\mathbf{q}\Big)_{\alpha\beta}
\Big(E.\big[t^A(\omega_{\alpha})t^B(\omega_{\beta})\big]\Big)\\
&=(d_A+d_B)t^A\big(\phi_{\mathbf{q}, t^B}\big)
+\sum_{i,j=1}^m\delta_{t^A,s^i}\delta_{t^B,s^j}\Big(\frac{1}{n_0+1}+\delta_{ij}\frac{1}{n_i+1}\Big).
\end{align*}
Let us now deal with the case where   $t^A\in \big\{t^{i,\epsilon},v^i,s^i,\rho^k,\xi^k\big\}$ and $t^B=\xi^l$. By studying  the analytic properties of the multivalued differential
\begin{equation*}
\lambda(P)\Phi_{\mathbf{q},A,l}(P):=\lambda(P)\frac{\phi_{\mathbf{q},t^A}(P)\phi_{\mathbf{q},\xi^l}(P)}{d\lambda(P)}
\end{equation*}
(its jumps and its poles) inside a fundamental polygon $F_g$ associated with the surface $C_g$, in order to arrive at (\ref{E-q-tAtB}), it suffices to use the equality
\begin{align*}
E.t^A\big(\phi_{\mathbf{q},\xi^l}\big)&=\frac{1}{2}\sum_{r=1}^N\lambda_j\phi_{\mathbf{q},t^A}(P_r)\phi_{\mathbf{q},\xi^l}(P_r)
=\sum_{r=1}^N\underset{P_r}{{\rm res}}\Big(\lambda(P)\Phi_{\mathbf{q},A,l}(P)\Big)
\end{align*}
(the first equality follows from Rauch variational formulas (\ref{q-W-Rauch})) and apply the residue theorem and proceed as in the proof of formula (\ref{u-k-sigma}).

\fd

More generally, if $\omega_{\mathbf{q}}$ is a quasi-homogeneous $\mathbf{q}$-differential of degree $d_0$ in the sense of Definition \ref{q-quasih} and $t^A(\omega_{\mathbf{q}})$ be one of the operations listed in (\ref{q-S-eta}), then the action of the vector field  $E=\sum_j\lambda_j\partial_{\lambda_j}$ on the function $t^A(\omega_{\mathbf{q}})$ is determined by
\begin{equation}\label{E-tA-q}
E.t^A(\omega_{\mathbf{q}})=(d_0+d_A)t^A(\omega_{\mathbf{q}})+r_{\omega_{\mathbf{q}},\phi_{\mathbf{q},t^A}},
\end{equation}
where $d_A$ is the degree of $\phi_{\mathbf{q},t^A}$ and $r_{\omega_{\mathbf{q}},\phi_{\mathbf{q},t^A}}$ is a constant depending on $\omega_{\mathbf{q}}$ and $\phi_{\mathbf{q},t^A}$.\\
As examples, when $\omega_{\mathbf{q}}$ is among the following quasi-homogeneous $\mathbf{q}$-differentials
$$
\underset{\infty^i}{\rm res}\ \lambda(Q)^{1+\frac{\beta}{n_i+1}}W_{\mathbf{q}}(P,Q),\quad \quad
\oint_{a_k}\big(\lambda(Q)\big)^{\sigma}W_{\mathbf{q}}(P,Q),\quad \quad \oint_{b_k}\big(\lambda(Q)\big)^{\nu}W_{\mathbf{q}}(P,Q),
$$
where $\beta,\sigma,\nu$ are positive integers,  then the constants $r_{\omega_{\mathbf{q}},\phi_{\mathbf{q},t^A}}$ in (\ref{E-tA-q}) are zero.

\subsection{Prepotentials of the $\mathbf{q}$-deformed Frobenius manifold structures}
Consider a quasi-homogeneous $\mathbf{q}$-differential $\omega_{\mathbf{q}}$ of degree $d_0\geq0$ (see Definition \ref{q-quasih}) such that $\omega_{\mathbf{q}}(P_j)\neq 0$ for all $j=1,\dots,N$. Then, the open set $\mathcal{M}_{\omega_{\mathbf{q}}}\subset \widehat{\mathcal{H}}_{g, L}(n_0,\dots,n_m)$ determined by  conditions (\ref{q-assumptions}) can be endowed with a Frobenius manifold structure (in the sense of Definition \ref{Frob Man def})  described  by
\begin{itemize}
\item the flat metric $\eta\big(\omega_{\mathbf{q}}\big)$ (\ref{q-eta}) and the corresponding system of flat coordinates (\ref{q-S-eta});
\item the multiplication low ``$\circ$'': $\partial_{\lambda_i}\circ\partial_{\lambda_j}=\delta_{ij}\partial_{\lambda_i}$;
\item the unit and Euler  vector fields $e=\sum_{j=1}^N\partial_{\lambda_j}$ and $E=\sum_j\lambda_j\partial_{\lambda_j}$.
\end{itemize}
Note that the symmetric 3-tensor $\mathbf{c}_{\omega_{\mathbf{q}}}$ defined by
$$
\mathbf{c}_{\omega_{\mathbf{q}}}(x,y,z):=\eta(\omega_{\mathbf{q}})(x\circ y,z)=\frac{1}{2}\sum_jx_jy_jz_j\omega_{\mathbf{q}}(P_j)^2,
$$
with $x=\sum_jx_j\partial_{\lambda_j}$, $y=\sum_jy_j\partial_{\lambda_j}$ and $z=\sum_jz_j\partial_{\lambda_j}$, has the following form with respect to flat coordinates (\ref{q-S-eta}):
\begin{equation}\label{c-FC-q}
\begin{split}
\mathbf{c}_{\omega_{\mathbf{q}}}\big(\partial_{t^A(\omega_{\mathbf{q}})},\partial_{t^B(\omega_{\mathbf{q}})},\partial_{t^C(\omega_{\mathbf{q}})}\big)
&=\frac{1}{2}\sum_{j=1}^N\frac{\phi_{\mathbf{q},t^{A^{\prime}}}(P_j)\phi_{\mathbf{q}, t^{B^{\prime}}}(P_j)\phi_{\mathbf{q}, t^{C^{\prime}}}(P_j)}
{\omega_{\mathbf{q}}(P_j)}
=\partial_{t^{C}(\omega_{\mathbf{q}})}t^{A^{\prime}}\big(\phi_{\mathbf{q}, t^{B^{\prime}}}\big),
\end{split}
\end{equation}
where the second equality is an immediate consequence of (\ref{tA-q-R1}) and Rauch formulas (\ref{q-W-Rauch}) for the $\mathbf{q}$-bidifferential $W_{\mathbf{q}}(P,Q)$.\\
As consequence,  in the system of flat coordinates (\ref{q-S-eta}), the multiplication $``\circ"$ is given in terms of the tensor  (\ref{c-FC-q}) as follows:
$$
\partial_{t^A(\omega_{\mathbf{q}})}\circ \partial_{t^B(\omega_{\mathbf{q}})}=\sum_{C}
\mathbf{c}_{\omega_{\mathbf{q}}}\big(\partial_{t^A(\omega_{\mathbf{q}})},\partial_{t^B(\omega_{\mathbf{q}})},\partial_{t^{C^{\prime}}(\omega_{\mathbf{q}})}\big)
\partial_{t^C(\omega_{\mathbf{q}})}.
$$

Now, by (\ref{degree-q}), (\ref{c-FC-q}) and a  direct adaptation of  the proof of formulas (\ref{Prep-eta}) and  (\ref{Prep-eta2}), we arrive at formula (\ref{Prep-q1}) below, which  offers a new approach to calculate the prepotential associated with a deformed Hurwitz-Frobenius manifold structure (where the unity is not necessarily flat). Note that the method suggested in \cite{Vasilisa2} uses an analogue of the Dubrovin bilinear pairing.
\begin{Thm} Let $\omega_{\mathbf{q}}$ be a quasi-homogeneous  $\mathbf{q}$-differential of degree $d_0\geq0$, $\big\{t^A(\omega_{\mathbf{q}})\big\}$ be the set of flat coordinates (\ref{q-S-eta}) of the flat metric $\eta(\omega_{\mathbf{q}})$ (\ref{q-eta}) and $\mathbf{c}_{\omega_{\mathbf{q}}}$ be the symmetric tensor (\ref{c-FC-q}).\\
Consider the function $\mathbf{F}_{\omega_{\mathbf{q}}}$ defined by
\begin{equation}\label{Prep-q1}
\begin{split}
&\mathbf{F}_{\omega_{\mathbf{q}}}
=\frac{1}{2(1+d_0)}\sum_{A,B}\frac{\big((d_0+d_A)t^A(\omega_{\mathbf{q}})+r_{\omega_{\mathbf{q}},\phi_{\mathbf{q},t^A}}\big)
\big((d_0+d_B)t^B(\omega_{\mathbf{q}})+r_{\omega_{\mathbf{q}},\phi_{\mathbf{q},t^B}}\big)}
{1+d_0+d_{A^{\prime}}}t^{A^{\prime}}(\phi_{\mathbf{q}, t^{B^{\prime}}})\\
&\quad+\frac{\log(-1)}{4}\sum_{i,j=1, i\neq j}^mv^i(\omega_{\mathbf{q}})v^j(\omega_{\mathbf{q}})
-\frac{3}{4(1+d_0)}\sum_{i,j=1}^m\bigg(\frac{1}{n_0+1}+\delta_{ij}\frac{1}{n_i+1}\bigg)v^i(\omega_{\mathbf{q}})v^j(\omega_{\mathbf{q}}),
\end{split}
\end{equation}
with  $v^i(\omega_{\mathbf{q}})$ being the function $v^i(\omega_{\mathbf{q}}):= \underset{\infty^i}{{\rm res}\ }\lambda(P)\omega_{\mathbf{q}}(P)$.\\
Then we have the following results:
\begin{description}
\item[1.] The function $\mathbf{F}_{\omega_{\mathbf{q}}}$ is a prepotential of  the Frobenius manifold structure induced by $\omega_{\mathbf{q}}$, i.e.
$$
\partial_{t^A(\omega_{\mathbf{q}})}\partial_{t^B(\omega_{\mathbf{q}})}\partial_{t^C(\omega_{\mathbf{q}})}\mathbf{F}_{\omega_{\mathbf{q}}}
=\mathbf{c}_{\omega_{\mathbf{q}}}\big(\partial_{t^A(\omega_{\mathbf{q}})},\partial_{t^B(\omega_{\mathbf{q}})},\partial_{t^C(\omega_{\mathbf{q}})}\big).
$$
\item[2.] The function $\mathbf{F}_{\omega_{\mathbf{q}}}$ is quasi-homogeneous of degree $2+2d_0$ in the sense of (\ref{quasihomog-def}), with
\begin{equation}\label{Prep-eta-homog-q}
E.\mathbf{F}_{\omega_{\mathbf{q}}}=2(1+d_0)\mathbf{F}_{\omega_{\mathbf{q}}}
+\frac{1}{2}\sum_{i,j=1}^m\bigg(\frac{1}{n_0+1}+\delta_{ij}\frac{1}{n_i+1}\bigg)v^i(\omega_{\mathbf{q}})v^j(\omega_{\mathbf{q}}),
\end{equation}
and  $E$ being the Euler vector field
\begin{equation}\label{Euler-q}
E=\sum_A\Big((d_0+d_A)t^A(\omega_{\mathbf{q}})+r_{\omega_{\mathbf{q}},\phi_{\mathbf{q},t^A}}\Big)\partial_{t^A(\omega_{\mathbf{q}})}.
\end{equation}
\item[3.] The Hessian matrix of  $\mathbf{F}_{\omega_{\mathbf{q}}}$ does not depend on the chosen  quasi-homogeneous $\mathbf{q}$-differential $\omega_{\mathbf{q}}$ and is determined by
\begin{equation}\label{q-hessian}
\partial_{t^A(\omega_{\mathbf{q}})}\partial_{t^B(\omega_{\mathbf{q}})}\mathbf{F}_{\omega_{\mathbf{q}}}=t^{A^{\prime}}(\phi_{\mathbf{q}, t^{B^{\prime}}})
-\frac{\log(-1)}{2}\sum_{i,j=1}^m(1-\delta_{ij})\delta_{t^A,v^i}\delta_{t^B,v^j}.
\end{equation}
\end{description}
\end{Thm}

\medskip

As in Section  4.1.5, we are going to pay  attention to quasi-homogeneous solutions to the WDVV equations induced by the normalized holomorphic $\mathbf{q}$-differentials
$\big\{\omega_{\mathbf{q},j}=\phi_{\mathbf{q},\rho^j}\big\}$, introduced in  (\ref{holomorphic-q}). \\
Basing on formula (\ref{Prep-q1}), we can extend all the results of Section  4.4.5. to the $\mathbf{q}$-deformation setting. However,  we  plan  to concentrate on the $\mathbf{q}$-analogues of the solutions in (\ref{Prep-holom3}) and (\ref{F-holo-g1}). \\
For each $l=1,\dots,g$, the holomorphic  $\mathbf{q}$-differential $\omega_{\mathbf{q},l}$ is $e$-primary (by Proposition \ref{primary-prop})  and quasi-homogeneous of degree 0. Thus, the flat coordinates $t^A\big(\omega_{\mathbf{q}, l}\big)$ are eigenfunctions of the Euler vector field $E$ and (\ref{E-q-tAtB}) becomes  as follows:
$$
E.t^A\big(\omega_{\mathbf{q}, l}\big)=d_At^A\big(\omega_{\mathbf{q}, l}\big),\quad \quad \forall\ t^A\in \big\{t^{i,\epsilon}, v^i, s^i, \rho^k, \xi^k\big\}.
$$
In addition, according to (\ref{Prep-q1}), the prepotential $\mathbf{F}_{\omega_{\mathbf{q},l}}$ is determined by
\begin{equation}\label{Prep-q2}
\begin{split}
\mathbf{F}_{\omega_{\mathbf{q},l}}
&=\frac{1}{2}\sum_{A,B}\frac{d_Ad_B}{1+d_{A^{\prime}}}t^A(\omega_{\mathbf{q},l})t^B(\omega_{\mathbf{q},l})t^{A^{\prime}}(\phi_{\mathbf{q}, t^{B^{\prime}}})
+\frac{\log(-1)}{4}\sum_{j,k=1, j\neq k}^mv^j(\omega_{\mathbf{q},l})v^k(\omega_{\mathbf{q},l})\\
&\quad-\frac{3}{4}\sum_{j,k=1}^m\bigg(\frac{1}{n_0+1}+\delta_{jk}\frac{1}{n_j+1}\bigg)v^j(\omega_{\mathbf{q},l})v^k(\omega_{\mathbf{q},l}).
\end{split}
\end{equation}

\medskip

In order to extend formulas (\ref{Prep-holom3}) and (\ref{F-holo-g1}) to the framework of $\mathbf{q}$-deformation,  let us start by expressing   the functions
$$
s^j\big(\phi_{\mathbf{q},s^i}\big):=p.v. \int_{\infty^0}^{\infty^j}\phi_{\mathbf{q},s^i},\quad \quad i,j=1,\dots,m,
$$
by means of flat coordinates of the flat metric $\eta(\omega_{\mathbf{q}, l})$ (\ref{q-eta}) induced by the normalized holomorphic $\mathbf{q}$-differentials. Here
$\phi_{\mathbf{q},s^i}$ is an Abelian  differential of the third kind normalized by (\ref{condition norm}), where the simple poles occur at $\infty^i$ and $\infty^0$ and the principal value is the same as in Theorem \ref{s-ij}.
The functions $s^j\big(\phi_{\mathbf{q},s^i}\big)$ are the $\mathbf{q}$-analogue of those obtained in (\ref{s-ij Formula}).
\begin{Prop} Let $\mathbb{B}_{\mathbf{q}}$ be the $\mathbf{q}$-Riemann matrix (\ref{B-q}) and $\mathcal{A}_{\mathbf{q}}(P)$ be the following $\mathbf{q}$-Abel map:
\begin{equation}\label{Abel-q0}
\mathcal{A}_{\mathbf{q}}(P):=\Big(\int^P\omega_{\mathbf{q},1},\dots,\int^P\omega_{\mathbf{q},g}\Big)^T,
\end{equation}
where $\omega_{\mathbf{q},1},\dots, \omega_{\mathbf{q},g}$ are the holomorphic $\mathbf{q}$-differentials defined by (\ref{holomorphic-q}). \\
If  $i\neq j$, then
\begin{equation}\label{s-ij-q}
\begin{split}
s^j\big(\phi_{\mathbf{q},s^i}\big)&=\log\Theta_{\Delta}\left(\big(I-\mathbb{B}_{\mathbf{q}}\mathbf{q}\big)^{-1}\big[\mathcal{A}_{\mathbf{q}}(\infty^j)
-\mathcal{A}\mathbf{_q}(\infty^i)]\Big|\mathbb{B}_{\mathbf{q}}\big(I-\mathbf{q}\mathbb{B}_{\mathbf{q}}\big)^{-1}\right)\\
&\quad -\log\Theta_{\Delta}\left((\big(I-\mathbb{B}_{\mathbf{q}}\mathbf{q}\big)^{-1}\big[\mathcal{A}_{\mathbf{q}}(\infty^j)-\mathcal{A}_{\mathbf{q}}(\infty^0)\Big]
\Big|\mathbb{B}_{\mathbf{q}}\big(I-\mathbf{q}\mathbb{B}_{\mathbf{q}}\big)^{-1}\right)\\
&\quad -\log\Theta_{\Delta}\left(\big(I-\mathbb{B}_{\mathbf{q}}\mathbf{q}\big)^{-1}\big[\mathcal{A}_{\mathbf{q}}(\infty^i)-\mathcal{A}_{\mathbf{q}}(\infty^0)]\Big|
\mathbb{B}_{\mathbf{q}}\big(I-\mathbf{q}\mathbb{B}_{\mathbf{q}}\big)^{-1}\right)\\
&\quad +\log\left(\sum_{k=1}^g\big(I-\mathbb{B}_{\mathbf{q}}\mathbf{q}\big)^{k\alpha}\omega_{\mathbf{q},\alpha}(\infty^0)\partial_{z_k}\Theta_{\Delta}
\Big(0\Big|\mathbb{B}_{\mathbf{q}}\big(I-\mathbf{q}\mathbb{B}_{\mathbf{q}}\big)^{-1}\Big)\right)\\
&\quad-{2{\rm i}\pi} \big(\mathcal{A}_{\mathbf{q}}(\infty^i)-\mathcal{A}_{\mathbf{q}}(\infty^0)\big)^{T}\mathbf{q}\big(I-\mathbb{B}_{\mathbf{q}}\mathbf{q}\big)^{-1}
\big(\mathcal{A}_{\mathbf{q}}(\infty^j)-\mathcal{A}_{\mathbf{q}}(\infty^0)\big)-\log(-1).
\end{split}
\end{equation}
When $i=j$, we have
\begin{equation}\label{s-ii-q}
\begin{split}
s^i\big(\phi_{\mathbf{q},s^i}\big)&=\log\left(\sum_{k=1}^g\big(I-\mathbb{B}_{\mathbf{q}}\mathbf{q}\big)^{k\alpha}\omega_{\mathbf{q},\alpha}(\infty^i)
\partial_{z_k}\Theta_{\Delta}\Big(0\Big|\mathbb{B}_{\mathbf{q}}\big(I-\mathbf{q}\mathbb{B}_{\mathbf{q}}\big)^{-1}\Big)\right)\\
&\quad +\log\left(\sum_{k=1}^g\big(I-\mathbb{B}_{\mathbf{q}}\mathbf{q}\big)^{k\alpha}\omega_{\mathbf{q},\alpha}(\infty^0)
\partial_{z_k}\Theta_{\Delta}\Big(0\Big|\mathbb{B}_{\mathbf{q}}\big(I-\mathbf{q}\mathbb{B}_{\mathbf{q}}\big)^{-1}\Big)\right)\\
&\quad -2\log\left(\Theta_{\Delta}\Big(\big(I-\mathbb{B}_{\mathbf{q}}\mathbf{q}\big)^{-1}\big[\mathcal{A}_{\mathbf{q}}(\infty^i)-\mathcal{A}_{\mathbf{q}}(\infty^0)\big]
\Big|\mathbb{B}_{\mathbf{q}}\big(I-\mathbf{q}\mathbb{B}_{\mathbf{q}}\big)^{-1}\Big)\right)\\
&\quad-{2{\rm i}\pi} \big(\mathcal{A}_{\mathbf{q}}(\infty^i)-\mathcal{A}_{\mathbf{q}}(\infty^0)\big)^{T}\mathbf{q}\big(I-\mathbb{B}_{\mathbf{q}}\mathbf{q}\big)^{-1}
\big(\mathcal{A}_{\mathbf{q}}(\infty^i)-\mathcal{A}_{\mathbf{q}}(\infty^0)\big)-\log(-1).
\end{split}
\end{equation}
Here, we have assumed the sum over the repeated  indices $\alpha=1,\dots,g$.
\end{Prop}
\emph{Proof:} Due to (\ref{B-q}) we have
\begin{equation}\label{B-B-q}
\mathbb{B}_{\mathbf{q}}=\big(I+\mathbb{B}\mathbf{q}\big)^{-1}\mathbb{B},\quad \quad\big(I+\mathbb{B}\mathbf{q}\big)^{-1}=I-\mathbb{B}_{\mathbf{q}}\mathbf{q}\quad \quad \text{and}\quad \quad
\mathbb{B}=\mathbb{B}_{\mathbf{q}}\big(I-\mathbf{q}\mathbb{B}_{\mathbf{q}}\big)^{-1}.
\end{equation}
This and (\ref{holomorphic-q}) enable us to write
\begin{equation}\label{h-h-q}
\omega_{\mathbf{q},k}(P)=\big(I+\mathbb{B}\mathbf{q}\big)^{k\alpha}\omega_{\alpha}(P)
=\big(I-\mathbb{B}_{\mathbf{q}}\mathbf{q}\big)_{k\alpha}\omega_{\alpha}(P).
\end{equation}
In particular, the usual Abel map and the introduced $\mathbf{q}$-Abel map (\ref{Abel-q0}) are related by
\begin{equation}\label{Abel-q}
\mathcal{A}_{\mathbf{q}}(P)=\big(I+\mathbb{B}\mathbf{q}\big)^{-1}\mathcal{A}(P)
=\big(I-\mathbb{B}_{\mathbf{q}}\mathbf{q}\big)\mathcal{A}(P).
\end{equation}
Finally, recalling expressions  (\ref{s-ij Formula}) for the function $s_{ij}$ obtained in Theorem \ref{s-ij}, we conclude that relations (\ref{s-ij-q}) and (\ref{s-ii-q}) follow from the following equality (due to (\ref{q-tAB-1}))
\begin{align*}
s^j\big(\phi_{\mathbf{q},s^i}\big)&=s_{ij}-{2{\rm i}\pi}\Big(\big(I+\mathbf{q}\mathbb{B}\big)^{-1}\mathbf{q}\Big)_{\alpha\beta}s^i(\omega_{\alpha})s^j(\omega_{\beta})\\
&=s_{ij}-{2{\rm i}\pi}\big(\mathcal{A}(\infty^i)-\mathcal{A}(\infty^0)\big)^T\Big(\big(I+\mathbf{q}\mathbb{B}\big)^{-1}\mathbf{q}\Big)
\big(\mathcal{A}(\infty^j)-\mathcal{A}(\infty^0)\big)
\end{align*}
and the preceding formulas (\ref{B-B-q}), (\ref{h-h-q}) and (\ref{Abel-q}).

\fd

\medskip

Let us now consider the $(2m+2g)$-dimensional Hurwitz space $\widehat{\mathcal{H}}_{g, m+1}(0,\dots,0)$ of ramified coverings $(C_g,\lambda)$ having  $m+1$ simple poles $\infty^0,\dots,\infty^m$. In this case, prepotential (\ref{Prep-q2}) takes the following form:
\begin{equation}\label{Prep-q3}
\begin{split}
\mathbf{F}_{\omega_{\mathbf{q},l}}
&=\frac{1}{2}\sum_{i,j=1}^mv^i(\omega_{\mathbf{q},l})v^j(\omega_{\mathbf{q},l})s^{j}(\phi_{\mathbf{q}, s^i})
+\sum_{i=1}^m\sum_{k=1}^gv^i(\omega_{\mathbf{q},l})\xi^k(\omega_{\mathbf{q},l})s^{i}(\omega_{\mathbf{q},k})\\
&\quad +\frac{1}{4{\rm i}\pi}\sum_{j,k=1}^g\xi^k(\omega_{\mathbf{q},l})\xi^j(\omega_{\mathbf{q},l})\big(\mathbb{B}_{\mathbf{q}}\big)_{jk}
 +\frac{\log(-1)}{4}\sum_{i,j=1, i\neq j}^mv^i(\omega_{\mathbf{q},l})v^j(\omega_{\mathbf{q},l})\\
&\quad-\frac{3}{4}\sum_{i,j=1}^m\big(1+\delta_{ij}\big)v^i(\omega_{\mathbf{q},l})v^j(\omega_{\mathbf{q},l}).
\end{split}
\end{equation}
In particular, this formula and expressions (\ref{s-ij-q}) and (\ref{s-ii-q}) for the functions  $s^{i}(\phi_{\mathbf{q}, s^j})$ allow us to arrive at the $\mathbf{q}$-analogue of formula (\ref{Prep-holom3}). Moreover, we observe that the function  $\mathbf{F}_{\omega_{\mathbf{q},j}}$ in (\ref{Prep-q3}) only depends on flat coordinates corresponding to the $g$ holomorphic $\mathbf{q}$-differentials (\ref{holomorphic-q}). Accordingly,   (\ref{Prep-q3}) turns out to be a new explicit  family of examples (parameterized by the values of the complex number $\mathbf{q}$) of genus one quasi-homogeneous solutions to the WDVV equations (\ref{WDVV equation}).\\
In the rest of this section, we are going to focus  on this example. Denote by $\omega_0$ be the (genus one) holomorphic differential normalized by
$$
\oint_a\omega_0=1,\quad \quad \oint_b\omega_0=\tau, \quad \text{with}\quad \Im\tau>0
$$
and $\phi_{\mathbf{q}}$ be the  normalized holomorphic  $\mathbf{q}$-differential (the genus one case of formula (\ref{holomorphic-q}))
\begin{equation}\label{q-phiq}
\phi_{\mathbf{q}}=\frac{\omega_0(P)}{1+\tau \mathbf{q}},\quad \quad \text{with}\quad \quad
\tau_{\mathbf{q}}:=\oint_b\phi_{\mathbf{q}}=\frac{\tau}{1+\tau \mathbf{q}}.
\end{equation}
Here  $\mathbf{q}$ is a complex number such that $1+\tau\mathbf{q}\neq 0$.
By using Proposition \ref{q-PropS}, we deduce that the system of flat coordinates (\ref{q-S-eta}) of the flat metric $\eta(\phi_{\mathbf{q}})$ is given by
\begin{equation}\label{FC-holo-q}
\begin{split}
&t_{\mathbf{q},0}=\frac{1}{2{\rm i}\pi}\oint_b\phi_{\mathbf{q}}=\frac{\tau_{\mathbf{q}}}{2{\rm i}\pi};\\
&t_{\mathbf{q},k}:=s^k(\phi_{\mathbf{q}})=\int_{\infty^0}^{\infty^k}\phi_{\mathbf{q}}=\mathcal{A}_{\mathbf{q}}(\infty^k)-\mathcal{A}_{\mathbf{q}}(\infty^0),\quad \quad k=1,\dots,m;\\
&t_{\mathbf{q},2m+1-k}:=v^k(\phi_{\mathbf{q}})=\underset{\infty^k}{{\rm res}}\ \lambda(P)\phi_{\mathbf{q}}(P)=\phi_{\mathbf{q}}(\infty^k),\quad k=1,\dots,m;\\
&t_{\mathbf{q},2m+1}:=\xi(\phi_{\mathbf{q}})=\oint_a\lambda(P)\phi_{\mathbf{q}}(P)+\mathbf{q}\oint_b\lambda(P)\phi_{\mathbf{q}}(P).
\end{split}
\end{equation}
Therefore, by substituting this and $g=1$ into  (\ref{Prep-q3}) we get
\begin{align*}
\mathbf{F}_{\phi_{\mathbf{q}}}
&=\frac{1}{2}\sum_{j,k=1}^mt_{\mathbf{q},2m+1-j}t_{\mathbf{q},2m+1-k}s^{j}(\phi_{\mathbf{q}, s^k})
+t_{\mathbf{q},2m+1}\sum_{k=1}^mt_{\mathbf{q},2m+1-k}t_{\mathbf{q},k}+\frac{(t_{\mathbf{q},2m+1})^2}{2}t_{\mathbf{q},0}\\
&\quad  +\frac{\log(-1)}{4}\sum_{j,k=1, j\neq k}^mt_{\mathbf{q},2m+1-j}t_{\mathbf{q},2m+1-k}
-\frac{3}{4}\sum_{i,j=1}^m\big(1+\delta_{ij}\big)t_{\mathbf{q},2m+1-j}t_{\mathbf{q},2m+1-k}.
\end{align*}
It remains to express functions $s^j(\phi_{\mathbf{q},s^i})$  in terms of the flat coordinates (\ref{FC-holo-q}). Making use of the link (\ref{theta1}) between the theta function $\Theta_{[\frac{1}{2},\frac{1}{2}]}$ with characteristics $[\frac{1}{2},\frac{1}{2}]$  and the $\theta_1$-Jacobi odd function, the equalities (\ref{FC-holo-q}) we can see that  the genus one case of the functions (\ref{s-ij-q}) and (\ref{s-ii-q}) are respectively given by
\begin{align*}
s^j\big(\phi_{\mathbf{q},s^k}\big)
&=\log\left(\frac{1}{1-{2{\rm i}\pi}\mathbf{q}t_{\mathbf{q},0}}\sum_{r=1}^mt_{\mathbf{q},2m+1-r}\right)
-\frac{2{\rm i}\pi\mathbf{q}}{1-{2{\rm i}\pi}\mathbf{q}t_{\mathbf{q},0}} t_{\mathbf{q},j}t_{\mathbf{q},k}\\
&\quad +\log\left(\frac{\theta_1'\left(0\Big|\frac{{2{\rm i}\pi}}{1-{2{\rm i}\pi}\mathbf{q}t_{\mathbf{q},0}}t_{\mathbf{q},0}\right)
\theta_1\left(\frac{t_{\mathbf{q},j}-t_{\mathbf{q},k}}{1-{2{\rm i}\pi}\mathbf{q}t_{\mathbf{q},0}}
\Big|\frac{{2{\rm i}\pi}t_{\mathbf{q},0}}{1-{2{\rm i}\pi}\mathbf{q}t_{\mathbf{q},0}}\right)}
{\theta_1\left(\frac{t_{\mathbf{q},j}}{1-2{\rm i}\pi\mathbf{q}t_{\mathbf{q},0}}
\Big|\frac{{2{\rm i}\pi}t_{\mathbf{q},0}}{1-{2{\rm i}\pi}\mathbf{q}t_{\mathbf{q},0}}\right)
\theta_1\left(\frac{t_{\mathbf{q},k}}{1-{2{\rm i}\pi}\mathbf{q}t_{\mathbf{q},0}}
\Big|\frac{{2{\rm i}\pi}t_{\mathbf{q},0}}{1-{2{\rm i}\pi}\mathbf{q}t_{\mathbf{q},0}}\right)}\right),\quad \quad \quad j\neq k
\end{align*}
and
\begin{align*}
s^k\big(\phi_{\mathbf{q},s^k}\big)
&=\log\left(\frac{t_{\mathbf{q},2m+1-k}}{1-2{\rm i}\pi\mathbf{q}t_{\mathbf{q},0}}\right)
+\log\left(\frac{1}{1-{2{\rm i}\pi}\mathbf{q}t_{\mathbf{q},0}}\sum_{r=1}^mt_{\mathbf{q},2m+1-r}\right)
-\frac{2{\rm i}\pi\mathbf{q}}{1-2{\rm i}\pi\mathbf{q}t_{\mathbf{q},0}}\big(t_{\mathbf{q},k}\big)^2\\
&\quad -2\log\left(\frac{\theta_1\left(\frac{t_{\mathbf{q},k}}{1-2{\rm i}\pi\mathbf{q}t_{\mathbf{q},0}}\Big|
\frac{{2{\rm i}\pi}t_{\mathbf{q},0}}{1-{2{\rm i}\pi}\mathbf{q}t_{\mathbf{q},0}}\right)}
{\theta_1'\left(0\Big|\frac{{2{\rm i}\pi}t_{\mathbf{q},0}}{1-{2{\rm i}\pi}\mathbf{q}t_{\mathbf{q},0}}\right)}\right).
\end{align*}

\medskip

\noindent Summarizing, we arrive at the following result.
\begin{Thm} Consider the genus one Hurwitz space $\widehat{\mathcal{H}}_{1, m+1}(0,\dots,0)$ of ramified coverings having  $m+1$ simple poles and $(2m+2)$-simple branch points.\\
Then  Frobenius manifold structure on $\widehat{\mathcal{H}}_{1, m+1}(0,\dots,0)$ induced by the normalized holomorphic $\mathbf{q}$-differential $\phi_{\mathbf{q}}$ (\ref{q-phiq}) is   determined  by the unit vector field $e=\partial_{t_{\mathbf{q},2m+1}}$, the Euler vector field $E=\sum_{k=0}^{m}t_{\mathbf{q},2m+1-k}\partial_{t_{\mathbf{q},2m+1-k}}$
and the prepotential
\begin{align*}
&\mathbf{F}_{\phi_{\mathbf{q}}}=\frac{(t_{\mathbf{q},2m+1})^2}{2}t_{\mathbf{q},0}+t_{\mathbf{q},2m+1}\sum_{k=1}^mt_{\mathbf{q},2m+1-k}t_{\mathbf{q},k}
-\frac{{\rm i}\pi\mathbf{q}}{1-2{\rm i}\pi\mathbf{q}t_{\mathbf{q},0}}\left(\sum_{k=1}^mt_{\mathbf{q},k}t_{\mathbf{q},2m+1-k}\right)^2\\
&\quad +\frac{1}{2}\sum_{k=1}^m\big(t_{\mathbf{q},2m+1-k}\big)^2\log\left(\frac{t_{\mathbf{q},2m+1-k}}{1-2{\rm i}\pi\mathbf{q}t_{\mathbf{q},0}}\right)
+\frac{1}{2}\left(\sum_{k=1}^mt_{\mathbf{q},2m+1-k}\right)^2\log\left(\sum_{r=1}^m\frac{t_{\mathbf{q},2m+1-r}}{1-2{\rm i}\pi\mathbf{q}t_{\mathbf{q},0}}\right)\\
&\quad -\sum_{k=1}^m\big(t_{\mathbf{q},2m+1-k}\big)^2\log\left(\frac{\theta_1\left(\frac{t_{\mathbf{q},k}}{1-2{\rm i}\pi\mathbf{q}t_{\mathbf{q},0}}\Big|
\frac{{2{\rm i}\pi}t_{\mathbf{q},0}}{1-2{\rm i}\pi\mathbf{q}t_{\mathbf{q},0}}\right)}
{\theta_1'\left(0\Big|\frac{{2{\rm i}\pi}t_{\mathbf{q},0}}{1-2{\rm i}\pi\mathbf{q}t_{\mathbf{q},0}}\right)}\right)\\
&\quad +\frac{1}{2}\sum_{j,k=1, j\neq k}^mt_{\mathbf{q},2m+1-j}t_{\mathbf{q},2m+1-k}
\log\left(\frac{\theta_1'\left(0\Big|\frac{{2{\rm i}\pi}t_{\mathbf{q},0}}{1-2{\rm i}\pi\mathbf{q}t_{\mathbf{q},0}}\right)
\theta_1\left(\frac{t_{\mathbf{q},j}-t_{\mathbf{q},k}}{1-2{\rm i}\pi\mathbf{q}t_{\mathbf{q},0}}
\Big|\frac{{2{\rm i}\pi}t_{\mathbf{q},0}}{1-2{\rm i}\pi\mathbf{q}t_{\mathbf{q},0}}\right)}
{\theta_1\left(\frac{t_{\mathbf{q},j}}{1-2{\rm i}\pi\mathbf{q}t_{\mathbf{q},0}}
\Big|\frac{{2{\rm i}\pi}t_{\mathbf{q},0}}{1-2{\rm i}\pi\mathbf{q}t_{\mathbf{q},0}}\right)
\theta_1\left(\frac{t_{\mathbf{q},k}}{1-2{\rm i}\pi\mathbf{q}t_{\mathbf{q},0}}
\Big|\frac{{2{\rm i}\pi}t_{\mathbf{q},0}}{1-2{\rm i}\pi\mathbf{q}t_{\mathbf{q},0}}\right)}\right)\\
&\quad  +\frac{\log(-1)}{4}\sum_{j,k=1, j\neq k}^mt_{\mathbf{q},2m+1-j}t_{\mathbf{q},2m+1-k}
-\frac{3}{4}\sum_{j,k=1}^m\big(1+\delta_{jk}\big)t_{\mathbf{q},2m+1-j}t_{\mathbf{q},2m+1-k}.
\end{align*}

\end{Thm}

\medskip

Let us mention the following remarks:  Let $\phi_{\mathbf{q}}$ be the holomorphic $\mathbf{q}$-differential defined by (\ref{q-phiq}) and  $\omega_0$ be the normalized holomorphic differential i.e. $\oint_a\omega_0=1$.
\begin{itemize}
\item Since by (\ref{q-limit-FC}),  for each $j=0,\dots,2m+1$,  the $j$th flat coordinate $t_{\mathbf{q},j}$ in (\ref{FC-holo-q}) coincides with the function $t_j$ when $\mathbf{q}=0$,  where $t_0,\dots,t_{2m+1}$ are the functions defined by (\ref{FC-holo-g1}), we have
$$
\mathbf{F}_{\phi_{\mathbf{q}}}\big(t_{\mathbf{q},0},\dots,t_{\mathbf{q},2m+1}\big)
\Big|_{\mathbf{q}=0}=\mathbf{F}_{\omega_0}(t_0,\dots,t_{2m+1})=\text{the prepotential given  by (\ref{F-holo-g1})}.
$$
\item When $m=1$, we have
\begin{align*}
\mathbf{F}_{\phi_{\mathbf{q}}}&=\frac{(t_{\mathbf{q},3})^2}{2}t_{\mathbf{q},0}+t_{\mathbf{q},1}t_{\mathbf{q},2}t_{\mathbf{q},3}
-\frac{{\rm i}\pi\mathbf{q}}{1-2{\rm i}\pi\mathbf{q}t_{\mathbf{q},0}}\left(t_{\mathbf{q},1}t_{\mathbf{q},2}\right)^2
 +\big(t_{\mathbf{q},2}\big)^2\log\left(\frac{t_{\mathbf{q},2}}{1-2{\rm i}\pi\mathbf{q}t_{\mathbf{q},0}}\right)\\
&\quad -\big(t_{\mathbf{q},2}\big)^2\log\left(\frac{\theta_1\left(\frac{t_{\mathbf{q},1}}{1-2{\rm i}\pi\mathbf{q}t_{\mathbf{q},0}}\Big|
\frac{{2{\rm i}\pi}t_{\mathbf{q},0}}{1-2{\rm i}\pi\mathbf{q}t_{\mathbf{q},0}}\right)}
{\theta_1'\left(0\Big|\frac{{2{\rm i}\pi}t_{\mathbf{q},0}}{1-2{\rm i}\pi\mathbf{q}t_{\mathbf{q},0}}\right)}\right)-\frac{3}{2}(t_{\mathbf{q},2})^2.
\end{align*}
Up to a change of variables and the (unimportant) term $-\frac{3}{2}(t_{\mathbf{q},2})^2$, this example of prepotential has been computed in \cite{Miguel}.
\end{itemize}

\section{Examples of solutions to the WDVV equations  in genus zero}
In this section, in Theorems \ref{g0-prep1}-\ref{g0-prep3}, we obtain explicit quasi-homogeneous solutions to the WDVV system depending on $2m$ variables.\\
Consider the genus zero  Hurwitz space of  meromorphic functions $\lambda:\P^1\longrightarrow\P^1$ given by
\begin{equation}\label{Superpotential-0}
\lambda(P)=\sum_{k=0}^m\frac{b_k}{z_P-a_k}=\frac{b_0}{z_P}+\sum_{k=1}^m\frac{b_k}{z_P-a_k}
\end{equation}
modulo automorphisms of $\P^1$, where $m\geq2$, $a_0=0,a_1,\dots,a_m$ are distinct points in $\C$ and  $b_0,\dots,b_m$ are nonzero complex numbers satisfying $b_0+b_1+\dots+b_m=1$. In particular $\lambda$ has exactly $m+1$ simple poles at $a_0,\dots,a_m$ and the point at infinity $\infty$ is one of its zeros. \\
We have
$$
\lambda'(P)=-\frac{f_{2m}(z_P)}{\prod_{k=0}^m(z_P-a_k)^2},\quad \text{with}\quad f_{2m}(z_P)=\sum_{k=0}^mb_k\prod_{j=0, j\neq k}^m\big(z_P-a_j\big)^2.
$$
We observe that the assumption  $b_0+b_1+\dots+b_m=1$ implies that  the leading coefficient of the $2m$-degree polynomial  $f_{2m}$ is 1. We also assume that $f_{2m}$ has $2m$ distinct roots, denoted by $\alpha_1,\dots, \alpha_{2m}$.  Let $P_j\in \P^1$ be the point represented by the complex number  $z_{P_j}=\alpha_j$ and $\lambda_j=\lambda(P_j)$ be its image by $\lambda$. Thus $P_1,\dots,P_{2m}$ (resp. $\lambda_1,\dots,\lambda_{2m}$)  are nothing but the  ramification points (resp. the branch points) of the branched covering $(\P^1,\lambda)$. \\
Since $a_0=0$ and the complex numbers $b_0,a_1,\dots,a_m$ are all nonzero (by assumption), we have
$$
f_{2m}(0)=\sum_{k=0}^mb_k\prod_{j=0, j\neq k}^ma_j^2=b_0\prod_{j=1}^ma_j^2\neq0.
$$
In particular, none of the ramification points $\alpha_1,\dots,\alpha_{2m}$ is zero.\\
Following the notation introduced in  Section  2.2, the simple Hurwitz space  consisting  of meromorphic functions $\lambda$ (\ref{Superpotential-0}), modulo automorphisms of $\P^1$, is denoted by $\mathcal{H}_{0,m+1}(0,\dots,0)$.
This space can be identified with the open subset of points $(a_1,\dots,a_m,b_1,\dots,b_m)\in \big(\C\setminus\{0\}\big)^{2m}$ determined by the conditions:
\begin{itemize}
\item[1)] $\alpha_1,\dots,\alpha_{2m}$ are distinct in $\C\setminus\{0\}$;
\item[2)] the branch points $\{\lambda_j\}_j$ satisfy  $\lambda_i\neq \lambda_j$ whenever $i\neq j$.
\end{itemize}

\noindent Let us now consider the family of Frobenius structures on the Hurwitz space $\mathcal{H}_{0,m+1}(0,\dots,0)$.
According to (\ref{Primary-e}),  we only have two types of Dubrovin's primary differentials:
\begin{equation}\label{g0-primary}
\begin{split}
&\phi_{k}(P)=\int_{0}^{a_k}W(P,Q)=\frac{dz_P}{z_P-a_k}-\frac{dz_P}{z_P}=\Omega_{0a_k}(P);\\
&\phi_{2m+1-k}(P)=\underset{a_k}{{\rm res}}\lambda(Q)W(P,Q)=\frac{b_k}{(z_P-a_k)^2}dz_P,
\end{split}
\quad \quad \quad \forall\ k=1,\dots,m,
\end{equation}
with $W(P,Q)$ being the genus zero bidifferential (\ref{g=0-W}).\\
As we know from Definition \ref{quasi-homo-diff} and Proposition \ref{primary-quasi},  primary  differentials in (\ref{g0-primary}) are also quasi-homogeneous and their degrees are given by:
\begin{equation}\label{g0-degree}
{\rm deg}(\phi_k)=0\quad \quad  \text{and}\quad \quad  {\rm deg}(\phi_{2m+1-k})=1,\quad \quad  \forall\ k=1,\dots,m.
\end{equation}
We shall also study  the Frobenius manifold structure associated with the following quasi-homogeneous differential $\phi_0$ of degree 0:
\begin{equation}\label{phi00}
\phi_0(P)=\int_{0}^{\infty}W(P,Q)=-\frac{dz_P}{z_P}.
\end{equation}
Besides leading to an example of non-normalized Frobenius manifold structure (i.e. where the unity is flat), considering differential (\ref{phi00}) turns out to be useful in calculating examples of prepotentials corresponding to the primary differentials (\ref{g0-primary}).
\begin{Prop} Let $\eta(\phi_k):=\displaystyle \frac{1}{2}\sum_j\big(\phi_k(P_j)\big)^2(d\lambda_j)^2$ be the Darboux-Egoroff metrics (\ref{eta-def}) induced by the Abelian differential $\phi_k$, $k=0,1,\dots,2m$, given by (\ref{g0-primary}) and (\ref{phi00}).\\
Then
\begin{align*}
&\phi_0(P_j)^2=-2\frac{\prod_{r=1}^m(\alpha_j-a_r)^2}{\prod_{s=1,s\neq j}^{2m}(\alpha_j-\alpha_s)};\\
&\phi_k(P_j)^2=-2a_k^2\frac{\prod_{r=1,r\neq k}^m(\alpha_j-a_r)^2}{\prod_{s=1,s\neq j}^{2m}(\alpha_j-\alpha_s)}, \hspace{2.5cm} k=1,\dots,m;\\
&\phi_{2m+1-k}(P_j)^2=-2b_k^2\frac{\alpha_j^2\prod_{r=1,r\neq k}^m(\alpha_j-a_r)^2}{(\alpha_j-a_k)^2\prod_{s=1,s\neq j}^{2m}(\alpha_j-\alpha_s)}, \quad  \quad k=1,\dots,m.
\end{align*}

\end{Prop}
\emph{Proof:} Due to (\ref{notation1-1}), we have
\begin{align*}
&\phi_0(P_j)^2=2\underset{\alpha_j}{{\rm res}}\frac{\phi_0(P)^2}{d\lambda(P)}=2\underset{\alpha_j}{{\rm res}}\frac{dz_P}{z_P^2\lambda'(P)};\\
&\phi_k(P_j)^2=2\underset{\alpha_j}{{\rm res}}\frac{\phi_k(P)^2}{d\lambda(P)}=2\underset{\alpha_j}{{\rm res}}\frac{a_k^2dz_P}{z_P^2(z_P-a_k)^2\lambda'(P)};\\
&\phi_{2m+1-k}(P_j)^2=2\underset{\alpha_j}{{\rm res}}\frac{b_k^2dz_P}{(z_P-a_k)^4\lambda'(P)}.
\end{align*}
Thus by simple calculations using the fact that
$$
\lambda'(P)=-\frac{\prod_{s=1}^{2m}\big(z_P-\alpha_s\big)}{\prod_{s=0}^m(z_P-a_s)^2},
$$
we obtain the desired results.

\fd

\medskip

\noindent  Let $k=0,\dots,2m$ be fixed. Then the system of flat coordinates (\ref{FBP-W}) of the metric $\eta(\phi_k)$  reduces to the following:
\begin{equation}\label{g0-FC}
\begin{split}
&\mathbf{t}^s(\phi_k)=p.v.\int_{0}^{a_s}\phi_k(P);\\
&\mathbf{t}^{2m+1-s}(\phi_k)=\underset{a_s}{{\rm res}}\lambda(P)\phi_{k}(P),
\end{split}
\quad \quad \quad \forall\ s=1,\dots,m.
\end{equation}
Here we would like to mention that the ordering of the functions $\mathbf{t}^j$, $j=1,\dots,2m$,  is chosen by respecting the correspondence (\ref{FF-model}) and (\ref{phi-t})
between flat coordinates and primary differentials.\\
Furthermore,   the $\eta(\phi_k)$-duality between flat coordinates (introduced in (\ref{duality}) and (\ref{duality-picture})) is described by the relation
\begin{equation}\label{g0-duality}
\mathbf{t}^{s^{\prime}}(\phi_k)=\mathbf{t}^{2m+1-s}(\phi_k), \quad \forall\ s=1,\dots,2m.
\end{equation}
and  thus the metric $\eta(\phi_k)$  becomes as follows (see formula (\ref{eta-tA})):
\begin{align*}
\eta(\phi_k)=\sum_{s=1}^{2m}d\mathbf{t}^s(\phi_k)\otimes d\mathbf{t}^{2m+1-s}(\phi_k).
\end{align*}
We first  calculate  flat  coordinates of the Darboux-Egoroff metric $\eta(\phi_0)$ by means of the parameters $a_1,\dots,a_m,b_0,b_1,\dots,b_m$ of the Hurwitz space $\mathcal{H}_{0,m+1}(0,\dots,0)$.
\begin{Prop} Let $\phi_0$ be the differential $\phi_0(P)=-dz_P/z_P$. Then, the flat coordinates (\ref{g0-FC}) of the metric $\eta(\phi_0)$ are:
\begin{equation}\label{g0-phi0-c}
\begin{split}
&t_{s}:=\mathbf{t}^s(\phi_0)=p.v.\int_{0}^{a_s}\phi_0(P)=\log(b_0)-\log(a_s);\\
&t_{2m+1-s}:=\mathbf{t}^{2m+1-s}(\phi_0)=\underset{a_s}{{\rm res}}\lambda(P)\phi_0(P)=-\frac{b_s}{a_s},
\end{split}
\quad \quad \quad \forall\ s=1,\dots,m.
\end{equation}
\end{Prop}
\emph{Proof:} Let $P^{\varepsilon}$ be the point of $\P^1$ such that  $z_{P^{\varepsilon}}=\varepsilon\in \C$ where $|\varepsilon|$ is small enough.
Since, by (\ref{Superpotential-0}),
$$
\varepsilon\lambda(P^{\varepsilon})=b_0+\varepsilon\sum_{j=1}^m\frac{b_j}{\varepsilon-a_j},
$$
it follows that
\begin{align*}
\int_{\varepsilon}^{a_s}\phi_0(P)&=-\int_{\varepsilon}^{a_s}\frac{dz_P}{z_P}=\log(\varepsilon)-\log(a_s)
=\log\bigg(b_0+\varepsilon\sum_{j=1}^m\frac{b_j}{\varepsilon-a_j}\bigg)-\log\big(\lambda(P^{\varepsilon})\big)-\log(a_s).
\end{align*}
Therefore, by definition of the principal value from Section 3.3, we obtain
\begin{align*}
t_s:=p.v.\int_0^{a_s}\phi_0(P)=\lim_{\varepsilon\to0}\bigg(\int_{\varepsilon}^{a_s}\phi_0(P)+\log\big(\lambda(P^{\varepsilon})\big)\bigg)
=\log(b_0)-\log(a_s).
\end{align*}
On the other hand, by straightforward  calculation we see that $t_{2m+1-s}:=\underset{a_s}{\rm res}\lambda(P)\phi_0(P)=-b_s/a_s$.

\fd

\medskip

The purpose of the next result is to find flat coordinates (\ref{g0-FC}) corresponding to the primary differentials listed in (\ref{g0-primary}). For later use, we express them by means of  flat coordinates $t_s,t_{2m+1-s}$  (\ref{g0-phi0-c}) corresponding to the differential $\phi_0(P)=-dz_P/z_P$.
\begin{Prop} Let $k=1,\dots,m$.
\begin{description}
\item[i)] Flat coordinates of the metric  $\eta(\phi_k)$ are given in terms of the flat coordinates $t_s,t_{2m+1-s}$ from (\ref{g0-phi0-c}) by:
\begin{equation}\label{g0-phik-c}
\begin{split}
&x_{k,s}:=\mathbf{t}^s(\phi_k)=\log\big(e^{t_s}-e^{t_k}\big),\quad\quad s\neq k;\\
&x_{k,k}:=\mathbf{t}^k(\phi_k)=t_k+\log(t_{2m+1-k});\\
&x_{k,2m+1-s}:=\mathbf{t}^{2m+1-s}(\phi_k)=t_{2m+1-s}\frac{e^{t_s}}{e^{t_s}-e^{t_k}},\quad \quad s\neq k;\\
&x_{k,2m+1-k}:=\mathbf{t}^{2m+1-k}(\phi_k)=e^{t_k}+\sum_{s=1}^mt_{2m+1-s}-\sum_{s=1,s\neq k}^mx_{k,2m+1-s}.
\end{split}
\end{equation}
\item[ii)] Flat coordinates of $\eta(\phi_{2m+1-k})$ are:
\begin{equation}\label{g0-phim+k-c}
\begin{split}
&y_{k,s}:=\mathbf{t}^s(\phi_{2m+1-k})=t_{2m+1-k}\frac{e^{t_k}}{e^{t_k}-e^{t_s}}=x_{s,2m+1-k},\quad \quad s\neq k;\\
&y_{k,k}:=\mathbf{t}^k(\phi_{2m+1-k})=e^{t_k}+\sum_{s=1}^mt_{2m+1-s}-\sum_{s=1,s\neq k}^my_{s,k}=x_{k,2m+1-k};\\
&y_{k,2m+1-s}:=\mathbf{t}^{2m+1-s}(\phi_{2m+1-k})=\frac{t_{2m+1-k}t_{2m+1-s}e^{t_k+t_s}}{\big(e^{t_k}-e^{t_s}\big)^2},\quad \quad s\neq k;\\
&y_{k,2m+1-k}:=\mathbf{t}^{2m+1-k}(\phi_{2m+1-k})=t_{2m+1-k}e^{t_k}-\sum_{s=1,s\neq k}^my_{k,2m+1-s}.
\end{split}
\end{equation}
\end{description}
\end{Prop}
\emph{Proof:} Let us first point out that, due to (\ref{g0-phi0-c}), the following  relations are satisfied:
\begin{equation}\label{g0-phi0-ab}
\begin{split}
&a_j=b_0e^{-t_j};\\
&b_j=-a_jt_{2m+1-j}=-b_0t_{2m+1-j}e^{-t_j},
\end{split}
\quad\quad \forall\ j=1,\dots,m.
\end{equation}
Let $P^{\varepsilon},P^{\delta}\in \P^1$ be such that  $z_{P^{\varepsilon}}=\varepsilon$ and $z_{P^{\delta}}=a_k+\delta$, where $\varepsilon,\delta\in \C$ are
such that $|\varepsilon|$ and $|\delta|$ are sufficiently small.\\
\textbf{i)} Assume that $s\neq k$. We have
\begin{align*}
\int_{\varepsilon}^{a_s}\phi_k(P)&=\log(a_s-a_k)-\log(\varepsilon-a_k)-\log(a_s)+\log(\varepsilon)\\
&=\textstyle\log(a_s-a_k)-\log(\varepsilon-a_k)-\log(a_s)-\log\big(\lambda(P^{\varepsilon})\big)+\log\Big(b_0+\varepsilon\sum_{j=1}^m\frac{b_j}{-a_j+\varepsilon}\Big).
\end{align*}
Therefore, using this and (\ref{g0-phi0-ab}) we arrive at
\begin{align*}
x_{k,s}&:=p.v.\int_{0}^{a_s}\phi_k(P)=\log(a_k-a_s)-\log(a_k)-\log(a_s)+\log(b_0)\\
&=\log\big(1-(a_s/a_k)\big)-\log(a_s)+\log(b_0)\\
&=\log\big(e^{t_s}-e^{t_k}\big).
\end{align*}
In  the same way we get
\begin{align*}
x_{k,k}=\log(b_k)+\log(b_0)-2\log(a_k)-\log(-1)=t_k+\log(t_{2m+1-k}).
\end{align*}
On the other hand,  (\ref{g0-phi0-ab}) implies that
\begin{align*}
\forall\ s\neq k,\quad x_{k,2m+1-s}&:=\underset{a_s}{{\rm res}}\lambda(P)\phi_k(P)=\frac{b_s}{a_s-a_k}-\frac{b_s}{a_s}\\
&=t_{2m+1-s}-\frac{t_{2m+1-s}e^{-t_s}}{e^{-t_s}-e^{-t_k}}\\
&=t_{2m+1-s}\frac{e^{t_s}}{e^{t_s}-e^{t_k}}.
\end{align*}
Furthermore, in view of  the residue theorem and (\ref{g0-phi0-ab}), we obtain
\begin{align*}
x_{k,2m+1-k}&:=\underset{a_k}{{\rm res}}\lambda(P)\phi_k(P)=-\underset{0}{{\rm res}}\lambda(P)\phi_k(P)-\sum_{s=1,s\neq k}^m\underset{a_s}{{\rm res}}\lambda(P)\phi_k(P)\\
&=\frac{b_0}{a_k}-\sum_{s=1}^m\frac{b_s}{a_s}-\sum_{s=1,s\neq k}^mx_{k,2m+1-s}\\
&=e^{t_k}+\sum_{s=1}^mt_{2m+1-s}-\sum_{s=1,s\neq k}^mx_{k,2m+1-s}.
\end{align*}
\textbf{ii)} When $s\neq k$, then
\begin{align*}
y_{k,s}&:=p.v.\int_0^{a_k}\phi_{2m+1-k}(P)=p.v.\int_0^{a_s}\frac{b_k}{(z_P-a_k)^2}dz_P=\int_0^{a_s}\frac{b_k}{(z_P-a_k)^2}dz_P\\
&=\frac{b_k}{a_k-a_s}-\frac{b_k}{a_k}=x_{s,2m+1-k}.
\end{align*}
If $s=k$, then
\begin{align*}
\int_{0}^{a_k+\delta}\frac{b_k}{(z_P-a_k)^2}dz_P&=-\frac{b_k}{\delta}-\frac{b_k}{a_k}
=-\frac{b_k}{a_k}+\sum_{j=0,j\neq k}^m\frac{b_j}{a_k-a_j+\delta}-\lambda(P^{\delta}).
\end{align*}
By subtracting the divergent term $\lambda(P^{\delta})$, letting $\delta\to0$ and employing (\ref{g0-phi0-ab}), we get
\begin{align*}
y_{k,k}&:=p.v.\int_0^{a_k}\phi_{2m+1-k}(P)=-\frac{b_k}{a_k}+\sum_{j=0,j\neq k}^m\frac{b_j}{a_k-a_j}\\
&=\frac{b_0}{a_k}-\sum_{s=1}^m\frac{b_k}{a_k}-\sum_{s=1,s\neq k}\bigg(\frac{b_s}{a_s-a_k}-\frac{b_s}{a_s}\bigg)\\
&=e^{t_k}+\sum_{s=1}^mt_{2m+1-s}-\sum_{s=1,s\neq k}^mx_{k,2m+1-s} =x_{k,2m+1-k}.
\end{align*}
Finally, let us deal with the functions $y_{k,2m+1-s}=\mathbf{t}^{2m+1-s}\big(\phi_{2m+1-k}\big)$, with $s=1,\dots,m$.
Due to (\ref{g0-phi0-ab}) we have
\begin{align*}
\forall\ s\neq k,\quad
y_{k,2m+1-s}&:=\underset{a_s}{{\rm res}}\lambda(P)\phi_{2m+1-k}(P)=\frac{b_sb_k}{(a_s-a_k)^2}
=\frac{t_{2m+1-k}t_{2m+1-s}e^{t_k+t_s}}{\big(e^{t_k}-e^{t_s}\big)^2}.
\end{align*}
In addition,  the residue theorem applied to the Abelian  differential $\lambda(P)\phi_{2m+1-k}(P)$ yields
\begin{align*}
y_{k,2m+1-k}&:=\underset{a_k}{{\rm res}}\lambda(P)\phi_{2m+1-k}(P)
=-\underset{0}{{\rm res}}\lambda(P)\phi_{2m+1-k}(P)-\sum_{s=1,s\neq k}^m\underset{a_s}{{\rm res}}\lambda(P)\phi_{2m+1-k}(P)\\
&=-\frac{b_0b_k}{a_k^2}-\sum_{s=1,s\neq k}y_{k,2m+1-s}=t_{2m+1-k}e^{t_k}-\sum_{s=1,s\neq k}y_{k,2m+1-s}.
\end{align*}

\fd

\medskip

The following corollary follows directly from  (\ref{g0-phik-c}) and (\ref{g0-phim+k-c}).
\begin{Cor} For any fixed $k=1,\dots, m$, we have
\begin{align}
\begin{split}\label{g0-R1}
&\sum_{s=1}^mx_{k,2m+1-s}=e^{t_k}+\sum_{s=1}^mt_{2m+1-s}=\sum_{s=1}^my_{s,k};
\end{split}\\
\begin{split}\label{g0-R2}
&\sum_{s=1}^my_{k,2m+1-s}=t_{2m+1-k}e^{t_k}=e^{x_{k,k}}.
\end{split}
\end{align}

\end{Cor}

Since the branch points $\lambda_1,\dots,\lambda_{2m}$ play the role of canonical coordinates of the  semi-simple Frobenius manifolds associated with the Abelian differentials $\phi_0,\phi_1,\dots,\phi_{2m}$, it follows that the unit and Euler vector fields are  respectively given  by
\begin{equation}\label{g0-e-E}
\mathbf{1}=\sum_{j=1}^{2m}\partial_{\lambda_j}\quad\quad \text{and}\quad \quad  \mathbf{E}=\sum_{j=1}^{2m}\lambda_j\partial_{\lambda_j}.
\end{equation}
We are going to give the expressions for these vector fields in  each system of flat coordinates attached to the considered differentials $\phi_k$, $k=0,\dots,2m$.

\begin{Prop} For $j=0,1,\dots,2m$, denote by $\mathbf{1}_{\phi_j}$ (resp. $\mathbf{E}_{\phi_j}$) the vector field $\mathbf{1}$ (resp. $\mathbf{E}$) (\ref{g0-e-E}) written  in the flat coordinates (\ref{g0-FC}) of the metric $\eta(\phi_j)$.
\begin{description}
\item[i)] Let $k=1,\dots,m$. In flat coordinates (\ref{g0-phik-c}), we have
\begin{equation}\label{g0-eE-phik}
\begin{split}
&\mathbf{1}_{\phi_k}=\partial_{x_{k,2m+1-k}};\\
&\mathbf{E}_{\phi_k}=\sum_{s=1}^m\bigg((1+\delta_{ks})\partial_{x_{k,s}}+x_{k,2m+1-s}\partial_{x_{k,2m+1-s}}\bigg).
\end{split}
\end{equation}
\item[ii)] Let $k=1,\dots,m$. Then in flat coordinates (\ref{g0-phim+k-c}), we have
\begin{equation}\label{g0-eE-phim+k}
\begin{split}
&\mathbf{1}_{\phi_{2m+1-k}}=\partial_{y_{k,k}};\\
&\mathbf{E}_{\phi_{2m+1-k}}=\sum_{s=1}^m\bigg(y_{k,s}\partial_{y_{k,s}}+2y_{k,2m+1-s}\partial_{y_{k,2m+1-s}}\bigg).
\end{split}
\end{equation}
\item[iii)] In flat coordinates (\ref{g0-phi0-c}), the vector fields (\ref{g0-e-E}) have respectively the following forms:
\begin{equation}\label{g0-eE-phi0}
\begin{split}
&\mathbf{1}_{\phi_0}=\frac{1}{1-\sum_{r=1}^mt_{2m+1-r}e^{-t_r}}\sum_{s=1}^me^{-t_s}\Big(\partial_{t_s}-t_{2m+1-s}\partial_{t_{2m+1-s}}\Big);\\
&\mathbf{E}_{\phi_0}=\sum_{s=1}^m\Big(\partial_{t_s}+t_{2m+1-s}\partial_{t_{2m+1-s}}\Big).
\end{split}
\end{equation}
\end{description}

\end{Prop}

Note that (\ref{g0-phi0-c}) and the assumption $1=b_0+\dots+b_m$ imply that
$$
1-\sum_{r=1}^mt_{2m+1-r}e^{-t_r}=1+\frac{1}{b_0}\sum_{r=1}^mb_r=\frac{1}{b_0}\neq 0.
$$
\noindent\emph{Proof:} \textbf{i)} and \textbf{ii)}  Given that $\phi_k$ and $\phi_{2m+1-k}$ (\ref{g0-primary}), with $k=1,\dots,m$,  are primary differentials, using formula (\ref{e-tA}), duality relations (\ref{g0-duality}) and the systems (\ref{g0-phik-c}) and (\ref{g0-phim+k-c}) of flat coordinates,  we conclude that
\begin{align*}
&\mathbf{1}_{\phi_k}=\frac{\partial}{\partial\mathbf{t}^{k^{\prime}}(\phi_k)}=\frac{\partial}{\partial\mathbf{t}^{2m+1-k}(\phi_k)}=\partial_{x_{k,2m+1-k}};\\
&\mathbf{1}_{\phi_{2m+1-k}}=\frac{\partial}{\partial\mathbf{t}^{(2m+1-k)^{\prime}}\big(\phi_{2m+1-k}\big)}
=\frac{\partial}{\partial\mathbf{t}^k\big(\phi_{2m+1-k}\big)}=\partial_{y_{k,k}}.
\end{align*}
Furthermore, formula (\ref{E-tA2}) reads in our setting as follows:
\begin{align*}
\mathbf{E}_{\phi_j}=\sum_{s=1}^{2m}\Big((d_j+d_s)\mathbf{t}^s(\phi_j)+r_{j,s}\Big)\frac{\partial}{\partial\mathbf{t}^s(\phi_j)},\quad \forall\ j=1,\dots,2m,
\end{align*}
where $d_j={\rm deg}(\phi_j)$ and the constants $r_{j,k}$ are the analogous of  (\ref{r0A}) (see also (\ref{r-AB})) and given by
$$
r_{j,s}=\sum_{l_1,l_2=1}^m\delta_{\phi_j,\phi_{l_1}}\delta_{\phi_s,\phi_{l_2}}\big(1+\delta_{l_1l_2}),\quad\quad  \forall\ j,s =1,\dots,2m.
$$
Here we used the fact the poles of the considered meromorphic functions (\ref{Superpotential-0}) are all simple.\\
This and  (\ref{g0-degree}) as well as  notation (\ref{g0-phik-c})   imply that
$$
\mathbf{E}_{\phi_k}=\sum_{s=1}^m\bigg((1+\delta_{ks})\partial_{x_{k,s}}+x_{k,2m+1-s}\partial_{x_{k,2m+1-s}}\bigg), \quad \forall\ k=1,\dots,m.
$$
By the same arguments, we also obtain the desired formula for $\mathbf{E}_{\phi_{2m+1-k}}$, $k=1,\dots,m$.

\medskip

\noindent \textbf{iii)} In view of (\ref{g0-phik-c}), we have
\begin{align*}
&x_{k,s}=\log\big(e^{t_s}-e^{t_k}\big)=x_{s,k}+\log(-1),\quad\quad s\neq k;\\
&x_{k,k}=t_k+\log(t_{2m+1-k});\\
&x_{k,2m+1-s}=t_{2m+1-s}e^{t_s-x_{k,s}},\quad \quad s\neq k;\\
&x_{k,2m+1-k}=e^{t_k}+\sum_{s=1}^mt_{2m+1-s}-\sum_{s=1,s\neq k}^mx_{k,2m+1-s}.
\end{align*}
Furthermore,  using the fact that $\mathbf{1}=\mathbf{1}_{\phi_k}=\partial_{x_{k,2m+1-k}}$  we deduce that the action of unit vector field $\mathbf{1}=\sum_{j=1}^{2m}\partial_{\lambda_j}$ on the coordinates $t_s$, $t_{2m+1-s}$, $x_{k,s}$, $x_{k,2m+1-s}$ (regarded as functions of branch points
 $\lambda_1,\dots,\lambda_{2m}$) leads to the following equations:
\begin{align*}
&\mathbf{1}\big(x_{k,s}\big)=0=\mathbf{1}(t_s)e^{t_s}-\mathbf{1}(t_k)e^{t_k},\quad\quad s\neq k;\\
&\mathbf{1}\big(x_{k,k}\big)=0=\mathbf{1}(t_k)+\frac{\mathbf{1}\big(t_{2m+1-k}\big)}{t_{2m+1-k}};\\
&\mathbf{1}\big(x_{k,2m+1-s}\big)=0=\mathbf{1}\big(t_{2m+1-s}\big)e^{t_s-x_{k,s}}+\mathbf{1}(t_s)t_{2m+1-s}e^{t_s-x_{k,s}},\quad \quad s\neq k;\\
&\mathbf{1}\big(x_{k,2m+1-k}\big)=1=\mathbf{1}(t_k)e^{t_k}+\sum_{s=1}^m\mathbf{1}\big(t_{2m+1-s}\big).
\end{align*}
From this we arrive at
\begin{equation*}
\begin{split}
&\bigg(1-\sum_{r=1}^mt_{2m+1-r}e^{-t_r}\bigg)\mathbf{1}(t_s)=e^{-t_s};\\
&\bigg(1-\sum_{r=1}^mt_{2m+1-r}e^{-t_r}\bigg)\mathbf{1}(t_{2m+1-s})=-t_{2m+1-s}e^{-t_s},
\end{split}
\quad \quad \forall\ s=1,\dots,m.
\end{equation*}
This shows the first stated  relation in (\ref{g0-eE-phi0}).\\
Alternatively, the expression for $\mathbf{1}_{\phi_0}$ can be proven  using the following facts (this method was indicated at the beginning of Section  4.3):
\begin{align*}
&\mathbf{1}\big(\mathbf{t}^k\big(\phi_0)\big)=\sum_{j=1}^{2m}\underset{P_j}{\rm res}\ \frac{\phi_0(P)\phi_k(P)}{d\lambda(P)}=-\phi_k(\infty)
=-\underset{\infty}{\rm res}\ \frac{\phi_k(P)}{\lambda(P)}=a_k;\\
&\mathbf{1}\big(\mathbf{t}^{2m+1-k}\big(\phi_0)\big)=\sum_{j=1}^{2m}\underset{P_j}{\rm res}\ \frac{\phi_0(P)\phi_{2m+1-k}(P)}{d\lambda(P)}=-\phi_{2m+1-k}(\infty)
=-\underset{\infty}{\rm res}\ \frac{\phi_{2m+1-k}(P)}{\lambda(P)}=b_k,
\end{align*}
where we have used the residue theorem, relation (\ref{notation3}) and expressions (\ref{Superpotential-0}) and (\ref{g0-primary}) for $\lambda$ and  differentials $\phi_k$, $\phi_{2m+1-k}$. Hence, in view of  (\ref{g0-phi0-ab}), we get the desired relation.  \\
On the other hand, we know that $\phi_0$ is the Abelian differential of the third kind $\phi_0=\Omega_{0\infty}$ (here  the point at infinity $\infty$ is a zero of $\lambda$ and $0$ is a simple pole of $\lambda$) and quasi-homogeneous of degree 0. Thus,   applying (\ref{E-tA2}) we get
\begin{align*}
\mathbf{E}_{\phi_0}=\sum_{s=1}^{2m}\Big(d_s\mathbf{t}^s(\phi_0)+r_{\phi_0,\phi_s}\Big)\frac{\partial}{\partial\mathbf{t}^s(\phi_0)}
=\sum_{s=1}^m\Big(\partial_{t_s}+t_{2m+1-s}\partial_{t_{2m+1-s}}\Big).
\end{align*}

\fd

\medskip

Let us now move on  to calculate  explicit solutions to the WDVV equations that are associated  with  the described $\phi_j$-semi-simple Frobenius manifold structures, $j=0,\dots,2m$.\\
In our setting, the generalized $\phi_j$-WDVV equations (\ref{WDVV equation}) take the following form:
$$
\left\{
\begin{array}{ll}
\displaystyle \big(\partial_{\mathbf{t}^a(\phi_j)}\mathbf{H}_{\phi_j}\big)\eta\big(\partial_{\mathbf{t}^b(\phi_j)}\mathbf{H}_{\phi_j}\big)
=\big(\partial_{\mathbf{t}^b(\phi_j)}\mathbf{H}_{\phi_j}\big)\eta\big(\partial_{\mathbf{t}^a(\phi_j)}\mathbf{H}_{\phi_j}\big),
& \hbox{}\quad a,b=1,\dots,2m;\\
\\
\eta=\sum_{a=1}^{2m} \mathbf{1}_{\phi_j}\big(\mathbf{t}^a(\phi_j)\big)\big(\partial_{\mathbf{t}^a(\phi_j)}\mathbf{H}_{\phi_j}\big).
\end{array}
\right.
$$
Here $\mathbf{H}_{\phi_j}$ is the Hessian matrix of the prepotential  $\mathcal{F}_{\phi_j}$ induced by $\phi_j$ and
the $2m\times 2m$ constant matrix $\eta$ is anti-diagonal with $\eta_{ab}=\delta_{2m+1,a+b}=\eta^{ab}$. \\
Note that according to (\ref{g0-eE-phik}), (\ref{g0-eE-phim+k}) and (\ref{g0-eE-phi0}), the second line of the preceding equations reads as:
$$
\eta=
\left\{
\begin{array}{lll}
\displaystyle\partial_{x_{k,2m+1-k}}\mathbf{H}_{\phi_k}, & \hbox{if}\quad k=1,\dots,m;\\
\\
\displaystyle\partial_{y_{k,k}}\mathbf{H}_{\phi_{2m+1-k}}, & \hbox{if}\quad k=1,\dots,m;\\
\\
\displaystyle\frac{1}{1-\sum_{r=1}^mt_{2m+1-r}e^{-t_r}}\sum_{s=1}^me^{-t_s}\Big(\partial_{t_s}\mathbf{H}_{\phi_{0}}-t_{2m+1-s}\partial_{t_{2m+1-s}}\mathbf{H}_{\phi_0}\Big),
& \hbox{if}\quad k=0.
\end{array}
\right.
$$
For our  purpose, we  calculate the $\phi_j$-WDVV solutions $\mathcal{F}_{\phi_j}$  applying formula (\ref{Prep-eta3}) (we choose to omit the quadratic terms to simplify formulas).\\
Employing notation (\ref{g0-FC}), formula (\ref{Prep-eta3})   tells us that the prepotential associated with the differential $\phi_j$ is as follows:
\begin{align*}
\mathcal{F}_{\phi_j}
&=\frac{1}{2(1+d_j)}\sum_{k,s=1}^{2m}\frac{\big(E.\mathbf{t}^k(\phi_j)\big)\big(E.\mathbf{t}^s(\phi_j)\big)}
{1+d_j+{\rm deg}(\phi_{2m+1-k})}\mathbf{t}^{2m+1-s}\big(\phi_{2m+1-k}\big)\\
&=\frac{1}{2(1+d_j)(2+d_j)}\sum_{k,s=1}^m\big(E.\mathbf{t}^k(\phi_j)\big)\big(E.\mathbf{t}^s(\phi_j)\big)\mathbf{t}^{2m+1-s}\big(\phi_{2m+1-k}\big)\\
&\quad + \frac{1}{2(1+d_j)(2+d_j)}\sum_{k,s=1}^m\big(E.\mathbf{t}^k(\phi_j)\big)\big(E.\mathbf{t}^{2m+1-s}(\phi_j)\big)\mathbf{t}^{s}\big(\phi_{2m+1-k}\big)\\
&\quad +\frac{1}{2(1+d_j)^2}\sum_{k,s=1}^m\big(E.\mathbf{t}^{2m+1-k}(\phi_j)\big)\big(E.\mathbf{t}^s(\phi_j)\big)\mathbf{t}^{2m+1-s}\big(\phi_k\big)\\
&\quad +\frac{1}{2(1+d_j)^2}\sum_{k,s=1}^m\big(E.\mathbf{t}^{2m+1-k}(\phi_j)\big)\big(E.\mathbf{t}^{2m+1-s}(\phi_j)\big)\mathbf{t}^s\big(\phi_k\big)\\
&=\frac{1}{2(1+d_j)(2+d_j)}\sum_{k,s=1}^m\big(E.\mathbf{t}^k(\phi_j)\big)\big(E.\mathbf{t}^s(\phi_j)\big)y_{k,2m+1-s}\\
&\quad + \frac{2d_j+3}{2(1+d_j)^2(2+d_j)}\sum_{k,s=1}^m\big(E.\mathbf{t}^s(\phi_j)\big)\big(E.\mathbf{t}^{2m+1-k}(\phi_j)\big)x_{k,2m+1-s}\\
&\quad +\frac{1}{2(1+d_j)^2}\sum_{k,s=1}^m\big(E.\mathbf{t}^{2m+1-k}(\phi_j)\big)\big(E.\mathbf{t}^{2m+1-s}(\phi_j)\big)x_{k,s},
\end{align*}
where we used duality relations (\ref{g0-duality}) in the first equality, relation (\ref{g0-degree}) giving the degree of $\phi_k$ in the second equality.
Moreover, the last equality  follows from the introduced notation in (\ref{g0-phik-c}) and (\ref{g0-phim+k-c}) as well as the relation
$$
y_{k,s}:=\mathbf{t}^{s}\big(\phi_{2m+1-k}\big)=\mathbf{t}^{2m+1-s}\big(\phi_k\big):=x_{s,2m+1-k}.
$$
Therefore, we arrive at
\begin{equation}\label{g0-prep}
\begin{split}
\mathcal{F}_{\phi_j}&=\frac{1}{2(1+d_j)(2+d_j)}\sum_{k,s=1}^m\big(E.\mathbf{t}^k(\phi_j)\big)\big(E.\mathbf{t}^s(\phi_j)\big)y_{k,2m+1-s}\\
&\quad + \frac{2d_j+3}{2(1+d_j)^2(2+d_j)}\sum_{k,s=1}^m\big(E.\mathbf{t}^s(\phi_j)\big)\big(E.\mathbf{t}^{2m+1-k}(\phi_j)\big)x_{k,2m+1-s}\\
&\quad +\frac{1}{2(1+d_j)^2}\sum_{k,s=1}^m\big(E.\mathbf{t}^{2m+1-k}(\phi_j)\big)\big(E.\mathbf{t}^{2m+1-s}(\phi_j)\big)x_{k,s}.
\end{split}
\end{equation}

In the next Theorems, we write (\ref{g0-prep}) for various concrete differentials $phi_j$, thus obtaining explicit prepotentials of the corresponding Frobenius manifolds.
\begin{Thm}\label{g0-prep1}
Let $\phi_0$ be the differential (\ref{phi00}). The $\phi_0$-prepotential is:
\begin{align*}
\mathcal{F}_{\phi_0}&=\sum_{k=1}^mt_{2m+1-k}e^{t_k}+\frac{1}{2}\sum_{k=1}^m\big(t_{2m+1-k}\big)^2\Big(t_k+\log\big(t_{2m+1-k}\big)\Big)
+\frac{1}{2}\sum_{k,s=1, s\neq k}^mt_{2m+1-k}t_{2m+1-s}\log\big(e^{t_s}-e^{t_k}\big).
\end{align*}
The function $\mathcal{F}_{\phi_0}$ is quasi-homogeneous of degree 2 with respect to the Euler vector field $\mathbf{E}_{\phi_0}$ (\ref{g0-eE-phi0}) and we have
$$
\mathbf{E}_{\phi_0}.\mathcal{F}_{\phi_0}=2\mathcal{F}_{\phi_0}+\sum_{k=1}^m\big(t_{2m+1-k}\big)^2+\frac{1}{2}\sum_{k,s=1, s\neq k}^mt_{2m+1-k}t_{2m+1-s}.
$$
Moreover, $\mathcal{F}_{\phi_0}$ provides a solution to the generalized WDVV equations (\ref{WDVV equation}).
\end{Thm}
\emph{Proof:} We have ${\rm deg}(\phi_0)=0$ and by (\ref{g0-eE-phi0}), the following   quasi-homogeneity property of the flat coordinates $t_k,t_{2m+1-k}$ (\ref{g0-phi0-c}) are satisfied:
$$
E.t_s=1,\quad E.t_{2m+1-s}=t_{2m+1-s},\quad \forall \ s=1,\dots,m.
$$
This and formula (\ref{g0-prep}) imply that
\begin{align*}
\mathcal{F}_{\phi_0}&=\frac{1}{4}\sum_{k,s=1}^my_{k,2m+1-s}+ \frac{3}{4}\sum_{k,s=1}^mt_{2m+1-k}x_{k,2m+1-s}+\frac{1}{2}\sum_{k,s=1}^mt_{2m+1-k}t_{2m+1-s}x_{k,s}\\
&=\frac{1}{4}\sum_{k=1}^m\bigg(\sum_{s=1}^my_{k,2m+1-s}\bigg)+ \frac{3}{4}\sum_{k=1}^mt_{2m+1-k}\bigg(\sum_{s=1}^mx_{k,2m+1-s}\bigg)
+\frac{1}{2}\sum_{k=1}^m\big(t_{2m+1-k}\big)^2x_{k,k}\\
&\quad +\frac{1}{2}\sum_{k,s=1, s\neq k}^mt_{2m+1-k}t_{2m+1-s}x_{k,s}\\
&=\frac{1}{4}\sum_{k=1}^mt_{2m+1-k}e^{t_k}+ \frac{3}{4}\sum_{k=1}^mt_{2m+1-k}\bigg(e^{t_k}+\sum_{s=1}^mt_{2m+1-s}\bigg)
+\frac{1}{2}\sum_{k=1}^m\big(t_{2m+1-k}\big)^2\Big(t_k+\log\big(t_{2m+1-k}\big)\Big)\\
&\quad +\frac{1}{2}\sum_{k,s=1, s\neq k}^mt_{2m+1-k}t_{2m+1-s}\log\big(e^{t_s}-e^{t_k}\big)\\
&=\sum_{k=1}^mt_{2m+1-k}e^{t_k}+\frac{1}{2}\sum_{k=1}^m\big(t_{2m+1-k}\big)^2\Big(t_k+\log\big(t_{2m+1-k}\big)\Big)
+\frac{1}{2}\sum_{k,s=1, s\neq k}^mt_{2m+1-k}t_{2m+1-s}\log\big(e^{t_s}-e^{t_k}\big).
\end{align*}
Here, in the third equality, we used (\ref{g0-R1}) and (\ref{g0-R2}) and replaced functions $x_{k,s}$ (\ref{g0-phik-c}), $s=1,\dots,m$, by their expressions in terms of the coordinates $t_s,t_{2m+1-s}$.

\fd

\begin{Thm}\label{g0-prep2}
Let $\phi_j$ be the Abelian differential of the third kind $\phi_j=\Omega_{0a_j}$ (\ref{g0-primary}), $j=1,\dots,m$. Then the corresponding prepotential is given by:
\begin{align*}
&\mathcal{F}_{\phi_j}=
\frac{x_{j,j}}{2}\big(x_{j,2m+1-j}\big)^2+ \big(x_{j,2m+1-j}\big)\sum_{k=1,k\neq j}^mx_{j,2m+1-k}x_{j,k}+
 e^{x_{j,j}}+\sum_{k=1,k\neq j}^mx_{j,2m+1-k}\Big(e^{x_{j,k}}-e^{x_{j,j}-x_{j,k}}\Big)\\
&\quad +\frac{1}{2}\sum_{k=1, k\neq j}^m\big(x_{j,2m+1-k}\big)^2\Big(x_{j,k}+\log(x_{j,2m+1-k})\Big)+\frac{1}{2}\sum_{k=1,k\neq j}^m\sum_{s=1, s\neq j,k}^m
x_{j,2m+1-k}x_{j,2m+1-s}\log\big(e^{x_{j,s}}-e^{x_{j,k}}\big).
\end{align*}
It satisfies:
$$
\mathbf{E}_{\phi_j}.\mathcal{F}_{\phi_j}=2\mathcal{F}_{\phi_j}+ \sum_{k=1}^m\big(x_{j,2m+1-k}\big)^2+\frac{1}{2}\sum_{k,s=1, k\neq s}^mx_{j,2m+1-k}x_{j,2m+1-s},
$$
with $\mathbf{E}_{\phi_j}$ being the Euler vector field (\ref{g0-eE-phik}).
\end{Thm}
\emph{Proof:} Applying formula (\ref{g0-prep}) and  using the fact that ${\rm deg}(\phi_j)=0$ and
$$
E.x_{j,r}=E.\mathbf{t}^r(\phi_j)=1+\delta_{jr},\quad \quad  E.x_{j,2m+1-r}=x_{j,2m+1-r},\quad \forall\ r=1,\dots,m,
$$
we can write
\begin{align*}
\mathcal{F}_{\phi_j}
&=\frac{1}{4}\sum_{k,s=1}^m\big(1+\delta_{jk}\big)\big(1+\delta_{js}\big)y_{k,2m+1-s}+ \frac{3}{4}\sum_{k,s=1}^m\big(1+\delta_{js}\big)x_{j,2m+1-k}x_{k,2m+1-s}\\
&\quad +\frac{1}{2}\sum_{k,s=1}^mx_{j,2m+1-k}x_{j,2m+1-s}x_{k,s}\\
&=\frac{1}{4}\sum_{k=1}^m\bigg(\sum_{s=1}^my_{k,2m+1-s}\bigg)+\frac{1}{2}\sum_{s=1}^my_{j,2m+1-s}+
\frac{1}{4}y_{j,2m+1-j}+ \frac{3}{4}\sum_{k=1}^mx_{j,2m+1-k}\bigg(x_{k,2m+1-j}+\sum_{s=1}^mx_{k,2m+1-s}\bigg)\\
&\quad +\frac{x_{j,j}}{2}\big(x_{j,2m+1-j}\big)^2+\big(x_{j,2m+1-j}\big)\sum_{k=1,k\neq j}^mx_{j,2m+1-k}\Big(x_{j,k}+\log(-1)/2\Big)\\
&\quad +\frac{1}{2}\sum_{k=1, k\neq j}^mx_{j,2m+1-k}\bigg(x_{j,2m+1-k}x_{k,k}+\sum_{s=1, s\neq j,k}^mx_{j,2m+1-s}x_{k,s}\bigg).
\end{align*}
Thus the crucial part remaining to do is to express the coordinates  $y_{k,2m+1-s}$, $y_{k,s}$ and $x_{k,s}$ by means of $x_{j,r}$ and $x_{j,2m+1-r}$, $r=1,\dots,m$. \\
Employing (\ref{g0-phik-c}), (\ref{g0-phim+k-c}), (\ref{g0-R1}) and (\ref{g0-R2}), we deduce the following:\\
\noindent $\bullet$ When $j,k,s$ are distinct, then
$$
x_{k,s}=\log(e^{t_s}-e^{t_k}\big)=\log\big(e^{x_{j,s}}-e^{x_{j,k}}\big).
$$
\noindent $\bullet$ For any $k\neq j$, we have
\begin{align*}
&x_{k,k}=\log(x_{j,2m+1-k})+x_{j,k};\\
&x_{k,2m+1-j}=-e^{x_{j,j}-x_{j,k}};\\
&\sum_{s=1}^mx_{k,2m+1-s}=\bigg(\sum_{s=1}^mx_{k,2m+1-s}-\sum_{s=1}^mx_{j,2m+1-s}\bigg)+\sum_{s=1}^mx_{j,2m+1-s}=e^{x_{j,k}}+\sum_{s=1}^mx_{j,2m+1-s};\\
&y_{j,2m+1-k}=-x_{j,2m+1-k}x_{k,2m+1-j}=x_{j,2m+1-k}e^{x_{j,j}-x_{j,k}}.
\end{align*}
\noindent $\bullet$ For all $k=1,\dots,m$, we have
\begin{align*}
&\sum_{s=1}^my_{k,2m+1-s}=e^{x_{k,k}}=\delta_{jk}e^{x_{j,j}}+(1-\delta_{jk})x_{j,2m+1-k}e^{x_{j,k}};\\
&y_{j,2m+1-j}=\sum_{s=1}^my_{j,2m+1-s}-\sum_{s=1, s\neq j}^my_{j,2m+1-s}=e^{x_{j,j}}-\sum_{s=1, s\neq j}^mx_{j,2m+1-s}e^{x_{j,j}-x_{j,s}}.
\end{align*}
Substituting these relations into the previous expression of $\mathcal{F}_{\phi_j}$, we  obtain
\begin{align*}
&\mathcal{F}_{\phi_j}
=\frac{1}{4}e^{x_{j,j}}+\frac{1}{4}\sum_{k=1, k\neq j}^mx_{j,2m+1-k}e^{x_{j,k}}+\frac{1}{2}e^{x_{j,j}}
+\frac{1}{4}e^{x_{j,j}}-\frac{1}{4}\sum_{s=1, s\neq j}^mx_{j,2m+1-s}e^{x_{j,j}-x_{j,s}}\\
&\quad + \frac{3}{4}x_{j,2m+1-j}\bigg(x_{j,2m+1-j}+\sum_{s=1}^mx_{j,2m+1-s}\bigg)+ \frac{3}{4}\sum_{k=1, k\neq j }^mx_{j,2m+1-k}
\bigg(-e^{x_{j,j}-x_{j,k}}+e^{x_{j,k}}+\sum_{s=1}^mx_{j,2m+1-s}\bigg)\\
&\quad +\frac{x_{j,j}}{2}\big(x_{j,2m+1-j}\big)^2+\big(x_{j,2m+1-j}\big)\sum_{k=1,k\neq j}^mx_{j,2m+1-k}\Big(x_{j,k}+\frac{\log(-1)}{2}\Big)\\
&\quad +\frac{1}{2}\sum_{k=1, k\neq j}^m\big(x_{j,2m+1-k}\big)^2\Big(x_{j,k}+\log(x_{j,2m+1-k})\Big)+\frac{1}{2}\sum_{k=1,k\neq j}^m\sum_{s=1, s\neq j,k}^m
x_{j,2m+1-k}x_{j,2m+1-s}\log\big(e^{x_{j,s}}-e^{x_{j,k}}\big)\\
&=\frac{x_{j,j}}{2}\big(x_{j,2m+1-j}\big)^2+ \big(x_{j,2m+1-j}\big)\sum_{k=1,k\neq j}^mx_{j,2m+1-k}x_{j,k}+
 e^{x_{j,j}}+\sum_{k=1,k\neq j}^mx_{j,2m+1-k}\Big(e^{x_{j,k}}-e^{x_{j,j}-x_{j,k}}\Big)\\
&\quad +\frac{1}{2}\sum_{k=1, k\neq j}^m\big(x_{j,2m+1-k}\big)^2\Big(x_{j,k}+\log(x_{j,2m+1-k})\Big)+\frac{1}{2}\sum_{k=1,k\neq j}^m\sum_{s=1, s\neq j,k}^m
x_{j,2m+1-k}x_{j,2m+1-s}\log\big(e^{x_{j,s}}-e^{x_{j,k}}\big)\\
&\quad +\frac{\log(-1)}{2}\big(x_{j,2m+1-j}\big)\sum_{k=1,k\neq j}^mx_{j,2m+1-k}+\frac{3}{4}\big(x_{j,2m+1-j}\big)^2
+\frac{3}{4}\sum_{k,s=1}^mx_{j,2m+1-k}x_{j,2m+1-s}.
\end{align*}
Finally,  subtracting the quadratic terms we arrive at the stated expression for the function $\mathcal{F}_{\phi_j}$.

\fd

\begin{Thm}\label{g0-prep3} For  $j=1,\dots,m$, let $\phi_{2m+1-j}$ be the Abelian differential (\ref{g0-primary}). Then the prepotential associated with the semi-simple Frobenius manifold structure induced by the differential $\phi_{2m+1-j}$ is determined by
\begin{align*}
&\mathcal{F}_{\phi_{2m+1-j}}=\frac{\big(y_{jj}\big)^2}{2}y_{j,2m+1-j}+y_{jj}\sum_{k=1,k\neq j}^my_{j,k}y_{j,2m+1-k}
-\frac{1}{2}\sum_{k=1,k\neq j}^m\big(y_{j,k}\big)^2y_{j,2m+1-k}\\
&\quad +\frac{1}{2}\bigg(\sum_{k=1}^my_{j,2m+1-k}\bigg)^2\log\Big(\sum_{r=1}^my_{j,2m+1-r}\Big)
+\frac{1}{2}\sum_{k=1, k\neq j}^m\big(y_{j,2m+1-k}\big)^2\log\big(y_{j,2m+1-k}\big)\\
&\quad -\sum_{k,s=1, s\neq j}^my_{j,2m+1-k}y_{j,2m+1-s}\log(y_{j,s})+\frac{1}{2}\sum_{k=1, k\neq j}^m\sum_{s=1,s\neq j,k}^m
y_{j,2m+1-k}y_{j,2m+1-s}\log\big(y_{j,s}-y_{j,k}\big)\\
&\quad -\frac{1}{12}\sum_{k=1,k\neq j}^m\sum_{s=1,s\neq j,k}^m\frac{y_{j,k}y_{j,2m+1-k}y_{j,2m+1-s}}{y_{j,k}-y_{j,s}}.
\end{align*}
The function $\mathcal{F}_{\phi_{2m+1-j}}$ is quasi-homogeneous of degree 4 with respect to the Euler vector field (\ref{g0-eE-phim+k}) and we have
$$
\mathbf{E}_{\phi_{2m+1-j}}.\mathcal{F}_{\phi_{2m+1-j}}=4\mathcal{F}_{\phi_{2m+1-j}}+\sum_{k=1}^m\big(y_{j,2m+1-k}\big)^2
+\frac{1}{2}\sum_{k,s=1, s\neq k}^my_{j,2m+1-k}y_{j,2m+1-s}.
$$
\end{Thm}
\emph{Proof:} Since ${\rm deg}(\phi_{2m+1-j})=1$,  formulas (\ref{g0-prep}) and (\ref{g0-eE-phim+k}) imply that
\begin{align*}
&\mathcal{F}_{\phi_{2m+1-j}}
=\frac{1}{12}\sum_{k,s=1}^my_{j,k}y_{j,s}y_{k,2m+1-s}+ \frac{5}{12}\sum_{k,s=1}^my_{j,s}y_{j,2m+1-k}y_{s,k}
+\frac{1}{2}\sum_{k,s=1}^my_{j,2m+1-k}y_{j,2m+1-s}x_{k,s}\\
&=\frac{\big(y_{jj}\big)^2}{12}y_{j,2m+1-j}+\frac{y_{jj}}{6}\sum_{k=1,k\neq j}^my_{j,k}y_{j,2m+1-k} +
\frac{1}{12}\sum_{k=1,k\neq j}^m\bigg((y_{j,k})^2y_{k,2m+1-k}+y_{j,k}\sum_{s=1, s\neq j,k}^my_{j,s}y_{k,2m+1-s}\bigg) \\
&\quad +\frac{5}{12}\big(y_{j,j}\big)^2y_{j,2m+1-j}+\frac{5}{12}y_{j,2m+1-j}\sum_{s=1,s\neq j}^my_{j,s}y_{s,j}+\frac{5}{6}y_{j,j}
\sum_{k=1, k\neq j}^my_{j,2m+1-k}y_{j,k}\\
&\quad + \frac{5}{12}\sum_{k=1,k\neq j}^my_{j,2m+1-k}\bigg(y_{j,k}\big(y_{k,k}-y_{jj}\big)+\sum_{s=1,s\neq j,k}^my_{j,s}y_{s,k}\bigg)\\
&\quad +\frac{1}{2}\big(y_{j,2m+1-j}\big)^2x_{j,j}+y_{j,2m+1-j}\sum_{k=1, k\neq j}^my_{j,2m+1-k}\Big(x_{k,j}+\frac{\log(-1)}{2}\Big)\\
&\quad +\frac{1}{2}\sum_{k=1, k\neq j}^m\bigg((y_{j,2m+1-k})^2x_{k,k}+y_{j,2m+1-k}\sum_{s=1,s\neq j,k}^my_{j,2m+1-s}x_{k,s}\bigg)\\
&=\frac{\big(y_{jj}\big)^2}{2}y_{j,2m+1-j}+y_{jj}\sum_{k=1,k\neq j}^my_{j,k}y_{j,2m+1-k} +
\frac{1}{12}\sum_{k=1,k\neq j}^m\bigg(\underbrace{(y_{j,k})^2y_{k,2m+1-k}+y_{j,k}\sum_{s=1, s\neq j,k}^my_{j,s}y_{k,2m+1-s}}_{Q_{j,k}}\bigg) \\
&\quad -\frac{5}{12}y_{j,2m+1-j}\sum_{s=1,s\neq j}^my_{j,2m+1-s}
+ \frac{5}{12}\sum_{k=1,k\neq j}^my_{j,2m+1-k}\bigg(\underbrace{y_{j,k}\big(y_{k,k}-y_{jj}\big)+\sum_{s=1,s\neq j,k}^my_{j,s}y_{s,k}}_{R_{j,k}}\bigg)\\
&\quad +\frac{1}{2}\big(y_{j,2m+1-j}\big)^2x_{j,j}+\frac{\log(-1)}{2}y_{j,2m+1-j}\sum_{k=1, k\neq j}^my_{j,2m+1-k}+
y_{j,2m+1-j}\sum_{k=1, k\neq j}^my_{j,2m+1-k}x_{k,j}\\
&\quad +\frac{1}{2}\sum_{k=1, k\neq j}^m\bigg((y_{j,2m+1-k})^2x_{k,k}+y_{j,2m+1-k}\sum_{s=1,s\neq j,k}^my_{j,2m+1-s}x_{k,s}\bigg).
\end{align*}
As above, we need some intermediate results to complete the calculation.
\begin{description}
\item[i)]For all  $k\neq j$ we have
\begin{align*}
&y_{k,j}=-\frac{y_{j,2m+1-k}}{y_{j,k}};\\
&x_{k,j}=x_{j,j}-\log(y_{j,k});\\
&x_{k,k}=x_{j,j}+\log(y_{j,2m+1-k})-2\log(y_{j,k}).
\end{align*}
\item[ii)]If $j,k,s$ are distinct, then
\begin{align*}
&x_{k,s}=x_{j,j}+\log\big(y_{j,s}-y_{j,k}\big)-\log(y_{j,k})-\log(y_{j,s});\\
&y_{k,s}=e^{x_{k,k}-x_{s,k}}=\frac{y_{j,2m+1-k}y_{j,s}}{y_{j,k}\big(y_{j,k}-y_{j,s}\big)};\\
&y_{k,2m+1-s}=-y_{k,s}y_{s,k}=\frac{y_{j,2m+1-k}y_{j,2m+1-s}}{\big(y_{j,k}-y_{j,s}\big)^2}.
\end{align*}
\end{description}
To see  these equalities we use expressions (\ref{g0-phik-c}) and (\ref{g0-phim+k-c}) involving the variables  $t_k,t_{m+k}$ .\\
In addition,  items \textbf{i)} and \textbf{ii)} and relations (\ref{g0-R1}) and (\ref{g0-R2}) lead to the following required assertions:
\begin{description}
\item[iii)] For all $k\neq j$:
\begin{align*}
Q_{j,k}&:=(y_{j,k})^2y_{k,2m+1-k}+y_{j,k}\sum_{s=1,s\neq j, k}^my_{j,s}y_{k,2m+1-s}\\
&=(y_{j,k})^2e^{x_{k,k}}-(y_{j,k})^2\sum_{s=1,s\neq k}^my_{k,2m+1-s}+y_{j,k}\sum_{s=1,s\neq j, k}^my_{j,s}y_{k,2m+1-s}\\
&=y_{j,2m+1-k}e^{x_{j,j}}-(y_{j,k})^2y_{j,2m+1-k}-\sum_{s=1,s\neq j,k}^m\frac{y_{j,k}y_{j,2m+1-k}y_{j,2m+1-s}}{y_{j,k}-y_{j,s}}.
\end{align*}
\item[vi)] For all $k\neq j$:
\begin{align*}
R_{j,k}&=y_{j,k}\big(y_{k,k}-y_{j,j}\big)+\sum_{s=1, s\neq j,k}^my_{j,s}y_{s,k}\\
&=y_{j,k}\bigg(-\frac{e^{x_{j,j}}}{y_{j,k}}+\sum_{s=1, s\neq j}^my_{s,j}-\sum_{s=1, s\neq k}^my_{s,k}\bigg)+\sum_{s=1, s\neq j,k}^my_{j,s}y_{s,k}\\
&=-e^{x_{j,j}}-y_{j,k}\sum_{s=1, s\neq j}^m\frac{y_{j,2m+1-s}}{y_{j,s}}-\big(y_{j,k}\big)^2+\sum_{s=1, s\neq j, k}^m\big(y_{j,s}-y_{j,k}\big)y_{s,k}\\
&=-e^{x_{j,j}}-y_{j,2m+1-k}-\big(y_{j,k}\big)^2.
\end{align*}
\end{description}
Finally,  omitting all the quadratic terms and employing  relations indicated in items \textbf{i)}-\textbf{vi)} and recalling that due to  (\ref{g0-R2}) we have $e^{x_{j,j}}=\sum_{r=1}^my_{j,2m+1-r}$, we arrive at
\begin{align*}
\mathcal{F}_{\phi_{2m+1-j}}&=\frac{\big(y_{jj}\big)^2}{2}y_{j,2m+1-j}+y_{jj}\sum_{k=1,k\neq j}^my_{j,k}y_{j,2m+1-k}\\
&\quad +\frac{1}{12}\sum_{k=1,k\neq j}^m\bigg(y_{j,2m+1-k}e^{x_{j,j}}-(y_{j,k})^2y_{j,2m+1-k}-\sum_{s=1,s\neq j,k}^m
\frac{y_{j,k}y_{j,2m+1-k}y_{j,2m+1-s}}{y_{j,k}-y_{j,s}}\bigg)\\
&\quad+ \frac{5}{12}\sum_{k=1,k\neq j}^my_{j,2m+1-k}\bigg(-e^{x_{j,j}}-y_{j,2m+1-k}-\big(y_{j,k}\big)^2\bigg)\\
&\quad +\frac{1}{2}\big(y_{j,2m+1-j}\big)^2x_{j,j}+y_{j,2m+1-j}\sum_{k=1, k\neq j}^my_{j,2m+1-k}\Big(x_{j,j}-\log(y_{j,k})\Big)\\
&\quad +\frac{1}{2}\sum_{k=1, k\neq j}^m\big(y_{j,2m+1-k}\big)^2\bigg(x_{j,j}+\log\big(y_{j,2m+1-k}\big)-2\log(y_{j,k})\bigg)\\
&\quad+\frac{1}{2}\sum_{k=1, k\neq j}^m\sum_{s=1,s\neq j,k}^my_{j,2m+1-k}y_{j,2m+1-s}\bigg(x_{j,j}+\log\big(y_{j,s}-y_{j,k}\big)-\log\big(y_{j,s}\big)
-\log\big(y_{j,k}\big)\bigg)
\end{align*}
which is equivalent to the expression claimed in the theorem.

\fd

\begin{Remark}
\end{Remark}
Let us consider the case $m=2$ of the obtained solutions to the WDVV equations stated in the above three theorems.  \\
For $j=1,2$, we introduce
$$
\begin{array}{cccc}
x_1=x_{j,j},\quad  & x_2=x_{j,k}, \quad j\neq k,\quad  & x_3=x_{j,5-k}, \quad k\neq j,\quad   & x_4=x_{j,5-j} \\
\\
y_1=y_{j,j},\quad  & y_2=y_{j,k}, \quad j\neq k,\quad  & y_3=y_{j,5-k}, \quad k\neq j,\quad   & y_4=y_{j,5-j} \\
\end{array}
$$
Then we have
\begin{align*}
F_1&=t_4e^{t_1}+t_3e^{t_2}+\frac{1}{2}\Big(t_4^2t_1+t_4^2\log(t_4)+ t_3^2t_2+t_3^2\log(t_3)\Big)+t_3t_4\log\big(e^{t_1}-e^{t_2}\big)+\frac{\log(-1)}{2}t_3t_4;\\
F_2&=\frac{x_1}{2}x_4^2+x_2x_3x_4+e^{x_1}+x_3e^{x_2}-x_3e^{x_1-x_2}+\frac{x_3^2}{2}x_2+\frac{x_3^2}{2}\log(x_3);\\
F_3&=\frac{y_1^2}{2}y_4+y_1y_2y_3-\frac{y_2^2}{2}y_3+\frac{1}{2}(y_3+y_4)^2\log\big(y_3+y_4\big)+\frac{y_3^2}{2}\log(y_3)-(y_3+y_4)y_3\log(y_2).
\end{align*}


\section{Examples of genus one solutions to the WDVV equations}
\subsection{Weierstrass functions}
We begin with recalling some useful formulas involving the three Weierstrass  functions $\wp,\zeta$ and $\sigma$ associated with the lattice $\mathbb{L}=2\omega_1\Z+2\omega_2\Z$, where $\omega_1, \omega_2$ are two complex numbers chosen in such a manner that  the imaginary part of the ratio $\tau:=\omega_2/{\omega_1}$ is  positive.  For more details on the properties of Weierstrass functions and related topics we refer to the monographs \cite{Akhiezer, Apostol, Chandra, Erdelyi, Du Val, Forsyth, Watson}.\\
The functions $\wp$, $\zeta$ and $\sigma$ are respectively defined  by:
\begin{align}\label{p-zeta functions}
\begin{split}
&\wp(u)=\wp(u;2\omega_1,2\omega_2):=\frac{1}{u^2}+\sum_{w\in \mathbb{L},w\neq0}\Big(\frac{1}{(u-w)^2}-\frac{1}{w^2}\Big);\\
&\zeta(u)=\zeta(u;2\omega_1,2\omega_2):=\frac{1}{u}+\sum_{w\in \mathbb{L},w\neq0}\Big(\frac{1}{u-w}+\frac{1}{w}+\frac{u}{w^2}\Big);\\
&\sigma(u)=\sigma(u;2\omega_1,2\omega_2):=u\prod_{w\in \mathbb{L},w\neq0}\big(1-\frac{u}{w}\big)\exp\Big(\frac{u}{w}+\frac{u^2}{2w^2}\Big).
\end{split}
\end{align}
Here is the list of properties of the three functions that will be important in what follows.
\begin{itemize}
\item[1.] Differential equations for the $\wp$-function:
\begin{align}
\begin{split}\label{p-diff equation}
&\wp'^2(u)=4\wp^3(u)-g_2\wp(u)-g_3=4\big(\wp(u)-e_1\big)\big(\wp(u)-e_2\big)\big(\wp(u)-e_3\big);
\end{split}\\
\begin{split}\label{pi-function equation}
&\wp''(u)=6\wp^2(u)-\frac{g_2}{2};\\
&\wp'''(u)=12\wp'(u)\wp(u);\\
&\wp^{(4)}(u)=120\wp^3(u)-18g_2\wp(u)-12g_3,
\end{split}
\end{align}
where $\omega_3:=\omega_1+\omega_2$, $e_j=\wp(\omega_j)$ and
\begin{equation}\label{g2-g3}
g_2:=g_2(2\omega_1,2\omega_2):=60\sum_{w\in \mathbb{L}, w\neq0}\frac{1}{w^4},\quad \quad\quad
g_3:=g_3(2\omega_1,2\omega_2):=140\sum_{w\in \mathbb{L}, w\neq0}\frac{1}{w^6}.
\end{equation}
\item[2.] The $\zeta$-Weierstrass function satisfies:
\begin{equation}\label{zeta-derivative}
\zeta'(u):=-\wp(u).
\end{equation}
\item[3.] The function $\zeta$  is the logarithmic derivative of $\sigma$:
\begin{equation}\label{zeta-sigma}
\sigma'(u)/\sigma(u)=\zeta(u).
\end{equation}
\item[4.] We have the homogeneity relations for arbitrary $\alpha\neq0$:
\begin{equation}\label{p-zeta-sigma homog}
\begin{split}
&\wp(\alpha u;2\alpha\omega_1,2\alpha\omega_2)=\alpha^{-2}\wp(u;2\omega_1,2\omega_2);\\
&\zeta(\alpha u;2\alpha\omega_1,2\alpha\omega_2)=\alpha^{-1}\zeta(u;2\omega_1,2\omega_2);\\
&\sigma(\alpha u;2\alpha\omega_1,2\alpha\omega_2)=\alpha\sigma(u;2\omega_1,2\omega_2).
\end{split}
\end{equation}
\item[5.] The functions $\zeta$ and $\sigma$ satisfy the following quasi-periodicity properties:
\begin{equation}\label{zeta-sigma periods}
\zeta(u+2\omega_j)=\zeta(u)+2\zeta(\omega_j),\quad \sigma(u+2\omega_j)=-\sigma(u)e^{2\zeta(\omega_j)u+2\omega_j\zeta(\omega_j)}.
\end{equation}
\item[6.] The periods $\omega_1,\omega_2$ and the quasi-periods of $\zeta$ are related by the Legendre relation:
\begin{equation}\label{Legendre}
\omega_2\zeta(\omega_1)-\omega_1\zeta(\omega_2)=\frac{{\rm{i}}\pi}{2}.
\end{equation}
\item[7.] Addition theorems for the functions $\wp$ and $\zeta$:
\begin{align}
&\label{addition-pi}\wp(u+v)=-\wp(u)-\wp(v)+\frac{1}{4}\bigg(\frac{\wp'(u)-\wp'(v)}{\wp(u)-\wp(v)}\bigg)^2;\\
&\label{addition-zeta}\zeta(u+v)=\zeta(u)+\zeta(v)+\frac{\wp'(u)-\wp'(v)}{2\big(\wp(u)-\wp(v)\big)}.
\end{align}

\end{itemize}
\subsection{Examples of 3-dimensional Hurwitz-Frobenius manifolds and Ramanujan identities}
We are interested in some examples of Frobenius manifold structures on  the Hurwitz space $\mathcal{H}_{1,2}(1)$ of genus one two-fold ramified coverings  $(\C/\mathbb{L},\lambda)$ of $\P^1$ where $\C/\mathbb{L}$ is the torus associated with the lattice $\mathbb{L}=2\omega_1\Z+2\omega_2\Z$ and $\lambda$ is the meromorphic function given by
\begin{equation}\label{Weierstrass1}
\lambda(P)=\wp(z_P)+c,\quad P\in \C/\mathbb{L}.
\end{equation}
Here $c$ is a nonzero  constant (with respect to $z_P$) and $z_P$ is a representative of the point $P$ in the fundamental parallelogram
$$
\mathcal{F}:=\Big\{2\alpha\omega_1+2\beta\omega_2:\quad 0\leq\alpha,\beta\leq1\Big\}.
$$
Due  to the  algebraic differential equation (\ref{p-diff equation}), the covering $(\C/\mathbb{L},\lambda)$ can be identified with the elliptic curve (known as the Weierstrass elliptic curve):
$$
\Big\{P=(\lambda,\mu): \quad \mu^2=4(\lambda-\lambda_1)(\lambda-\lambda_2)(\lambda-\lambda_3)\Big\},
$$
where $\lambda_j=\wp(\omega_j)+c=e_j+c$ and $\mu=\wp'(z_P)$.
It has four simple ramification points $P_j=(\lambda_j,0)$ for $j=1,2,3$ and $\infty^0=(\infty,\infty)$. Note that the parameter $c$ makes the branch points $\lambda_j$ independent of each other (if $c=0$, they become dependent since $e_1+e_2+e_3=0$) and this permits us to use them as local coordinates on the
Hurwitz space $\mathcal{H}_{1,2}(1)$. \\
Using the homogeneity relation (\ref{p-zeta-sigma homog}) we can see that the equivalence class (as defined in Section 2.2) of the pair $(\C/\mathbb{L},\lambda)$ consists of ramified coverings $(\C/\widetilde{\mathbb{L}},\widetilde{\lambda})$ such that
$$
\widetilde{\mathbb{L}}=\alpha\mathbb{L}\quad \quad  \text{and}\quad \quad\widetilde{\lambda}(P)=\alpha^2\wp(z_P,2\alpha\omega_1,2\alpha\omega_2)+c
$$
for some nonzero complex number $\alpha$. \\
In order to construct Hurwitz-Frobenius manifold structures in genus one, we restrict ourselves to an open neighborhood in $\mathcal{H}_{1,2}(1)$ of the generic covering $(\C/\mathbb{L},\lambda)$ with a fixed Torelli marking $\{a,b\}$. Here the cycles $a,b$ are fixed to be the segments $a=[x,x+2\omega_1]$ and $b=[x,x+2\omega_2]$.\\
We will apply formula (\ref{Prep-eta4}) to compute the prepotentials of the semi-simple Hurwitz-Frobenius manifolds structures determined by the following  differentials:
\begin{align*}
\phi_1(P)&:=\frac{1}{2{\rm{i}}\pi}\oint_bW(P,Q);\\
\phi_2(P)&:=\sqrt{2}{\rm res}_{\infty^0}\sqrt{\lambda(Q)}W(P,Q);\\
\phi_3(P)&:=\oint_a\lambda(Q)W(P,Q).
\end{align*}
According to (\ref{Primary-e}) and Corollary \ref{Rmk-primary}, the  three differentials $\phi_1,\phi_2$ and $\phi_3$ are primary with respect to the unit vector field $e=\sum_{j=1}^3\partial_{\lambda_j}$. Furthermore, the differentials $\phi_j$ are all quasi-homogeneous and their  quasi-homogeneous degrees are:
$$
deg(\phi_1)=0,\quad deg(\phi_2)=1/2,\quad deg(\phi_3)=deg(\phi_4)=1.
$$
As a first step, we start  with the following explicit formula for the symmetric bidifferential $W$ (see Section 2.3) in genus one:
\begin{Lemma}
The canonical meromorphic bidifferential on the torus $\C/{\mathbb{L}}$, equipped with a fixed canonical homology basis $\{a,b\}$,  is given by
\begin{equation}\label{W-g=1}
W(P,Q)=\Big(\wp(z_P-z_Q)+\frac{\zeta(\omega_1)}{\omega_1}\Big)dz_Pdz_Q.
\end{equation}
\end{Lemma}
\emph{Proof:} The right hand side is clearly a symmetric bidifferential and has a second-order pole on the diagonal $P=Q$  with biresidue 1. Moreover,  due to (\ref{zeta-derivative}) and the quasi-periodicity (\ref{zeta-sigma periods}) of the function zeta, we see that  its $a$-period is zero.

\fd

It is more convenient to work with the following explicit formulas for the four differentials $\phi_1, \dots,\phi_4$. The results  are obtained with the help of  (\ref{W-g=1}).
\begin{Prop}
\begin{description}
\item[1)] Normalized holomorphic differential:
\begin{equation}\label{phi1}
\phi_1(P)=\frac{dz_P}{2\omega_1}.
\end{equation}
\item[2)] Abelian differential of the second kind:
\begin{equation}\label{phi2}
\phi_2(P)=\sqrt{2}\Big(\wp(z_P)+\frac{\zeta(\omega_1)}{\omega_1}\Big)dz_P.
\end{equation}
\item[3)] Multivalued differential with jumps along the cycle $b$:
\begin{equation}\label{phi3}
\phi_3(P)=\bigg(4\omega_1\wp^2(z_P)-4\zeta(\omega_1)\wp(z_P)+2\wp'(z_P)\big(\omega_1\zeta(z_P)-\zeta(\omega_1)z_P\big)
-\frac{g_2}{2}\omega_1-2\frac{\zeta^2(\omega_1)}{\omega_1}\bigg)dz_P.
\end{equation}
\end{description}
\end{Prop}
\begin{Remark}
In view of (\ref{phi3}) and  the Legendre identity (\ref{Legendre}), we can  observe that the jumps of the differential $\phi_3$ are such that:
$$
\phi_3(P+a)-\phi_3(P)=0,\quad \quad \phi_3(P+b)-\phi_3(P)=-2{\rm{i}}\pi \wp'(z_P)dz_P.
$$
On the other hand,  introducing  the (holomorphic)  function
\begin{equation}\label{V-j}
\mathcal{V}(u):=\omega_1\wp'(u)+2\zeta(\omega_1)\Big(\zeta(u)-\frac{\zeta(\omega_1)}{\omega_1}u\Big)
+2\wp(u)\Big(\omega_1\zeta(u)-\zeta(\omega_1)u\Big), \quad u\in \C,
\end{equation}
and employing the differential equation $2\wp''=12\wp^2-g_2$ and expression (\ref{phi3}) for the differential $\phi_3$ we can write
\begin{equation}\label{V-j-der}
\phi_3(P)=d\mathcal{V}(z_P)=\mathcal{V}'(z_P)dz_P.
\end{equation}
\end{Remark}
\noindent\emph{Proof:} \textbf{1)} Formula (\ref{phi1}) can be easily checked  using  (\ref{W-g=1}), (\ref{zeta-derivative}), the quasi-periodicity property (\ref{zeta-sigma periods}) of the $\zeta$-function and Legendre's relation (\ref{Legendre}).\\
\textbf{2)} In order to show (\ref{phi2}), let us first claim that near the point $\infty^0$, the holomorphic differential $dz_P=2\omega_1\phi_1(P)$ behaves as follows
\begin{equation}\label{phi1-behavior}
dz_P\underset{P\sim \infty^0}=\Big(1+\frac{3c}{2}z_{\infty}(P)^2+\big(\frac{g_2}{8}+\frac{15}{8}c^2\big)z_{\infty}(P)^4+\dots\Big)dz_{\infty}(P),
\end{equation}
where $z_{\infty}(P)$ is the standard  local parameter near $\infty^0$ defined by $\lambda(P)=z_{\infty}(P)^{-2}$.\\
Indeed, since  the principal part near the origin  of $\wp'(z)$ is $-2z^{-3}$ and $d\lambda(P)=\wp'(z_P)dz_P$,  the holomorphic differential $dz_P$ can be rewritten as
\begin{equation}\label{phi1-lambda}
dz_P=\frac{d\lambda(P)}{\wp'(z_p)}=-\frac{d\lambda(P)}{2\sqrt{(\lambda(P)-\lambda_1)(\lambda(P)-\lambda_2)(\lambda(P)-\lambda_3)}}.
\end{equation}
In particular
$$
dz_P\underset{P\sim \infty^0}=\frac{dz_{\infty}(P)}{f\big(z_{\infty}(P)\big)},
$$
where $f$ is the function
\begin{align*}
f(x)&=\Big(1-(\lambda_1+\lambda_2+\lambda_3)x^2+(\lambda_1\lambda_2+\lambda_2\lambda_3+\lambda_3\lambda_1)x^4-(\lambda_1\lambda_2\lambda_3)x^6\Big)^{1/2}\\
&=\Big(1-3cx^2+\big(3c^2-\frac{g_2}{4}\big)x^4+\big(\frac{g_3}{4}-c\frac{g_2}{4}+c^3\big)x^6\Big)^{1/2}.
\end{align*}
Hence the behavior of $f$ near zero gives (\ref{phi1-behavior}).\\
Now (\ref{phi1-behavior})  and (\ref{W-g=1}) yield that
$$
W(P,Q)\underset{Q\sim \infty^0}=\Big(\wp(z_P-z_Q)+\frac{\zeta(\omega_1)}{\omega_1}\Big)
\Big(1+\frac{3c}{2}z_{\infty}(Q)^2+\dots\Big)dz_{\infty}(Q)dz_P.
$$
Accordingly, we obtain
$$
\phi_2(P)=\sqrt{2}W(P,\infty^0)=\sqrt{2}\frac{W(P,Q)}{dz_{\infty}(Q)}\Big|_{Q=\infty^0}=\sqrt{2}\Big(\wp(z_P)+\frac{\zeta(\omega_1)}{\omega_1}\Big)dz_P.
$$
\textbf{3)} Let us look at (\ref{phi3}). We have
$$
\phi_3(P)=\oint_a(\lambda(Q)-c)W(P,Q)=\left(\oint_a\wp(z_Q)\wp(z_Q-z_P)dz_Q\right)dz_P-2\frac{\zeta^2(\omega_1)}{\omega_1}dz_P.
$$
Thus it suffices  to establish that
\begin{equation}\label{p-key relation}
\oint_a\wp(z_Q)\wp(z_Q-z_P)dz_Q=4\omega_1\wp^2(z_P)-4\zeta(\omega_1)\wp(z_P)+2\wp'(z_P)\big(\omega_1\zeta(z_P)-\zeta(\omega_1)z_P\big)
-\frac{g_2}{2}\omega_1.
\end{equation}
In order to show (\ref{p-key relation}), let us claim that:
\begin{equation*}
\Big(\wp(z_Q)-\wp(z_P)\Big)\wp(z_Q-z_P)
=2\wp^2(z_P)+\wp(z_P)\wp(z_Q)+\wp'(z_P)\Big(\zeta(z_Q-z_P)-\zeta(z_Q)+\zeta(z_P)\Big)-\frac{g_2}{4}.
\end{equation*}
Indeed, we can write
\begin{align*}
&\Big(\wp(z_Q)-\wp(z_P)\Big)\wp(z_Q-z_P)\\
&=\wp^2(z_P)-\wp^2(z_Q)+\frac{\wp'^2(z_P)+\wp'^2(z_Q)+2\wp'(z_Q)\wp'(z_P)}{4\big(\wp(z_Q)-\wp(z_P)\big)}\\
&=\wp^2(z_P)-\wp^2(z_Q)+\frac{\wp'^2(z_P)}{4\big(\wp(z_Q)-\wp(z_P)\big)}+\frac{\wp'(z_P)\wp'(z_Q)}{2\big(\wp(z_Q)-\wp(z_P)\big)}
+\frac{4\wp^3(z_Q)-g_2\wp(z_Q)-g_3}{4\big(\wp(z_Q)-\wp(z_P)\big)}\\
&=\wp^2(z_P)-\wp^2(z_Q)+\frac{\wp'^2(z_P)}{4\big(\wp(z_Q)-\wp(z_P)\big)}+\frac{\wp'(z_P)\wp'(z_Q)}{2\big(\wp(z_Q)-\wp(z_P)\big)}\\
&\quad+\frac{4\big(\wp^3(z_Q)-\wp^3(z_P)\big)-g_2\big(\wp(z_Q)-\wp(z_P)\big)+4\wp^3(z_P)-g_2\wp(z_P)-g_3}{4\big(\wp(z_Q)-\wp(z_P)\big)}\\
&=2\wp^2(z_P)+\wp(z_P)\wp(z_Q)+\frac{\wp'^2(z_P)}{2\big(\wp(z_Q)-\wp(z_P)\big)}+\frac{\wp'(z_P)\wp'(z_Q)}{2\big(\wp(z_Q)-\wp(z_P)\big)}-\frac{g_2}{4}\\
&=2\wp^2(z_P)+\wp(z_P)\wp(z_Q)+\wp'(z_P)\Big(\zeta(z_Q-z_P)-\zeta(z_Q)+\zeta(z_P)\Big)-\frac{g_2}{4},
\end{align*}
where we have used\\
- the addition theorem (\ref{addition-pi}) for the $\wp$-function in the first equality;\\
- the differential equation (\ref{p-diff equation}) in the second and forth equalities;\\
- the addition theorem (\ref{addition-zeta}) for the $\zeta$-function in the last one. \\
Finally, the claimed relation, the  $\sigma$-quasi-periodicity property $\sigma(u+2\omega_1)=-\sigma(u)e^{2\zeta(\omega_1)u+2\omega_1\zeta(\omega_1)}$ and a straightforward calculation based on (\ref{zeta-derivative}) and (\ref{zeta-sigma periods}) enable us to  arrive at (\ref{p-key relation}).

\fd

\medskip

Our next goal is to calculate the $\phi_j$-Darboux-Egoroff metrics:
$$
\eta(\phi_j)=\frac{1}{2}\sum_{k=1}^3\big(\phi_j(P_k)\big)^2(d\lambda_k)^2, \quad j=1,2,3,
$$
and  the corresponding flat coordinates $\mathbf{t}^A(\phi_j)$. Recall that $\phi_j(P_k)$ denotes the evaluation  of the differential $\phi_j$ at the ramification point $P_k$ with respect to the local parameter $x_k(P)=\sqrt{\lambda(P)-\lambda_k}$ (see (\ref{notation1})).\\
From (\ref{FBP-W}) we know that the three flat coordinates of the flat metric $\eta(\phi_j)$ are chosen as follows:
\begin{align*}
\mathbf{t}^1(\phi_j):=\frac{1}{2{\rm{i}}\pi}\oint_b\phi_j(P); \quad \mathbf{t}^2(\phi_j):=\sqrt{2}{\rm res}_{\infty^0}\sqrt{\lambda(P)}\phi_j(P);\quad \mathbf{t}^3(\phi_j):=\oint_a\lambda(P)\phi_j(P).
\end{align*}

\begin{Prop}\label{Flat m-c}
\textbf{\emph{1)}} The flat metric induced by the normalized holomorphic differential $\phi_1$ is given by
$$
\eta(\phi_1)=\frac{(d\lambda_1)^2}{8\omega_1^2(\lambda_1-\lambda_2)(\lambda_1-\lambda_3)}+\frac{(d\lambda_2)^2}{8\omega_1^2(\lambda_2-\lambda_1)(\lambda_2-\lambda_3)}+
\frac{(d\lambda_3)^2}{8\omega_1^2(\lambda_3-\lambda_1)(\lambda_3-\lambda_2)}.
$$
The flat coordinates of $\eta(\phi_1)$ are
\begin{equation}\label{phi1-coord}
\begin{split}
&t_1:=\mathbf{t}^1(\phi_1)=\frac{\omega_2}{2{\rm{i}}\pi\omega_1}=\frac{\tau}{2{\rm{i}}\pi};\\
&t_2:=\mathbf{t}^2(\phi_1)=\frac{1}{\sqrt{2}\omega_1};\\
&t_3:=\mathbf{t}^3(\phi_1)=-\frac{\zeta(\omega_1)}{\omega_1}+c.
\end{split}
\end{equation}
\textbf{\emph{2)}} The flat metric  $\eta(\phi_2)$ is given by
$$
\eta(\phi_2)=\frac{(e_1\omega_1+\zeta(\omega_1))^2}{\omega_1^2(\lambda_1-\lambda_2)(\lambda_1-\lambda_3)}(d\lambda_1)^2
+\frac{(e_2\omega_1+\zeta(\omega_1))^2}{\omega_1^2(\lambda_2-\lambda_1)(\lambda_2-\lambda_3)}(d\lambda_2)^2+
\frac{(e_3\omega_1+\zeta(\omega_1))^2}{\omega_1^2(\lambda_3-\lambda_1)(\lambda_3-\lambda_2)}(d\lambda_3)^2.
$$
The flat coordinates of $\eta(\phi_2)$ are
\begin{equation}\label{phi2-coord}
\begin{split}
&x_1:=\mathbf{t}^1(\phi_2)=t_2;\\
&x_2:=\mathbf{t}^2(\phi_2)=2\frac{\zeta(\omega_1)}{\omega_1}+c;\\
&x_3:=\mathbf{t}^3(\phi_2)=\frac{\sqrt{2}}{6}g_2\omega_1-2\sqrt{2}\frac{\zeta^2(\omega_1)}{\omega_1}.
\end{split}
\end{equation}
\textbf{\emph{3)}} The flat metric $\eta(\phi_3)$ is explicitly given by
\begin{align*}
\eta(\phi_3)&=\frac{\big(\mathcal{V}'(\omega_1)\big)^2}{2(\lambda_1-\lambda_2)(\lambda_1-\lambda_3)}(d\lambda_1)^2
+\frac{\big(\mathcal{V}'(\omega_2)\big)^2}{2(\lambda_2-\lambda_1)(\lambda_2-\lambda_3)}(d\lambda_2)^2
+\frac{\big(\mathcal{V}'(\omega_3)\big)^2}{2(\lambda_3-\lambda_1)(\lambda_3-\lambda_2)}(d\lambda_3)^2,
\end{align*}
with $\mathcal{V}$ being the function (\ref{V-j}).
The corresponding flat coordinates  are
\begin{equation}\label{phi3-coord}
\begin{split}
&y_1:=\mathbf{t}^1(\phi_3)=c-\frac{\zeta(\omega_1)}{\omega_1}-\lambda(P_0);\\
&y_2:=\mathbf{t}^2(\phi_3)=\frac{\sqrt{2}}{6}g_2\omega_1-2\sqrt{2}\frac{\zeta^2(\omega_1)}{\omega_1}=x_3;\\
&y_3:=\mathbf{t}^3(\phi_3)=g_3\omega_1^2-g_2\omega_1\zeta(\omega_1)+4\frac{\zeta^3(\omega_1)}{\omega_1},
\end{split}
\end{equation}
where $P_0$ is an arbitrary  starting point of the contour $b$.
\end{Prop}

\noindent\emph{Proof:} \textbf{1)}
Since the standard local parameter near the simple ramification point $P_1$ is $x_1(P)=\sqrt{\lambda(P)-\lambda_1}$,  it follows that $d\lambda(P)=2x_1(P)dx_1(P)$. This, (\ref{phi1}) and (\ref{phi1-lambda}) yield that
$$
\phi_1(P)\underset{P\sim P_1}{=}-\frac{dx_1(P)}{2\omega_1\sqrt{(\lambda(P)-\lambda_2)(\lambda(P)-\lambda_3)}}.
$$
Therefore
$$
\phi_1(P_1):=\frac{\phi_1(P)}{dx_1(P)}\Big|_{P=P_1}=-\frac{1}{2\omega_1\sqrt{(\lambda_1-\lambda_2)(\lambda_1-\lambda_3)}}.
$$
In the same way we get $\phi_1(P_2)$ and $\phi_1(P_3)$.\\
Using  (\ref{phi1}), (\ref{phi1-behavior}) and (\ref{zeta-derivative}) we obtain
\begin{align*}
&t_1=\frac{1}{2{\rm{i}}\pi}\int_{x}^{x+2\omega_2}\frac{dz_P}{2\omega_1}=\frac{\omega_2}{2{\rm{i}}\pi\omega_1};\\
&t_2=\sqrt{2}{\rm res}_{0}\frac{1}{z_{\infty}(P)}\Big(1+\frac{3c}{2}z^2_{\infty}(P)+\dots\Big)\frac{dz_{\infty}(P)}{2\omega_1}=\frac{1}{\sqrt{2}\omega_1};\\
&t_3=\frac{1}{2\omega_1}\int_x^{x+2\omega_1}\Big(\wp(z_P)+c\Big)dz_p=-\frac{\zeta(\omega_1)}{\omega_1}+c.
\end{align*}
\textbf{2)} From (\ref{phi1}) and (\ref{phi2}), we observe that $\phi_2$ can be expressed in terms of the normalized holomorphic differential $\phi_1$ as follows
$$
\phi_2(P)=2\sqrt{2}\big(\omega_1\wp(z_P)+\zeta(\omega_1)\big)\phi_1(P).
$$
Accordingly for $k=1,2,3$ we have
$$
\phi_2(P_k)=2\sqrt{2}\big(\omega_1\wp(\omega_k)+\zeta(\omega_1)\big)\phi_1(P_k)=2\sqrt{2}\big(\omega_1e_k+\zeta(\omega_1)\big)\phi_1(P_k)
$$
which gives the stated formula for the metric $\eta(\phi_2)$.\\
Furthermore, by (\ref{phi2}), (\ref{zeta-derivative}), (\ref{phi1-behavior}) and Legendre's relation (\ref{Legendre}) we get
\begin{align*}
x_1&=\frac{\sqrt{2}}{2{\rm{i}}\pi}\int_x^{x+2\omega_2}\Big(\wp(z_P)+\frac{\zeta(\omega_1)}{\omega_1}\Big)dz_P
=\frac{\sqrt{2}}{2{\rm{i}}\pi}\Big(-2\zeta(\omega_2)+2\frac{\omega_2}{\omega_1}\zeta(\omega_1)\Big)=\frac{1}{\sqrt{2}\omega_1};\\
x_2&=2\underset{\infty^0}{{\rm res}}\bigg(\sqrt{\lambda(P)}\Big(\wp(z_P)+\frac{\zeta(\omega_1)}{\omega_1}\Big)dz_P\bigg)
=2\underset{\infty^0}{{\rm res}}\bigg(\sqrt{\lambda(P)}\Big(\lambda(P)-c+\frac{\zeta(\omega_1)}{\omega_1}\Big)dz_P\bigg)\\
&=2\underset{0}{{\rm res}}\bigg(\frac{1}{z_{\infty}(P)}\Big(\frac{1}{z^2_{\infty}(P)}-c+\frac{\zeta(\omega_1)}{\omega_1}\Big)
\Big(1+\frac{3c}{2}z^2_{\infty}(P)+o\left(z^2_{\infty}(P)\right)\Big)dz_{\infty}(P)\Big)\bigg)\\
&=c+2\frac{\zeta(\omega_1)}{\omega_1};\\
x_3&:=\oint_a\lambda(P)\phi_2(P)=\oint_a\big(\lambda(P)-c\big)\phi_2(P)
=\sqrt{2}\int_x^{x+2\omega_1}\Big(\wp^2(z_P)+\frac{\zeta(\omega_1)}{\omega_1}\wp(z_P)\Big)dz_p\\
&=-2\sqrt{2}\frac{\zeta^2(\omega_1)}{\omega_1}+\sqrt{2}\int_x^{x+2\omega_1}\Big(\frac{1}{6}\wp''(z_P)+\frac{g_2}{12}\Big)dz_P
=-2\sqrt{2}\frac{\zeta^2(\omega_1)}{\omega_1}+\frac{\sqrt{2}}{6}g_2\omega_1.
\end{align*}
\textbf{3)} Because of (\ref{V-j-der}) and (\ref{phi1}), we can write  $\phi_3(P)=2\omega_1\mathcal{V}'(z_P)\phi_1(P)$. This implies that $\phi_3(P_j)=2\omega_1\mathcal{V}(\omega_j)\phi(P_j)$ and then we  arrive at the desired expression of the metric $\eta(\phi_3)$. On the other hand, we have
\begin{align*}
y_1&:=\frac{1}{2{\rm{i}}\pi}\oint_b\phi_3(P)=\frac{2}{{\rm{i}}\pi}\zeta(\omega_1)\Big(\zeta(\omega_2)-\frac{\omega_2}{\omega_1}\zeta(\omega_1)\Big)-\wp(x)
=-\frac{\zeta(\omega_1)}{\omega_1}-\wp(x);\\
y_2&:=\sqrt{2}{\rm res}_{\infty^0}\sqrt{\lambda(P)}\phi_3(P)=x_3;\\
y_3&:=\oint_a\lambda(P)\phi_3(P)=\oint_a\big(\lambda(P)-c\big)\phi_3(P)\\
&=\int_x^{x+2\omega_1}\bigg(4\omega_1\wp^3(z_P)-4\zeta(\omega_1)\wp^2(z_P)+2\wp'(z_P)\wp(z_P)\big(\omega_1\zeta(z_P)-\zeta(\omega_1)z_P\big)\bigg)dz_P
+g_2\omega_1\zeta(\omega_1)+4\frac{\zeta^3(\omega_1)}{\omega_1}\\
&=5\omega_1\int_x^{x+2\omega_1}\wp^3(z_P)dz_P-3\zeta(\omega_1)\int_x^{x+2\omega_1}\wp^2(z_P)dz_P+g_2\omega_1\zeta(\omega_1)+4\frac{\zeta^3(\omega_1)}{\omega_1}\\
&=g_3\omega_1^2-g_2\omega_1\zeta(\omega_1)+4\frac{\zeta^3(\omega_1)}{\omega_1}.
\end{align*}

\fd
\begin{Remark} Observe that  if we assume that the contours $a$ and $b$ start at a point $P_0$ such that $\lambda(P_0)=0$, then we  have
$$
y_1=\mathbf{t}^1(\phi_3)=c-\zeta(\omega_1)/{\omega_1}=\mathbf{t}^3(\phi_1)=t_3.
$$
As a consequence, the following matrix is symmetric:
\begin{equation}\label{Hessian dual g=1}
H:=(H_{AB})=\big(\mathbf{t}^B(\phi_A)\big)
=\begin{pmatrix}
t_1 & t_2 & t_3 \\
x_1 & x_2 & x_3 \\
y_1 & y_2 & y_3
\end{pmatrix}
\end{equation}
In the next results of this subsection, this additional assumption is considered to be fulfilled.
\end{Remark}

Note that, according to (\ref{Entries eta}), the common constant matrix of the dual metrics $\eta^*(\phi_j)$ w.r.t. the flat coordinates
$\big\{\mathbf{t}^1(\phi_j),\mathbf{t}^2(\phi_j), \mathbf{t}^3(\phi_j)\big\}$ is anti-diagonal:
$$
\big(\eta^{\alpha\beta}\big):=
\begin{pmatrix}
0 & 0 & 1 \\
0 & 1 & 0 \\
1 & 0 & 0
\end{pmatrix},
\quad \quad\quad  \eta^{\alpha\beta}=\eta^*(\phi_j)\big(d\mathbf{t}^{\alpha}(\phi_j),d\mathbf{t}^{\beta}(\phi_j)\big), \quad \alpha,\beta=1,2,3.
$$
In particular, this implies the following duality relations between flat coordinates of the metric $\eta(\phi_j)$:
\begin{equation}\label{duality1-g=1}
\mathbf{t}^{A^{\prime}}(\phi_j)=\mathbf{t}^{4-A}(\phi_j),\quad A=1,2,3.
\end{equation}
Now,  we use these relations and apply formula (\ref{Prep-eta4}) (which coincides with formulas (\ref{Prep-eta}) and (\ref{Prep-eta2}) in this case), we deduce that  the prepotential associated with the semi-simple  Hurwitz-Frobenius manifold induced by the differential $\phi_j$ is given by:
\begin{equation}\label{Prep Ex1}
\begin{split}
\mathbf{F}_{\phi_j}&=\frac{1}{2(1+d_j)}\sum_{A,B=1}^3\frac{\big(d_j+d_{A^{\prime}}\big)\big(d_j+d_{B^{\prime}}\big)}
{1+d_j+d_A}\mathbf{t}^{A^{\prime}}(\phi_j)\mathbf{t}^{B^{\prime}}(\phi_j)H_{AB}\\
&=\frac{1}{2(1+d_j)}\sum_{A,B=1}^3\frac{\big(d_j+d_{4-A}\big)\big(d_j+d_{4-B}\big)}
{1+d_j+d_A}\mathbf{t}^{4-A}(\phi_j)\mathbf{t}^{4-B}(\phi_j)H_{AB}\\
&=\frac{1}{2}\mathbf{t}^{3}(\phi_j)^2H_{11}+\frac{(d_j+1/2)^2}{2(d_j+1)(d_j+3/2)}\mathbf{t}^2(\phi_j)^2H_{22}
+\frac{d_j^2}{2(d_j+1)(d_j+2)}\mathbf{t}^1(\phi_j)^2H_{33}\\
&\quad+\frac{(d_j+1/2)(2d_j+5/2)}{2(d_j+1)(d_j+3/2)}\mathbf{t}^{2}(\phi_j)\mathbf{t}^{3}(\phi_j)H_{12}
+\frac{d_j(2d_j+3)}{2(d_j+1)(d_j+2)}\mathbf{t}^1(\phi_j)\mathbf{t}^3(\phi_j)H_{13}\\
&\quad+\frac{d_j(d_j+1/2)(2d_j+7/2)}{2(d_j+1)(d_j+2)(d_j+3/2)}\mathbf{t}^{1}(\phi_j)\mathbf{t}^{2}(\phi_j)H_{23},
\end{split}
\end{equation}
where $d_j$ is the quasi-homogeneous  degree of the differential $\phi_j$ and $(H_{AB})$ is the  symmetric matrix (\ref{Hessian dual g=1}).
In Theorems \ref{Prep-Chazy} and \ref{phi23}, we consider differentials $\phi_1$, $\phi_2$ and $\phi_3$ one by one and rewrite (\ref{Prep Ex1}) explicitly in each case.

\subsubsection{Prepotential associated with the Chazy equation}
\begin{Thm}\label{Prep-Chazy} The semi-simple Hurwitz-Frobenius manifold induced by the normalized holomorphic  differential $\phi_1$ is described by the Darboux-Egoroff metric $\eta(\phi_1)$ and its flat coordinates $t_1,t_2,t_3$  (\ref{phi1-coord}), the unit vector field $e_{\phi_1}=\partial_{t_3}$, the Euler vector field $\mathcal{E}_{\phi_1}=\frac{t_2}{2}\partial_{t_2}+t_3\partial_{t_3}$ and the prepotential
\begin{equation}\label{Prep-phi1-g1}
\mathbf{F}_{\phi_1}(t_1,t_2,t_3)=\frac{t_3^2}{2}t_1+\frac{t_3}{2}t_2^2+\frac{\pi^2}{24}t_2^4E_2(2{\rm{i}}\pi t_1),
\end{equation}
where $E_2$ is the Eisenstein series defined by
\begin{equation}\label{Eisenstein2}
E_2(\tau)=P(q):=1-24\sum_{n=1}^{\infty}\frac{nq^n}{1-q^n},\quad \quad q=e^{2{\rm{i}}\pi\tau},\quad\quad \Im(\tau)>0.
\end{equation}
Moreover, the function $\mathbf{F}_{\phi_1}$ is quasi-homogeneous of degree $2$: $\mathcal{E}_{\phi_1}.\mathbf{F}_{\phi_1}=2\mathbf{F}_{\phi_1}$.
\end{Thm}

\begin{Remark}
If we make the change of variables
$$
(t_1,t_2,t_3)\longleftrightarrow (u_1,u_2,u_3):=\big(\frac{t_3}{2{\rm{i}}\pi},t_2,2{\rm{i}}\pi{t_1}\big),
$$
then
$$
\mathbf{F}_{\phi_1}(t_1,t_2,t_3)=2{\rm{i}}\pi\bigg(\frac{(u_1)^2}{2}u_3+\frac{u_1}{2}(u_2)^2-\frac{\rm{i}\pi}{48}(u_2)^4E_2(u_3)\bigg)
=2{\rm{i}}\pi\mathrm{F}(u_1,u_2,u_3),
$$
where $\mathrm{F}(u_1,u_2,u_3)$ is the solution to the WDVV equations obtained by Dubrovin (see \cite{Dubrovin2D}, formula (C.87)).
\end{Remark}

\noindent\emph{Proof:} Applying  (\ref{e-tA}) and taking into account  duality relations (\ref{duality1-g=1}), we see that the unit vector field $e=\sum_{j=1}^3\partial_{\lambda_j}$ is represented by
$$
e=\partial_{\mathbf{t}^{1^{\prime}}(\phi_1)}=\partial_{\mathbf{t}^3(\phi_1)}=\partial_{t_3}.
$$
Moreover, the expression of the Euler vector field in flat coordinates of $t_1,t_2,t_3$ follows directly from the relation
$$
E.\mathbf{t}^i(\phi_j)=(d_i+d_j)\mathbf{t}^i(\phi_j),\quad d_i=deg(\phi_j).
$$
Now, since the holomorphic differential $\phi_1$ is of degree $d_1=0$, formula (\ref{Prep Ex1}) implies that the prepotential $\mathbf{F}_{\phi_1}$ reduces to
\begin{align*}
\mathbf{F}_{\phi_1}&=\frac{t_3^2}{2}H_{11}+\frac{1}{12}t_2^2H_{22}+\frac{5}{12}t_2t_3H_{12}\\
&=\frac{t_3^2}{2}t_1+\frac{1}{12}t_2^2x_2+\frac{5}{12}t_2^2t_3.
\end{align*}
In addition, by (\ref{phi1-coord}), (\ref{phi2-coord}) and the homogeneity property of the $\zeta$-function  we have
$$
x_2=t_3+3\frac{\zeta(\omega_1;2\omega_1,2\omega_2)}{\omega_1}=t_3+3\frac{\zeta(1/2|\tau)}{2\omega_1^2}=t_3+3t_2^2\zeta(1/2|\tau).
$$
On the other hand, it is known that the function $\zeta(1/2|\tau)$ (sometimes called Weierstrass eta function) satisfies
\begin{equation}\label{zeta-E2}
2\zeta(1/2|\tau)=\zeta(z+1|\tau)-\zeta(z|\tau)=\frac{\pi^2}{3}E_2(\tau),
\end{equation}
where $\zeta(z|\tau)$ is the zeta function associated with the lattice $\Z+\tau\Z$ (see for instance \cite{Akhiezer}-Section 20 and \cite{Dubrovin2D}).
Substituting this into the expression of the function $x_2$ we arrive at
\begin{equation}\label{x2}
x_2=t_3+\frac{\pi^2}{2}t_2^2E_2(\tau)=t_3+\frac{\pi^2}{2}t_2^2E_2({2{\rm i}\pi}t_1)
\end{equation}
and this shows formula (\ref{Prep-phi1-g1}).

\fd

The following result appeared in \cite{Ablowitz} and (\cite{Dubrovin2D}-Appendix C). It allows to view the 3-dimensional semi-simple Frobenius manifold induced by the holomorphic differential $\phi_1$ as a geometric framework of the Chazy equation (\ref{E2-Chazy}).
\begin{Cor} The Eisenstein series $E_2$ satisfies the Chazy equation:
\begin{equation}\label{E2-Chazy}
E_2'''-2{\rm i}\pi E_2E_2''+3{\rm i}\pi\big(E_2'\big)^2=0.
\end{equation}
\end{Cor}
\emph{Proof:}  By the above theorem and Dubrovin's theory of Frobenius manifolds, we know that the prepotential $\mathbf{F}_{\phi_1}$ obeys the WDVV equations:
\begin{align*}
F_1F_3F_2=F_2F_3F_1,
\end{align*}
where $F_j$ is the Hessian matrix of the $\partial_{t_j}\mathbf{F}_{\phi_1}$ and hence $F_3=F_3^{-1}=(\eta^{\alpha\beta})$ is the anti-diagonal constant matrix of the metric $\eta(\phi_1)$.\\
By calculating the left and right hand sides of this equation, we see that the associativity condition is equivalent to:
$$
\partial_{t_1}^3\mathbf{F}_{\phi_1}+\big(\partial_{t_1}^2\partial_{t_2}\mathbf{F}_{\phi_1}\big)\big(\partial_{t_2}^3\mathbf{F}_{\phi_1}\big)=
\big(\partial_{t_1}\partial_{t_2}^2\mathbf{F}_{\phi_1}\big)^2
$$
Substituting the function $\mathbf{F}_{\phi_1}$ (\ref{Prep-phi1-g1}) into this equation we conclude that $E_2$ solves the Chazy equation.

\fd

\subsubsection{Ramanujan identities}

In the next result, we show how Ramanujan's identities \cite{Ramanujan1916}, involving the Eisenstein series $E_2,E_4$ and $E_6$, can be easily derived from the formalism we have developed for prepotentials of Hurwitz-Frobenius manifolds, namely from our formula (\ref{Prep-eta}) in Theorem \ref{Prep-eta-thm} and the observation that all the Frobenius structures of the same family share the Hessian matrix (\ref{Prep-eta-H}).
Let $q=e^{2{\rm{i}}\pi\tau}$, with $\Im(\tau)>0$. The Eisenstein series $E_2$ is given by (\ref{Eisenstein2}) and the series $E_4$ and $E_6$ are respectively defined  by
\begin{equation}\label{E4-E6}
\begin{split}
&E_{4}(\tau)=Q(q):=1+240\sum_{n=1}^{\infty}\frac{n^3q^n}{1-q^n};\\
&E_{6}(\tau)=R(q):=1-504\sum_{n=1}^{\infty}\frac{n^5q^n}{1-q^n}.
\end{split}
\end{equation}
\begin{Prop} Let $x_3=\mathbf{t}^3(\phi_2)$ and $y_3=\mathbf{t}^3(\phi_3)$ be the functions in (\ref{phi2-coord}) and (\ref{phi3-coord}), respectively.
Then we have
\begin{equation}\label{Ramanujan1}
x_3=y_2=\frac{\rm{i}\pi^3}{3}t_2^3E_2'(\tau),\quad \quad\quad  y_3=-\frac{\pi^4}{6}t_2^4E_2''(\tau),\quad\quad  \text{with}\quad t_2=\frac{1}{\sqrt{2}\omega_1}.
\end{equation}
In particular, the following Ramanujan identities hold:
\begin{align}
&\label{R-E2}\frac{1}{2{\rm{i}}\pi}E_2'=\frac{1}{12}\big(E_2^2-E_4\big);\\
&\label{R-E4}\frac{1}{2{\rm{i}}\pi}E_4'=\frac{1}{3}\big(E_2E_4-E_6\big);\\
&\label{R-E6}\frac{1}{2{\rm{i}}\pi}E_6'=\frac{1}{2}\big(E_2E_6-E_4^2\big).
\end{align}

\end{Prop}
\emph{Proof:} Using (\ref{Prep-eta-H}), we deduce that the Hessian matrix of the function $\mathbf{F}_{\phi_1}$ is connected to  matrix $H$
(\ref{Hessian dual g=1}) by:
$$
\partial_{t_A}\partial_{t_B}\mathbf{F}_{\phi_1}=\mathbf{t}^{A^{\prime}}\big(\phi_{B^{\prime}}\big)=H_{A^{\prime}B^{\prime}}, \quad \text{for all $A,B$}.
$$
Note that this expression of the Hessian matrix follows from our formalism. On the other hand, we can derive it from expression (\ref{Prep-phi1-g1}) for prepotential $\mathbf{F}_{\phi_1}$ by differentiation.  Comparing these two ways produces the three Ramanujan identities. Namely, using the duality of flat coordinates from
(\ref{duality1-g=1}), we obtain
$$
x_3=H_{23}=\partial_{t_1}\partial_{t_2}\mathbf{F}_{\phi_1}=\frac{\rm{i}\pi^3}{3}t_2^3E_2'(2{\rm{i}}\pi{t_1})
=\frac{\rm{i}\pi^3}{3}t_2^3E_2'(\tau)
$$
and
$$
y_3=H_{33}=\partial_{t_1}\partial_{t_1}\mathbf{F}_{\phi_1}=-\frac{\pi^4}{6}t_2^4E_2''(2{\rm{i}}\pi{t_1})=-\frac{\pi^4}{6}t_2^4E_2''(\tau).
$$
On the other hand, in (\ref{phi2-coord}) and (\ref{phi3-coord}), we have already obtained an expression for the flat coordinates $x_3$ and $y_3$ by direct calculation in terms of the Weierstrass invariants $g_2$, $g_3$ and the zeta-function. Let us now recall that $g_2$ and $g_3$ can also be expressed via the Eisenstein  series $E_4$ and $E_6$, respectively. Specifically, we have (see for instance \cite{Apostol}, Theorem 1.18)
\begin{equation}\label{g2-g3-E46}
\begin{split}
&g_2(2\omega_1,2\omega_2)=\frac{1}{16\omega_1^4}g_2(1,\tau)=\frac{\pi^4}{12\omega_1^4}E_4(\tau);\\
&g_3(2\omega_1,2\omega_2)=\frac{1}{64\omega_1^6}g_3(1,\tau)=\frac{\pi^6}{216\omega_1^6}E_6(\tau).
\end{split}
\end{equation}
Now, equating the two expressions for $x_3$, in (\ref{phi2-coord}) and (\ref{Ramanujan1}) and using
 the first relation in (\ref{g2-g3-E46}) and formula (\ref{zeta-E2}), yields the first Ramanujan differential equation (\ref{R-E2}):
\begin{align*}
E_2'(\tau)&=\frac{6\sqrt{2}\omega_1^3}{\rm{i}\pi^3}x_3
=\frac{6\sqrt{2}\omega_1^3}{\rm{i}\pi^3}\Big(\frac{\sqrt{2}}{6}\omega_1g_2(2\omega_1; 2\omega_2)-2\sqrt{2}\frac{\zeta^2(\omega_1; 2\omega_1,2\omega_2)}{\omega_1}\Big)\\
&=\frac{6\sqrt{2}\omega_1^3}{\rm{i}\pi^3}\Big(\frac{\pi^4\sqrt{2}}{72\omega_1^3}E_4(\tau)-\frac{2\sqrt{2}}{4\omega_1^3}\zeta^2(1/2|\tau)\Big)\\
&=\frac{\pi}{6\rm{i}}E_4(\tau)-\frac{\pi}{6\rm{i}}E_2^2(\tau).
\end{align*}
The second Ramanujan identity (\ref{R-E4}) is obtained in the same way by comparing two expressions for $y_3$. By virtue of (\ref{phi3-coord}) and (\ref{zeta-E2}) and (\ref{g2-g3-E46}), we have
\begin{align*}
y_3&=\omega_1^2g_3(2\omega_1,2\omega_2)-\omega_1g_2(2\omega_1,2\omega_2)\zeta(\omega_1;2\omega_1,2\omega_2)+\frac{4}{\omega_1}\zeta^3(\omega_1;2\omega_1,2\omega_2)\\
&=\frac{\pi^6}{216\omega_1^4}E_6(\tau)-\frac{\pi^4}{24\omega_1^4}E_4(\tau)\zeta(1/2|\tau)+\frac{1}{2\omega_1^4}\zeta^3(1/2|\tau)\\
&=\frac{\pi^6}{216\omega_1^4}E_6(\tau)-\frac{\pi^6}{144\omega_1^4}E_2(\tau)E_4(\tau)+\frac{\pi^6}{432\omega_1^4}E_2^3(\tau).
\end{align*}
This and (\ref{Ramanujan1}) yield that
\begin{equation}\label{Ramanujan3}
E_2''=-\frac{24\omega_1^4}{\pi^4}y_3=-\frac{\pi^2}{9}E_6+\frac{\pi^2}{6}E_2E_4-\frac{\pi^2}{18}E_2^3.
\end{equation}
On the other hand, if we use the first Ramanujan identity (\ref{R-E2}), we also get
$$
E_2''=\frac{\pi}{6i}E_4'+\frac{\pi^2}{18}E_4E_2-\frac{\pi^2}{18}E_2^3.
$$
Thus, by comparing this with (\ref{Ramanujan3}) we deduce that (\ref{R-E4}) holds true.\\
It remains to prove the third identity (\ref{R-E4}).
Making use of the Chazy equation (\ref{E2-Chazy}), (\ref{Ramanujan3})  as well as the first and second Ramanujan relations (\ref{R-E2})-(\ref{R-E4}), we get
\begin{align*}
0&=E_2'''-2{\rm i}\pi E_2E_2''+3{\rm i}\pi\big(E_2'\big)^2\\
&=-\frac{2}{3}E_6'+E_2'E_4+E_2E_4'-E_2'E_2^2-2{\rm i}\pi E_2\Big(-\frac{2}{3}E_6+E_2E_4-\frac{1}{3}E_2^3\Big)-\frac{2{\rm i}\pi}{4}\big(E_2^4-2E_2^2E_4+E_4^2)\\
&=-\frac{2}{3}E_6'+\frac{2{\rm i}\pi}{3}E_2E_6-\frac{2{\rm i}\pi}{3}E_4^2.
\end{align*}
From this we arrive at (\ref{R-E6}).

\fd

\begin{Remark}  Assume that $2\omega_1=1$ and $2\omega_2=\tau$ and consider the Weierstrass functions $\wp=\wp(\cdot|\tau)$ and $\zeta=\zeta(\cdot|\tau)$ associated with the lattice $\Z+\tau\Z$.   Then from  explicit formulas (\ref{phi2}) and (\ref{phi3}) for the differentials $\phi_2$ and $\phi_3$, the two relations in (\ref{Ramanujan1}) (where we take $\omega_1=1/2$) and the fact that $x_3=\mathbf{t}^3(\phi_2)$ and $y_3=\mathbf{t^3}(\phi_3)$, we obtain  that  the Eisenstein series $E_2$ and its first and second derivatives can be viewed  as the $a$-periods of some specific differentials as follows:
\begin{align*}
&E_2=\frac{6}{\pi^2}\zeta(1/2|\tau)=-\frac{3}{\pi^2}\oint_a\wp(u|\tau)du;\\
&E_2'=\frac{3}{2\rm{i}\pi^3}\oint_{a}\bigg(\wp^2(u)+2\zeta(1/2)\wp(u)\bigg)du;\\
&E_2''=-\frac{3}{2\pi^4}\oint_{a}\bigg(2\wp^2(u)-4\zeta(1/2)\wp(u)+\wp'(u)\big(\zeta(u)-2\zeta(1/2)u\big)-\frac{g_2}{4}-4\zeta^2(1/2)\bigg)\wp(u)du.
\end{align*}
\end{Remark}

\medskip

As an application, we shall employ  the first Ramanujan differential equation (\ref{R-E2}) and a special case of  relation (\ref{derivatives lambda-j})
to derive a nonlinear differential equation satisfied  by the three functions
$$
e_1(\tau):=\wp(1/2|\tau),\quad e_2(\tau):=\wp(\tau/2|\tau),\quad e_3(\tau):=\wp((1+\tau)/2|\tau)
$$
with $\wp(u|\tau):=\wp(u;1,\tau)$. This  differential equation is a particular case of the rule for the derivative of the $\wp$-function with respect to the modulus $\tau$, recently proved  in \cite{Brezhnev} (Section 10).\\
Recall that (\ref{derivatives lambda-j})  gives a way to calculate   derivatives of the critical values of the Weierstrass covering (\ref{Weierstrass1}):
$$
\lambda_j=\wp(\omega_j; 2\omega_1,2\omega_2)+c=e_j(2\omega_1,2\omega_2)+c,\quad \quad j=1,2,3,
$$
with respect to flat coordinates $t_1, t_2,t_3$ (\ref{phi1-coord}) of the metric $\eta(\phi_1)$.
\begin{Cor} For $j=1,2,3$, the function $e_j(\tau)$ satisfies the equation:
$$
\frac{1}{2{\rm{i}}\pi}e_j'(\tau)=-\frac{1}{2\pi^2}e_j^2(\tau)+\frac{1}{6}E_2e_j(\tau)+\frac{\pi^2}{9}E_4.
$$

\end{Cor}
\emph{Proof:} Using (\ref{derivatives lambda-j}) and relation $\tau=2\rm{i}\pi \mathbf{t}^1(\phi_1)=2\rm{i}\pi t_1$ from (\ref{phi1-coord}), we have
\begin{equation}\label{pde-ej}
\frac{1}{2{\rm{i}}\pi}\partial_{\tau}\lambda_j=-\frac{1}{4\pi^2}\frac{\partial\lambda_j}{\partial \mathbf{t}^1(\phi_1)}=-\frac{1}{4\pi^2}\frac{\phi_3(P_j)}{\phi_1(P_j)}.
\end{equation}
Due to relations (\ref{phi1-coord}), (\ref{zeta-E2}), $e_j(\tau)=(2\omega_1)^2e_j$ and the first Ramanujan identity (\ref{R-E2}), the left hand side of (\ref{pde-ej}) can be written as
\begin{align*}
\frac{1}{2{\rm{i}}\pi}\partial_{\tau}\lambda_j&=\frac{1}{2{\rm{i}}\pi}\partial_{\tau}\Big(\frac{1}{4\omega_1^2}e_j(\tau)+c\Big)
=\frac{1}{4\omega_1^2}\frac{e_j'(\tau)}{2{\rm{i}}\pi}+\frac{1}{2{\rm{i}}\pi}\partial_{\tau}\Big(t_3+\frac{1}{2\omega_1^2}\zeta(1/2|\tau)\Big)\\
&=\frac{1}{4\omega_1^2}\Big(\frac{1}{2{\rm{i}}\pi}e_j'(\tau)+\frac{\pi^2}{3}\frac{1}{2{\rm{i}}\pi}E'_2(\tau)\Big)\\
&=\frac{1}{4\omega_1^2}\Big(\frac{1}{2{\rm{i}}\pi}e_j'(\tau)+\frac{\pi^2}{36}E_2^2-\frac{\pi^2}{36}E_4\Big).
\end{align*}
Now, using explicit form  (\ref{phi3}) of the differential $\phi_3$ and expression (\ref{g2-g3-E46}) for $g_2$ in terms of $E_4$, we deduce that the right
 hand side of (\ref{pde-ej}) is
\begin{align*}
-\frac{1}{4\pi^2}\frac{\phi_3(P_j)}{\phi_1(P_j)}
&=-\frac{1}{4\pi^2}\Big(8\omega_1^2e_j^2(2\omega_1,2\omega_2)-8\omega_1\zeta(\omega_1)e_j(2\omega_1,2\omega_2)-g_2\omega_1^2-4\zeta^2(\omega_1)\Big)\\
&=-\frac{1}{4\omega_1^2}\Big(\frac{1}{2\pi^2}e_j^2-\frac{1}{6}E_2e_j-\frac{\pi^2}{12}E_4-\frac{\pi^2}{36}E_2^2\Big).
\end{align*}
This shows the desired differential equations.

\fd

\begin{Remark}
A link between the Ramanujan identities and the Gromov-Witten correlation functions was recently studied in \cite{Shen-Zhou}.
\end{Remark}
\subsubsection{Further examples in dimension 3}
In the next theorem we deal with prepotentials associated the differentials $\phi_2$ (\ref{phi2}) and $\phi_3$ (\ref{phi3}).
\begin{Thm}\label{phi23} The prepotentials corresponding to  the semi-simple Hurwitz-Frobenius manifold induced by the differential $\phi_2$ and $\phi_3$ are respectively given by
\begin{align*}
\mathbf{F}_{\phi_2}
&=\frac{x_2^3}{6}+x_1x_2x_3+\frac{x_3^2}{4{\rm{i}}\pi}(E_2')^{-1}\Big(\frac{3x_3}{{\rm{i}}\pi^3x_1^3}\Big)
-\frac{2}{15}x_1^3x_3E_2\Big((E_2')^{-1}\Big(\frac{3x_3}{{\rm{i}}\pi^3x_1^3}\Big)\Big)
-\frac{\pi^4}{180}x_1^6E_2''\Big((E_2')^{-1}\Big(\frac{3x_3}{{\rm{i}}\pi^3x_1^3}\Big)\Big);\\
\mathbf{F}_{\phi_3}&=\frac{y_1}{2}y_2^2+\frac{y_1^2}{2}y_3+\frac{y_3^2}{4{\rm i}\pi}\chi^{-1}\Big(-\frac{8}{3}y_3^3y_2^{-4}\Big)
+\frac{27}{40\pi}y_2y_3\big(-6y_3\big)^{1/4}\Big(E_2''\Big[\chi^{-1}\Big(-\frac{8}{3}y_3^3y_2^{-4}\Big)\Big]\Big)^{-1/4}\\
&\quad +\frac{9}{80\pi^2}y_2^2\big(-6y_3\big)^{1/2}\Big(E_2''\Big[\chi^{-1}\Big(-\frac{8}{3}y_3^3y_2^{-4}\Big)\Big]\Big)^{-1/2}
E_2\Big(\chi^{-1}\Big(-\frac{8}{3}y_3^3y_2^{-4}\Big)\Big),
\end{align*}
where $(E_2')^{-1}$ denotes the inverse function  of the Eisenstein series (\ref{Eisenstein2}) and $\chi^{-1}$ is the inverse function of
$$
\chi(\tau):=\frac{\big(E_2''(\tau)\big)^3}{\big(E_2'(\tau)\big)^4}.
$$
The functions $\mathbf{F}_{\phi_2}$ and $\mathbf{F}_{\phi_3}$ are respectively quasi-homogeneous with respect to the Euler vector fields:
\begin{align*}
E_{\phi_2}&=\frac{x_1}{2}\partial_{x_1}+x_2\partial_{x_2}+\frac{3}{2}x_3\partial_{x_3};\\
E_{\phi_3}&=y_1\partial_{y_1}+\frac{3}{2}y_2\partial_{y_2}+2y_3\partial_{y_3},
\end{align*}
with  $E_{\phi_2}.\mathbf{F}_{\phi_2}=3\mathbf{F}_{\phi_2}$ and $E_{\phi_3}.\mathbf{F}_{\phi_3}=4\mathbf{F}_{\phi_3}$.

\end{Thm}

\noindent\emph{Proof}: In the two cases,  using (\ref{Entries-g-eta}) we obtain the precise action of the Euler vector field.
Let us now prove the stated formulas for prepotentials $\mathbf{F}_{\phi_2}$ and $\mathbf{F}_{\phi_3}$.\\
\noindent\textbf{1)} When $\phi_j=\phi_2$ we have $d_2=1/2$ and thus by (\ref{Prep Ex1}) we have
\begin{align*}
\mathbf{F}_{\phi_2}&=\frac{x_3^2}{2}H_{11}+\frac{x_2^2}{6}H_{22}+\frac{x_1^2}{30}H_{33}+\frac{7}{12}x_2x_3H_{12}+\frac{4}{15}x_1x_3H_{13}+\frac{3}{20}x_1x_2H_{23}\\
&=\frac{x_3^2}{2}t_1+\frac{x_2^3}{6}+\frac{x_1^2}{30}y_3+\frac{7}{12}x_1x_2x_3+\frac{4}{15}x_1x_3t_3+\frac{3}{20}x_1x_2x_3\\
&=\frac{x_3^2}{4{\rm{i}}\pi}\tau+\frac{x_2^3}{6}+\frac{x_1^2}{30}y_3+\frac{11}{15}x_1x_2x_3+\frac{4}{15}x_1x_3t_3.
\end{align*}
We need to express the coordinates  $\tau$, $t_3$ and $y_3$ in terms of $x_1,x_2,x_3$.  Recall from (\ref{phi2-coord}) that $x_1=1/{\sqrt{2}\omega_1}$ and then relation (\ref{Ramanujan1}) implies that
$$
x_3= \frac{\rm{i}\pi^3}{6\sqrt{2}\omega_1^3}E_2'(\tau)=\frac{{\rm{i}}\pi^3}{3}x_1^3E_2'(\tau).
$$
The function $E_2''$ cannot vanish identically and then there is an open subset of $U\subset \mathbb{H}$ (where $\mathbb{H}$ is the upper half plane) such that  $E'_2$ is a biholomorphic function on $U$. This implies that
\begin{align*}
&\tau=(E_2')^{-1}\Big(\frac{3x_3}{{\rm{i}}\pi^3x_1^3}\Big);\\
&y_3=-\frac{\pi^4}{6}x_1^4E_2''\Big((E_2')^{-1}\Big(\frac{3x_3}{{\rm{i}}\pi^3x_1^3}\Big)\Big);\\
&t_3=x_2-\frac{x_1^2}{2}E_2(\tau)=x_2-\frac{x_1^2}{2}E_2\Big((E_2')^{-1}\Big(\frac{3x_3}{{\rm{i}}\pi^3x_1^3}\Big)\Big),
\end{align*}
which gives us the stated formula for $\mathbf{F}_{\phi_2}$.\\
\textbf{2)} Here we treat the case where  $\phi_j=\phi_3$.  Since $d_3=1$ and
$$
x_2=y_1+3\frac{\zeta(\omega_1)}{\omega_1}=y_1+\frac{t_2^2}{2}E_2(\tau),
$$
it follows that
\begin{align*}
\mathbf{F}_{\phi_3}&=\frac{y_3^2}{2}H_{11}+\frac{9}{40}y_2^2H_{22}+\frac{y_1^2}{12}H_{33}+\frac{27}{40}y_2y_3H_{12}+\frac{5}{12}y_1y_3H_{13}+\frac{11}{40}y_1y_2H_{23}\\
&=\frac{y_3^2}{2}t_1+\frac{9}{40}y_2^2x_2+\frac{y_1^2}{2}y_3+\frac{27}{40}y_2y_3t_2+\frac{11}{40}y_1y_2^2\\
&=\frac{y_1}{2}y_2^2+\frac{y_1^2}{2}y_3+\frac{y_3^2}{4{\rm i}\pi}\tau +\frac{27}{40}y_2y_3t_2+\frac{9}{80}y_2^2t_2^2E_2(\tau).
\end{align*}
By (\ref{Ramanujan1}) we can write
$$
\chi(\tau):=\frac{\big(E_2''(\tau)\big)^3}{\big(E_2'(\tau)\big)^4}=-\frac{8y_3^3}{3y_2^4}
$$
and in particular we have
$$
\chi'(\tau)\partial_{y_3}\tau=-8\frac{y_3^2}{y_2^4}.
$$
Hence $\chi$ is biholomorphic on some open subset of the upper half plane and then we can write
$$
\tau=\chi^{-1}\Big(-\frac{8}{3}y_3^3y_2^{-4}\Big).
$$
Moreover,  given that  $y_3=-\frac{\pi^4}{6}t_2^4E_2''(\tau)$, we get
$$
t_2=\frac{\big(-6y_3\big)^{1/4}}{\pi}\Big(E_2''\Big[\chi^{-1}\Big(-\frac{8}{3}y_3^3y_2^{-4}\Big)\Big]\Big)^{-1/4}.
$$
Accordingly we conclude that
\begin{align*}
\mathbf{F}_{\phi_3}
&=\frac{y_1}{2}y_2^2+\frac{y_1^2}{2}y_3+\frac{y_3^2}{4{\rm i}\pi}\tau +\frac{27}{40}y_2y_3t_2+\frac{9}{80}y_2^2t_2^2E_2(\tau)\\
&=\frac{y_1}{2}y_2^2+\frac{y_1^2}{2}y_3+\frac{y_3^2}{4{\rm i}\pi}\chi^{-1}\Big(-\frac{8}{3}y_3^3y_2^{-4}\Big)
+\frac{27}{40\pi}y_2y_3\big(-6y_3\big)^{1/4}\Big(E_2''\Big[\chi^{-1}\Big(-\frac{8}{3}y_3^3y_2^{-4}\Big)\Big]\Big)^{-1/4}\\
&\quad +\frac{9}{80\pi^2}y_2^2\big(-6y_3\big)^{1/2}\Big(E_2''\Big[\chi^{-1}\Big(-\frac{8}{3}y_3^3y_2^{-4}\Big)\Big]\Big)^{-1/2}
E_2\Big(\chi^{-1}\Big(-\frac{8}{3}y_3^3y_2^{-4}\Big)\Big)
\end{align*}
which is the stated formula.



\subsection{Examples of 3-dimensional deformed Hurwitz-Frobenius manifolds}
Let $q$ be a complex number such that $1+q\tau\neq 0$, with $\tau=\frac{\omega_2}{\omega_1}$. In this subsection we briefly describe the  ingredients of  the deformed Frobenius manifold structures on the three dimensional Hurwitz space of Weierstrass elliptic curves (\ref{Weierstrass1}).
\begin{description}
\item $\bullet$ \emph{Genus one $q$-bidifferential.}
\end{description}
 From (\ref{q-W-def}) and (\ref{W-g=1}) we deduce that the genus one  $q$-bidifferential is given by
\begin{equation}\label{q-W-g1}
W_q(P,Q)=\Big(\wp(z_P-z_Q)+\frac{\zeta(\omega_1)}{2\omega_1}-\frac{2{\rm i}\pi{q}}{4\omega_1^2(1+q\tau)}\Big)dz_Pdz_Q.
\end{equation}
\begin{description}
\item $\bullet$ \emph{Primary $q$-bidifferentials and flat coordinates of the corresponding flat metrics.}
\end{description}
Using the intimate connection between the Dubrovin primary differentials and their $\mathbf{q}$-deformations (\ref{q-phi-tA}) and (\ref{q-phi-xik}), we see that the $q$-analogue of the primary differentials (\ref{phi1}), (\ref{phi2}) and (\ref{phi3}) are explicitly determined by
\begin{equation}\label{q-primary}
\begin{split}
\phi_{q,1}(P)&:=\frac{1}{2{\rm{i}}\pi}\oint_bW_q(P,Q)=\frac{1}{1+q\tau}\phi_1(P);\\
\phi_{q,2}(P)&:=\sqrt{2}{\rm res}_{\infty^0}\sqrt{\lambda(Q)}W_q(P,Q)\\
&=\phi_2(P)-{2{\rm i}\pi}\frac{qt_2}{1+q\tau}\phi_1(P);\\
\phi_{q,3}(P)&:=\oint_a\lambda(Q)W_q(P,Q)+q\oint_b\lambda(Q)W_q(P,Q)\\
&=\phi_3(P)+q\phi_4(P)-{2{\rm i}\pi}q\frac{t_3+qs_1}{1+q\tau}\phi_1(P).
\end{split}
\end{equation}
Here $t_1,t_2,t_3$ are the functions (\ref{phi1-coord}), $s_1:=\frac{1}{2{\rm{i}}\pi}\oint_b\phi_4$ and $\phi_4$ denotes the following multivalued differential
\begin{equation}\label{phi4}
\begin{split}
&\phi_4(P)=\oint_b\lambda(Q)W(P,Q)=\mathcal{U}'(z_P)dz_P=2\omega_1\mathcal{U}'(z_P)\phi_1(P),\quad \text{with}\\
&\mathcal{U}(u):=\omega_2\wp'(u)+2\zeta(\omega_2)\Big(\zeta(u)-\frac{\zeta(\omega_1)}{\omega_1}u\Big)
+2\wp(u)\Big(\omega_2\zeta(u)-\zeta(\omega_2)u\Big)+ {\rm{i}}\pi\frac{c}{\omega_1}u.
\end{split}
\end{equation}
We obtain (\ref{phi4}) by arguing as in (\ref{phi3}) and (\ref{V-j-der}).\\
The three relations in  (\ref{q-primary}) and the precise values of the evaluations $\phi_{k}(P_j)$ that were computed in Proposition \ref{Flat m-c} allow us to determine  explicit formulas for  the Darboux-Egoroff metrics
$$
\eta(\phi_{q,k})=\frac{1}{2}\sum_{j=1}^3\big(\phi_{q,k}(P_j)\big)^2(d\lambda_j)^2,\quad \quad k=1,2,3
$$
induced by the $q$-differentials $\phi_{q,1}$, $\phi_{q,2}$ and $\phi_{q,3}$. Let us denote by $\mathbf{t}^j(\phi_{q,k})$ the flat coordinates (\ref{q-S-eta}) of the metric
$\eta(\phi_{q,k})$, where
\begin{align*}
&\mathbf{t}^1(\phi_{q,k}):=\frac{1}{2{\rm{i}}\pi}\oint_b\phi_{q,k};\\
&\mathbf{t}^2(\phi_{q,k}):=\sqrt{2}{\rm res}_{\infty^0}\sqrt{\lambda(P)}\phi_{q,k}(P);\\ &\mathbf{t}^3(\phi_{q,k}):=\oint_a\lambda(P)\phi_{q,k}(P)+q\oint_b\lambda(P)\phi_{q,k}(P).
\end{align*}

In order to determine explicit formulas for the functions $\big\{\mathbf{t}^k(\phi_{q,s}):\ k,s=1,2,3\big\}$, we need the following lemma.
\begin{Lemma}
Let $\phi_4$ be the multivalued differential (\ref{phi4}) and $\tau={2{\rm{i}}\pi}t_1, t_2,t_3$ be the coordinates  (\ref{phi1-coord}). We have
\begin{equation}\label{phi4-coord}
\begin{split}
s_1&:=\frac{1}{2{\rm{i}}\pi}\oint_b\phi_4=t_3\tau+{\rm{i}}\pi t_2^2;\\
s_2&:=\sqrt{2}\underset{\infty^0} {\rm res}\sqrt{\lambda(P)}\phi_4(P)
=\frac{\pi^4}{18}t_2^3\bigg({\tau} E_4(\tau)-{\tau}E_2^2(\tau)+\frac{12{\rm{i}}}{\pi}E_2(\tau)\bigg)+{2{\rm{i}}\pi}t_2t_3;\\
s_3&:=\oint_a\lambda(P)\phi_4(P)=\frac{\pi^6}{108}{\tau}t_2^4\Big(E_2^3(\tau)-3E_2(\tau)E_4(\tau)+2E_6(\tau)\Big)-\frac{{\rm i}\pi^5}{12}t_2^4
\Big(E_2^2(\tau)-E_4(\tau)\Big);\\
s_4&:=\oint_b\lambda(P)\phi_4(P)=2{{\rm{i}}\pi}t_3^2\tau -4\pi^2t_3t_2^2-\pi^4t_2^4 E_2+\frac{\pi^6}{54}t_2^4\tau^2\big(E_6-E_2E_4)\\
&\quad\hspace{2.3cm} +\frac{\pi^6}{108}t_2^4\tau^2E_2\big(E_2^2-E_4\big)-\frac{{\rm{i}}\pi^5}{6}t_2^4\tau (E_2^2-E_4).
\end{split}
\end{equation}
Here $E_2,E_4,E_6$ denote  the Eisenstein series defined in (\ref{Eisenstein2}) and (\ref{E4-E6}).
\end{Lemma}
\emph{Proof:} We obtain  $s_1$  using (\ref{phi4}), the Legendre identity (\ref{Legendre}), (\ref{phi1-coord}).  By a simple calculation based on  (\ref{phi4}) and (\ref{Weierstrass1}) and  properties of Weierstrass's functions, we arrive at
\begin{align*}
&s_2=\frac{\sqrt{2}}{6}g_2\tau\omega_1-2\sqrt{2}\frac{\zeta(\omega_1)}{\omega_1}\zeta(\omega_2)+\frac{2{\rm{i}}\pi}{\sqrt{2}\omega_1}c;\\
&s_3=g_3\omega_1\omega_2-g_2\omega_2\zeta(\omega_1)+\frac{{\rm{i}}\pi}{4}g_2+{{\rm{i}}\pi}c^2-2{\rm{i}}\pi c\frac{\zeta(\omega_1)}{\omega_1}
+4\frac{\zeta^2(\omega_1)}{\omega_1}\zeta(\omega_2);\\
&s_4=2{{\rm{i}}\pi}cs_1+g_3\omega_2^2-g_2\omega_2\zeta(\omega_2)-2{\rm{i}}\pi\frac{c}{\omega_1}\zeta(\omega_2)
+4\frac{\zeta(\omega_1)}{\omega_1}\zeta^2(\omega_2).
\end{align*}
Thus, the desired  expressions for $s_2$, $s_3$ and $s_4$ in terms of $\tau,t_2,t_3$ follow from the Legendre identity (\ref{Legendre}), (\ref{g2-g3-E46}), (\ref{zeta-E2}) and the following fact that
$$
c=t_3+\zeta(\omega_1)/\omega_1=t_3+\frac{\pi^2}{6}t_2^2E_2(\tau)
$$
which is an immediate consequence of (\ref{phi1-coord}).
\fd

\medskip

In the next proposition, we express flat coordinates of the metrics $\eta(\phi_{q,k})$ in terms of the functions $s_1,s_2,s_3,s_4$ from (\ref{phi4-coord}) as well as the systems of flat coordinates (\ref{phi1-coord})-(\ref{phi3-coord}).
\begin{Prop}
\begin{description}
\item[i)] Flat coordinates of $\eta(\phi_{q,1})$:
\begin{equation}\label{phi1q-coord}
\begin{split}
&t_{q,1}:=\frac{1}{2{\rm{i}}\pi}\oint_b\phi_{q,1}(P)=\frac{\tau_q}{2{\rm{i}}\pi}=\frac{t_1}{1+q\tau};\\
&t_{q,2}:=\sqrt{2}{\rm res}_{\infty^0}\sqrt{\lambda(P)}\phi_{q,1}(P)=\frac{t_2}{1+q\tau};\\
&t_{q,3}:=\oint_a\lambda(P)\phi_{q,1}(P)+q\oint_b\lambda(P)\phi_{q,1}(P)=\frac{t_3+qs_1}{1+q\tau}.
\end{split}
\end{equation}
\item[ii)] Flat coordinates of $\eta(\phi_{q,2})$:
\begin{equation}\label{phi2q-coord}
\begin{split}
&x_{q,1}:=\mathbf{t}^1(\phi_{q,2})=t_{q,2};\\
&x_{q,2}:=\mathbf{t}^2(\phi_{q,2})=x_2-{2{\rm i}\pi}\frac{q}{1+q\tau}(t_2)^2;\\
&x_{q,3}:=\mathbf{t}^3(\phi_{q,2})=x_3-{2{\rm i}\pi}\frac{q}{1+q\tau}t_2t_3+qs_2-{2{\rm i}\pi}\frac{q^2}{1+q\tau}t_2s_1.
\end{split}
\end{equation}
\item[iii)] Flat coordinates of $\eta(\phi_{q,3})$:
\begin{equation}\label{phi3q-coord}
\begin{split}
y_{q,1}&:=\mathbf{t}^1(\phi_{q,3})=t_{q,3};\\
y_{q,2}&:=\mathbf{t}^2(\phi_{q,3})=x_{q,3};\\
y_{q,3}&:=\mathbf{t}^3(\phi_{q,3})
=y_3+qs_3-{2{\rm i}\pi}\frac{q(t_3+s_1)}{1+q\tau}t_3+qs_3+qs_4-{2{\rm i}\pi}\frac{q^2(t_3+s_1)}{1+q\tau}s_1.
\end{split}
\end{equation}
\end{description}
\end{Prop}
\emph{Proof:} In view of (\ref{q-primary}), we obtain (\ref{phi1q-coord})-(\ref{phi3q-coord}).

\fd

\begin{description}
\item $\bullet$\emph{Prepotential associated with the holomorphic $q$-differential $\phi_{q,1}$.}
\end{description}
Applying (\ref{Prep-q1}) for a prepotential of the deformed Frobenius manifold, we deduce that the prepotential corresponding to the deformed $\phi_{q,j}$-Frobenius manifold structure  is given by
\begin{equation}\label{Prep Ex1-q}
\begin{split}
\mathbf{F}_{\phi_{q,j}}&=\frac{1}{2(1+d_j)}\sum_{k,s=1}^3\frac{\big(d_j+d_{4-k}\big)\big(d_j+d_{4-s}\big)}
{1+d_j+d_k}\mathbf{t}^{4-k}(\phi_{q,j})\mathbf{t}^{4-s}(\phi_{q,j})\mathbf{t}^k(\phi_{q,s}).
\end{split}
\end{equation}
This formula is the $q$-analogue of  (\ref{Prep Ex1}).\\
We are going to apply formula (\ref{Prep Ex1-q}) to calculate the prepotential $\mathbf{F}_{\phi_{q,1}}$ corresponding to the deformed Frobenius manifold structure induced by the $q$-differential $\phi_{q,1}$. Let us mention that this example appeared in \cite{Vasilisa2}, where a generalized Dubrovin bilinear pairing was employed.\\
Let  $E_2$ be the Eisenstein series (\ref{Eisenstein2}) and $E_{q,2}$ be the $q$-Eisenstein series defined by
\begin{equation}\label{E2-q}
E_{q,2}(\tau_q):=\frac{1}{(1-q\tau_q)^2}E_2\Big(\frac{\tau_q}{1-q\tau_q}\Big)-\frac{6{\rm i}q}{\pi(1-q\tau_q)},\quad \quad \tau_q:=\frac{\tau}{1+q\tau}.
\end{equation}
\begin{Remark}
Note that when $q$ is an integer, we have $\Im(\tau_q)>0$ and thus, due to the following  quasi-modular property of $E_2$ \cite{Apostol, Zagier}:
$$
E_2\Big(\frac{az+b}{cz+d}\Big)=(cz+d)^2E_2(z)-\frac{6{\rm i}}{\pi}c(cz+d),\quad \quad
\begin{pmatrix}
  a & b \\
  c & d
\end{pmatrix}\in SL(2,\Z),
$$
we see  that (\ref{E2-q}) becomes as:
$$
E_{q,2}(\tau_q)=E_2(\tau_q).
$$
\end{Remark}
Let $t_{q,1}$, $t_{q,2}$, $t_{q,3}$ be the flat coordinates (\ref{phi1q-coord}) of the flat metric $\eta(\phi_{q,1})$.
Then the  WDVV-quasi-homogeneous solution $\mathbf{F}_{\phi_{q,1}}$ induced by the holomorphic $q$-differential $\phi_{q,1}$ is of the form:
\begin{equation}\label{F-phiq1}
\begin{split}
\mathbf{F}_{\phi_{q,1}}&=\frac{(t_{q,3})^2}{2}t_{q,1}+\frac{(t_{q,2})^2}{2}t_{q,3}+\frac{\pi^2}{24}(t_{q,2})^4E_{q,2}(\tau_q).
\end{split}
\end{equation}
The prepotential $\mathbf{F}_{\phi_{q,1}}$ is quasi-homogeneous of degree 2 with respect to the Euler vector field $\frac{t_{q,2}}{2}\partial_{t_{q,2}}+t_{q,3}\partial_{t_{q,3}}$.
Indeed,  using (\ref{Prep Ex1-q}) and bearing in mind the fact that $\phi_{q,1}$ is a quasi-homogeneous $q$-differential of degree 0 (see Section  5.1),  we have
\begin{align*}
\mathbf{F}_{\phi_{q,1}}
&=\frac{(t_{q,3})^2}{2}\mathbf{t}^1(\phi_{q,1})+\frac{1}{12}(t_{q,2})^2\mathbf{t}^2(\phi_{q,2})+\frac{5}{12}t_{q,2}t_{q,3}\mathbf{t}^1(\phi_{q,2})\\
&=\frac{(t_{q,3})^2}{2}t_{q,1}+\frac{1}{12}(t_{q,2})^2\mathbf{t}^2(\phi_{q,2})+\frac{5}{12}(t_{q,2})^2t_{q,3}.
\end{align*}
Therefore, we only need to prove that
\begin{equation}\label{x-q2}
x_{q,2}:=\mathbf{t}^2(\phi_{q,2})=t_{q,3}+\frac{\pi^2}{2}(t_{q,2})^2E_{q,2}(\tau_q),
\end{equation}
in order to obtain the prepotential as a function of the variables $t_{q,1},t_{q,2}$ and $t_{q,3}$. \\
Let us first observe that (\ref{q-primary}) and (\ref{x2}) imply that
$$
x_{q,2}=x_2-{2{\rm i}\pi}\frac{q}{1+q\tau}(t_2)^2=t_3+\frac{\pi^2}{2}t_2^2E_2(\tau)-{2{\rm i}\pi}\frac{q}{1+q\tau}(t_2)^2.
$$
On the other hand,  (\ref{phi1q-coord}) implies that  the inverse of the map $(\tau={2{\rm{i}}\pi}t_1,t_2,t_3)\longmapsto (\tau_q={2{\rm{i}}\pi}t_{q,1},t_{q,2},t_{q,3})$ is determined by
\begin{equation}\label{phi1-phi1q}
\begin{split}
&\tau=\frac{\tau_q}{1-q\tau_q};\\
&t_2=\frac{t_{q,2}}{1-q\tau_q};\\
&t_3=t_{q,3}-\frac{i\pi {q}}{1-q\tau_q}(t_{q,2})^2.
\end{split}
\end{equation}
Plugging this into the above expression for $x_{q,2}=x_{q,2}(\tau,t_2,t_3)$ and using (\ref{E2-q}) we obtain (\ref{x-q2}).
\begin{description}
\item $\bullet$ \emph{Chazy equation.}
\end{description}
Similarly to (\ref{E2-Chazy}), the WDVV equations for the prepotential (\ref{F-phiq1}) is equivalent to the Chazy equation for the function $E_{q,2}$ defined  by (\ref{E2-q}):
$$
E_{q,2}'''-2{\rm i}\pi E_{q,2}E_{q,2}''+3{\rm i}\pi\big(E_{q,2}'\big)^2=0.
$$
\begin{description}
\item $\bullet$ \emph{Generalized Ramanujan identities.}
\end{description}
Let $E_{q,2}$ be given by (\ref{E2-q}) and $E_{q,4}$ and $E_{q,6}$ be the following $q$-Eisenstein series:
\begin{equation}\label{E4-E6-q}
\begin{split}
&E_{q,4}(\tau_q):=\frac{1}{(1-q\tau_q)^4}E_{4}\Big(\frac{\tau_q}{1-q\tau_q}\Big);\\
&E_{q,6}(\tau_q):=\frac{1}{(1-q\tau_q)^6}E_{6}\Big(\frac{\tau_q}{1-q\tau_q}\Big),
\end{split}
\quad \quad\quad \quad \quad  \tau_q:=\frac{\tau}{1+q\tau},
\end{equation}
with $E_4$ and $E_6$ being the Eisenstein series (\ref{E4-E6}).\\
Then the following $q$-Ramanujan identities are satisfied:
\begin{equation}\label{q-Ramanujan}
\begin{split}
&\frac{1}{2{\rm{i}}\pi}E_{q,2}'(\tau_q)=\frac{1}{12}\big(E_{q,2}^2(\tau_q)-E_{q,4}(\tau_q)\big);\\
&\frac{1}{2{\rm{i}}\pi}E_{q,4}'(\tau_q)=\frac{1}{3}\big(E_{q,2}(\tau_q)E_{q,4}(\tau_q)-E_{q,6}(\tau_q)\big);\\
&\frac{1}{2{\rm{i}}\pi}E_{q,6}'(\tau_q)=\frac{1}{2}\big(E_{q,2}(\tau_q)E_{q,6}(\tau_q)-E_{q,4}^2(\tau_q)\big).
\end{split}
\end{equation}
The differential equations (\ref{q-Ramanujan}) can be proved  using the usual Ramanujan identities (\ref{R-E2})-(\ref{R-E6}) and  expressions (\ref{E2-q}) and (\ref{E4-E6-q}) for the functions $E_{q,2}$, $E_{q,4}$ and $E_{q,6}$. \\
Alternatively, they can be obtained  by  adapting the scheme of the proof of (\ref{R-E2})-(\ref{R-E2}) where the main idea relies on the following result:
\begin{align*}
&x_3=\mathbf{t}^{2}(\phi_{q,3})=\mathbf{t}^{3}(\phi_{q,2})=\partial_{t_{q,1}}\partial_{t_{q,2}}\mathbf{F}_{\phi_{q,1}}=\frac{{\rm i}\pi^3}{3}(t_{q,2})^3E'_{q,2}(\tau_q);\\
&y_3=\mathbf{t}^{3}(\phi_{q,3})=\partial_{t_{q,1}}^2\mathbf{F}_{\phi_{q,1}}=-\frac{\pi^4}{6}(t_{q,2})^4E''_{q,2}(\tau_q).
\end{align*}
These equalities  are  direct consequence of the fact that the three prepotentials $\mathbf{F}_{\phi_{q,j}}$ in (\ref{Prep Ex1-q}) have the same Hessian matrix:
$$
\partial_{\mathbf{t}^k(\phi_{q,j})}\partial_{\mathbf{t}^s(\phi_{q,j})}\mathbf{F}_{\phi_{q,j}}=\mathbf{t}^{4-k}(\phi_{q,4-s}),\quad \quad \forall\ j,k,s=1,2,3.
$$
\begin{description}
\item $\bullet$ \emph{Prepotentials associated with the $q$-differentials $\phi_{q,2}$ and $\phi_{q,3}$.}
\end{description}
Let $\phi_{q,2}$ and $\phi_{q,3}$ be differential given by
(\ref{q-primary}) and  $x_{q,k}:=\mathbf{t}^{k}(\phi_{q,2})$ (\ref{phi2q-coord}) and $y_{q,k}:=\mathbf{t}^{k}(\phi_{q,3})$ (\ref{phi3q-coord}) be the flat coordinates of the flat metrics
$\eta(\phi_{q,2})$ and $\eta(\phi_{q,3})$, respectively.
Let us denote by  $(E_{q,2}')^{-1}$  the inverse function  of the $q$-Eisenstein series (\ref{E2-q}) and  $\chi_q^{-1}$ that of
$$
\chi_q(\tau_q):=\frac{\big(E_{q,2}''(\tau_q)\big)^3}{\big(E_{q,2}'(\tau_q)\big)^4}.
$$
Then  using  arguments similar to those in the proof of  Theorem \ref{phi23}, we conclude that the prepotentials  $\mathbf{F}_{\phi_{q,2}}$ and $\mathbf{F}_{\phi_{q,3}}$ take the following forms:
\begin{align*}
\mathbf{F}_{\phi_{q,2}}
&=\frac{(x_{q,2})^3}{6}+x_{q,1}x_{q,2}x_{q,3}+\frac{(x_{q,3})^2}{4{\rm{i}}\pi}\big(E_{q,2}'\big)^{-1}\Big(\frac{3x_{q,3}}{{\rm{i}}\pi^3(x_{q,1})^3}\Big)
-\frac{2}{15}(x_{q,1})^3x_{q,3}E_{q,2}\Big(\big(E_{q,2}'\big)^{-1}\Big(\frac{3x_{q,3}}{{\rm{i}}\pi^3(x_{q,1})^3}\Big)\Big)\\
&\quad -\frac{\pi^4}{180}(x_{q,1})^6E_{q,2}''\Big(\big(E_{q,2}'\big)^{-1}\Big(\frac{3x_{q,3}}{{\rm{i}}\pi^3(x_{q,1})^3}\Big)\Big)
\end{align*}
and
\begin{align*}
\mathbf{F}_{\phi_{q,3}}&=\frac{y_{q,1}}{2}(y_{q,2})^2+\frac{(y_{q,1})^2}{2}y_{q,3}+\frac{(y_{q,3})^2}{4{\rm i}\pi}\chi_q^{-1}\Big(-\frac{8}{3}(y_{q,3})^3(y_{q,2})^{-4}\Big)\\
&\quad +\frac{27}{40\pi}y_{q,2}y_{q,3}\big(-6y_{q,3}\big)^{1/4}\Big(E_{q,2}''\Big[\chi_q^{-1}\Big(-\frac{8}{3}(y_{q,3})^3(y_{q,2})^{-4}\Big)\Big]\Big)^{-1/4}\\
&\quad +\frac{9}{80\pi^2}(y_{q,2})^2\big(-6y_{q,3}\big)^{1/2}\Big(E_{q,2}''\Big[\chi_q^{-1}\Big(-\frac{8}{3}(y_{q,3})^3(y_{q,2})^{-4}\Big)\Big]\Big)^{-1/2}
E_{q,2}\Big(\chi_q^{-1}\Big(-\frac{8}{3}(y_{q,3})^3(y_{q,2})^{-4}\Big)\Big).
\end{align*}
The functions $\mathbf{F}_{\phi_{q,2}}$ and $\mathbf{F}_{\phi_{q,3}}$ satisfy the following quasi-homogeneity  property:
\begin{align*}
&\Big(\frac{x_{q,1}}{2}\partial_{x_{q,1}}+x_{q,2}\partial_{x_{q,2}}+\frac{3}{2}x_{q,3}\partial_{x_{q,3}}\Big).\mathbf{F}_{\phi_{q,2}}=3\mathbf{F}_{\phi_{q,2}};\\
&\Big(y_{q,1}\partial_{y_{q,1}}+\frac{3}{2}y_{q,2}\partial_{y_{q,2}}+2y_{q,3}\partial_{y_{q,3}}\Big).\mathbf{F}_{\phi_{q,3}}=4\mathbf{F}_{\phi_{q,3}}.
\end{align*}

\section{Appendix}
\subsection{Appendix 1: Diagonal flat metrics}

The goal of this Appendix is to prove the flatness property of the  metric $\mathbf{ds^2_{\beta_1,\beta_2}}(\omega)$ given by (\ref{ds}).

\begin{Thm}\label{flatness-W} Let $(\beta_1,\beta_2)\neq (0,0)$ and $\omega$ be a differential satisfying (\ref{Diff model}) and (\ref{Assymptions}). Then the diagonal  metric $\mathbf{ds^2_{\beta_1,\beta_2}}(\omega)$ defined in (\ref{ds}) is flat.
\end{Thm}
We need the following lemma which investigates some useful  relations involving  the Christoffel symbols (\ref{CS}) of the diagonal metric (\ref{ds}).
\begin{Lemma} Let $i,j,s \in\big\{1,\dots,N\big\}$ be distinct. Then the Christoffel symbols of the metric $\mathbf{ds^2_{\beta_1,\beta_2}}(\omega)$ satisfy the following relations:
\begin{align}
&\label{CS-aba-W}\partial_{\lambda_{s}}\Gamma_{ij}^i=\Gamma_{ij}^i\Gamma_{js}^j
-\Gamma_{ij}^i\Gamma_{is}^i+\Gamma_{is}^i\Gamma_{sj}^{s},\\
&\label{CS-aab-W}\partial_{\lambda_{s}}\Gamma_{ii}^j
=\Gamma_{ii}^j\Gamma_{is}^i-\Gamma_{ii}^j\Gamma_{js}^j-\Gamma_{ii}^{s}\Gamma_{ss}^j,\\
&\label{CS-partial-a-aba-W}\partial_{\lambda_i}\Gamma_{ij}^i=\frac{\omega(P_j)}{2\omega(P_i)}\partial_{\lambda_i}W(P_i,P_j)+
\Gamma_{ij}^i\Gamma_{ij}^j-\frac{\partial_{\lambda_i}\omega(P_i)}{\omega(P_i)}\Gamma_{ij}^i,\\
&\label{CS-partial-b-aba-W}\partial_{\lambda_j}\Gamma_{ij}^i=\frac{\omega(P_j)}{2\omega(P_i)}\partial_{\lambda_j}W(P_i,P_j)
-(\Gamma_{ij}^i)^2+\frac{\partial_{\lambda_j}\omega(P_j)}{\omega(P_j)}\Gamma_{ij}^i,\\
&\label{CS-aaa-W}
\partial_{\lambda_j}\Gamma_{ii}^i=\partial_{\lambda_i}\Gamma_{ij}^i,\\
\begin{split}\label{CS sum-W}
\begin{array}{ll}
\displaystyle\sum_{\underset{s\neq i,j}{s=1,}}^N \Gamma_{jj}^{s}\Gamma_{is}^i&=\displaystyle\frac{\beta_1}{\beta_1\lambda_j+\beta+2}\Gamma_{ij}^i
+\frac{\omega(P_j)}{2\omega(P_i)}\partial_{\lambda_j}W(P_i,P_j)\\
&\quad \displaystyle+\frac{\beta_1\lambda_i+\beta_2}{\beta_1\lambda_j+\beta_2}\times\frac{\omega(P_j)}{2\omega(P_i)}\partial_{\lambda_i}W(P_i,P_j).
\end{array}
\end{split}
\end{align}
\end{Lemma}
\emph{Proof:}
Taking the partial derivative with respect to a  branch point $\lambda_{s}$, $s\neq i,j$,  in the first relation in (\ref{CS}) and using  Rauch formulas
(\ref{Rauch}) and (\ref{Rauch-2}), we obtain (\ref{CS-aba-W}). Relation (\ref{CS-aab-W}) follows immediately from the second equality in (\ref{CS}) and (\ref{CS-aba-W}). We get the two  equalities (\ref{CS-partial-a-aba-W}) and (\ref{CS-partial-b-aba-W})  using (\ref{CS}) and (\ref{Rauch-2}). \\
By virtue of  Rauch formulas (\ref{Rauch-2}), expressions (\ref{CS}) for the Christoffel symbols and (\ref{CS-partial-b-aba-W}) we obtain
\begin{align*}
\partial_{\lambda_j}\Gamma_{ii}^i&=\frac{\partial_{\lambda_i}\partial_{\lambda_j}\omega(P_i)}{\omega(P_i)}-
\frac{\partial_{\lambda_i}\omega(P_i)}{(\omega(P_i))^2}\partial_{\lambda_j}\omega(P_i)\\
&=\frac{\omega(P_j)}{2\omega(P_i)}\partial_{\lambda_i}W(P_i,P_j)+\frac{1}{4}W(P_i,P_j)^2
-\frac{\partial_{\lambda_i}\omega(P_i)}{\omega(P_i)}\Gamma_{ij}^i\\
&=\partial_{\lambda_i}\Gamma_{ij}^i,
\end{align*}
which establishes (\ref{CS-aaa-W}).\\
Finally, note that from (\ref{CS}) and the Rauch formulas, we can write
$$
\forall\ s\neq i,j,\quad \Gamma_{jj}^{s}\Gamma_{is}^i=-\frac{\beta_1\lambda_s+\beta_2}{\beta_1\lambda_j+\beta_2}\frac{\omega(P_j)}{2\omega(P_i)}
\partial_{\lambda_s}W(P_i,P_j).
$$
We obtain (\ref{CS sum-W})  using this relation together with  the action (\ref{Auxi1}) of the  vector field $\sum_k(\beta_1\lambda_k+\beta_2)\partial_{\lambda_k}$ on $W(P_i,P_j)$.

\fd

\medskip

\noindent We  now move on to prove  Theorem \ref{flatness-W}.

\noindent \emph{Proof of Theorem \ref{flatness-W}:} We are going to show that the Riemann curvature tensor of the metric $\mathbf{ds^2_{\beta_1,\beta_2}}(\omega)$ vanishes identically.
Recall that the Riemann curvature tensor is defined  by
$$
R_{ijk}^t:=\partial_j\Gamma_{ik}^t-\partial_k\Gamma_{ij}^t+\textstyle\sum_{s}\big(\Gamma_{ik}^s\Gamma_{sj}^t-\Gamma_{ij}^s\Gamma_{sk}^t \big).
$$
We  distinguish four cases.
\medskip

\noindent\emph{First case:}  $i, j, k\in \{1,\dots,N\}$ are distinct. We have
\begin{align*}
R_{ijk}^t
&=\partial_j\Gamma_{ik}^t-\partial_k\Gamma_{ij}^t+\Gamma_{ik}^i\Gamma_{ij}^t+\Gamma_{ik}^k\Gamma_{kj}^t-\Gamma_{ij}^i\Gamma_{ik}^t-\Gamma_{ij}^j\Gamma_{jk}^t.
\end{align*}
If $t\notin\{i,j,k\}$, clearly we have $R_{ijk}^t=0$. By (\ref{CS-aba-W}) we deduce that if $t=i$,
\begin{align*}
R_{ijk}^i&=\partial_j\Gamma_{ik}^i-\partial_k\Gamma_{ij}^i
=\Gamma_{ik}^{i}\Gamma_{kj}^{k}-\Gamma_{ik}^{i}\Gamma_{ij}^{i}+\Gamma_{jk}^{j}\Gamma_{ij}^{i}-\Gamma_{ij}^{i}\Gamma_{jk}^{j}+\Gamma_{ij}^{i}\Gamma_{ik}^{i}
-\Gamma_{kj}^{k}\Gamma_{ik}^{i}=0
\end{align*}
and if $t=j$,
\begin{align*}
R_{ijk}^j&=-\partial_k\Gamma_{ij}^j+\Gamma_{ik}^i\Gamma_{ij}^j+\Gamma_{ik}^k\Gamma_{kj}^j-\Gamma_{ij}^i\Gamma_{ik}^j-\Gamma_{ij}^j\Gamma_{jk}^j\\
&=-\Gamma_{ji}^j\Gamma_{ik}^i+\Gamma_{ji}^j\Gamma_{jk}^j-\Gamma_{jk}^j\Gamma_{ki}^j+\Gamma_{ik}^i\Gamma_{ij}^j+\Gamma_{ik}^k\Gamma_{kj}^j-\Gamma_{ij}^j\Gamma_{jk}^j
=0.
\end{align*}
Again (\ref{CS-aba-W}) implies the vanishing of $R_{ijk}^k$. Indeed,
\begin{align*}
R_{ijk}^k&=\partial_j\Gamma_{ik}^k-\partial_k\Gamma_{ij}^k+\Gamma_{ik}^i\Gamma_{ij}^k+\Gamma_{ik}^k\Gamma_{kj}^k-\Gamma_{ij}^i\Gamma_{ik}^k
-\Gamma_{ij}^j\Gamma_{jk}^k\\
&=\Gamma_{ik}^{k}\Gamma_{ij}^{i}-\Gamma_{ik}^{k}\Gamma_{kj}^{k}+\Gamma_{kj}^{k}\Gamma_{ji}^{i}+\Gamma_{ik}^k\Gamma_{kj}^k
-\Gamma_{ij}^i\Gamma_{ik}^k-\Gamma_{ij}^j\Gamma_{jk}^k=0.
\end{align*}
\medskip

\noindent\emph{Second case:}\ $i=j\neq k$. In this case, the Riemann curvature tensor takes the following  form
\begin{align*}
R_{iik}^t
&=\partial_i\Gamma_{ik}^t-\partial_k\Gamma_{ii}^t+\Gamma_{ik}^i\Gamma_{ii}^t+\Gamma_{ik}^k\Gamma_{ki}^t-\sum_s\Gamma_{ii}^s\Gamma_{sk}^t.
\end{align*}
If $t\notin \{i,k\}$ then by  (\ref{CS-aab-W}), we get
\begin{align*}
R_{iik}^t&=-\partial_k\Gamma_{ii}^t+\Gamma_{ik}^i\Gamma_{ii}^t+\Gamma_{ik}^k\Gamma_{ki}^t-\Gamma_{ii}^k\Gamma_{kk}^t-\Gamma_{ii}^t\Gamma_{tk}^t\\
&=-\Gamma_{ii}^{t}\Gamma_{ik}^{i}+\Gamma_{ii}^{t}\Gamma_{kt}^{t}+\Gamma_{ii}^{k}\Gamma_{kk}^{t}+
\Gamma_{ik}^i\Gamma_{ii}^t-\Gamma_{ii}^k\Gamma_{kk}^t-\Gamma_{ii}^t\Gamma_{tk}^t=0.
\end{align*}
If $t=i$, then (\ref{CS-aaa-W}) implies that
\begin{align*}
R_{iik}^i&=\partial_i\Gamma_{ik}^i-\partial_k\Gamma_{ii}^i+\Gamma_{ik}^i\Gamma_{ii}^i+\Gamma_{ik}^k\Gamma_{ki}^i-\sum_s\Gamma_{ii}^s\Gamma_{sk}^i\\
&=\Gamma_{ik}^i\Gamma_{ii}^i+\Gamma_{ik}^k\Gamma_{ki}^i-\Gamma_{ii}^i\Gamma_{ik}^i-\Gamma_{ii}^k\Gamma_{kk}^i\\
&=\Gamma_{ik}^k\Gamma_{ki}^i-\Gamma_{ii}^k\Gamma_{kk}^i=0.
\end{align*}
If $t=k$, according to (\ref{CS}), (\ref{CS-partial-a-aba-W}), (\ref{CS-partial-b-aba-W}) and (\ref{CS sum-W}) we can write
\begin{align*}
R_{iik}^k&=\partial_i\Gamma_{ik}^k-\partial_k\Gamma_{ii}^k+\Gamma_{ik}^i\Gamma_{ii}^k+(\Gamma_{ik}^k)^2-\sum_s\Gamma_{ii}^s\Gamma_{sk}^k\\
&=\partial_i\Gamma_{ik}^k+ \frac{\beta_1}{\beta_1\lambda_i+\beta_2}\Gamma_{ik}^k+\frac{\beta_1\lambda_k+\beta_2}{\beta_1\lambda_i+\beta_2}
\partial_k\Gamma_{ik}^k+\Gamma_{ik}^i\Gamma_{ii}^k
+(\Gamma_{ik}^k)^2-\Gamma_{ii}^i\Gamma_{ik}^k-\Gamma_{ii}^k\Gamma_{kk}^k-\sum_{s\neq i,k}\Gamma_{ii}^s\Gamma_{sk}^k\\
&=\frac{\omega(P_i)}{2\omega(P_k)}\partial_{\lambda_i}W(P_k,P_i)-(\Gamma_{ik}^k)^2+\frac{\partial_{\lambda_i}\omega(P_i)}{\omega(P_i)}\Gamma_{ik}^k
+\frac{\beta_1}{\beta_1\lambda_i+\beta_2}\Gamma_{ik}^k\\
&\quad+\frac{\beta_1\lambda_k+\beta_2}{\beta_1\lambda_i+\beta_2}\bigg(\frac{\omega(P_i)}{2\omega(P_k)}\partial_{\lambda_k}W(P_k,P_i)+
\Gamma_{ik}^k\Gamma_{ik}^i-\frac{\partial_{\lambda_k}\omega(P_k)}{\omega(P_k)}\Gamma_{ik}^k\bigg)\\
&\quad-\frac{\beta_1\lambda_k+\beta_2}{\beta_1\lambda_i+\beta_2}\Gamma_{ik}^k\Gamma_{ik}^i+(\Gamma_{ik}^k)^2
-\bigg(-\frac{\beta_1}{2\beta_1\lambda_i+2\beta_2}+\frac{\partial_{\lambda_{i}}\omega(P_i)}{\omega(P_i)}\bigg)\Gamma_{ik}^k\\
&\quad+\frac{\beta_1\lambda_k+\beta_2}{\beta_1\lambda_i+\beta_2}\bigg(-\frac{\beta_1}{2\beta_1\lambda_k+2\beta_2}
+\frac{\partial_{\lambda_k}\omega(P_k)}{\omega(P_k)}\bigg)\Gamma_{ik}^k-\sum_{s\neq i,k}\Gamma_{ii}^s\Gamma_{sk}^k\\
&=\frac{\omega(P_i)}{2\omega(P_k)}\partial_{\lambda_i}W(P_{k},P_{i})+  \frac{\beta_1\lambda_k+\beta_2}{\beta_1\lambda_i+\beta_2}\frac{\omega(P_i)}{2\omega(P_k)}\partial_{\lambda_k}W(P_k,P_i)
+\frac{\beta_1}{\beta_1\lambda_i+\beta_2}\Gamma_{ik}^k-\sum_{s\neq i,k}\Gamma_{ii}^s\Gamma_{sk}^k\\
&=0.
\end{align*}
\noindent\emph{Third case:}\ $i=k\neq j$. The vanishing of $R_{iji}^t$  follows from the second case and the properties of the Riemann tensor:
\begin{align*}
\forall\ t,\quad R_{iji}^t=-R_{iij}^t=0.
\end{align*}
\noindent\emph{Fourth case:}  If $j=k$, then  we always  have  $R_{ijj}^t=R_{iii}^t=0$.

\fd

\subsection{Appendix 2: Multivalued differentials and quasi-homogeneity properties}
In this appendix we explore some results related to multivalued differentials defined on surface $C_g$ of genus $g\geq 1$, where $\big\{(C_g,\lambda),\{a_k,b_k\}_k\big\}$ is a fixed point in the simple Hurwitz space $\widehat{\mathcal{H}}_{g, L}(n_0,\dots,n_m)$.\\
Let $\sigma$ and $\nu$ be two  positive integers. For $k=1,\dots,g$, consider the multivalued differentials
\begin{equation}\label{multiva-def}
\phi_{u^{k,\sigma}}(P):=\oint_{a_k}\big(\lambda(Q)\big)^{\sigma}W(P,Q),\quad \quad \phi_{\widetilde{u}^{k,\nu}}(P):=\oint_{b_k}\big(\lambda(Q)\big)^{\nu}W(P,Q).
\end{equation}
\begin{Prop}
Let $j=1,\dots,g$.  The differentials $\phi_{u^{k,\sigma}}$ and $\phi_{\widetilde{u}^{k,\nu}}$ (\ref{multiva-def}) have the following jumps:
\begin{equation}\label{sigma-nu-jump}
\begin{split}
&\phi_{u^{k,\sigma}}(P+b_j)-\phi_{u^{k,\sigma}}(P)=-2{\rm i}\pi \delta_{jk}d\big(\lambda(P)\big)^{\sigma};\\
&\phi_{u^{k,\sigma}}(P+a_j)-\phi_{u^{k,\sigma}}(P)=0;\\
&\phi_{\widetilde{u}^{k,\nu}}(P+a_j)-\phi_{\widetilde{u}^{k,\nu}}(P)=2{\rm i}\pi \delta_{jk}d\big(\lambda(P)\big)^{\nu};\\
&\phi_{\widetilde{u}^{k,\nu}}(P+b_j)-\phi_{\widetilde{u}^{k,\nu}}(P)=0.
\end{split}
\end{equation}
Moreover,  the periods of the  differentials $\phi_{u^{k,\sigma}}$ and  $\phi_{\widetilde{u}^{k,\nu}}$ are as follows:
\begin{equation}\label{sigma-nu-periods}
\begin{split}
&\oint_{a_j}\phi_{u^{k,\sigma}}=0;\\
&\oint_{b_j}\phi_{u^{k,\sigma}}=-2{\rm{i}}\pi\big(\lambda(P_0)\big)^{\sigma}\delta_{jk}+2{\rm{i}}\pi \oint_{a_k}\big(\lambda(P)\big)^{\sigma}\omega_j(P);\\
&\oint_{a_j}\phi_{\widetilde{u}^{k,\nu}}=2{\rm i}\pi\delta_{jk}\big(\lambda(P_0)\big)^{\nu};\\
&\oint_{b_j}\phi_{\widetilde{u}^{k,\nu}}=2{\rm i}\pi\oint_{b_k}\big(\lambda(P)\big)^{\nu}\omega_j(P).
\end{split}
\end{equation}
 Here $P_0$ denotes an arbitrary  starting point of the cycles $a_k, b_k$.
\end{Prop}
\emph{Proof:} We obtain the jumps and the periods of the multivalued differential $\phi_{u^{k,\sigma}}$ by a direct adaptation of  the proofs of (\ref{type5}) and (\ref{Fubini1}). \\
Let us deal with the differential $\phi_{\widetilde{u}^{k,\nu}}$.
We have
\begin{align*}
&\big(\lambda(Q)\big)^{\nu}W(P,Q)=\big(\lambda(Q)\big)^{\nu}d_Pd_Q\log\Big(\Theta_{\Delta}\big(\mathcal{A}(P)-\mathcal{A}(Q)\big)\Big)\\
&=d_Q\Big[\big(\lambda(Q)\big)^{\nu}d_P\log\Big(\Theta_{\Delta}\big(\mathcal{A}(P)-\mathcal{A}(Q)\big)\Big)\Big]
 -\Big[d_P\log\Big(\Theta_{\Delta}\big(\mathcal{A}(P)-\mathcal{A}(Q)\big)\Big)\Big]d\big(\lambda(Q)\big)^{\nu}.
\end{align*}
Integrating both sides with respect to $Q$ along the contour $b_k$ while  choosing  a starting point $P_0$ of the contour $b_k$ and keeping in
mind the quasi-periodicity properties (\ref{theta-quasi}) of the Riemann Theta function, we see that the differential $\phi_{\widetilde{u}^{k,\nu}}$ can
 be represented by
$$
\phi_{\widetilde{u}^{k,\nu}}(P)=2{\rm i}\pi\big(\lambda(P_0)\big)^{\nu}\omega_k(P)-
\oint_{b_k}d_P\log\Big(\Theta_{\Delta}\big(\mathcal{A}(P)-\mathcal{A}(Q)\big)\Big)d\big(\lambda(Q)\big)^{\nu}.
$$
Let us define
$$
F_{k,\nu}(P):=\int^P\phi_{\widetilde{u}^{k,\nu}}=2{\rm i}\pi\big(\lambda(P_0)\big)^{\nu}\int^P\omega_k-
\oint_{b_k}\log\Big(\Theta_{\Delta}\big(\mathcal{A}(P)-\mathcal{A}(Q)\big)\Big)d\big(\lambda(Q)\big)^{\nu}.
$$
Then, due to the quasi-periodicity  property  (\ref{theta-quasi}) of the Riemann Theta function, we see that $F_{k,\nu}$ is a multivalued function with
\begin{align*}
F_{k,\nu}(P+a_j)-F_{k,\nu}(P)&=0=\oint_{a_j}\phi_{\widetilde{u}^{k,\nu}}, \quad \forall\ j\neq k;\\
F_{k,\nu}(P+a_k)-F_{k,\nu}(P)&=2{\rm i}\pi\big(\lambda(P_0)\big)^{\nu}=\oint_{a_k}\phi_{\widetilde{u}^{k,\nu}};\\
F_{k,\nu}(P+b_j)-F_{k,\nu}(P)
&=2{\rm i}\pi\big(\lambda(P_0)\big)^{\nu}\mathbb{B}_{jk}-\oint_{b_k}\Big(-{\rm i}\pi \mathbb{B}_{jj}-2{\rm{i}}\pi\prs{\beta}{e_j}
-2{\rm i}\pi\int_Q^P\omega_j\Big)d\big(\lambda(Q)\big)^{\nu}\\
&=2{\rm i}\pi\big(\lambda(P_0)\big)^{\nu}\mathbb{B}_{jk}+2{\rm i}\pi \oint_{b_k}\Big(\int_Q^P\omega_j\Big)d\big(\lambda(Q)\big)^{\nu}\\
&=2{\rm i}\pi\big(\lambda(P_0)\big)^{\nu}\mathbb{B}_{jk}+2{\rm i}\pi\Big(\big(\lambda(Q)\big)^{\nu}\int_Q^P\omega_j\Big)\bigg|_{Q=P_0}^{Q=P_0+b_k}
+2{\rm i}\pi \oint_{b_k}\big(\lambda(Q)\big)^{\nu}\omega_j(Q)\\
&=2{\rm i}\pi \oint_{b_k}\big(\lambda(Q)\big)^{\nu}\omega_j(Q).
\end{align*}
This, given that $\phi_{\widetilde{u}^{k,\nu}}(P)=dF_{k,\nu}(P)$, yields
$$
\phi_{\widetilde{u}^{k,\nu}}(P+b_j)-\phi_{\widetilde{u}^{k,\nu}}(P)=0=\phi_{\widetilde{u}^{k,\nu}}(P+a_l)-\phi_{\widetilde{u}^{k,\nu}}(P)
$$
 for all $j,l=1,\dots,g$ and $l\neq k$. \\
On the other hand,  the expression $F_{k,\nu}(P+a_k)-F_{k,\nu}(P)$ assumes  that the point $P$ is considered  as the starting point of the
 contour $a_k$. Therefore, since  the point $P_0$ can be chosen as the intersection point of the contours $a_k$ and $b_k$, i.e. $P_0=P$, we get
$$
\phi_{\widetilde{u}^{k,\nu}}(P+a_k)-\phi_{\widetilde{u}^{k,\nu}}(P)=dF_{k,\nu}(P+a_k)-dF_{k,\nu}(P)=2\rm{i}\pi d\big(\lambda(P)\big)^{\nu}.
$$

\fd

\begin{Prop} Let $\phi_{u^{k,\sigma}}$ and $\phi_{\widetilde{u}^{k,\nu}}$ be the multivalued differentials  defined by (\ref{multiva-def}) and $t^A$ be an operation from the set $\big\{t^{i,\alpha},\ v^i,\ s^i,\ \rho^k,\  u^k\Big\}$ described in (\ref{FBP-W})-(\ref{S-eta}). \\
Then the action of the vector field $e=\sum_{j=1}^N\partial_{\lambda_j}$ on the functions $t^A\big(\phi_{u^{k,\sigma}}\big)$ and $t^A\big(\phi_{\widetilde{u}^{k,\nu}}\big)$ is given by
\begin{equation}\label{e-sigma-nu}
\begin{split}
&e.t^A\big(\phi_{u^{k,\sigma}}\big)=\sigma \oint_{a_k}\lambda(P)^{\sigma-1}\phi_{t^A}(P);\\
&e.t^A\big(\phi_{\widetilde{u}^{k,\nu}}\big)=2{\rm i}\pi\big(\lambda(P_0)\big)^{\nu}+\nu\oint_{b_k}\big(\lambda(P)\big)^{\nu-1}\phi_{t^A}(P).
\end{split}
\end{equation}

\end{Prop}
\emph{Proof:} Let us denote by $F_g$ the fundamental polygon associated with the compact surface $C_g$. The differential
\begin{equation}\label{Phi-A-sigma}
\Phi_{A,k,\sigma}(P):=\frac{\phi_{t^A}(P)}{d\lambda(P)}\phi_{u^{k,\sigma}}(P)
\end{equation}
is multivalued and, by (\ref{sigma-nu-jump}), its jumps  are such that $\Phi_{A,k,\sigma}(P+a_j)-\Phi_{A,k,\sigma}(P)=0$ and
$$
\Phi_{A,k,\sigma}(P+b_j)-\Phi_{A,k,\sigma}(P)=-2{\rm i}\pi\Big(\delta_{jk}\sigma\big(\lambda(P)\big)^{\sigma-1}\phi_{t^A}(P)+\delta_{\phi_{t^A},\phi_{u^j}}
\phi_{u^{k,\sigma}}(P)-2{\rm i}\pi\delta_{jk}\delta_{\phi_{t^A},\phi_{u^j}}d\big(\lambda(P)\big)^{\sigma}\Big).
$$
Therefore, using  Rauch formulas (\ref{partial-t-W}) and  the residue theorem we conclude that
\begin{align*}
e.t^A\big(\phi_{u^{k,\sigma}}\big)&=\frac{1}{2}\sum_{j=1}^N\phi_{t^A}(P_j)\phi_{u^{k,\sigma}}(P_j)=\sum_{j=1}^N\underset{P_j}{{\rm res}}\Phi_{A,k,\sigma}(P)
=\frac{1}{2{\rm i}\pi}\oint_{F_g}\Phi_{A,k,\sigma}(P)\\
&=\frac{1}{2{\rm i}\pi}\sum_{j=1}^g\bigg(\oint_{a_j}\Phi_{A,k,\sigma}(P)-\oint_{a_j}\Phi_{A,k,\sigma}(P+b_j)\bigg)\\
&=\sigma\oint_{a_k}\lambda(P)^{\sigma-1}\phi_{t^A}(P).
\end{align*}
Now, due to (\ref{type5}) and (\ref{sigma-nu-jump}), the jumps of  the multivalued  differential
$$
\widetilde{\Phi}_{A,k,\nu}(P):=\frac{\phi_{t^A}(P)}{d\lambda(P)}\phi_{\widetilde{u}^{k,\nu}}(P)
$$
are as follows:
\begin{align*}
&\widetilde{\Phi}_{A,k,\nu}(P+a_j)-\widetilde{\Phi}_{A,k,\nu}(P)=2{\rm i}\pi \delta_{jk}\nu \big(\lambda(P)\big)^{\nu-1}\phi_{t^A}(P);\\
&\widetilde{\Phi}_{A,k,\nu}(P+b_j)-\widetilde{\Phi}_{A,k,\nu}(P)=-2{\rm i}\pi\delta_{\phi_{t^A},\phi_{u^j}}\phi_{\widetilde{u}^{k,\nu}}(P).
\end{align*}
Using this and proceeding as in the previous case, we deduce that
\begin{align*}
&e.t^A\big(\phi_{\widetilde{u}^{k,\nu}}\big)=\sum_{j=1}^N\underset{P_j}{{\rm res}}\widetilde{\Phi}_{A,k,\nu}(P)
=\frac{1}{2{\rm i}\pi}\oint_{F_g}\widetilde{\Phi}_{A,k,\nu}(P)\\
&=\frac{1}{2{\rm i}\pi}\sum_{j=1}^g\bigg(\oint_{a_j}\widetilde{\Phi}_{A,k,\nu}(P)-\oint_{a_j}\widetilde{\Phi}_{A,k,\nu}(P+b_j)\bigg)
+\frac{1}{2{\rm i}\pi}\sum_{j=1}^g\bigg(\oint_{b_j}\widetilde{\Phi}_{A,k,\nu}(P)-\oint_{b_j}\widetilde{\Phi}_{A,k,\nu}(P+a_j^{-1})\bigg)\\
&=\sum_{j=1}^g\bigg(\delta_{\phi_{t^A},\phi_{u^j}}\oint_{a_j}\phi_{\widetilde{u}^{k,\nu}}(P)\bigg)+
\nu\oint_{b_k}\big(\lambda(P)\big)^{\nu-1}\phi_{t^A}(P)\\
&=2{\rm i}\pi\big(\lambda(P_0)\big)^{\nu}+\nu\oint_{b_k}\big(\lambda(P)\big)^{\nu-1}\phi_{t^A}(P).
\end{align*}

\fd

\begin{Prop} Consider the multivalued differentials $\phi_{u^{k,\sigma}}$ and $\phi_{\widetilde{u}^{k,\nu}}$ defined by (\ref{multiva-def}) and assume that all the cycles $\{a_k,b_k\}$ start at a marked point $P_0$ whose  projection by $\lambda$ is zero. Then,  for any operation  $t^A$ among those defining  flat coordinates  in (\ref{FBP-W})-(\ref{S-eta}), the functions $t^A\big(\phi_{u^{k,\sigma}}\big)$ and $t^A\big(\phi_{\widetilde{u}^{k,\nu}}\big)$ are eigenfunctions of the vector field $E=\sum_{s=1}^N\lambda_s\partial_{\lambda_s}$ with
\begin{align}
&\label{u-k-sigma}E.t^A\big(\phi_{u^{k,\sigma}}\big)=(d_A+\sigma)t^A\big(\phi_{u^{k,\sigma}}\big);\\
&\label{v-k-nu} E.t^A\big(\phi_{\widetilde{u}^{k,\nu}}\big)=(d_A+\nu)t^A\big(\phi_{\widetilde{u}^{k,\nu}}\big).
\end{align}
\end{Prop}
\emph{Proof:} Let $(C_g,\lambda)$ be a generic  branched covering representing an equivalence  class in the simple Hurwitz space
$\widehat{\mathcal{H}}_{g, L}(n_0,\dots,n_m)$. Denote by $F_g$ the fundamental polygon associated with the compact surface $C_g$. \\
Using the properties of the multivalued  differential  $\Phi_{A,k,\sigma}$ defined by (\ref{Phi-A-sigma}), we deduce that the differential $\lambda(P)\Phi_{A,k,\sigma}(P)$
has no jumps along the cycles $a_j$ and its  jumps along the cycle $b_j$ are as follows:
\begin{align*}
&\lambda(P)\Phi_{A,k,\sigma}(P+b_j)-\lambda(P)\Phi_{A,k,\sigma}(P)\\
&=-2{\rm i}\pi\Big(\delta_{jk}\sigma\big(\lambda(P)\big)^{\sigma}\phi_{t^A}(P)+\delta_{\phi_{t^A},\phi_{u^j}}\lambda(P)
\phi_{u^{k,\sigma}}(P)-2{\rm i}\pi\delta_{jk}\delta_{\phi_{t^A},\phi_{u^j}}\sigma\big(\lambda(P)\big)^{\sigma}d\lambda(P)\Big).
\end{align*}
Moreover, the restriction of the differential $\phi_{u^{k,\sigma}}$ to the fundamental polygon $F_g$  is holomorphic inside $F_g$.  This enables us
to apply  identity (\ref{g-eta-trick}) with $\phi_0=\phi_{u^{k,\sigma}}$.  Therefore, using  Rauch formulas (\ref{partial-t-W}) and formula (\ref{g-eta-trick}) while
bearing  in mind the jumps of $\lambda(P)\Phi_{A,k,\sigma}$, we get
\begin{align*}
&E.t^A\big(\phi_{u^{k,\sigma}}\big)=\frac{1}{2}\sum_{j=1}^N\lambda_j\phi_{t^A}(P_j)\phi_{u^{k,\sigma}}(P_j)
=\sum_{j=1}^N\underset{P_j}{{\rm res}}\bigg(\frac{\lambda(P)\phi_{t^A}(P)}{d\lambda(P)}\phi_{u^{k,\sigma}}(P)\bigg)\\
&=\frac{1}{2{\rm{i}}\pi}\oint_{\partial{F_g}}\lambda(P)\Phi_{A,k,\sigma}(P)
-\sum_{i=0}^{m}\sum_{\alpha=1}^{n_i}\delta_{\phi_{t^A},\phi_{t^{i,\alpha}}}\underset{\infty^i}{{\rm res}}
\bigg(\frac{\lambda(P)\phi_{t^A}(P)}{d\lambda(P)}\phi_{u^{k,\sigma}}(P)\bigg)\\
&\quad -\sum_{i=1}^{m}\delta_{\phi_{t^A},\phi_{v^{i}}}\underset{\infty^i}{{\rm res}}\bigg(\frac{\lambda(P)\phi_{t^A}(P)}{d\lambda(P)}\phi_{u^{k,\sigma}}(P)\bigg)\\
&=\frac{1}{2{\rm{i}}\pi}\sum_{j=1}^g\oint_{a_j}\bigg(\Phi_{A,k,\sigma}(P)-\Phi_{A,k,\sigma}(P+b_j)\bigg)
+\sum_{i=0}^{m}\sum_{\alpha=1}^{n_i}\delta_{\phi_{t^A},\phi_{t^{i,\alpha}}}\frac{\alpha}{n_i+1}t^{i,\alpha}\big(\phi_{u^{k,\sigma}}\big)
+\sum_{i=1}^{m}\delta_{\phi_{t^A},\phi_{v^{i}}}v^i\big(\phi_{u^{k,\sigma}}\big)\\
&=\sum_{j=1}^g\oint_{a_j}\bigg(\delta_{jk}\sigma\big(\lambda(P)\big)^{\sigma}\phi_{t^A}(P)+\delta_{\phi_{t^A},\phi_{u^j}}
\lambda(P)\phi_{u^{k,\sigma}}(P)-2{\rm i}\pi\delta_{jk}\delta_{\phi_{t^A},\phi_{u^j}}\sigma \big(\lambda(P)\big)^{\sigma}d\lambda(P)\bigg)\\
&\quad +\sum_{i=0}^{m}\sum_{\alpha=1}^{n_i}\delta_{\phi_{t^A},\phi_{t^{i,\alpha}}}\frac{\alpha}{n_i+1}t^{i,\alpha}\big(\phi_{u^{k,\sigma}}\big)
+\sum_{i=1}^{m}\delta_{\phi_{t^A},\phi_{v^{i}}}v^i\big(\phi_{u^{k,\sigma}}\big)\\
&=\sigma\oint_{a_k}\big(\lambda(P)\big)^{\sigma}\phi_{t^A}(P)+\sum_{j=1}^g\delta_{\phi_{t^A},\phi_{u^j}}\oint_{a_j}\lambda(P)\phi_{u^{k,\sigma}}(P)
 +\sum_{i=0}^{m}\sum_{\alpha=1}^{n_i}\delta_{\phi_{t^A},\phi_{t^{i,\alpha}}}\frac{\alpha}{n_i+1}t^{i,\alpha}\big(\phi_{u^{k,\sigma}}\big)
+\sum_{i=1}^{m}\delta_{\phi_{t^A},\phi_{v^{i}}}v^i\big(\phi_{u^{k,\sigma}}\big)\\
&=\sigma t^A\big(\phi_{u^{k,\sigma}}\big)+\sum_{j=1}^g\delta_{\phi_{t^A},\phi_{u^j}}u^j\big(\phi_{\sigma,k}\big)
+\sum_{i=0}^{m}\sum_{\alpha=1}^{n_i}\delta_{\phi,\phi_{t^{i,\alpha}}}\frac{\alpha}{n_i+1}t^{i,\alpha}\big(\phi_{\sigma,k}\big)
+\sum_{i=1}^{m}\delta_{\phi,\phi_{v^{i}}}v^i\big(\phi_{\sigma,k}\big).
\end{align*}
In the last equality we have interchanged the order of integration over the cycles $a_k$ and $b_k$ when $\phi_{t^A}=\phi_{\rho^k}=\omega_k$. More precisely, we have used the following relation:
$$
\oint_{a_k}\big(\lambda(P)\big)^{\sigma}\omega_k(P)=\big(\lambda(P_0)\big)^{\sigma}+\frac{1}{2{\rm{i}}\pi}\oint_{b_j}\phi_{u^{k,\sigma}}
=\frac{1}{2{\rm{i}}\pi}\oint_{b_j}\phi_{u^{k,\sigma}},
$$
where the first equality follows from (\ref{sigma-nu-periods}) and the second is an immediate consequence of the  fact that the cycles $a_k$ and $b_k$ start at a point $P_0$ such  that $\lambda(P_0)=0$.\\
Formula (\ref{v-k-nu}) can be obtained in a similar way.
\fd

\medskip

The next proposition addresses  the quasi-homogeneity property of the  functions
\begin{equation}\label{q-sigma}
q^j\big(\phi_{u^{k,\sigma}}\big):=\oint_{a_j}\log\big(\lambda(P)\big)\phi_{u^{k,\sigma}}(P), \quad j=1,\dots,g,
\end{equation}
with  $\phi_{u^{k,\sigma}}$ being the differential defined by (\ref{multiva-def}). Note that these functions
are among the family of  flat coordinates (\ref{flat basis-g}) of the intersection form $\mathbf{g}\big(\phi_{u^{k,\sigma}}\big)$ (\ref{Inter-form}).
\begin{Prop} For each $j=1,\dots,g$,  the function $q^j\big(\phi_{u^{k,\sigma}}\big)$ (\ref{q-sigma})
is an eigenfunction of the vector field $E=\sum_{s=1}^N\lambda_s\partial_{\lambda_s}$ with
\begin{equation}\label{E-q-sigma}
E.q^j\big(\phi_{u^{k,\sigma}}\big)=\sigma q^j\big(\phi_{u^{k,\sigma}}\big).
\end{equation}

\end{Prop}
\emph{Proof:}  Consider the multivalued differential
$$
\Psi_{u^{k,\sigma},q^j}(P):=\frac{\lambda(P)\psi_{q^j}(P)}{d\lambda(P)}\phi_{u^{k,\sigma}}(P),\quad \text{with}\quad \psi_{q^k}(P):=\oint_{a_k}\log(\lambda(Q))W(P,Q).
$$
Due to (\ref{jump-E}) and  (\ref{sigma-nu-jump}), the jumps of $\Psi_{u^{k,\sigma},q^j}(P)$ are as follows:
\begin{align*}
&\Psi_{u^{k,\sigma},q^j}(P+a_r)-\Psi_{u^{k,\sigma},q^j}(P)=0;\\
&\Psi_{u^{k,\sigma},q^j}(P+b_r)-\Psi_{u^{k,\sigma},q^j}(P)\\
&=-2{\rm i}\pi \delta_{jr}\phi_{u^{k,\sigma}}(P)-2{\rm i}\pi\sigma\delta_{kr}\big(\lambda(P)\big)^{\sigma}\psi_{q^j}(P)
+(2{\rm i}\pi)^2\sigma\delta_{jr}\delta_{kr}d\big(\lambda(P)\big)^{\sigma}.
\end{align*}
Thus,  observing that the ramification points $P_1,\dots,P_N$ are the only simple poles of the meromorphic differential $\Psi_{u^{k,\sigma},q^j}$  within a fixed fundamental polygon $F_g$  and using its jumps, we get
\begin{align*}
&E.q^j\big(\phi_{u^{k,\sigma}}\big)=\sum_{s=1}^N\underset{P_s}{{\rm res}}\Psi_{u^{k,\sigma},q^j}(P)=\frac{1}{2{\rm{i}}\pi}\oint_{\partial{F_g}}\Psi_{u^{k,\sigma},q^j}(P)\\
&=\frac{1}{2{\rm{i}}\pi}\sum_{r=1}^g\oint_{a_r}\bigg(\Psi_{u^{k,\sigma},q^j}(P)-\Psi_{u^{k,\sigma},q^j}(P+b_r)\bigg)\\
&=\oint_{a_j}\phi_{u^{k,\sigma}}(P)+\sigma\oint_{a_k}\big(\lambda(P)\big)^{\sigma}\psi_{q^j}(P)
+2{\rm i}\pi\sigma\sum_{r=1}^g\delta_{jr}\delta_{kr}\oint_{a_r}d\big(\lambda(P)\big)^{\sigma}\\
&=\sigma q^j\big(\phi_{u^{k,\sigma}}\big).
\end{align*}

\fd

\subsection{Appendix 3: Dubrovin's bilinear pairing  and prepotentials}
Let $\omega^{(1)}$ and $\omega^{(2)}$ be two differentials on the surface $C_g$ holomorphic outside the points $\infty^0,\dots,\infty^m$ with the following properties:
\begin{align*}
&\omega^{(\alpha)}(P)\underset{P\sim \infty^i}=\sum_{n=-n^{(\alpha)}}^{\infty}c_{n,i}^{(\alpha)}z_i(P)^ndz_i(P)+\frac{1}{n_i+1}d\left(
\sum_{n=1}^{\infty}r_{n,i}^{(\alpha)}\big(\lambda(P)\big)^n\log\big(\lambda(P)\big)\right);\\
&\oint_{a_k}\omega^{(\alpha)}=A_{k}^{(\alpha)};\\
&\omega^{(\alpha)}(P+a_k)-\omega^{(\alpha)}(P)=dp_k^{(\alpha)}\big(\lambda(P)\big),\quad \quad p_k^{(\alpha)}=\sum_{j>0}p_{kj}^{(\alpha)}\big(\lambda(P)\big)^j;\\
&\omega^{(\alpha)}(P+b_k)-\omega^{(\alpha)}(P)=dq_k^{(\alpha)}\big(\lambda(P)\big),\quad \quad q_k^{(\alpha)}=\sum_{j>0}q_{kj}^{(\alpha)}\big(\lambda(P)\big)^j,
\end{align*}
where $z_i(P)$ is the local parameter near the pole $\infty^i$  defined by $\lambda(P)=z_i(P)^{-n_i-1}$, $n^{(\alpha)}\in \Z$ and $c_{n,i}^{(\alpha)}$, $r_{n,i}^{(\alpha)}$,
$p_{kj}^{(\alpha)}$ and $q_{kj}^{(\alpha)}$ are some coefficients.\\
The Dubrovin bilinear pairing of $\omega^{(1)}$ and $\omega^{(2)}$ is defined by
\begin{align*}
\mathfrak{F}[\omega^{(1)}, \omega^{(2)}]&=\sum_{i=0}^m\left(\sum_{n=0}^{\infty}\frac{c_{-n-2,i}^{(1)}}{n+1}c_{n,i}^{(2)}+c_{-1,i}^{(1)}p.v.\int_{P_0}^{\infty^i}\omega^{(2)}
-p.v. \int_{P_0}^{\infty^i}\sum_{n=1}^{\infty}r_{n,i}^{(1)}\big(\lambda(P)\big)^n\omega^{(2)}\right)\\
&\quad +\frac{1}{2{\rm i}\pi}\sum_{k=1}^g\left(-\oint_{a_k}q_k^{(1)}(\lambda)\omega^{(2)}+\oint_{b_k}p_k^{(1)}(\lambda)\omega^{(2)}+A_{k}^{(1)}\oint_{b_k}\omega^{(2)} \right),
\end{align*}
where $P_0$ is a marked point on $C_g$ such that $\lambda(P_0)=0$.

\begin{Thm} Let $\phi$ be one of the primary differential (\ref{Primary-e}) and $\Psi_{\phi}$ be the following multivalued differential
$$
\Psi_{\phi}(P):=\left( p.v. \int_{\infty^0}^{P}\phi\right)d\lambda(P).
$$
Then the Dubrovin prepotential associated with the Frobenius manifold structure determined by $\phi$ is given by means of the bilinear pairing $\mathfrak{F}[\cdot, \cdot]$ as follows:
\begin{equation}
F_{\phi}=\frac{1}{2}\mathfrak{F}[\Psi_{\phi}, \Psi_{\phi}].
\end{equation}

\end{Thm}


\bigskip

\textbf{Acknowledgements:}
The author is grateful to Vasilisa Shramchenko for   several valuable comments and discussions during the realization of  this work.
The author thanks  Universit\'{e} de Sherbrooke and Institut des Sciences math\'{e}matiques (ISM) for their financial support.


\end{document}